\documentclass[a4paper,fleqn,usenatbib]{mnras}

\usepackage[T1]{fontenc}
\usepackage{ae,aecompl}

\usepackage{graphicx}	
\usepackage{amsmath}	
\usepackage{amssymb}	
\usepackage[update,prepend]{epstopdf}
\usepackage{morefloats}
\usepackage{hyperref}
\usepackage{booktabs}
\usepackage{color}
\usepackage{ulem}

\title[Phot. of Barred LSB Discs]{Bar properties and photometry of barred low surface brightness disc galaxies}

\author[W.\ Peters \& R.\ Kuzio de Naray]{
Wesley Peters$^{1}$\thanks{E-mail: peters@astro.gsu.edu}
and
Rachel Kuzio de Naray$^{1}$\thanks{E-mail: kuzio@astro.gsu.edu}
\\
$^{1}$Department of Physics \& Astronomy, Georgia State University, PO Box 5060, Atlanta, GA 30302-5060, USA\\\\
}

\date{Accepted XXX. Received YYY; in original form ZZZ}

\pubyear{2018}

\begin{document}
\label{firstpage}
\pagerange{\pageref{firstpage}--\pageref{lastpage}}
\maketitle

\begin{abstract}
We measure the bar properties (length, strength, and corotation radius) of a sample of barred low surface brightness (LSBs) galaxies and compare to previous results for both LSBs and high surface brightness galaxy (HSBs) samples. In addition, we present new optical \textit{B}- and \textit{I}-band surface brightness profiles, magnitudes, and colours. We find that bars in LSBs are shorter and weaker when compared with those in HSBs galaxies. Based on analysis of four different bar length measures on simulated galaxy images, we find our bar length measure based on azimuthal light profiles is the most accurate out of the four tested here when seeing effects are taken into account. We also find that bars in LSBs have comparable relative bar pattern speeds to those in HSBs, the majority of which we find to be `fast' (i.e. $\mathcal{R} <$\ 1.4). In general, we find that our barred LSBs have comparable central surface brightnesses and total magnitudes to unbarred LSBs, albeit slightly brighter. Our barred LSBs fall on a shallower disc scale length (h) vs $M_{B}^{T}$ relation than unbarred LSBs. Finally, we find our barred LSBs to be just as gas rich as unbarred LSBs (i.e. $M_{gas}/(M_{gas} + M_{*}) > 0.5$).
\end{abstract}




\section{Introduction}

Low surface brightness galaxies (LSBs) are disc galaxies typically defined as having central surface brightness $\mu_{0}(B) > 22.0$\ mag arcsec$^{-2}$\ \citep[e.g.][]{mcgaugh1995b} or $\mu_{0}(B) > 23.0$\ mag arcsec$^{-2}$\ \citep[e.g.][]{impey1997}. Although LSBs occupy a large portion of the galaxy population, they are often biased against in large surveys due to their faint nature \citep{mcgaugh1995b,oneil2000, trachternach2006}. LSBs are not exclusively low-mass galaxies, but instead span all mass ranges \citep{mcgaugh1995a, oneil2000}. LSBs are also HI rich (i.e. $M_{\mathrm{HI}}/M_{*} > 1$) when compared with HSBs \citep{deblok1996,pahwa2018}, suggesting they may be relatively unevolved systems. 

LSBs are vital to our galaxy formation and evolution theories. In particular, LSBs serve as ideal dark matter laboratories, as analysis of their rotation curves have indicated they are dominated by dark matter at nearly all radii \citep[e.g.][]{deblok1996, deblok1997, mcgaugh2000, swaters2003, haghi2018}. As such, the dynamics of unbarred LSBs are particularly well understood. However, given the inherent noncircular motions of \textit{barred} LSBs, these galaxies and their bars are less well studied. Nonetheless, these systems could shed further light on the nature of dark matter.

Due to their gas rich nature \citep[e.g.][]{mcgaugh1997, cervantes2017b}, dark matter domination \citep[e.g.][]{deblok1997, bothun1997}, and isolated environments \citep[e.g][]{bothun1993, mo1994, rosenbaum2004, du2015, honey2018}, LSBs are expected to be quite stable against bar formation. Indeed, the bar fraction for LSBs has historically been quite low, $\sim$8$\%$, \citep{mihos1997, honey2016} when compared with the HSB bar fraction, $\sim$60$\%$\ \citep[e.g.][]{eskridge2000, marinova2007}. However, large-scale LSB studies have begun increasing this bar fraction to $\sim$20$\%$\ \citep[see][]{cervantes2017a, pahwa2018}. \textcolor{black}{Finally, while LSBs are relatively isolated, there is recent work to suggest that galaxy isolation has less of an effect on bar formation than previously thought \citep{casteels2013, lin2014, zana2018}.}

In a previous paper \citep[][hereafter Paper I]{peters2018}, we measured the photometric bar properties of four well-studied LSBs: UGC 628, F563-V2, F568-1, and F568-3. We found that the bars in these galaxies had comparable lengths and relative bar pattern speeds to those in HSBs, albeit weaker. Most interestingly, we found that three of the four galaxies were hosts to `fast' bars, contrary to expectations for \textcolor{black}{objects in non-rotating dark matte haloes.}

Here, we use the same methods as in Paper I for 11 new targets that are comparably less well studied, i.e. no existing surface brightness profiles, mass modeling, or rotation curves. We have used the same measurement techniques as those in Paper I, with an additional bar length measure used here. Finally, we have created mock galaxy images in order to fully test our various bar length measurements, specifically to quantify how accurately we can determine the other two parameters we are interested in: strength and corotation radius. 

We have also obtained broadband $B$- and $I$-band photometry of our full sample, and derived surface brightness profiles, magnitudes, and colours. In addition, we have used published HI fluxes to obtain HI mass estimates, and population synthesis models to obtain stellar mass estimates in order to examine the gas fractions of a subset of our sample.

The organization of this paper is as follows. We detail our sample, data acquisition and reduction in Sec.~\ref{sec:sam_dat}. Our methods and results for the bar properties of our sample are presented in Sec.~\ref{sec:barProps}, and we describe and show results of our mock galaxy image procedure in Sec.~\ref{sec:photmodel}. We outline our methods for measuring photometry and present the results in Sec.~\ref{sec:phot}. We use published HI fluxes and obtain stellar mass estimates for a subset of our sample in Sec.~\ref{sec:gasFraction} to examine gas fractions. Finally, our discussion and conclusions are in Sec.~\ref{sec:discussion}.

\section{Sample and Data}
\label{sec:sam_dat}

In this section we briefly outline our sample selection and observations of barred LSBs. We also detail our data reduction.

\subsection{Sample}
\label{ssec:sample}

As discussed in Paper I, we have assembled a sample of 15 barred LSBs drawn from the catalogs of \citet{schombert1992} and \citet{impey1996}. Not much is known about the 11 new targets in this sample, with few of the galaxies being observed since appearing in \citet{impey1996}. We show the entire observed sample in Table~\ref{sample}. Here, we list the R.A. (col. 1) and Dec. (col. 2) coordinates, date of observation (col. 3), photometric bands observed (col. 4), average seeing for observations (col. 5), derived position angles (col. 6) and inclinations (col. 7).

\begin{table*}
  \centering
  \caption{Sample of observed barred LSBs. The R.A. (col. 1) and Dec. (col. 2) coordinates, date of observation (col. 3), photometric bands observed (col. 4), average seeing for observations, $B$\ then $I$, (col. 5), derived photometric disc position angles (col. 6) and inclinations (col. 7), distances (col. 8), and morphologies (col. 9). Position angles and inclinations for UGC 628, F563-V2, F568-1, and F568-3 are taken from Paper I, with the remainder derived in this work (Sec.~\ref{ssec:data}). The Tully-Fisher distance for UGC 628 is taken from \citet{tully2016}; the remaining distances are derived using available spectroscopic distances from NED.}
  \label{sample}
  \begin{tabular}{lccccccccc}
    \hline
    Galaxy                 &    R.A.    &    Dec.   &      Date       & Bands     &   Seeing    &    P.A.    &    $i$      &  D     & Morph. \\
                           &  (J2000)   &  (J2000)  &                 &           & ($\arcsec$) &($^{\circ}$) & ($^{\circ}$) &  (Mpc) &  \\
    \hline
    UGC 628                & 01:00:51.9 & +19:28:33 & 7 August 2016   & $B,I$     &  1.74, 1.50  &  -42.80    &   58.20    &  86    & SBc \\
    LEDA 135682            & 03:39:34.2 & -00:30:43 & 29 October 2017 & $B,I$     &  1.06, 0.92  &   42.66    &   32.75    & 104    & SBbc \\
    LEDA 135684            & 03:41:44.5 & -02:00:09 & 29 October 2017 & $B,I$     &  1.12, 0.91  &   30.07    &   44.35    &  46    & SBd \\
    LEDA 135693            & 03:53:26.3 & +00:50:32 & 29 October 2017 & $B,I$     &  0.88, 0.87  &   37.84    &   22.12    & 154    & SB(r)b \\
    UGC 2925               & 04:01:02.5 & -00:43:03 & 29 October 2017 & $B,I$     &  1.00, 0.79  &  -36.64    &   31.79    &  57    & SB(r)c \\
    F563-V2                & 08:53:03.8 & +18:26:09 & 24, 28 February 2017   & $B,I$     &  1.50, 1.54  &  -32.00    &   29       &  60    & Irr \\
    F568-1                 & 10:26:06.3 & +22:26:01 & 24, 28 February 2017   & $B,I$     &  1.43, 1.08  &  -86.00    &   24.90    &  92    & SBbc \\
    F568-3                 & 10:27:20.2 & +22:14:24 & 24, 28 February 2017   & $B,I$     &  1.59, 1.15  &  -11.40    &   39.60    &  83    & SBcd \\
    LEDA 135782            & 11:28:29.6 & +00:08:40 & 13 May 2018     & $B,I$     &  1.27, 1.21  &    5.27    &   35.91    & 208    & SBc \\
    UGC 8066               & 12:57:00.3 & +01:01:43 & 13 May 2018     & $B,I$     &  1.11, 0.91  &   -8.46    &   50.90    &  40    & SB(r)d \\
    LEDA 135867            & 14:48:56.4 & -00:43:38 & 13 May 2018     & $B,I$     &  1.32, 1.25  &   73.40    &   38.30    & 119    & SBd \\
    F602-1                 & 22:34:46.0 & +22:33:48 & 25 August 2017  & $B,V,R,I$ &  0.84, 0.97  &   -6.00    &   28.58    & 105    & SBc \\
    PGC 70352              & 23:03:21.5 & +01:53:12 & 25 August 2017  & $B,R,I$   &  1.11, 1.10  &   60.90    &   31.09    &  73    & SBd \\
    ASK 25131              & 23:12:21.0 & -01:05:42 & 25 August 2017  & $B,V,R,I$ &  1.04, 0.83  &   -2.96    &   38.00    & 107    & SBcd \\
    $[$ISI96$]$ 2329-0204  & 23:31:40.8 & -01:48:01 & 25 August 2017  & $B,V,R,I$ &  0.87, 0.74  &   20.29    &   31.43    & \dotso & SBcd \\
    \hline
  \end{tabular}
\end{table*}

In col. 8 we list distances (Mpc). Only one of our galaxies, UGC 628, has a redshift independent distance \citep{tully2016}. We estimated the distances for the remainder of the galaxies by averaging the derived spectroscopic redshift distances on the NASA Extragalactic Database (NED)\footnote{The NASA/IPAC Extragalactic Database (NED) is operated by the Jet Propulsion Laboratory, California Institute of Technology, under contract with the National Aeronautics and Space Administration.}, assuming $H_{0}$ = 73 km s$^{-1}$\ Mpc$^{-1}$. There are no available spectra or distance measurements for $[$ISI96$]$ 2329-0204. For those galaxies with distances less than $\sim$100 Mpc, these distance estimates likely have large errors. However, we use the distances here to obtain estimates of physical bar lengths and disc-scale lengths from our arcsec measurements, allowing the physical values to be scaled later on if a redshift independent distance is determined at a later time.

In col. 9 we list the morphologies of the sample, following the same method as for HSBs \citep[i.e.][]{devaucouleurs1959}. As with LSBs in general, our targets are all late- to very late-type galaxies. With the exception of F563-V2, all our galaxies have a spiral structure, some much more pronounced than others. 

\subsection{Observations and Reduction}
\label{ssec:data}

To observe our targets, we used the ARCTIC imager on the 3.5-m telescope at Apache Point Observatory\footnote{Based on observations obtained with the Apache Point Observatory 3.5-meter telescope, which is owned and operated by the Astrophysical Research Consortium.}. ARCTIC has a field of view of 7.5$\arcmin \times$7.5$\arcmin$. We used ARCTIC in single read out mode with 2$\times$2 binning, giving a plate scale of 0.228 arcsec pix$^{-1}$. We obtained $B$- and $I$-band images of 13 new galaxies. In addition, we obtained \textit{V}- and \textit{R}-band images for a subset of four galaxies.

We observed each galaxy in all bands with 3$\times$600 sec. exposures, with 15$\arcsec$\ dithering between each image. In addition, we used a circular dithering pattern for the \textit{I}-band images in order to create a master fringe pattern for reduction. We reduced the data using standard packages and routines in \texttt{IRAF}\footnote{IRAF is distributed by the National Optical Astronomy Observatory, which is operated by the Association of Universities for Research in Astronomy (AURA) under a cooperative agreement with the National Science Foundation}: \texttt{CCDPROC}, \texttt{IMCOMBINE}, etc.. For photometric calibration, stars near our targets from \citet{landolt1992} were observed to obtain photometric zero points. These zero points were found to be accurate to 0.2 mags by checking with stars in the field from Pan-STARRS1 \citep{chambers2016} and TASS Mark III \citep{richmond2000}. Sky values were determined by subtracting the average value of six 50$\times$50 pixel star-free boxes in each image \citep[see Paper I and][]{schombert2014}.

As with Paper I, we deproject our galaxy images for our analysis. To do this, we use the \texttt{IRAF} task \texttt{ELLIPSE} to determine the disc position angle and inclination of each galaxy in each photometric band. We then take the average value between all available bands to be the P.A. and inclination of the galaxy, with the largest deviation from the average assigned as the error. Errors in P.A. and inclination are typically $\sim\pm$5$^{\circ}$\ and $\sim\pm$2$^{\circ}$, respectively. The deprojection is then performed using \texttt{GEOTRAN} to rotate and `stretch' the image. Our derived P.A. and inclinations are listed in Table~\ref{sample}.

\begin{figure*}
  \centering
  \includegraphics[scale=0.63]{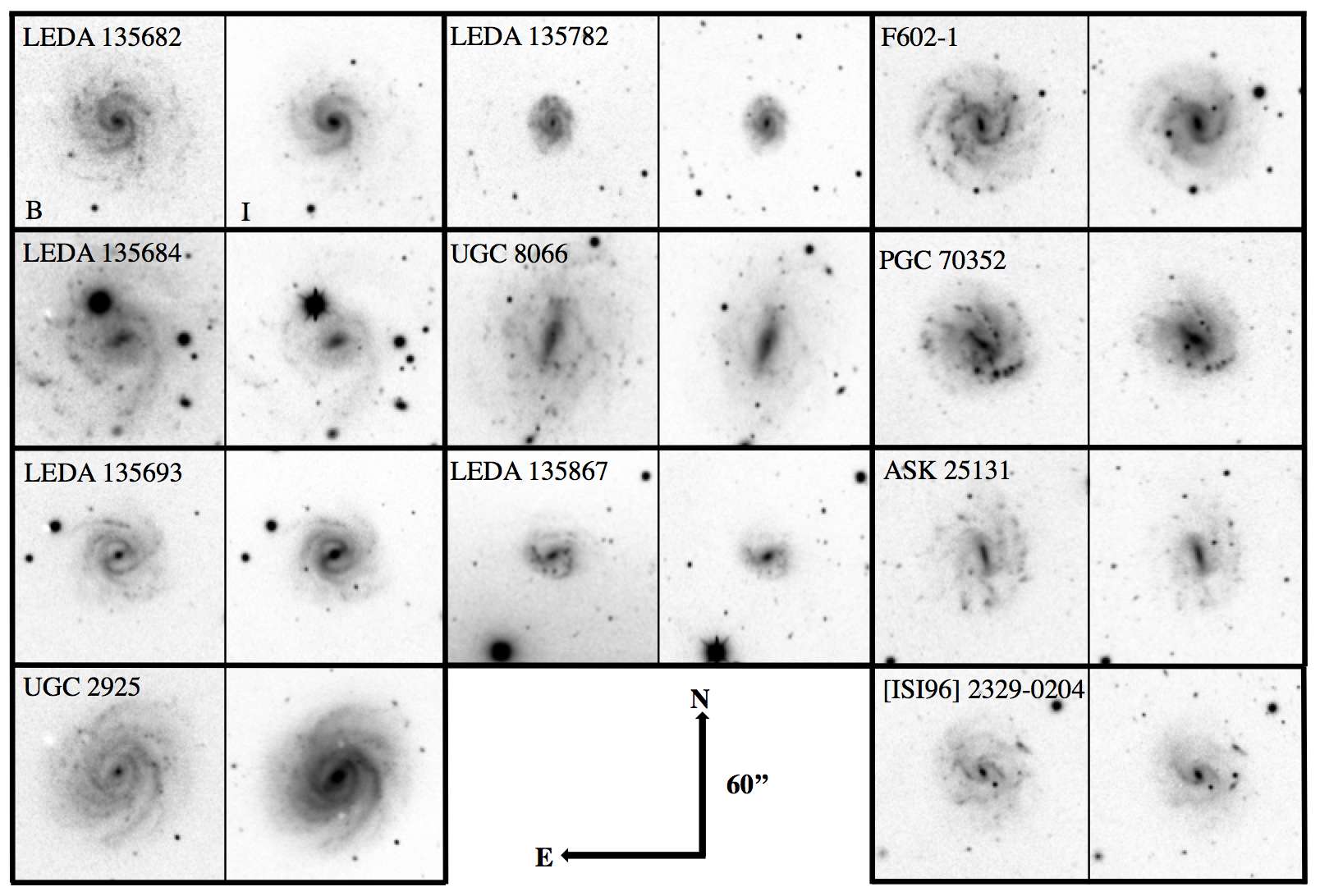}
  \caption{\textit{B}- and \textit{I}-band ARCTIC images of our sample. Galaxies are separated by thick black lines, with the \textit{B}-band image on the left and the \textit{I}-band image on the right for each galaxy. Each image is 400$\times$400 pixels, or 1.52$\arcmin \times$1.52$\arcmin$, with each \textit{B}-band image scaled identically. All \textit{I}-band images are scaled identically with the exception of UGC 2925. All images are shown with \texttt{asinh} scaling to display structure more easily. North and East are indicated by the arrows in the bottom center, each arrow being 60$\arcsec$\ long. Images for UGC 628, F563-V2, F568-1, and F568-3 are in Paper I.}
  \label{galimages}
\end{figure*}

Our final, reduced, on-sky \textit{B}- and \textit{I}-band images for our new galaxies are presented in Fig.~\ref{galimages}. Here, images of the same photometric band have been scaled the same way for easy comparison (except for the \textit{I}-band image of UGC 2925 which has a larger scaling range), and every image uses \texttt{asinh} scaling to best show the structure in each galaxy. Each image is 400$\times$400 pixels, or 1.52$\arcmin \times$1.52$\arcmin$.

\section{Bar Properties}
\label{sec:barProps}

Here we describe the methods used to characterise each of the bar properties for the full sample. In addition, we present and discuss the results in relation to previous LSB and HSB results.

\subsection{Methods}
\label{ssec:barMethods}

We follow the methods outlined in Paper I for determining the bar length, strength, and corotation radius but we also employ a new bar length measure with the aim of improving the elliptical isophotal bar length measure for our sample. For a more detailed description of each of these methods, see Sec. 3 in Paper I. All of our measurements use the deprojected $I$-band galaxy images, since bars are stellar features and are more prominent and well-defined in redder photometric bands. For the sake of example, we will use LEDA 135782 as a walkthrough for each meaasurement. The remaining galaxies are shown in the Appendix figures. 

\subsubsection{Bar Lengths}
\label{sssec:methodbarlength}

Since the bar strength and corotation radius are dependent on the bar radius, it is important that the bar radius be accurately determined. Therefore, as in Paper I, we use multiple methods for measuring the bar length: (1) elliptical isophotes \citep[e.g.,][]{wozniak1995,aguerri1998,aguerri2000a,aguerri2000b,aguerri2009}; (2) Fourier analysis of azimuthal light profiles \citep[e.g.][]{ohta1990,aguerri1998,aguerri2000a}; (3) azimuthal behaviour of the bar \citep{ohta1990}. We briefly describe each method below. We select the best measurement of the bar length by visually comparing the various measures on the deprojected galaxy images (see Paper I). For a few targets, the bars are no longer than 2$\arcsec$\ ($\sim$8.77 pixels), so some methods are unable to find a bar length. 

\paragraph{Elliptical Isophotes}
\label{ssssec:isophotes}

When fitting the light distribution of a galaxy with elliptical isophotes, a bar can be found by analyzing the behaviour of both the eccentricity and position angles of the ispohotes. For example, a traditional approach is to use the radius of maximum ellipticity as the bar length \citep[e.g.][]{wozniak1995}.

However, the bar length can be underestimated when using the radius of maximum ellipticity \citep{michel2006}. In addition, because the bar strength and relative bar pattern speed depend on the bar length, choosing the most accurate measure of the bar length is crucial. Due to this, we have employed an additional bar length measure using the elliptical isophotes that was not used in Paper I. This method defines the bar length to be where the position angle of the isophotes diverges a certain amount away from the position angle of the bar \citep{wozniak1995,aguerri2009}. We therefore have two measures using the elliptical isophotes: (1) $R_{e}$, the radius of maximum ellipticity, and (2) $R_{\mathrm{PA}}$, the radius where the P.A. of the isophotes differs by more than 5 degrees from the value at $R_{e}$.

\begin{figure}
  \centering
  \includegraphics[scale=0.38]{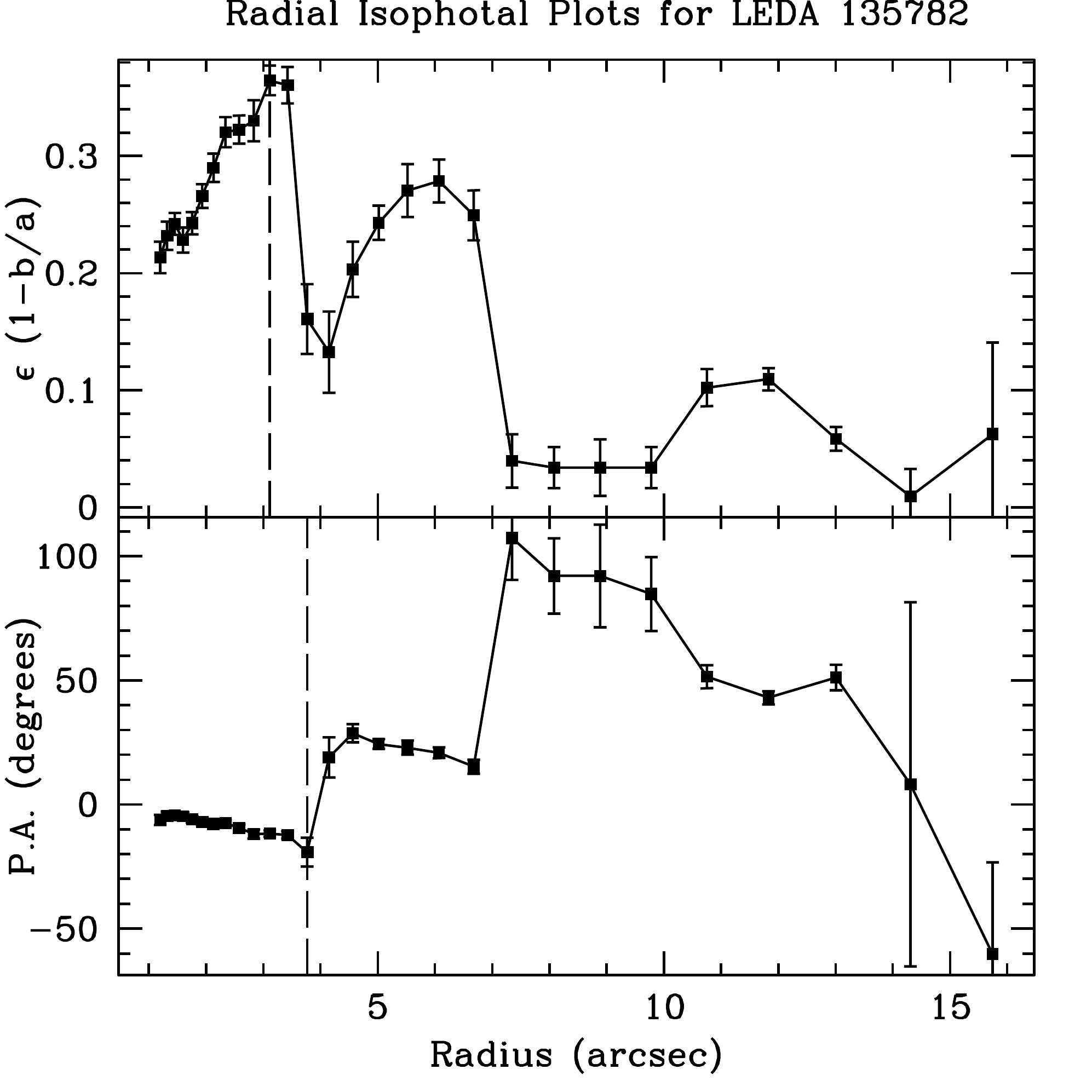}
  \caption{Radial plots of eccentricity (top panel) and P.A. (bottom panel) for LEDA 135782. The vertical dashed lines indicate the bar lengths based on maximum eccentricity (top panel) and position angle discontinuity (bottom panel).}
  \label{leda135782_elipBars}
\end{figure}

The deprojected radial plots of eccentricity and P.A. for LEDA 135782 are shown in Fig.~\ref{leda135782_elipBars}. In each panel, the vertical dashed line indicates the bar length from the respective method. Using the radius of maximum eccentricity, we find a bar length of $3.11\arcsec \pm 0.30\arcsec$, and the bar length from the discontinuous P.A. is $3.77\arcsec \pm 0.36\arcsec$. The errors for both these methods come from the radial spacing between the points. We find that discontinuous P.A. radius method fails for one of our galaxies, F563-V2.

The bar lengths for each galaxy are listed in Columns 1 and 2 in Table~\ref{barlens}. Plots similar to Fig.~\ref{leda135782_elipBars} for the remaining galaxies are in shown in the online Appendix in Fig.~1.

\paragraph{Fourier Analysis}
\label{ssssec:fourier}

The Fourier analysis is done by decomposing our deprojected images into azimuthal light profiles $I_{r}(\theta)$, and then decomposing these via a Fourier transform:

\begin{equation}
  \mathcal{F}(r) = \int_{-\pi}^{\pi} I_{r}(\theta) \exp{(-2i \theta)} d\theta.
\end{equation}
The Fourier coefficients are
\begin{equation}
  A_{m}(r) = \frac{1}{\pi} \int_{0}^{\pi} I_{r}(\theta) \cos{(m\theta)} d\theta
\end{equation}
\begin{equation}
  B_{m}(r) = \frac{1}{\pi} \int_{0}^{\pi} I_{r}(\theta) \sin{(m\theta)} d\theta
\end{equation}
and the amplitudes are
\begin{equation}
  I_{0}(r) = \frac{A_{0}(r)}{2}
\end{equation}
\begin{equation}
  I_{m}(r) = \sqrt{A_{m}^{2}(r) + B_{m}^{2}(r)}
\end{equation}
In order to determine the bar length, we use the bar/interbar ($I_{b}/I_{ib}$) Fourier intensities method from \citet{aguerri2000a}, where the bar region is defined as
\begin{equation}
  \frac{I_{b}}{I_{ib}} > \frac{1}{2} \left[ \left(\frac{I_{b}}{I_{ib}}\right)_{max} - \left(\frac{I_{b}}{I_{ib}}\right)_{min}\right] + \left( \frac{I_{b}}{I_{ib}} \right)_{min}.
\end{equation}
The last radius at which the above is satisfied is taken to be the bar radius. The bar and interbar intensities are defined as 
\begin{equation}
  I_{b} = I_{0} + I_{2} + I_{4} + I_{6}
\end{equation}
\begin{equation}
  I_{ib} = I_{0} - I_{2} + I_{4} - I_{6}
\end{equation}

While \citet{aguerri2000a} state Equation 6 describes the bar region better than $(I_{b}/I_{ib}) > 2$, both our results from Paper I and the results from \citet{aguerri2009} show that this method can tend to overestimate the true bar length. Regardless, we use this method to determine how effective (or not) it is when applied to LSBs.

\begin{figure}
  \centering
  \includegraphics[scale=0.38]{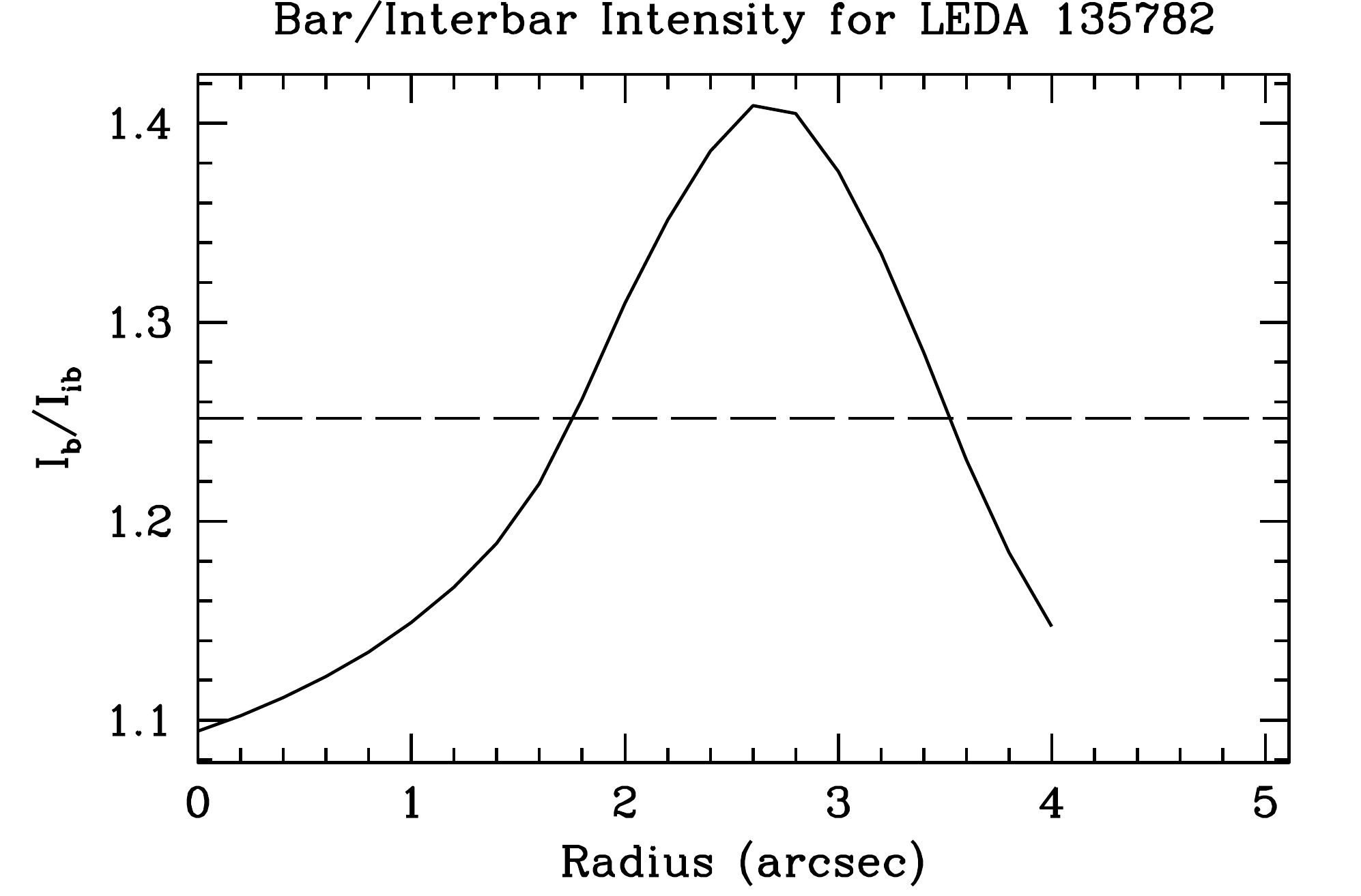}
  \caption{Bar/Interbar Fourier intensities for LEDA 135782. The dashed horizontal line denotes Equation 6, or the bar region. As the bar region is defined as where the ratio is \textit{greater} than Equation 6, only the second crossing denotes the bar length.}
  \label{fourierBar}
\end{figure}

The bar/interbar Fourier intensity plot for LEDA 135782 is shown in Fig.~\ref{fourierBar}, where the dashed horizontal line denotes Equation 6. We only consider data near the bar region as including radii in the disc can bias the method due to spiral arms, hence why the profile ends near 4$\arcsec$. Using this method, we find a Fourier bar length of $3.5\arcsec \pm 0.46\arcsec$. The error for this method comes from the radial spacing of the azimuthal light profiles.

\begin{figure}
  \centering
  \includegraphics[scale=0.38]{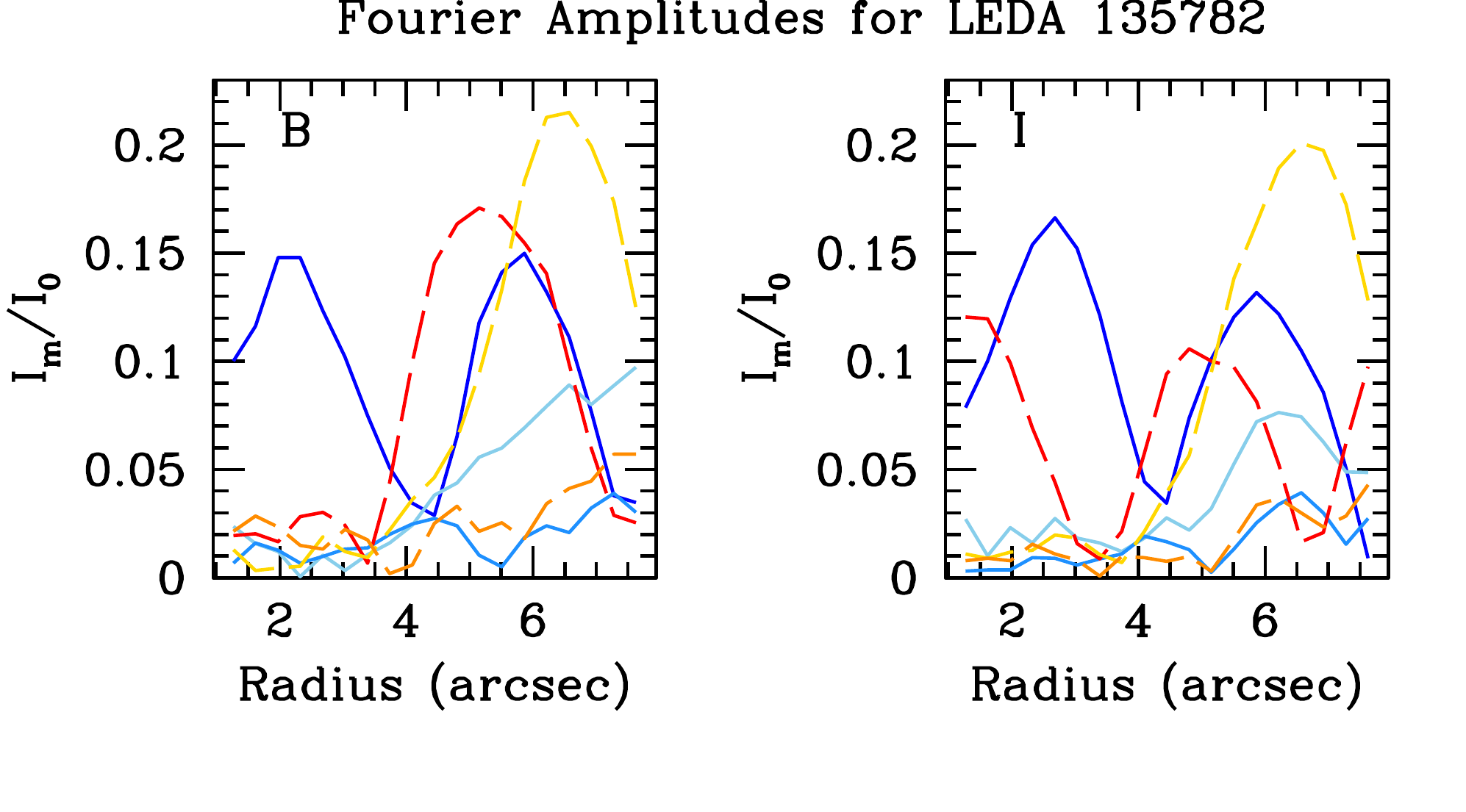}
  \caption{Fourier amplitudes for LEDA 135782: $B$-band shown in the left panel and $I$-band in the right panel. Bluer colours show even modes with deep blue being the $m=2$\ mode, and redder colours show odd modes with deep red being the $m=1$\ mode. Here we can see strong $m=2$\ modes in the bar region and disc region. Interestingly, we see strong $m=1$\ and $m=3$\ (yellow) modes in the disc region.}
  \label{fouramp}
\end{figure}

LEDA 135782 has interesting Fourier amplitudes, shown in Fig.~\ref{fouramp}. Here we see strong $m=2$\ (deep blue) modes in the bar and disc region, which is not surprising for a barred spiral. However, we also see strong $m=1$\ (deep red) and $m=3$\ (yellow) modes in the disc. When looking at Fig.~\ref{galimages}, LEDA 135782 appears to have three spiral arms, consistent with the strong odd modes.

We find that the strength of the spiral arm pattern greatly affects the accuracy of this method. For instance, for those galaxies with a strong dual-arm pattern (see UGC 628 and F568-1 in Paper I and LEDA 135682 in Fig.~\ref{galimages}), the bar length is consistently overpredicted. The method finds the bar length more easily when the spiral arms are weaker.

In addition, we find the Fourier method fails to find a bar length for \textcolor{black}{three} of our galaxies, most likely due to the spiral arms: LEDA 135867, PGC 70352, and $[$ISI96$]$ 2329-0204 (middle center, middle right, and bottom right in Fig.~\ref{galimages}). When examining the Fourier amplitudes for these three galaxies, we find even modes that are stronger in the disc than in the bar, preventing the method from accurately determining a bar region. 

The bar lengths for each galaxy using the Fourier Analysis method are listed in Column 3 of Table~\ref{barlens}. Figures similar to Fig.~\ref{fourierBar} and Fig.~\ref{fouramp} for the remaining galaxies are shown in the online Appendix in Fig.~2 and Fig.~3.

\paragraph{Azimuthal Light Profile}
\label{ssssec:azimuthal}

Finally, we use azimuthal light profiles to characterize the azimuthal behaviour of the bar. Following the procedure outlined in Paper I, we determine the azimuthal centroid of the bar and track \textcolor{black}{either where the humps in the azimuthal light profiles no longer trace a constant azimuthal centroid, or} where the azimuthal profiles begin to trace the spiral arms at the end of the bar. To do this, we make the following assumptions: (1) the bar is constant in azimuthal angle, and (2) the (two) spiral arms, \textcolor{black}{if present}, are separated by 180$^{\circ}$\ azimuthally. While these two assumptions may not always be true (e.g. the bar may have internal structure that causes it to be warped), they allow us to analyze the behaviour of the bar in order to determine the length.
 
Due to the different morphologies of our sample and sizes of our galaxies on the sky, we varied the azimuthal spacing and starting radius when constructing the azimuthal light profiles, as opposed to leaving all galaxies with the same radial spacing of 0.46$\arcsec$\ and starting radii of 2$\arcsec$\ as was the case in Paper I. This allowed bars that were quite short, $<$2$\arcsec$, to be measured.

\begin{figure}
  \centering
  \includegraphics[scale=0.2]{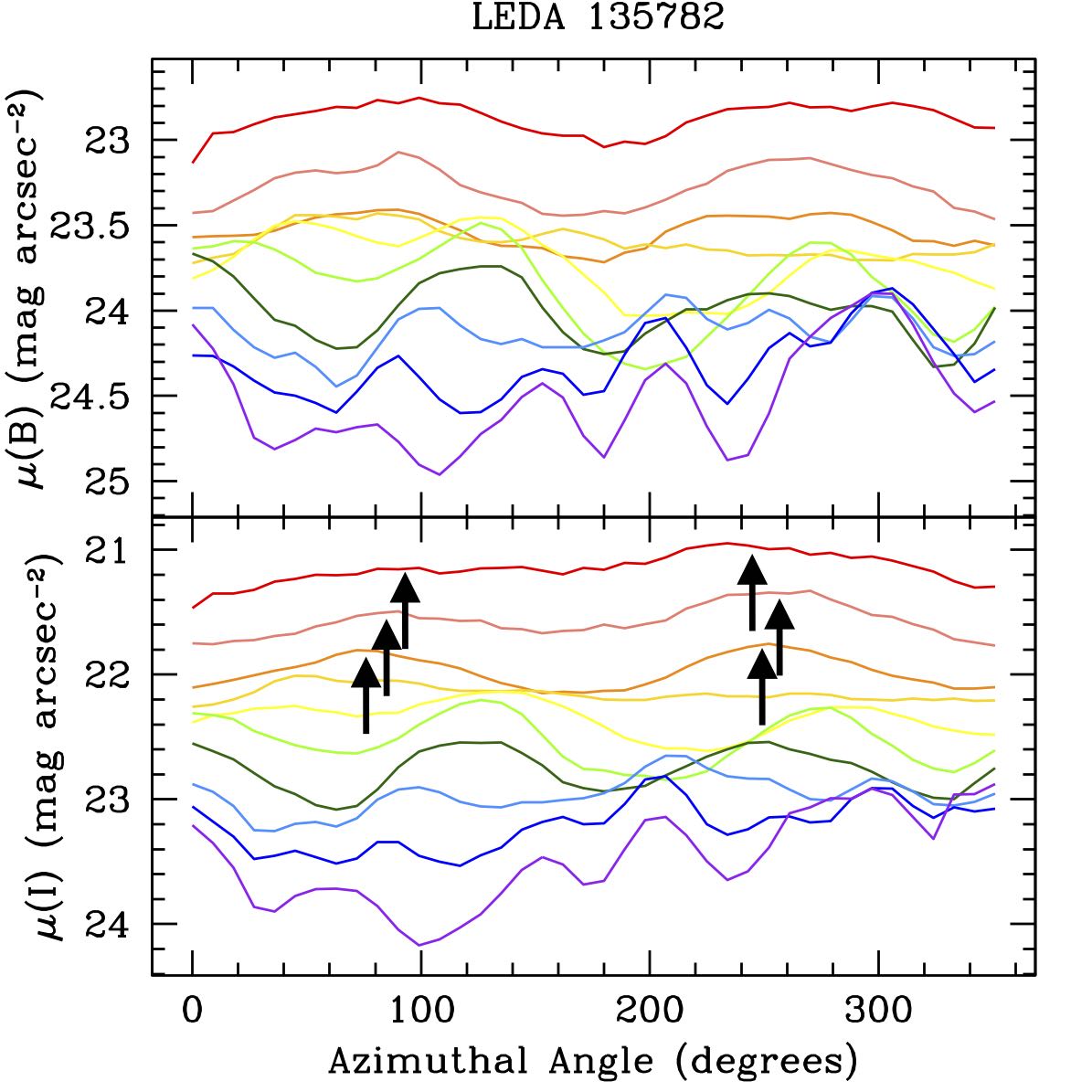}
  \caption{$B$-band (top) and $I$-band (bottom) azimuthal light profiles (in mag arcsec$^{-2}$) for LEDA 135782. Profiles are plotted every $\sim$0.7$\arcsec$\ for clarity. Colour scheme indicates radial position, with redder colours at smaller radii. \textcolor{black}{The black arrows are placed to assist the reader in identifying the humps in the inner radii.}}
  \label{azProfs}
\end{figure}

The $B$-band and $I$-band azimuthal light profiles for LEDA 135782 are shown in Fig.~\ref{azProfs}. Here, the profiles are plotted every $\sim$0.7$\arcsec$\ for clarity, and the colour scheme indicates the radial position with redder colours at smaller radii. We can see the clear presence of a bar in both bands, as indicated by the \textcolor{black}{black arrows} at $\sim$90$^{\circ}$\ and $\sim$270$^{\circ}$\ \textcolor{black}{in the inner radii}, as well as a three arm pattern present in the outer radii (green profiles) in both bands.

\begin{figure}
  \centering
  \includegraphics[scale=0.38]{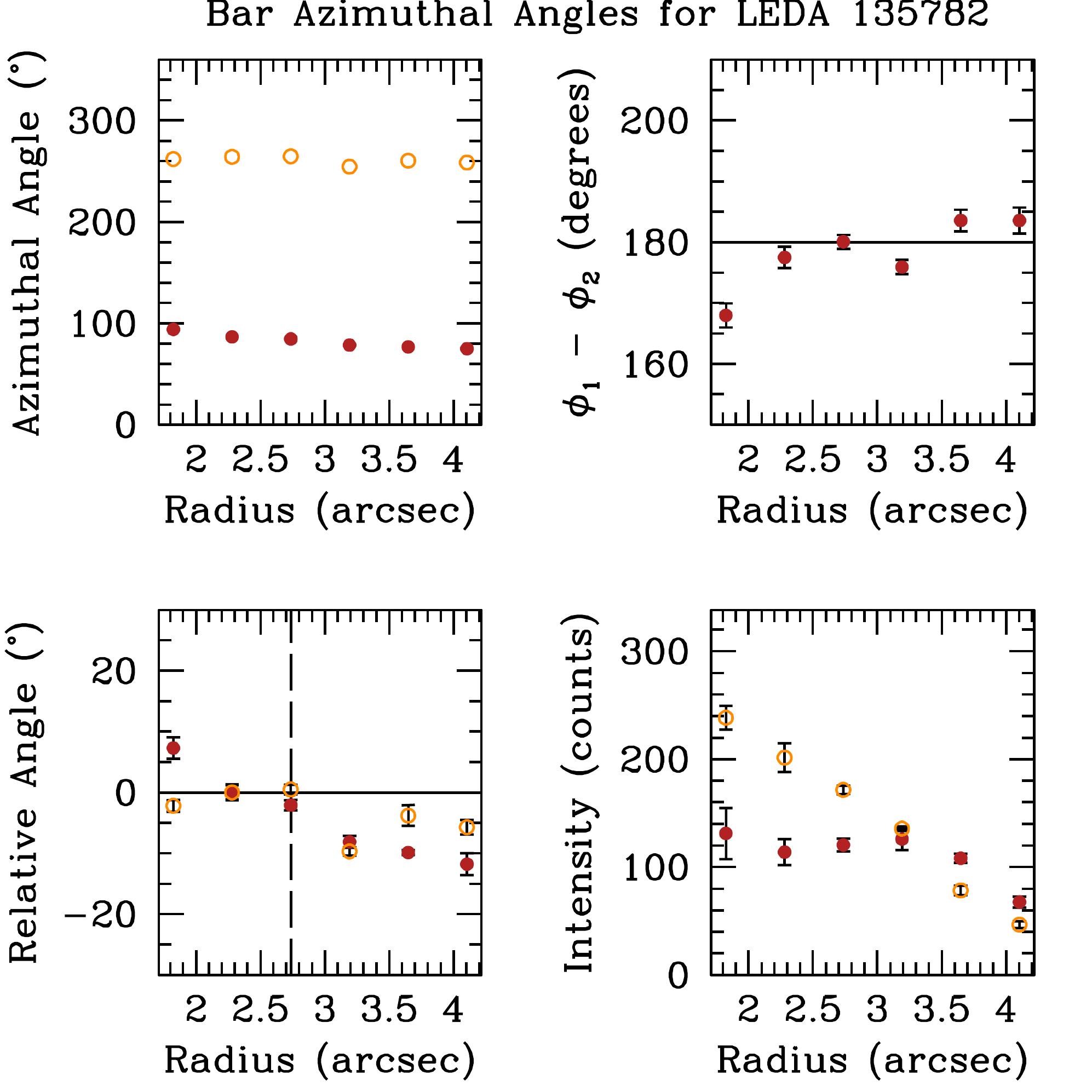}
  \caption{Bar azimuthal angle information for LEDA 135782. \textit{Top Left}: azimuthal position of the two bar humps; \textit{Top Right}: difference between azimuthal angles for two humps; \textit{Bottom Left}: azimuthal difference from the bar centroid; \textit{Bottom Right}: ADU intensity of the bar humps. The dashed vertical line in the bottom left panel denotes the azimuthal bar length, or where the azimuthal position of the humps has diverged from the bar azimuthal centroid.}
  \label{azBar}
\end{figure}

We show the azimuthal information for the bar in LEDA 135782 in Fig.~\ref{azBar}, \textcolor{black}{deriving the azimuthal positions and errors of the humps by fitting gaussians to the profiles.} Here we show the azimuthal positions of the bar `humps' (top left), the azimuthal difference between the two humps (top right), the azimuthal difference from the bar centroid (bottom left), and the ADU intensity of the bar humps (bottom right). The dashed vertical line in the bottom left denotes the bar length, or where the azimuthal difference from the bar centroid has diverged. LEDA 135782 is a prime case of a very small bar (on the sky), which can pose a challenge when using this method. However, it is clear from looking at the bottom left panel that the first three points are relatively constant, best seen in the open yellow points, and show a dramatic change after the dashed line. We find a bar length of 2.66$\arcsec \pm 0.68\arcsec$. \textcolor{black}{The errors are determined based on the behaviour of the azimuthal light profiles, as discussed in Paper I.} We assign the larger errors due to the small bar in this galaxy.

For LEDA 135682 and $[$ISI96$]$ 2329-0204, the bar is quite small on the sky ($\sim$2$\arcsec$) and hard to measure with this method. In order to get as accurate a measurement as possible, we decreased the starting radius of our azimuthal light profiles to 1$\arcsec$\ for each galaxy, and lowered the radial spacing for LEDA 135682 to 1.5 pixels (0.34$\arcsec$) and to 1 pixel (0.228$\arcsec$) for $[$ISI96$]$ 2329-0204. To ensure we maintained at least one pixel in each azimuthal division, we increased the size of the azimuthal bins to 9$^{\circ}$\ for both galaxies. This is a loss of resolution on these azimuthal light profiles, but the azimuthal locations of the bar remained unchanged.

The bar lengths for each galaxy measured using azimuthal light profiles are listed in Column 4 of Table~\ref{barlens}. Figures similar to Fig.~\ref{azProfs} and Fig.~\ref{azBar} for the remaining galaxies are shown in the online Appendix in Fig.~4 and Fig.~5 respectively.

\subsubsection{Bar Strengths}
\label{sssec:methodbarstrength}

\textcolor{black}{Instead of using the same bar strength measure as in Paper I \citep[i.e. from][]{aguerri2000a}, we modify the method to include the higher order even components, as these can be significant \citep{ohta1990}:}

\begin{equation}
   S_{b} = \frac{1}{r_{bar}} \sum_{m=2,4,6} \int_{0}^{r_{bar}} \frac{I_{m}}{I_{0}} dr, 
\end{equation}
where $m=2,4,6$\ are the even modes, and $r_{bar}$\ is the bar radius from the azimuthal method. We note that our measure is only a lower limit, as we cannot probe down to $r=0$, since this would result in azimuthal bins with no pixels when constructing the azimuthal light profiles.

\subsubsection{Corotation Radii}
\label{sssec:methodcorotation}

We use the method put forth in \citet{puerari1997} to measure the corotation radius ($R_{\mathrm{CR}}$) of the bar, the radius where disc orbits are equal to the pattern speed of the bar. This method determines $R_{\mathrm{CR}}$\ via the intersection of phase profiles of the Fourier transforms of the \textit{B}- and \textit{I}-band images. The phase is given by

\begin{equation}
  \Theta(r) = \arctan{ \left( \frac{\mathrm{Re}(\mathcal{F}(r))}{\mathrm{Im}(\mathcal{F}(r))} \right) },
\end{equation}
The idea behind this method is based on using the two different photometric bands, $B$ and $I$, as proxies for two different stellar populations, young and old. We take the \textit{first} phase intersection after the bar length to be the corotation radius. Corotation should not occur within the bar region \citep{contopoulos1980, bureau1999}, and multiple intersections after the bar may be representative of the pattern speed of the disc \citep{puerari1997}.

\begin{figure}
  \centering
  \includegraphics[scale=0.4]{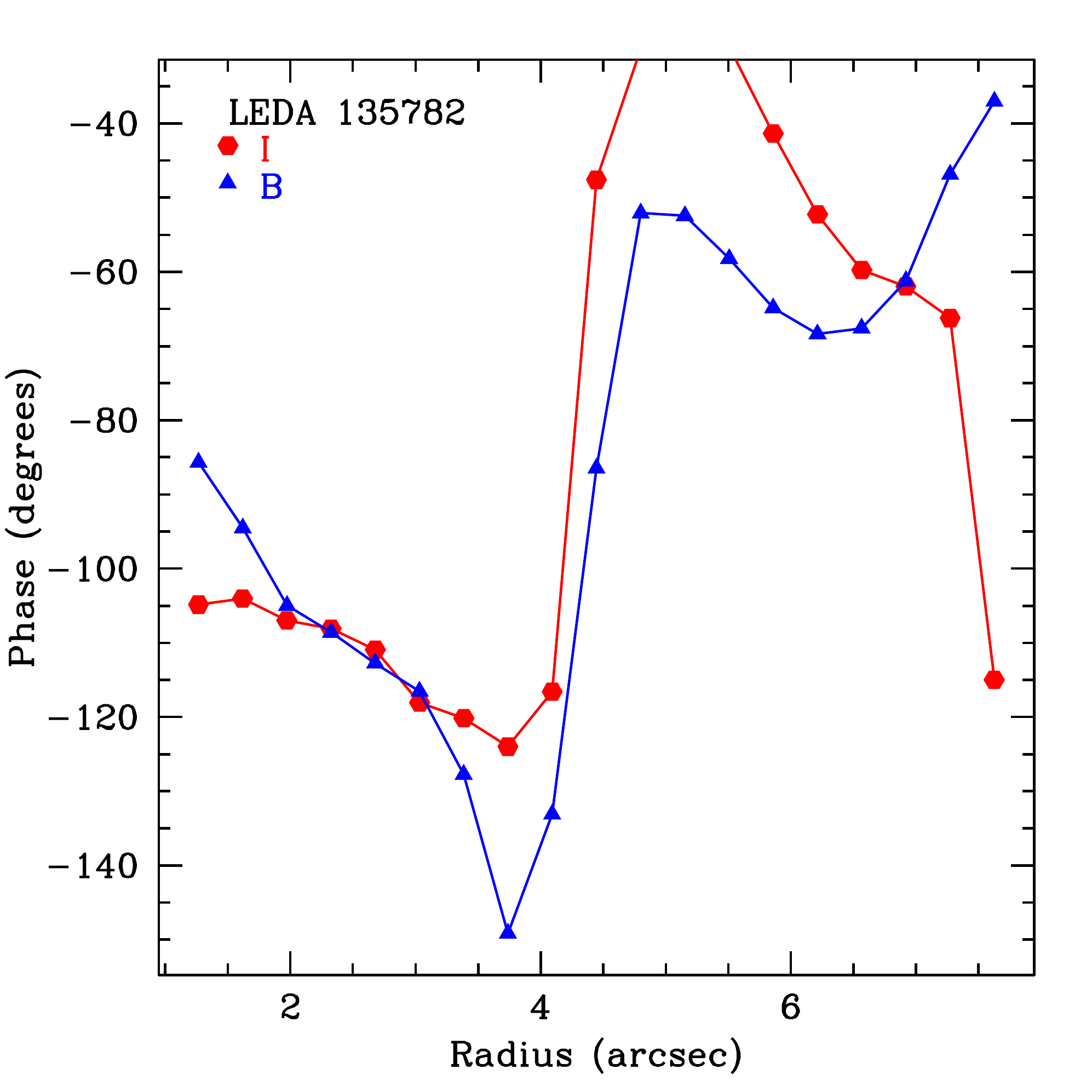}
  \caption{$B$-band (blue triangles) and $I$-band (red squares) phase profiles for LEDA 135782. Intersections of the phase profiles denote corotation radii, with the first intersection \textit{after} the bar length being the bar corotation radius.}
  \label{phase}
\end{figure}

Again, we use LEDA 135782 as an example for this process. The phase profiles for LEDA 135782 are shown in Fig.~\ref{phase}. Here we see there are multiple phase intersections near the bar region: 2.2$\arcsec$, 2.8$\arcsec$, and 3.1$\arcsec$. Using the azimuthal bar length from Table~\ref{barlens}, this means the bar corotation radius is at 2.8$\arcsec$, the first intersection after the bar radius. In addition, we find a phase intersection farther out, at 6.8$\arcsec$. This is likely a disc corotation radius. 

Plots similar to Fig.~\ref{phase} for the remaining galaxies are shown in the online Appendix in Fig.~6.

\subsection{Results}
\label{ssec:barResults}

Here, we discuss the three bar properties (length, strength, corotation radius) for the entire sample. We do not discuss each galaxy individually, but we do focus on interesting or problematic galaxies when applicable (for more detailed discussion on these methods and individual results on our first sample, see Paper I). The bar lengths from each of the four techniques for each galaxy are listed in Table~\ref{barlens}, and our final derived bar properties are listed in Table~\ref{barprops}.

\subsubsection{Bar Lengths}
\label{sssec:resbarlength}

\begin{figure}
  \centering
  \includegraphics[scale=0.17]{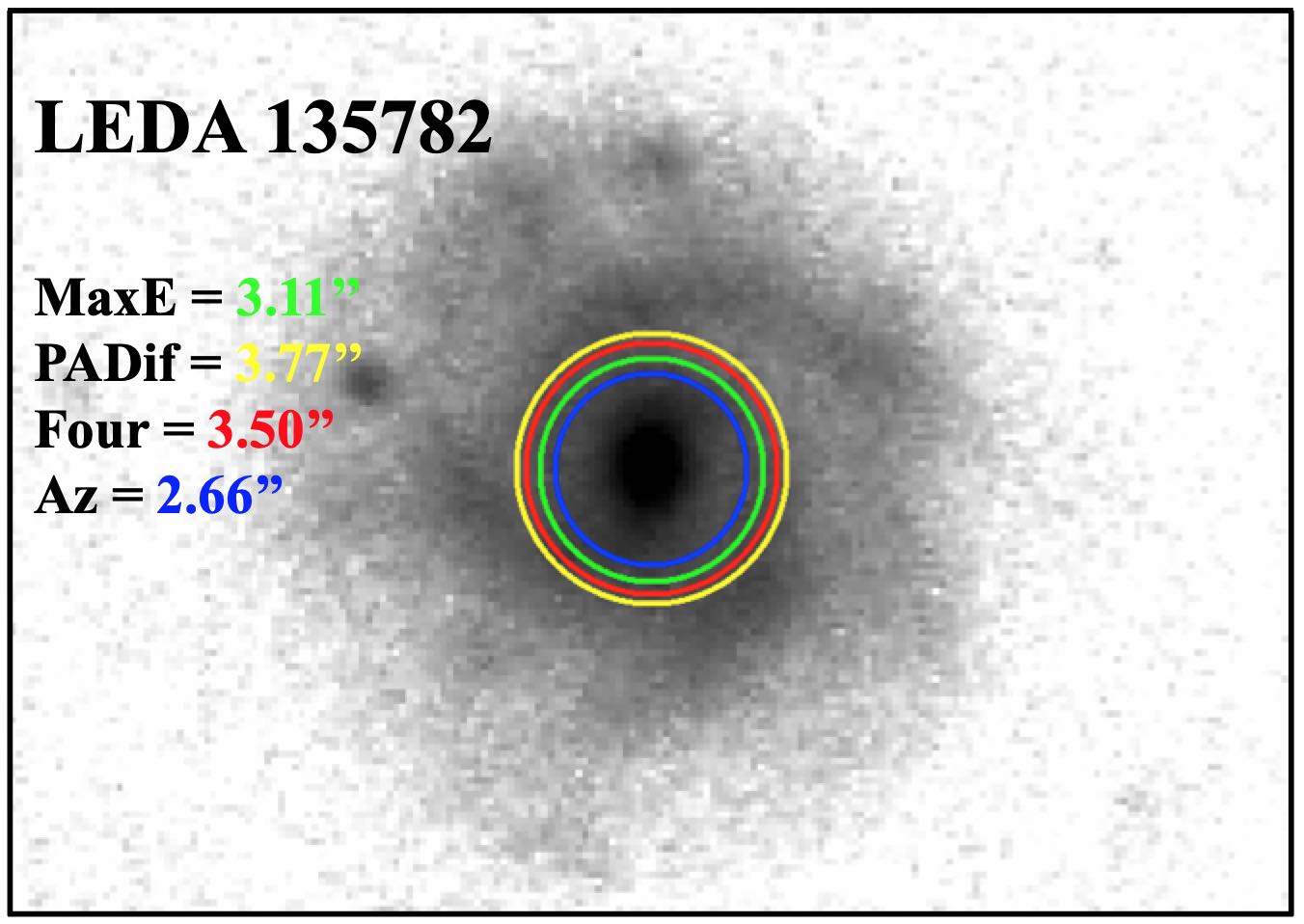}
  \caption{The four bar length measures plotted over the deprojected $I$-band image of LEDA 135782: $R_{e}$\ (green), $R_{\mathrm{PA}}$\ (yellow), $R_{\mathcal{F}}$\ (red), and $R_{az}$\ (blue). Here we can also see the three spiral arm pattern.}
  \label{barVis}
\end{figure}

The bar lengths measured using each technique for each galaxy are listed in Table~\ref{barlens}. We also include the bar length measurements from Paper I, as well as apply the $R_{\mathrm{PA}}$\ measure to the four galaxies from Paper I. We determine the final bar length for all galaxies by visually plotting the various bar length measures over the deprojected $I$-band images (see Fig.~\ref{barVis}, Fig.~7 in the online Appendix, and Paper I). We report these values as $R_{\mathrm{bar}}$\ in Table~\ref{barprops}. In general, we find an average bar length of 2.5 kpc for our sample.

\begin{table*}
  \centering
  \caption{Comparison of our various \textit{I}-band bar length measures, all in arcsec: radius of maximum ellipticity ($R_{e}$), radius of position angle discontinuity ($R_{\mathrm{PA}}$), Fourier method ($R_{\mathcal{F}}$), and the azimuthal method ($R_{az}$). `$\dotso$' denotes where a given method failed for a given galaxy. Values for UGC 628, F563-V2, F568-1, and F568-3 are taken from Paper I.}
  \label{barlens}
  \begin{tabular}{lcccc}
    \hline
    Galaxy                  & $R_{e}$           & $R_{\mathrm{PA}}$  & $R_{\mathcal{F}}$   & $R_{az}$           \\
    \hline
    UGC 628                 & 11.63 $\pm$ 0.68 & 14.36 $\pm$ 0.68 & 16.96 $\pm$ 0.46 & 11.21 $\pm$ 0.92 \\
    LEDA 135682             &  2.05 $\pm$ 0.2  &  2.73 $\pm$ 0.26 &  2.90 $\pm$ 0.23 &  2.05 $\pm$ 0.46 \\
    LEDA 135684             &  2.87 $\pm$ 0.27 &  9.01 $\pm$ 0.86 &  6.30 $\pm$ 0.46 &  7.07 $\pm$ 0.92 \\
    LEDA 135693             &  7.91 $\pm$ 0.76 &  8.70 $\pm$ 0.83 &  6.10 $\pm$ 0.46 &  5.70 $\pm$ 0.46 \\
    UGC 2925                & 11.60 $\pm$ 1.11 & 12.76 $\pm$ 1.22 &  9.10 $\pm$ 0.46 &  8.44 $\pm$ 0.92 \\
    F563-V2                 &  5.02 $\pm$ 0.91 &  $\dotso$        &  7.96 $\pm$ 0.46 &  6.65 $\pm$ 0.46 \\
    F568-1                  &  8.44 $\pm$ 0.91 & 10.26 $\pm$ 0.91 &  7.76 $\pm$ 0.46 &  4.37 $\pm$ 0.46 \\
    F568-3                  &  7.75 $\pm$ 0.68 & 16.64 $\pm$ 0.68 & 13.96 $\pm$ 0.46 &  8.93 $\pm$ 0.92 \\
    LEDA 135782             &  3.11 $\pm$ 0.3  &  3.77 $\pm$ 0.36 &  3.50 $\pm$ 0.46 &  2.66 $\pm$ 0.68 \\
    UGC 8066                &  6.07 $\pm$ 0.58 & 15.75 $\pm$ 1.50 & 16.50 $\pm$ 0.34 &  7.07 $\pm$ 0.68 \\
    LEDA 135867             &  6.77 $\pm$ 0.65 &  8.19 $\pm$ 0.78 & $\dotso$         &  4.39 $\pm$ 0.34 \\
    F602-1                  &  1.19 $\pm$ 0.27 &  2.80 $\pm$ 0.27 & 4.20 $\pm$ 0.34  &  4.11 $\pm$ 0.34 \\
    PGC 70352               &  5.28 $\pm$ 0.50 &  7.03 $\pm$ 0.67 & $\dotso$         &  5.02 $\pm$ 0.68 \\
    ASK 25131               &  3.28 $\pm$ 0.31 &  9.36 $\pm$ 0.89 & 8.70 $\pm$ 0.34  &  7.76 $\pm$ 0.92 \\
    $[$ISI96$]$ 2329-0204   &  1.64 $\pm$ 0.16 &  2.64 $\pm$ 0.25 & $\dotso$         &  2.05 $\pm$ 0.46 \\
    \hline
  \end{tabular}
\end{table*}

We plot the four different bar length measures over the deprojected $I$-band image of LEDA 135782 in Fig.~\ref{barVis}. Here we see that the azimuthal bar length (blue) is the best measure of the bar in this galaxy, as the other three extend into the spiral arms. We find that this holds for the other galaxies in our sample (see Fig.~7 in the online Appendix), suggesting that the azimuthal method is the most accurate of the four used here, consistent with our findings in Paper I.

\begin{figure}
  \centering
  \includegraphics[scale=0.4]{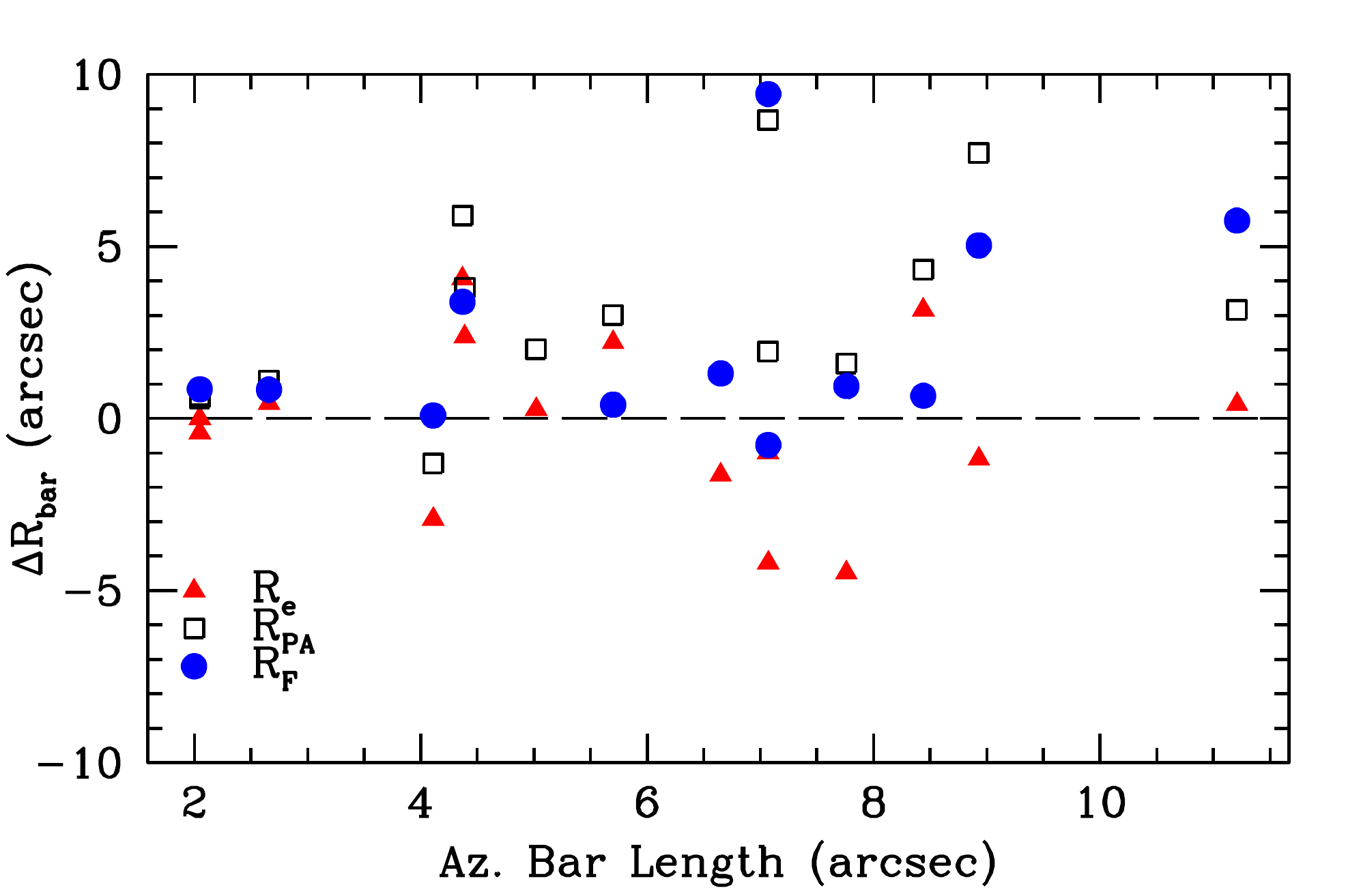}
  \caption{Comparison of our bar length measures relative to the azimuthal method (i.e. $\Delta R_{\mathrm{bar}} = R_{method} - R_{az}$): $R_{e}$\ shown as red triangles, $R_{\mathrm{PA}}$\ shown as open black squares, and $R_{\mathcal{F}}$\ shown as blue circles. By taking $R_{az}$\ as the true bar length, we find the other three measures overpredict the bar length, sometimes significantly.}
  \label{barComp}
\end{figure}

With this in mind, in Fig.~\ref{barComp} we show the various measures in relation to the azimuthal method. Here, $\Delta R_{\mathrm{bar}}$\ denotes the difference between either $R_{e}$\ (red triangles), $R_{\mathrm{PA}}$\ (open black squares), or $R_{\mathcal{F}}$\ (blue circles) and $R_{az}$. We discuss each measure in relation to $R_{az}$\ below.

We find that the two bar length measures based on the behaviour of the elliptical isophotes do not produce consistent results. $R_{e}$\ equally over and under predicts the bar length compared to $R_{az}$. This is because the measure can be biased towards longer values due to to a very elliptical bar or the presence of spiral arms, as well as biased towards shorter bars due to any highly elliptical feature in the inner regions.

We find that $R_{\mathrm{PA}}$\ almost consistently overpredicts the bar length, often quite extremely (i.e. $>$\ 2$\arcsec$). This is not too surprising, given that this measure is dependent on $R_{e}$. Since $R_{\mathrm{PA}}$\ can only be larger than $R_{e}$, this method can only obtain an accurate measurement if $R_{e}$\ underpredicts the bar length.

In general, the Fourier method overpredicts the bar length compared to $R_{az}$. However, we find better results for this full sample when compared with our results from Paper I. In fact, $R_{\mathcal{F}}$\ appears to be a better predictor of bar length than $R_{e}$\ or $R_{\mathrm{PA}}$\ when looking at Fig.~\ref{barComp} (at least when $R_{az} < 9\arcsec$). 

The azimuthal method does not rely on prior knowledge about the size of the bar, or the bar region. The three previous methods, however, can be biased towards shorter or longer values depending on how they are used. For example, if there is inner structure within the bar, the radius of maximum ellipticity ($R_{e}$) can be pushed inwards towards shorter values (see LEDA 135684, F602-1 and ASK 25131 in Table~\ref{barlens}). This can in turn bias $R_{\mathrm{PA}}$, as this is determined via a change in P.A. relative to the value at $R_{e}$. To mitigate this effect, one can begin fitting isophotes at a larger initial radius, but this can lead to missing structure and requires knowledge about the bar length, possibly introducing bias. Our various bar length measures will be explored in more detail in Sec.~\ref{sec:photmodel}.

In order to obtain final bar lengths, we have visually examined the bar lengths over the deprojected $I$-band images for all galaxies (see Fig.~\ref{barVis}, Fig.~\ref{barPlots}, and Paper I). We have selected the best bar length from this examination and list them as the bar length in Table~\ref{barprops}. \textcolor{black}{In all cases, the best bar length measure comes from our azimuthal method.}

\subsubsection{Bar Strengths}
\label{sssec:resbarstrength}

Using Equation 9 and $R_{az}$\ as $R_{bar}$, we find bar strengths that range from 0.07 (LEDA 135682) to \textcolor{black}{0.49} (ASK 25131), with an average value of \textcolor{black}{0.25}. As with Paper I, we examined the Fourier amplitudes of each galaxy and decreased the starting location of the azimuthal light profiles in order to gauge the confidence of our bar strength measure. We found that this did not significantly change the values, suggesting that our bar strengths, while lower limits, are accurate. \textcolor{black}{We find that accounting for the higher order even modes gives a better picture of the bar strength, as some galaxies have significant contribution from these modes. For example, we find a bar strength of 0.25 for F568-3 here, and a bar strength of 0.19 in Paper I where we only accounted for the $m=2$\ mode.}

\subsubsection{Corotation Radii}
\label{sssec:rescorotation}

We find that the majority of the bar corotation radii in our sample occur very close to the end of the bars in our sample. This is consistent with results from HSBs \citep[see e.g.][]{perez2012, aguerri2015, sierra2015}, and from the LSBs in Paper I. We address a problem galaxy, $[$ISI96$]$ 2329-0204, below.

\paragraph{Special Note on $[$ISI96$]$ 2329-0204}
\label{sssec:specialnote2} 

Measuring a corotation radius for $[$ISI96$]$ 2329-0204 proves to be quite difficult. For this galaxy, we have \textit{BVRI}\ images, and thus create phase profiles for all four bands (see Fig~\ref{phases} in the Appendix). When examining the phase intersections, we find that the \textit{B}-band behaves quite differently from the other three bands. So much so that it does not intersect the \textit{I}-band phase profile until $\sim$11$\arcsec$. This would leave us with a relative bar pattern speed that is $\sim$10, not a realistic number as this is more likely indicating the location of a disc corotation. This deviation from the other three bands is most pronounced at radii greater than $\sim$3$\arcsec$, near the end of the bar. 

Because this bar is quite small on the sky, we decreased the radial spacing of the azimuthal light profiles to one pixel, or 0.228$\arcsec$. This did not produce better results, and no phase intersection between the \textit{B}- and \textit{I}-bands are observed at radii less than 11$\arcsec$.

To add to this conundrum, the \textit{B}-band phase profile intersects the \textit{R}-band phase profile at 2.8$\arcsec$\ and 3$\arcsec$, both at the end of the bar. Whether or not this is the radius of corotation is not entirely clear, as the method from \citet{puerari1997} needs photometric bands separated by a large enough wavelength range. However, there has been previous work that uses `bluer' bands than \textit{I}- to determine the corotation radius \citep[see ][]{sierra2015}. With this in mind, we will take the phase intersection between the \textit{B}- and \textit{R}-bands at 2.8$\arcsec$\ to be the corotation radius of the bar, as using the phase intersections between \textit{B}- and \textit{I}- is most likely not due to the bar.

\subsection{Discussion of Bar Properties}
\label{ssec:barDiscuss}

Our final bar properties for the whole sample are listed in Table~\ref{barprops}. Our sample has an average bar length of 2.5 kpc. This is shorter than what is typical for HSBs. When looking at surveys of HSBs, bars in these galaxies tend to be in the range $3$\ kpc $\leq R_{bar} \leq$\ 5 kpc \citep[see e.g.][]{erwin2005, marinova2007, aguerri1998, aguerri2009}.

\begin{table*}
  \centering
  \caption{Final bar properties of the whole sample. Bar lengths are in arcsec ($R_{\mathrm{bar}}$) and kpc ($R_{\mathrm{bar}}^{'}$), and corotation radii are in arcsec ($R_{\mathrm{CR}}$) and kpc ($R_{\mathrm{C}}^{'}$), relative bar pattern speeds ($\mathcal{R}$), and lower limits on bar strength (S$_{b}$) are unitless. Lengths in kpc assume the distances in Table~\ref{sample}.}
  \label{barprops}
  \begin{tabular}{lcccccc}
    \hline
    Galaxy                  & $R_{\mathrm{bar}}$  & $R_{\mathrm{bar}}^{'}$ & $R_{\mathrm{CR}}$   & $R_{\mathrm{CR}}^{'}$ & $\mathcal{R}$   & S$_{\mathrm{b}}$ \\
                            & (arcsec)         & (kpc)              & (arcsec)         & (kpc) & & \\
    \hline
    UGC 628                 & 11.21 $\pm$ 0.92 &  4.67 $\pm$ 0.38   & 13.96 $\pm$ 0.46 & 5.81 $\pm$ 0.19    & 1.25 $\pm$ 0.11 &    0.32       \\
    LEDA 135682             &  2.05 $\pm$ 0.46 &  1.04 $\pm$ 0.23   &  2.70 $\pm$ 0.34 & 1.36 $\pm$ 0.17    & 1.32 $\pm$ 0.34 &    0.07       \\
    LEDA 135684             &  7.07 $\pm$ 0.92 &  1.57 $\pm$ 0.20   &  7.90 $\pm$ 0.46 & 1.75 $\pm$ 0.10    & 1.12 $\pm$ 0.16 &    0.30       \\
    LEDA 135693             &  5.70 $\pm$ 0.46 &  4.25 $\pm$ 0.69   &  7.30 $\pm$ 0.46 & 5.45 $\pm$ 0.34    & 1.28 $\pm$ 0.22 &    0.36       \\
    UGC 2925                &  8.44 $\pm$ 0.92 &  2.34 $\pm$ 0.26   & 12.50 $\pm$ 0.46 & 3.47 $\pm$ 0.13    & 1.48 $\pm$ 0.17 &    0.18       \\
    F563-V2                 &  6.65 $\pm$ 0.46 &  1.95 $\pm$ 0.13   & 15.86 $\pm$ 0.46 & 4.64 $\pm$ 0.13    & 2.38 $\pm$ 0.18 &    0.29       \\
    F568-1                  &  4.37 $\pm$ 0.46 &  1.94 $\pm$ 0.20   &  5.86 $\pm$ 0.46 & 2.61 $\pm$ 0.20    & 1.34 $\pm$ 0.18 &    0.13       \\
    F568-3                  &  8.93 $\pm$ 0.46 &  3.61 $\pm$ 0.37   & 10.06 $\pm$ 0.46 & 4.07 $\pm$ 0.19    & 1.13 $\pm$ 0.13 &    0.25       \\
    LEDA 135782             &  2.66 $\pm$ 0.68 &  2.69 $\pm$ 0.69   &  2.80 $\pm$ 0.34 & 2.83 $\pm$ 0.34    & 1.05 $\pm$ 0.30 &    0.11       \\
    UGC 8066                &  7.07 $\pm$ 0.68 &  1.39 $\pm$ 0.13   &  7.50 $\pm$ 0.34 & 1.47 $\pm$ 0.07    & 1.06 $\pm$ 0.11 &    0.38       \\
    LEDA 135867             &  4.39 $\pm$ 0.34 &  2.53 $\pm$ 0.20   &  5.40 $\pm$ 0.34 & 3.11 $\pm$ 0.20    & 1.23 $\pm$ 0.12 &    0.37       \\
    F602-1                  &  4.11 $\pm$ 0.34 &  2.09 $\pm$ 0.17   &  5.30 $\pm$ 0.34 & 2.70 $\pm$ 0.17    & 1.29 $\pm$ 0.14 &    0.21       \\
    PGC 70352               &  5.02 $\pm$ 0.68 &  1.79 $\pm$ 0.24   &  6.60 $\pm$ 0.46 & 2.35 $\pm$ 0.16    & 1.32 $\pm$ 0.20 &    0.13       \\
    ASK 25131               &  7.76 $\pm$ 0.92 &  4.02 $\pm$ 0.48   & 10.60 $\pm$ 0.46 & 5.49 $\pm$ 0.24    & 1.37 $\pm$ 0.17 &    0.49       \\
    $[$ISI96$]$ 2329-0204   &  2.05 $\pm$ 0.46 &  $\dotso$          &  2.80 $\pm$ 0.28 & $\dotso$           & 1.37 $\pm$ 0.34 &    0.18       \\

    \hline
  \end{tabular}
\end{table*}

With the corotation radii of our sample, we can calculate the relative bar pattern speeds for our sample, $\mathcal{R} = R_{\mathrm{CR}}/R_{\mathrm{bar}}$, also listed in Table~\ref{barprops}. Bars are considered `fast' rotators if $\mathcal{R} < 1.4$\ and `slow' rotators if $\mathcal{R} > 1.4$ \citep{athanassoula1992, elmegreen1996, debattista2000}. \textcolor{black}{While historically} the almost unanimous result for HSBs are fast bars \citep[e.g.][]{perez2012,aguerri2015,sierra2015}, \textcolor{black}{recent work has found an increasing number of `slow' bars \citep{font2017, guo2019}}. The results for dark matter dominated galaxies, LSBs specifically, are much more unclear due to the lack of pattern speed measurements. 

Centrally dense, \textcolor{black}{nonrotating}, dark matter halos are expected to dynamically slow down bars over time \citep{weinberg1985, debattista2000}. Indeed, very slow bars in dark matter dominated galaxies have been reported in the literature \citep{bureau1999, chemin2009, banerjee2013}. \textcolor{black}{In order to examine possible trends between $\mathcal{R}$\ and galaxy properties, \citet{guo2019} used a sample of 53 barred galaxies and found a significant number of very slow bars. Interestingly, they found no correlation between a very slow bar and dark matter content.} Our previous results in Paper I indicated that three out of the four galaxies we analyzed were hosts to \textit{fast} bars. Here we find that 13 out of the 15 galaxies in this work are hosts to fast bars.

Noteably, only one galaxy \textcolor{black}{in our sample} has a \textit{very} slow bar (i.e. $\mathcal{R} > 2$), F563-V2, and one galaxy, UGC 2925, has a slow bar that is comparable to previous results on HSBs \citep{aguerri1998, sierra2015, font2017, guo2019}. The remainder of the bars in our sample are \textit{fast}. \textcolor{black}{While this may seem to indicate that the bars in our sample have not been slowed down, dark matter halos are \textit{not} static and can have significant angular momentum, arising from tidal exchanges during the formation of the galaxy \citep[see][]{peebles1969}. This in turn means that the angular momentum exchange between the disc and the halo can be mitigated, sometimes significantly \citep{long2014}. Thus, halos cannot serve as a pure angular momentum sink, and the angular momentum of the halo, often characterised by the halo spin parameter $\lambda$, can dictate important aspects of a bar.}

\textcolor{black}{For example, \citet{fujii2019} found that for high $\lambda$\ halos, bars actually \textit{speed up}, while also becoming shorter and weaker. This is especially of note, as LSB discs are expected to form in high $\lambda$\ halos \citep[e.g.][]{dalcanton1997, jimenez1998, kim2013}. Thus, without knowing $\lambda$, we cannot truly make a definitive statement on the implications of our relative bar pattern speeds.}

\begin{figure}
  \centering
  \includegraphics[scale=0.4]{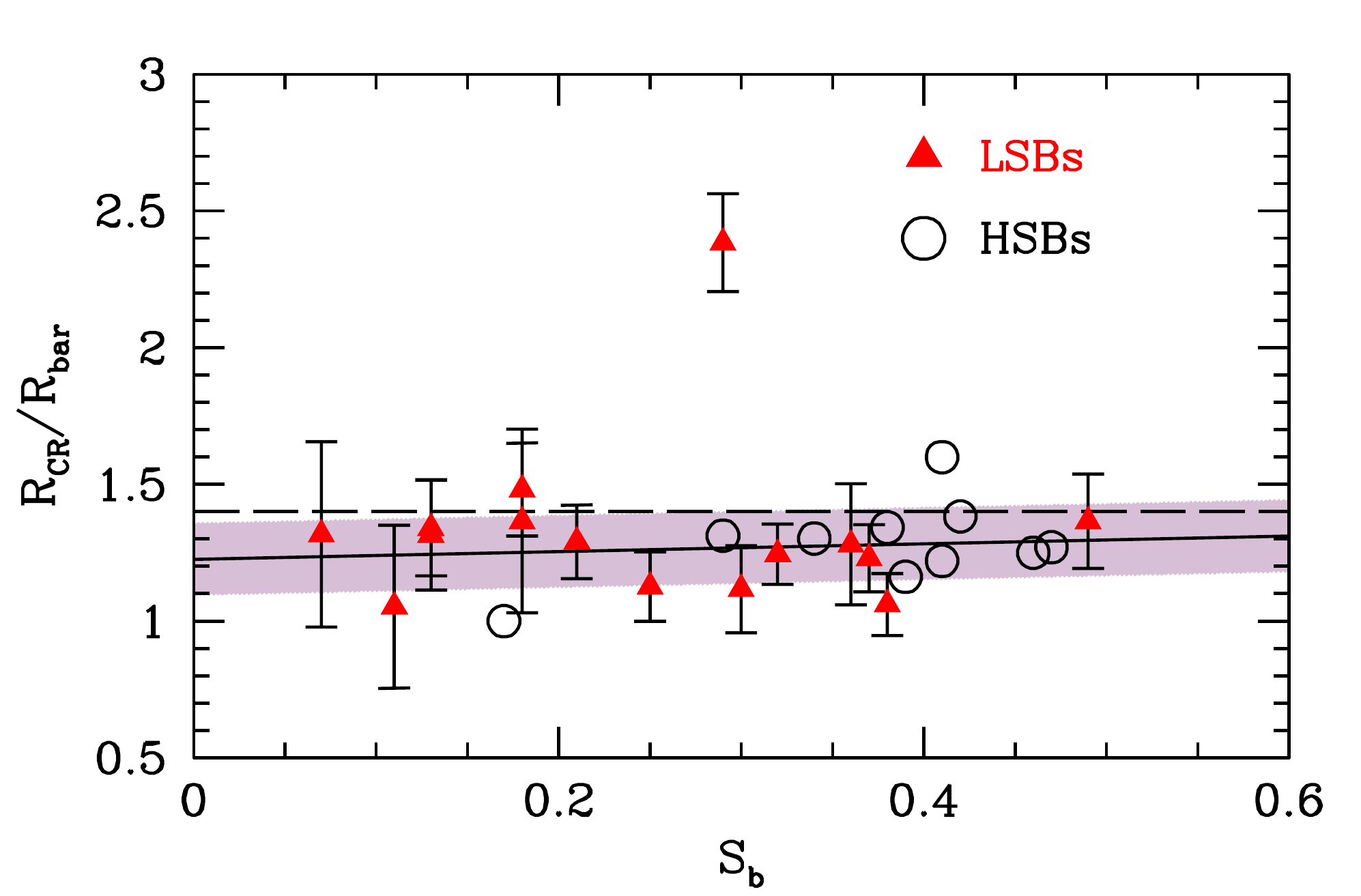}
  \caption{Relative bar pattern speed ($\mathcal{R} = R_{\mathrm{CR}}/R_{\mathrm{bar}}$) as a function of bar strength (S$_{b}$) for our sample (red triangles) and HSBs from \citet{aguerri1998} (open circles). The solid line indicates the fit to the HSBs and LSBs, excluding the outlier discussed in the text: $\mathcal{R} = 1.23 +0.14S_{b}$. The shaded region shows the scatter in the relation: $\sigma = 0.13$. The horizontal dashed line is the separator between fast and slow bars (i.e. $\mathcal{R} = 1.4$).}
  \label{speedStr}
\end{figure}

We show $\mathcal{R}$ versus bar strength in Fig.~\ref{speedStr}, including the HSBs from \citet{aguerri1998}, as this work uses the same measure as used here for corotation radius and bar strength. This is an updated plot of Fig. 14 from Paper I. Here we also show the fit to both LSBs and HSBs, excluding the outlier of F563-V2 (the point far above the rest) and find $\mathcal{R} = 1.23 + 0.14S_{b}$\ with a scatter of 0.13. This is \textcolor{black}{close to identical to the} relation from Paper I, with a near identical scatter, \textcolor{black}{and is consistent with the lack of relation found by \citet{guo2019}}. Our relation is plotted in Fig.~\ref{speedStr} as the solid line, with the shaded region indicating the scatter.

When looking at Fig.~\ref{speedStr}, we find that LSBs and HSBs form a continuum, with the LSBs clustered on the weaker strength end (average $\sim0.25$) transitioning into the stronger HSBs (0.37 for the HSBs shown here). This also then translates to surface brightness, with bars being fast rotators across all surface brightnesses.

As both the strength and relative bar pattern speed are dependent on the bar length, it is crucial that this property be measured as accurately as possible. For this reason, we have employed the four different measurement techniques here and have found that our $R_{az}$\ method performs the best for all the galaxies in our sample. However, in order to be as confident in this method as possible, we have tested the four methods used here on simulated galaxy images consisting of a disc and bar. This is therefore an idealized case for each method, as no spiral arms or disc features can affect the measurement. We spend the entirety of the next section discussing this process. Because the azimuthal bar length performs so well on our data, we expect this method to perform just as well on the simulated images. However, the lack of spiral arms or disc features could affect the accuracy of the other three methods as well.

\section{Analysis of Mock Galaxy Images}
\label{sec:photmodel}

In order to determine how well we can determine the bar properties of our galaxies, we constructed fake galaxy images with randomised structural and photometric parameters. We used observational parameters similar to those of our observed sample and those of the ARCTIC imager in order to compare with our observations. We broadly follow the procedure in \citet{aguerri2009}.

\subsection{Creating the Images}
\label{ssec:photmodimage}

For each galaxy, we assume two components: an exponential disc and a Ferrers bar \citep[see][]{laurikainen2005}. We do not consider a bulge component since LSBs do not typically have a bulge \citep{pahwa2018}, and because our sample lacks significant bulges. \textcolor{black}{In addition, we do not include spiral arms in our images in order to provide an ideal scenario for each bar length fitting method.} The intensity profile of an exponential disc is given by
\begin{equation}
  I(r) = I_{0}e^{-r/h},
\end{equation}
where $I_{0}$\ is the central disc intensity and $h$\ is the disc scale length. The intensity profile of a Ferrers bar is given by
\begin{equation}
  I(r) = I_{0} \left( 1 - \left( \frac{r}{r_{bar}} \right)^{2} \right)^{n_{bar}+0.5},
\end{equation}
where $I_{0}$\ is the central bar intensity, $r_{bar}$\ is the length of the bar, and $n_{bar}$\ is the bar shape parameter. The intensity is defined to be zero for $r > r_{bar}$. The radial coordinate is given by
\begin{equation}
  r = \left( \left| x \right|^{c} + \left| \frac{y}{1 - \epsilon_{bar}} \right|^{c} \right)^{1/c},
\end{equation}
where $\epsilon_{bar}$\ is the ellipticity of the bar, and $c$\ is a parameter that controls the shape of the bar isophotes \citep{athanassoula1990}. Pure elliptical isophotes have $c = 2$, boxy isophotes have $c > 2$, and discy isophotes have $c < 2$.

In order to construct observationally accurate galaxies for our purposes, we selected galaxy properties that match the general LSB population, with bar values similar to our results from Paper I and from Sec.~\ref{ssec:barResults}. We begin by randomly assigning a disc central surface brightness 
\begin{equation}
  22.5 < \mu_{0,B} < 24 
\end{equation} 
and bar central surface brightness
\begin{equation}
  22.0 < \mu_{0,B} < 23.0
\end{equation}
with the condition that the central bar surface brightness may not exceed the disc central surface brightness by more than 1 magnitude. \textcolor{black}{This condition was chosen in order to closely match the photometric properties of our sample (see Sec. \ref{sec:phot}).} We then convert these to intensity values via
\begin{equation}
  I = tp^{2}10^{(Z - \mu)/2.5},
\end{equation}
where $t$\ is the exposure time, $p$\ is the plate scale of the detector, and $Z$ is the photometric zeropoint. We used values that matched our observations to obtain images similar to our observed galaxies: $t = 600$\ sec., $p = 0.228$\ arcsec pix$^{-1}$, and $Z_{B} = 25$. In order to produce images similar to our observations, we select distances 
\begin{equation}
70\ \mathrm{Mpc} < D < 200\ \mathrm{Mpc}
\end{equation}
For the other parameters, such as bar ellipticity and disc inclination, we broadly follow the methods laid out in \citet{aguerri2009}. For the bar length, we select from a range representative of our results in Sec.~\ref{ssec:barResults}. We also force the bar length to be shorter than the disc scale length in order to prevent constructing very unrealistic galaxies (i.e. a bar length of 15$\arcsec$\ and a disc scale length of 3$\arcsec$\ is not considered realistic).

After the galaxy images were created, we added noise to each pixel in order to closely match our observations. The gain and read-out noise of ARCTIC are 1.98 e$^{-}$\ ADU$^{-1}$\ and 3.8 e$^{-1}$\ respectively. Sky noise was added in assuming a 600 sec. exposure in \textit{B}. Finally, to account for seeing, we convolved the images with a 2D Moffat PSF
\begin{equation}
  PSF(r) = \frac{\beta - 1}{\pi \alpha^{2}} \left( 1 + \frac{r^{2}}{\alpha} \right)^{-\beta},
\end{equation}
where $\alpha = \mathrm{FWHM}/(2\sqrt{2^{1/\beta} - 1})$\ is the seeing parameter, and $\beta$\ is typically taken to be $\sim$3.5 \citep[see ][]{trujillo2001}. To test a wide range of observations, we randomly select our seeing to be between 0.8$\arcsec$ and 2$\arcsec$, consistent with the average seeing values of our observations (Table~\ref{sample}).

\begin{figure}
  \centering
  \includegraphics[scale=0.3]{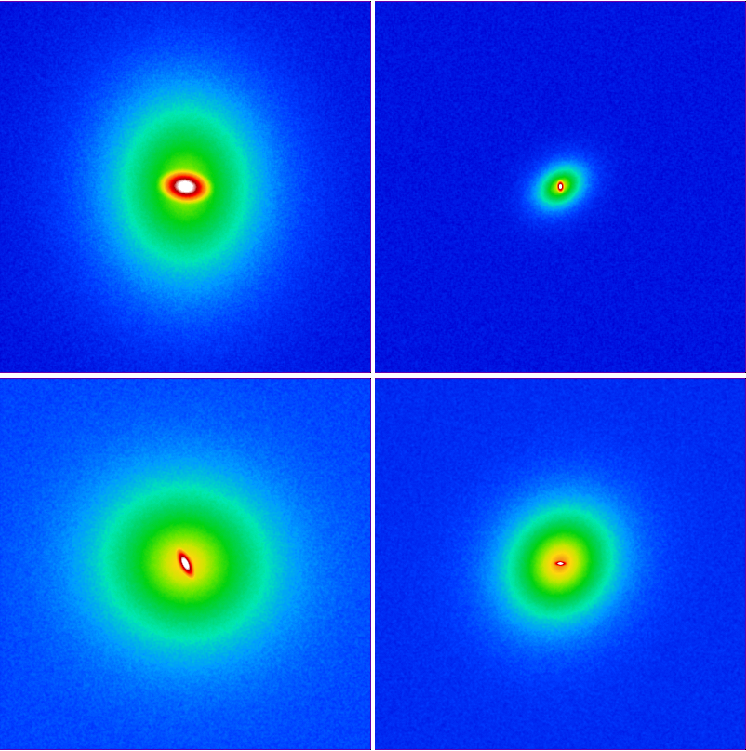}
  \caption{Examples of the mock images showcasing various sized bars and discs. Each image is 200$\times$200 pixels (45.6$\arcsec \times$45.6$\arcsec$) and is scaled the same way.}
  \label{mockImgs}
\end{figure}

\begin{figure}
  \centering
  \includegraphics[scale=0.38]{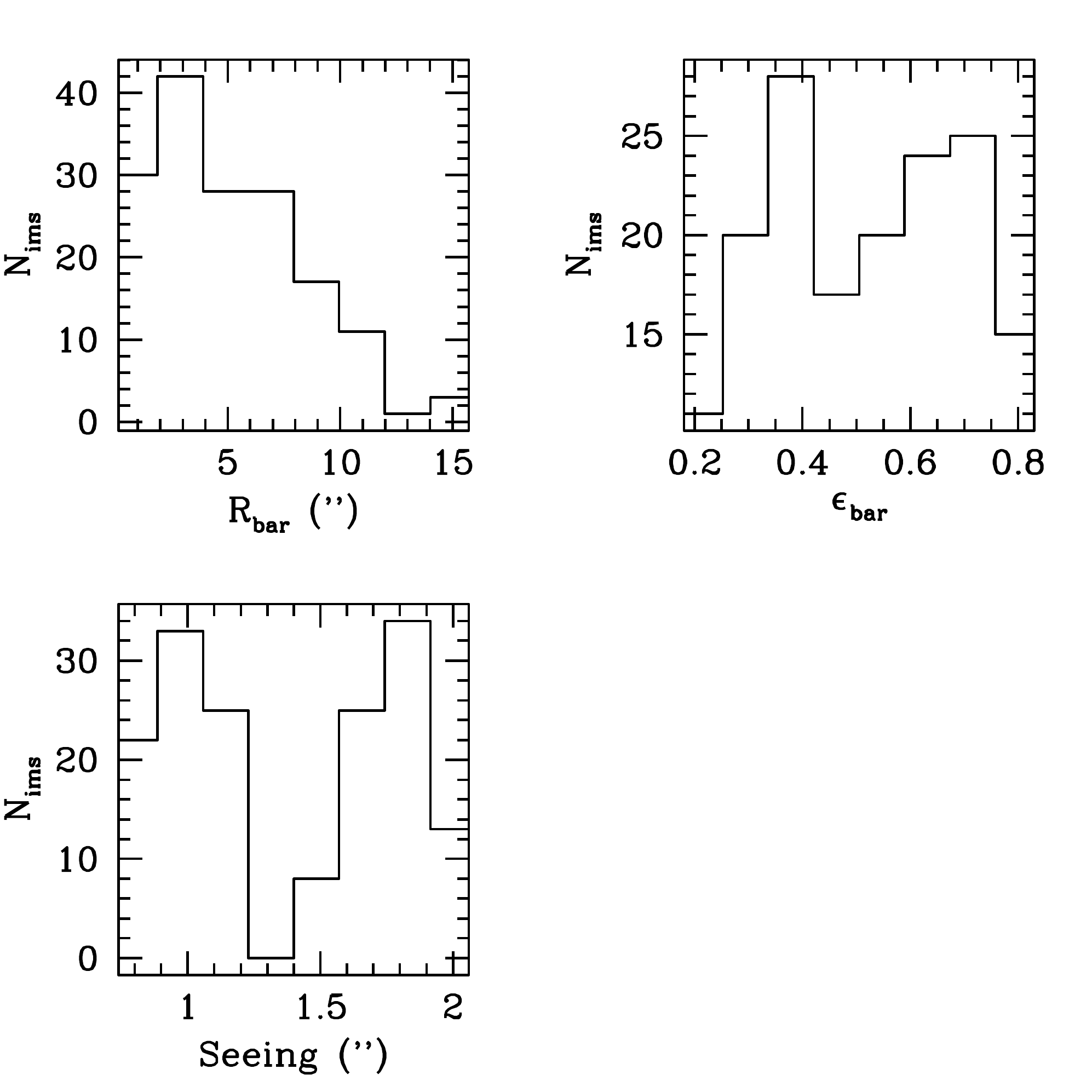}
  \caption{Histograms of bar length (top left), ellipticity (top right), and seeing (bottom left) for our mock images.}
  \label{imStats}
\end{figure}

Examples of the images we created are shown in Fig.~\ref{mockImgs}. These images were randomly selected, and are representative of the types of galaxy images we have created. In total we have created 200 images. Histograms of bar lengths, ellipticity, and seeing are shown in Fig.~\ref{imStats}. Here we can see that we are sampling a large range of bar lengths, but that we have a large over sampling of shorter bar lengths. This is because the bar lengths in our sample are quite short (Sec.~\ref{ssec:barResults}) and we are hoping to gauge how well the various methods work for our sample. We see that we are sampling a large range of bar ellipticities, $\sim0.2 \leq \epsilon_{bar} \leq \sim0.8$. The two peaks in the seeing histogram arise from the distributions around `good' and `bad' seeing, which we have defined here to be $\sim$1$\arcsec$\ and $\sim$1.8$\arcsec$\ respectively.

\subsection{Measurements of Mock Images}
\label{ssec:photmodmeasure}

To maintain consistency, we follow the same reduction process as our real observations. We determine the sky value by using the box method of \citet{schombert2014} and use \texttt{ELLIPSE} and \texttt{GEOTRAN} to deproject our images. We then use our four methods described in Sec.~\ref{ssec:barMethods} to determine the bar length. 

As opposed to our real galaxies (see Sec.~\ref{ssec:barResults}), we do not have concerns over spiral arms influencing our isophotal bar length measures, as we have explicitly not included them. Therefore, we have an ideal scenario for each measurement technique.

Due to the large number of images tested (200), we automated the azimuthal light profile measurement by standardizing each set of profiles. We did this by finding the azimuthal centroid of the bar, usually located at the maximum value of the profile. We then adjusted the entire set of profiles so that the bar centroid is aligned with either 90$^{\circ}$\ or 270$^{\circ}$, depending on which is closest to the centroid. It is then assumed that the two humps in the azimuthal light profiles will remain at 90$^{\circ}$\ and 270$^{\circ}$. As there are no spiral arms in our images, we instead use the relative intensity of the bar hump to the disc to determine where the bar ends.

\subsection{Results of  Measurements}
\label{ssec:photmodresults}

\begin{figure}
  \centering
  \includegraphics[scale=0.38]{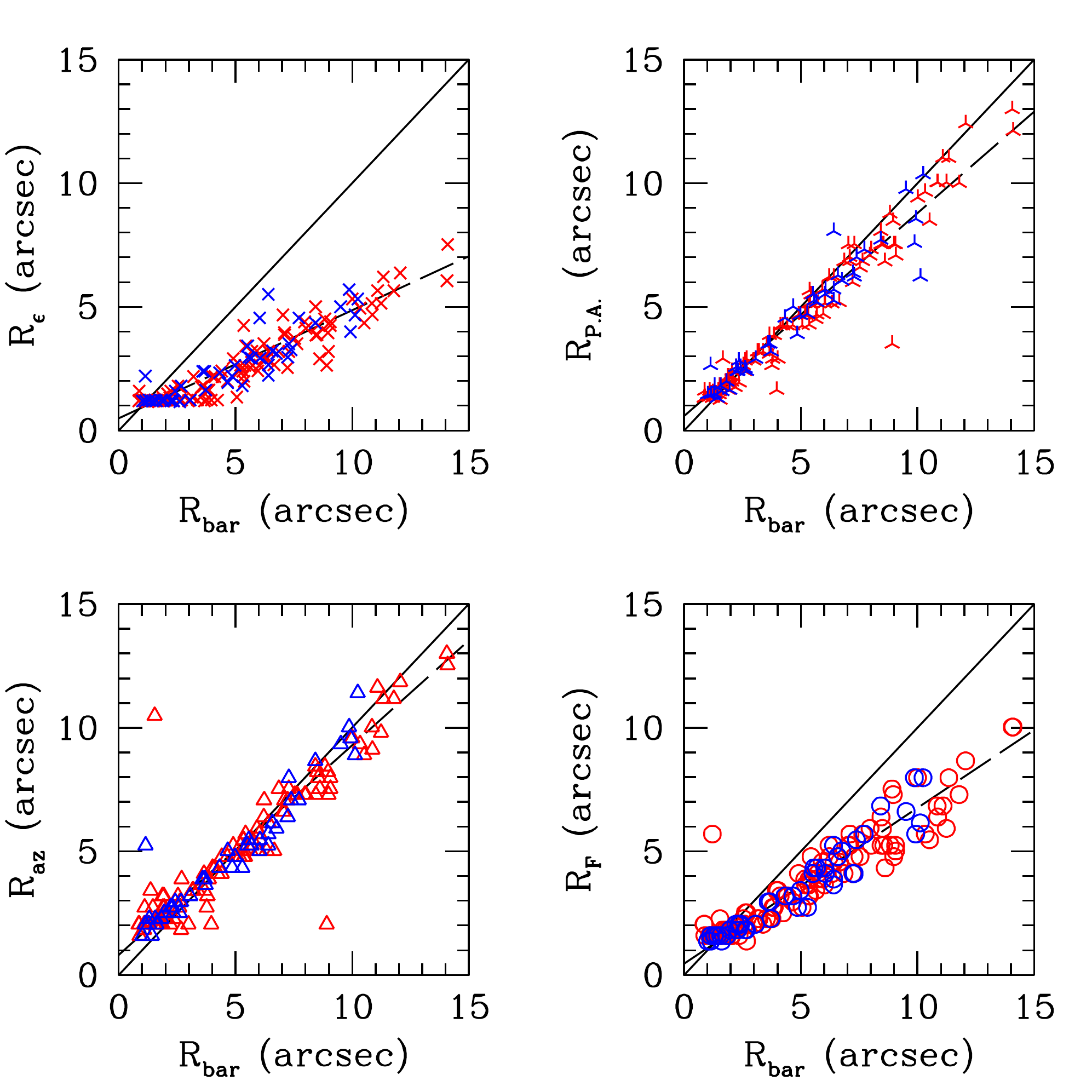}
  \caption{Our four bar length measures when applied to fake galaxy images: $R_{e}$\ (top left), $R_{\mathrm{PA}}$\ (top right), $R_{az}$\ (bottom left), and $R_{\mathcal{F}}$\ (bottom right). Images with seeing less than 1$\arcsec$\ are shown as blue points, and images with seeing greater than 1$\arcsec$\ are shown as red points. The dashed lines in each panel show the linear fits to the blue points.}
  \label{photMod_measures}
\end{figure}

We show the comparison between the true and measured bar length for our mock images in Fig.~\ref{photMod_measures}: $R_{\epsilon}$\ (top left), $R_{\mathrm{P.A.}}$\ (top right), $R_{az}$\ (bottom left), and $R_{\mathcal{F}}$\ (bottom right). Blue points denote those images with less than 1$\arcsec$\ seeing, and red points denote images with greater than 1$\arcsec$\ seeing. In each of these subfigures the solid line denotes unity, and the dashed lines are fits to the blue points (i.e. those with good seeing). It is clear from Fig.~\ref{photMod_measures} that the seeing in each image greatly increases the scatter of the measurement.

We find that the radius of maximum ellipticity (top left panel) does a poor job of measuring the bar length, often measuring a length 50$\%$\ shorter than the true length. This is an expected result, especially for Ferrers bars \citep{michel2006,aguerri2009}. The fit for this measure is $R_{e} = 0.44R_{bar} + 0.49$. Even though $R_{\mathrm{P.A.}}$\ (top right panel) depends on $R_{e}$, we find that it does an excellent job of finding the bar length, only really failing at larger radii (i.e. $r > 5\arcsec$). The fit for this measure is $R_{\mathrm{P.A.}} = 0.82R_{bar} + 0.59$. Our bar length measure using the azimuthal light profiles (bottom left), $R_{az}$, performs the best out of the four methods tested, albeit only slightly better than $R_{\mathrm{P.A}}$. The fit for this measure is $R_{az} = 0.85R_{bar} + 0.80$. \textcolor{black}{Lastly, we find that $R_{\mathcal{F}}$\ under predicts the bar length, contrary to our results using real data (see UGC 628 and F568-1 in Paper I and LEDA 135682 in Fig.~\ref{galimages}, for example) and previous works \citep{aguerri2009}, suggesting that spiral arms can heavily bias this method to longer bar lengths.} The fit for this measure is $R_{\mathcal{F}} = 0.63R_{bar} + 0.46$.

Based on this modeling, we find that the bar length measure based on the behaviour of the azimuthal light profiles to be the best measure out of the four used here. While $R_{\mathrm{P.A.}}$\ almost performs as well, the dependence on $R_{\epsilon}$\ for this measure is concerning, and can result in wrong measurements. That $R_{\mathrm{P.A}}$\ performs so much better here than with our real data is surprising, although this could simply be that $R_{\epsilon}$\ underpredicts the bar length so significantly here. In addition, the change in P.A. used here is simply following what has been used before \citep[i.e. from][]{aguerri2009}. Contrary to this, there is strong motivation for the azimuthal method, as it traces the location of the bar and any spiral arms in the galaxy, only being limited by the resolution and seeing. We do, however, note the large amount of scatter at smaller bar lengths for $R_{az}$. In additon, we see that this technique fails to find an accurate bar length for three bars, seen as the wildly innacurate points in the bottom left panel of Fig.~\ref{photMod_measures}. \textcolor{black}{When examining these images, we found that these failures were caused by a combination of factors. First, all three images contain bars and discs that have central surface brightnesses within $\sim$0.4 mag arcsec$^{-2}$\ of each other, making it somewhat hard to distinguish the two components in the azimuthal light profiles. Second, the position angles of the bars and discs are all within $\sim$5$^{\circ}$\ of each other, which causes the bar to become slightly lost in the inclined disc. Finally, two of the failures have poor seeing, which further amplifies the issues caused by the first two points.}

This suggests that our bar properties reported in Sec.~\ref{ssec:barResults} are accurate. More importantly, this means that the bars in our LSBs are indeed \textit{fast} \textcolor{black}{(but see Sec.~\ref{sec:discussion})}.

\section{Photometry}
\label{sec:phot}

In this section, we detail our methods for measuring the photometry, and present our results. The surface brightness profiles, magnitudes, and colours for the 11 targets not in Paper I are new measurements: as to the best of our knowledge no such measurements have been previously published.

\subsection{Methods}

Here we detail how we construct surface brightness profiles, as well as obtaining total magnitudes and colours.

\subsubsection{Surface Brightness Profiles}
\label{sssec:methodsSurfBright}

We first obtain intensity profiles ($I(r)$) of our galaxies using the \texttt{IRAF} routine \texttt{ELLIPSE}. This method of constructing surface brightness profiles for LSBs has been used succesfully in the literature \citep[e.g.][]{wittman2017}. We then convert these to surface brightness values via:
\begin{equation}
  \mu(r) = -2.5\log{\left( \frac{I(r)}{t\ p^{2}} \right)} + Z_{p},
\end{equation}
where $t$\ is the exposure time in seconds, $p$\ is the plate scale of the instrument, and $Z_{p}$\ is the photometric zeropoint of the observation (Sec.~\ref{ssec:data}). We then fit for the central surface brightness ($\mu_{0}$) and disc scale length ($h$), starting beyond the bar region, $r > R_{\mathrm{bar}}$\ (see Sec.~\ref{ssec:barResults}). 

In order to correct for the likely low internal dust extinction in our galaxies \citep[e.g.][]{bothun1997, matthews2005, wyder2009, hinz2007, honey2016}, we assume the discs are optically thin slabs. The correction for this assumption is simply:

\begin{equation}
  \mu_{0} = \mu_{0}^{f} - 2.5\log{(a/b)}
\end{equation}
where $\mu_{0}$\ is the observed central surface brightness, $\mu_{0}^{f}$\ is the face-on central surface brightness, and a and b are the semi-major and -minor axes respectively (Table~\ref{sample}). This correction increases with the inclination, making magnitudes brighter for highly inclined discs.

\subsubsection{Magnitudes}
\label{sssec:methodsmags}

Due to the faint nature of LSBs, using a typical $m_{25}$\ magnitude is not an adequate measure of the total luminosity. Instead, we follow \citet{deblok1995} and use the total magnitude, $m_{T}$, of the disc, found via
\begin{equation}
  m_{T} = \mu_{0} - 2.5\log{(2\pi h^{2})} - 2.5\log{(\cos{i})}
\end{equation}
where $i$\ is the inclination of the disc (Table~\ref{sample}). Integrating out to infinity involves extrapolating the disc scale length, which can introduce error. To counter this, \citet{deblok1995} also use another magnitude measure, $m_{apt}$, which takes the entire data of the galaxy disc into account. As their data probe down to $\sim$28 mag arcsec$^{-2}$\ in $B$, almost all the light of the galaxy is being probed. Our observations are not quite as deep, down to $\sim$26 mag arcsec$^{-2}$\ in $B$\ for most targets, so we do not use $m_{apt}$\ here.

\subsubsection{Colours}
\label{sssec:methodsColors}

Colour profiles were obtained by subtracting the $B$- and $I$-band surface brightness profiles from each other. 

We obtain three different total colours for each galaxy: 
\begin{enumerate}
  \item a \textbf{bar colour} consisting of the average colour within the bar region ($r < R_{bar}$)
  \item a \textbf{disc colour} consisting of the average colour outside the bar region ($r > R_{bar}$)
  \item an \textbf{area-weighted colour} within the 25.5 $B$-mag arcsec$^{-2}$\ isophote \citep[e.g.][]{mcgaugh1994,deblok1995} ($r < R_{25.5}$)
\end{enumerate}

We measure these three colours in the same manner. First, we create a ($B-I$) colour map for each galaxy by rebinning each image by a factor of 2 to reduce the noise, and subtracting the new $I$-band image from the $B$-band image, obtaining a colour in each pixel \citep[see][]{deblok1995, schombert2011}. We finally obtain colours by taking the mean value in each of the defined regions, rejecting divergent pixels and those with large errors.

\subsection{Results}

We list central surface brightnesses, magnitudes, disc scale lengths and other information for our whole sample in Table~\ref{photometry}. All surface brightnesses and magnitudes reported are corrected for Galactic extinction using $A_{\lambda}$\ values from \citet{schlafly2011}, also listed in Table~\ref{photometry}. Data for F563-V2, F568-1, and F568-3 are taken from \citet{mcgaugh1994} and \citet{deblok1995}. Although LSBs are thought to have low dust content, we correct for any internal dust by assuming the discs are optically thin slabs (see Sec.~\ref{sssec:methodsSurfBright}).

With the exception of UGC 628, F563-V2, F568-1, and F568-3, the surface brightness profiles, magnitudes, and colours are all new. In addition, this is one of the few samples of exclusively barred LSB photometry.

\begin{table*}
  \centering
  \caption{\textit{B}- and \textit{I}-band central surface brightness ($\mu_{0}$), disc scale lengths in arcsec ($h$) and kpc ($h^{'}$), and total apparent magnitudes ($m_{T}$) and absolute magnitudes ($M_{T}$) for our LSBs. Data for F568-1, and F568-3 are taken from \citet{deblok1995}. Data for F563-V2 are taken from \citet{mcgaugh1994} (no \textit{I}-band photometry available). Central surface brightness and magnitudes are corrected using the $A_{\lambda}$\ values in Column 8 \citep{schlafly2011}. As a reminder, there is no distance available for $[$ISI96$]$ 2329-0204. Other distance-dependent values use the distances listed in Table~\ref{sample}.}
  \label{photometry}
  \begin{tabular}{lccccccc}
    \hline
    Galaxy                  & Band &     $\mu_{0}$       &   $h$    & $h^{'}$  & $m_{T}$ & $M_{T}$   & $A_{\lambda}$  \\
                            &      & (mag arcsec$^{-2}$) & (arcsec) &  (kpc)   & (mag)  & (mag)    &   (mag)      \\
    \hline
    UGC 628                 &   B  &       22.36        &   13.87  &   5.78   &   15.35  & -19.32   &    0.158     \\
                            &   I  &       20.15        &    8.33  &   3.47   &   14.25  & -20.42   &    0.065     \\
    LEDA 135682             &   B  &       22.31        &   10.09  &   5.10   &   15.49  & -19.60   &    0.380     \\
                            &   I  &       20.97        &    7.27  &   3.67   &   14.86  & -20.23   &    0.158     \\
    LEDA 135684             &   B  &       22.37        &   10.61  &   2.13   &   15.61  & -17.69   &    0.344     \\
                            &   I  &       21.58        &    9.58  &   1.40   &   15.05  & -18.25   &    0.143     \\
    LEDA 135693             &   B  &       21.42        &    6.30  &   4.70   &   15.51  & -20.43   &    1.086     \\
                            &   I  &       20.17        &    5.72  &   4.27   &   14.47  & -21.47   &    0.451     \\
    UGC 2925                &   B  &       20.71        &   10.18  &   2.83   &   13.86  & -19.93   &    1.781     \\
                            &   I  &       18.75        &    8.99  &   2.50   &   12.19  & -21.60   &    0.739     \\
    F563-V2                 &   B  &       21.95        &    7     &   2.05   &   15.87  & -18.18   &  $\dotso$    \\
                            &   I  &      $\dotso$      & $\dotso$ & $\dotso$ & $\dotso$ & $\dotso$ &  $\dotso$    \\
    F568-1                  &   B  &       23.65        &   12.86  &   5.72   &   16.23  & -18.58   &  $\dotso$    \\
                            &   I  &       22.32        &   12.17  &   5.41   &   15.02  & -19.79   &  $\dotso$    \\
    F568-3                  &   B  &       22.79        &   10.76  &   4.35   &   15.91  & -18.70   &  $\dotso$    \\
                            &   I  &       21.04        &    8.37  &   3.38   &   14.70  & -19.91   &  $\dotso$    \\
    LEDA 135782             &   B  &       22.22        &    4.58  &   4.63   &   17.16  & -19.43   &    0.094     \\
                            &   I  &       20.41        &    3.35  &   3.38   &   16.01  & -20.58   &    0.039     \\
    UGC 8066                &   B  &       23.34        &   14.04  &   2.76   &   16.10  & -16.94   &    0.063     \\
                            &   I  &       21.83        &   11.34  &   2.23   &   15.07  & -17.97   &    0.026     \\
    LEDA 135867             &   B  &       22.38        &    4.92  &   2.84   &   17.19  & -18.19   &    0.179     \\
                            &   I  &       21.07        &    4.15  &   2.39   &   16.24  & -19.14   &    0.074     \\
    F602-1                  &   B  &       22.47        &   11.87  &   4.92   &   15.24  & -19.87   &    0.169     \\
                            &   I  &       20.92        &    9.66  &   4.65   &   14.14  & -20.97   &    0.070     \\
    PGC 70352               &   B  &       21.58        &    7.64  &   2.72   &   15.33  & -19.00   &    0.261     \\
                            &   I  &       20.18        &    6.64  &   2.37   &   14.24  & -20.09   &    0.108     \\
    ASK 25131               &   B  &       23.09        &    9.14  &   4.74   &   16.55  & -18.59   &    0.135     \\
                            &   I  &       21.16        &    6.49  &   3.36   &   15.36  & -19.78   &    0.056     \\
    $[$ISI96$]$ 2329-0204   &   B  &       23.14        &    9.38  & $\dotso$ &   16.48  & $\dotso$ &    0.162     \\
                            &   I  &       21.62        &    9.04  & $\dotso$ &   15.02  & $\dotso$ &    0.067     \\
    \hline
  \end{tabular}
\end{table*}

\subsubsection{Surface Brightness Profiles}
\label{sssec:surfbright}

Our \textit{B}- and \textit{I}-band surface brightness profiles and ($B-I$) colour profiles are shown in Fig.~\ref{surfBright}. We do not have photometrically calibrated ARCTIC data for F563-V2, F568-1, or F568-3. The majority of our galaxies exhibit very nearly Freeman Type I profiles \citep{freeman1970}, or pure exponential disks. For some galaxies, the bar is not noticeable in these surface brightness profiles, LEDA 135782 or PGC 70352 for example. For others, it very clearly dominates the inner light profile, as in the case of LEDA 135693 or $[$ISI96$]$\ 2329-0204. 

The mean $B$-band and $I$-band central surface brightnesses are 22.39 and 20.28 mag arcsec$^{-2}$\ respectively, consistent with the findings in \citet{zhong2008}, who looked at roughly 12000 LSBs from the SDSS catalog. We find a mean \textit{B}-band disc scale length of 3.68 kpc and a mean \textit{I}-band disc scale length of 2.03 kpc. We find no correlation between scale length and central surface brightness, consistent with previous works \citep{mcgaugh1994, deblok1995, zhong2008, pahwa2018}, as shown in the top panel of Fig.~\ref{scalePlot}. Here, black points are our data, blue squares are \citet{deblok1995}, and red triangles are \citet{mcgaugh1994}, both of which use comparable methods for both barred and unbarred LSBs. We can see that roughly one dex in scale length spans roughly 4 dex in central surface brightness.

\begin{figure*}
  \centering
  \includegraphics[scale=0.38]{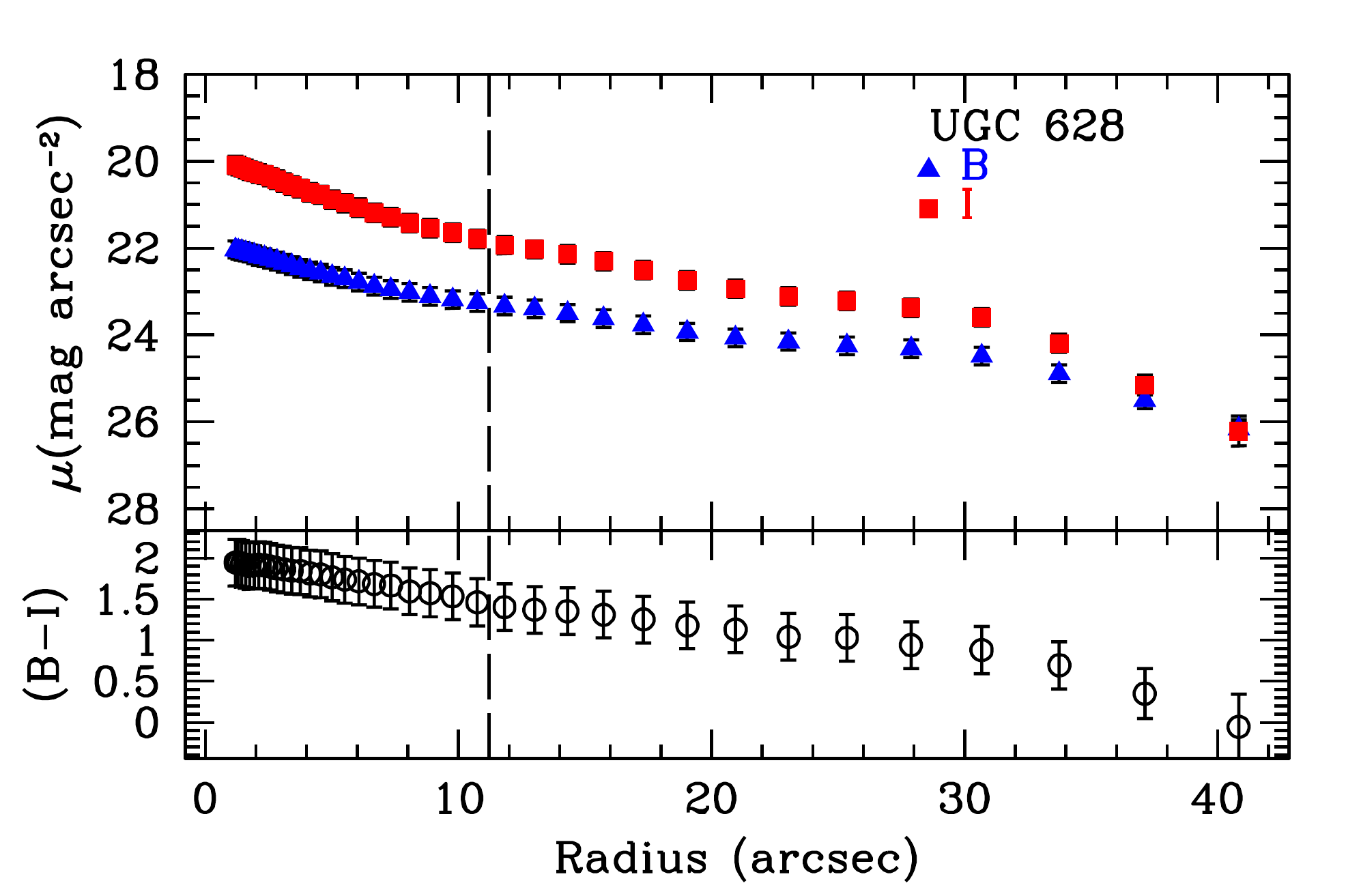} \includegraphics[scale=0.38]{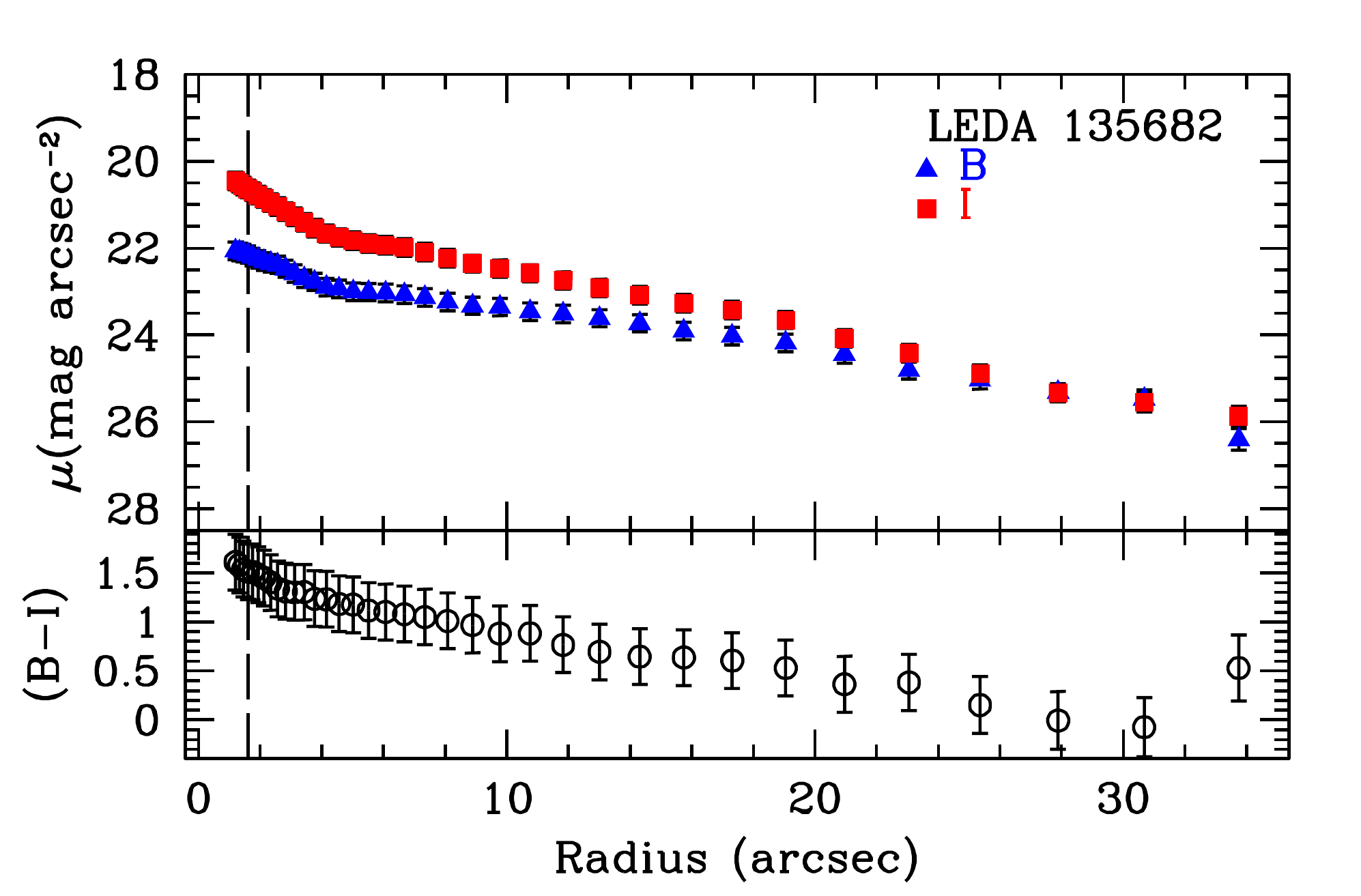} 
  \\
  \includegraphics[scale=0.38]{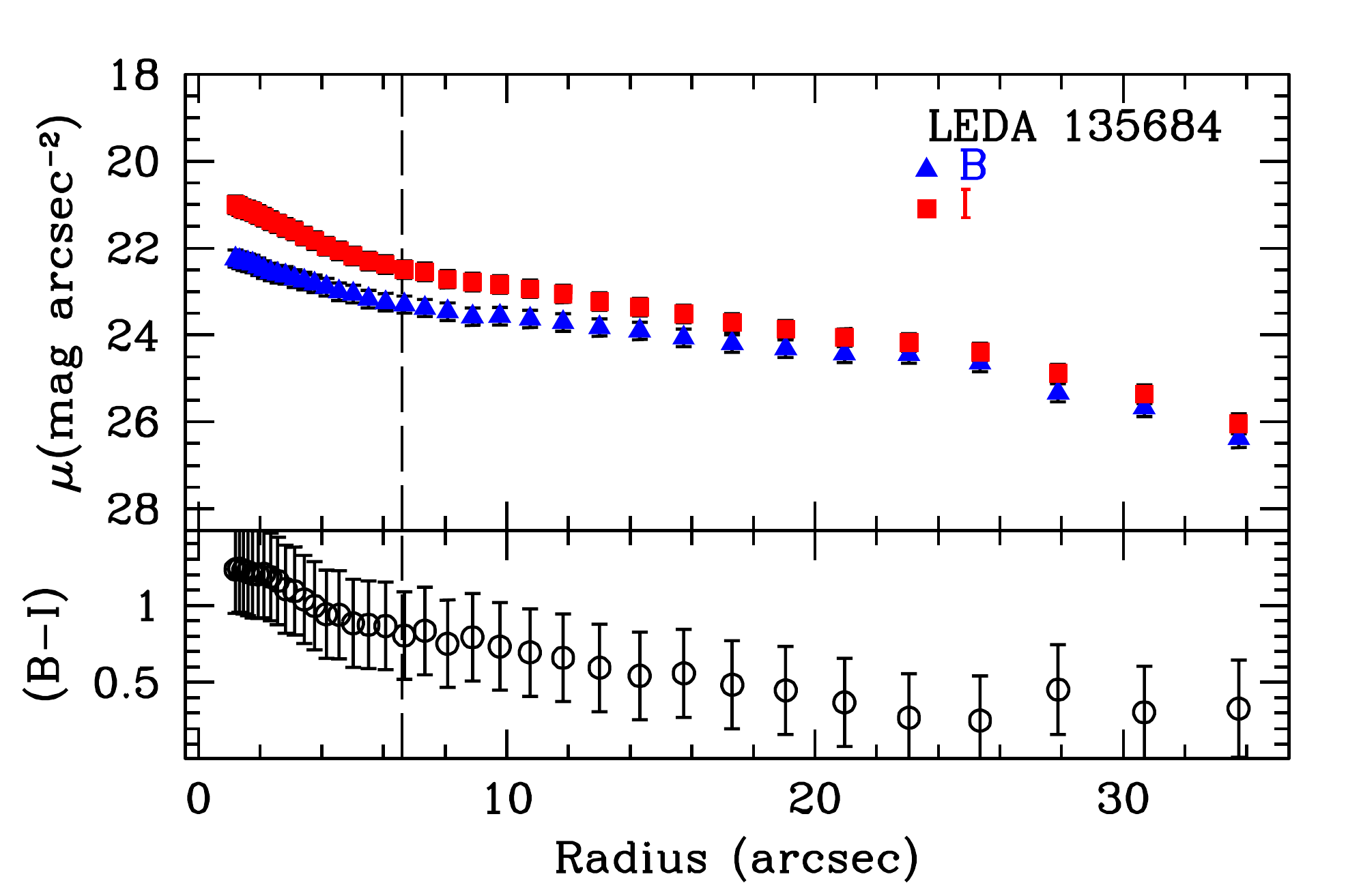} \includegraphics[scale=0.38]{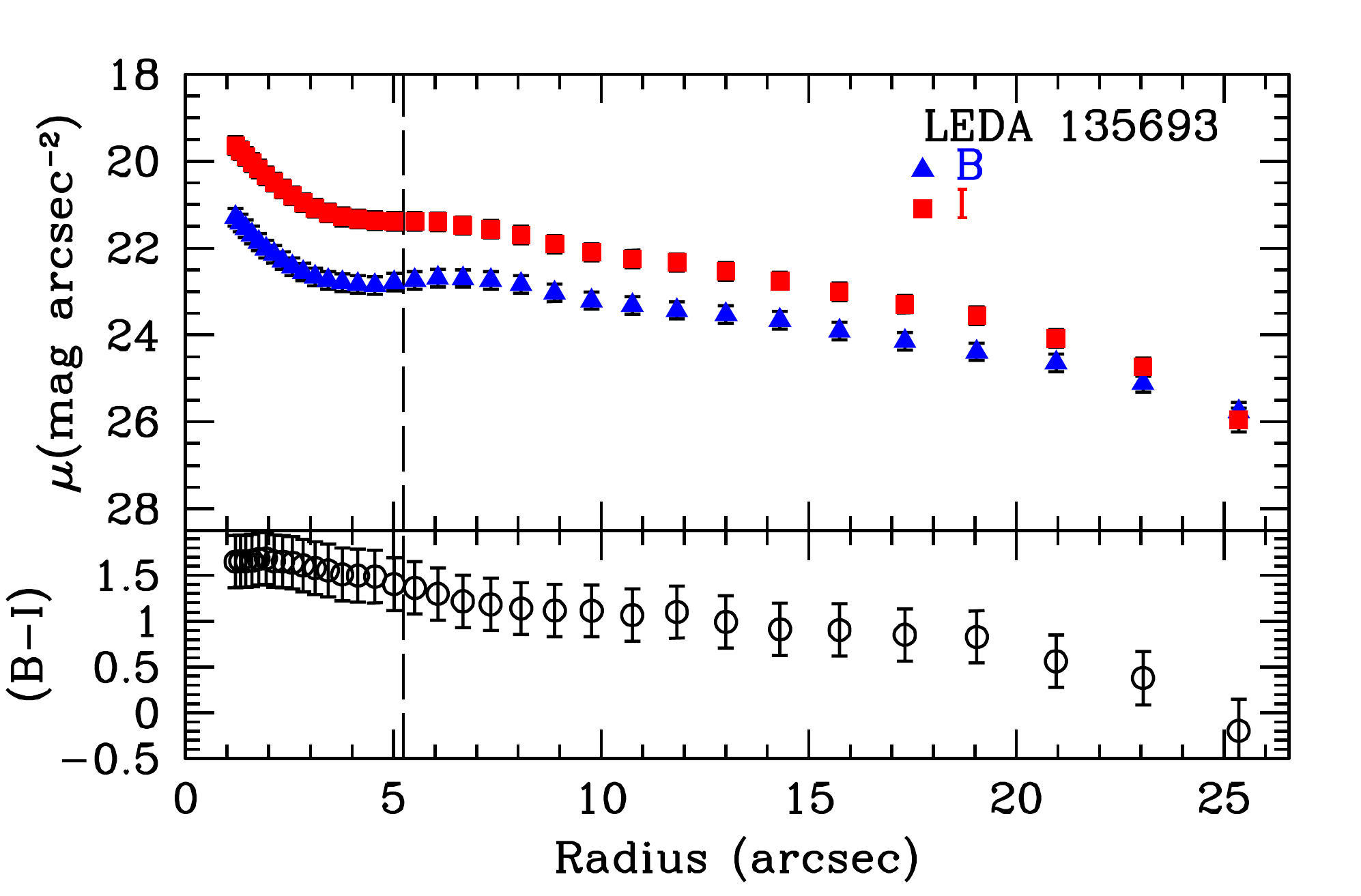} 
  \\
  \includegraphics[scale=0.38]{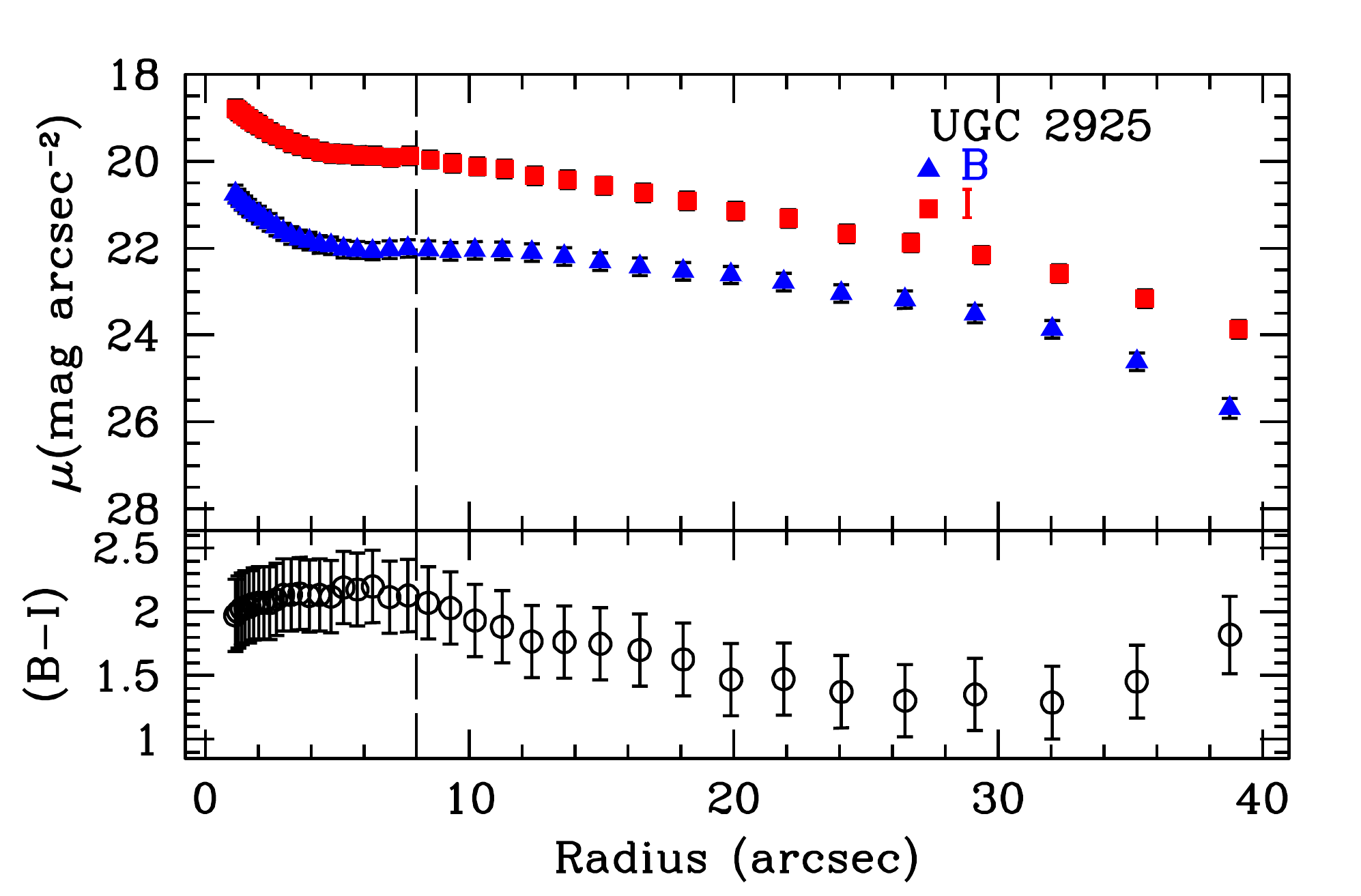} \includegraphics[scale=0.38]{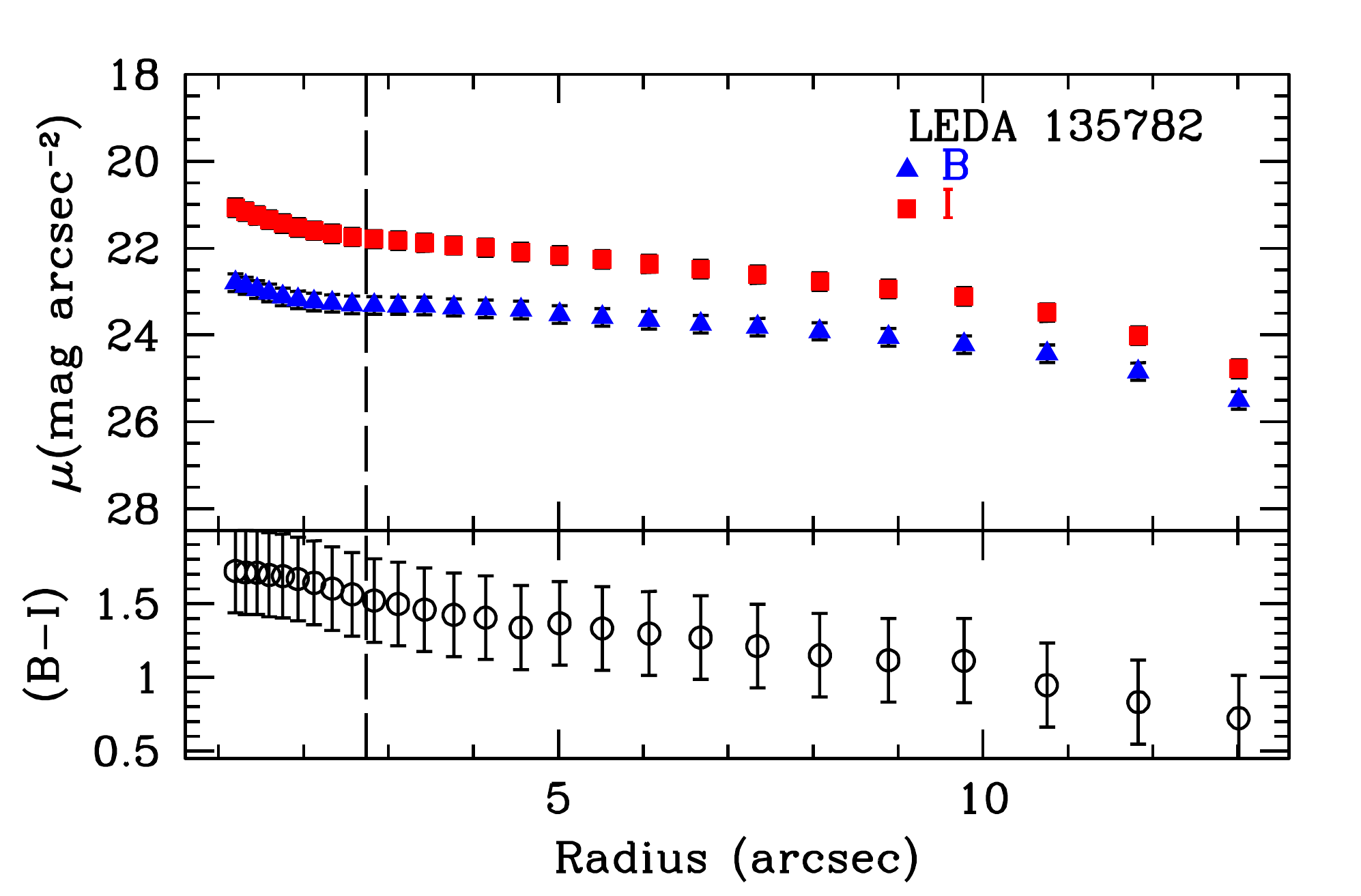} 
  \\
  \includegraphics[scale=0.38]{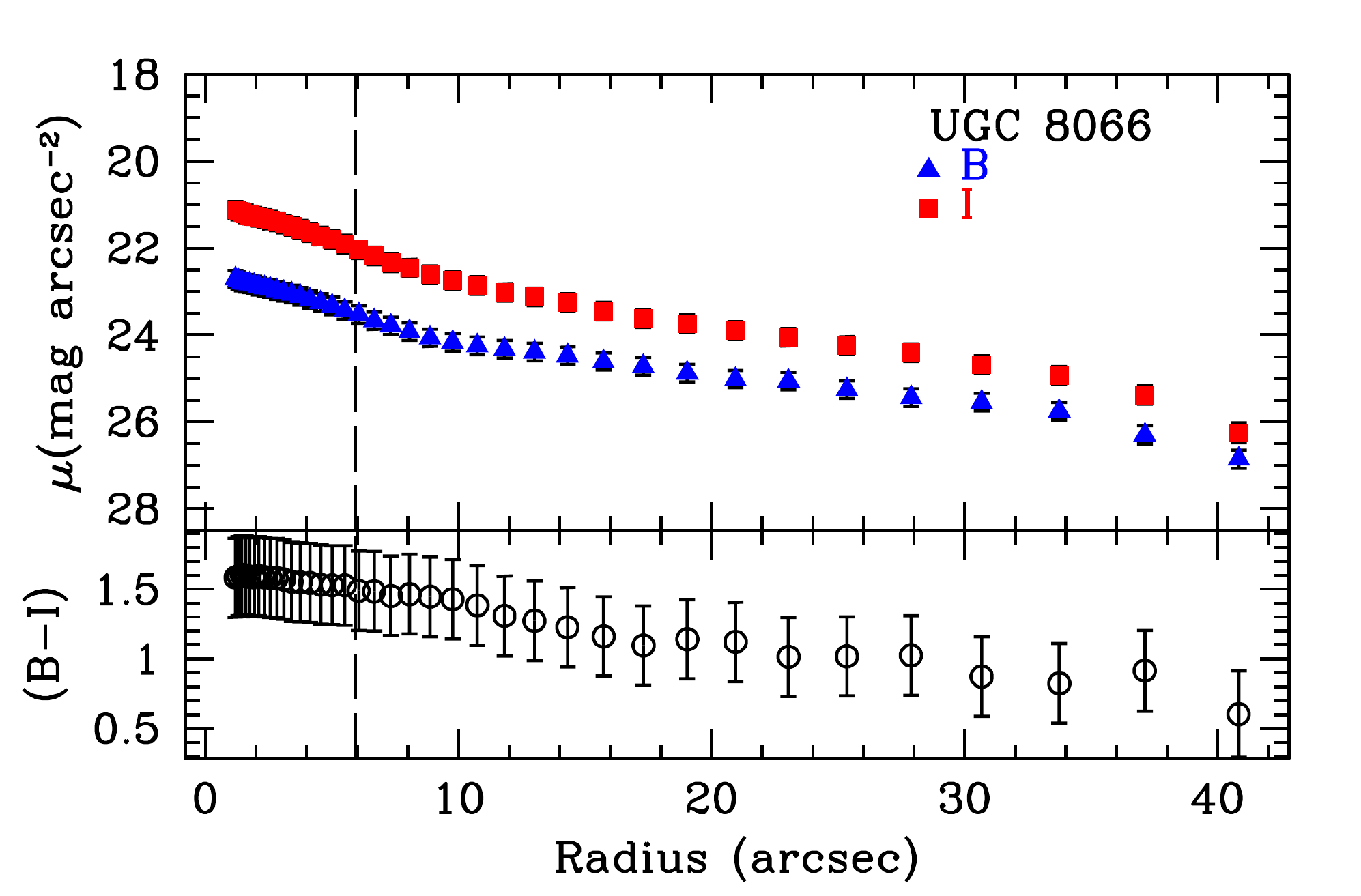} \includegraphics[scale=0.38]{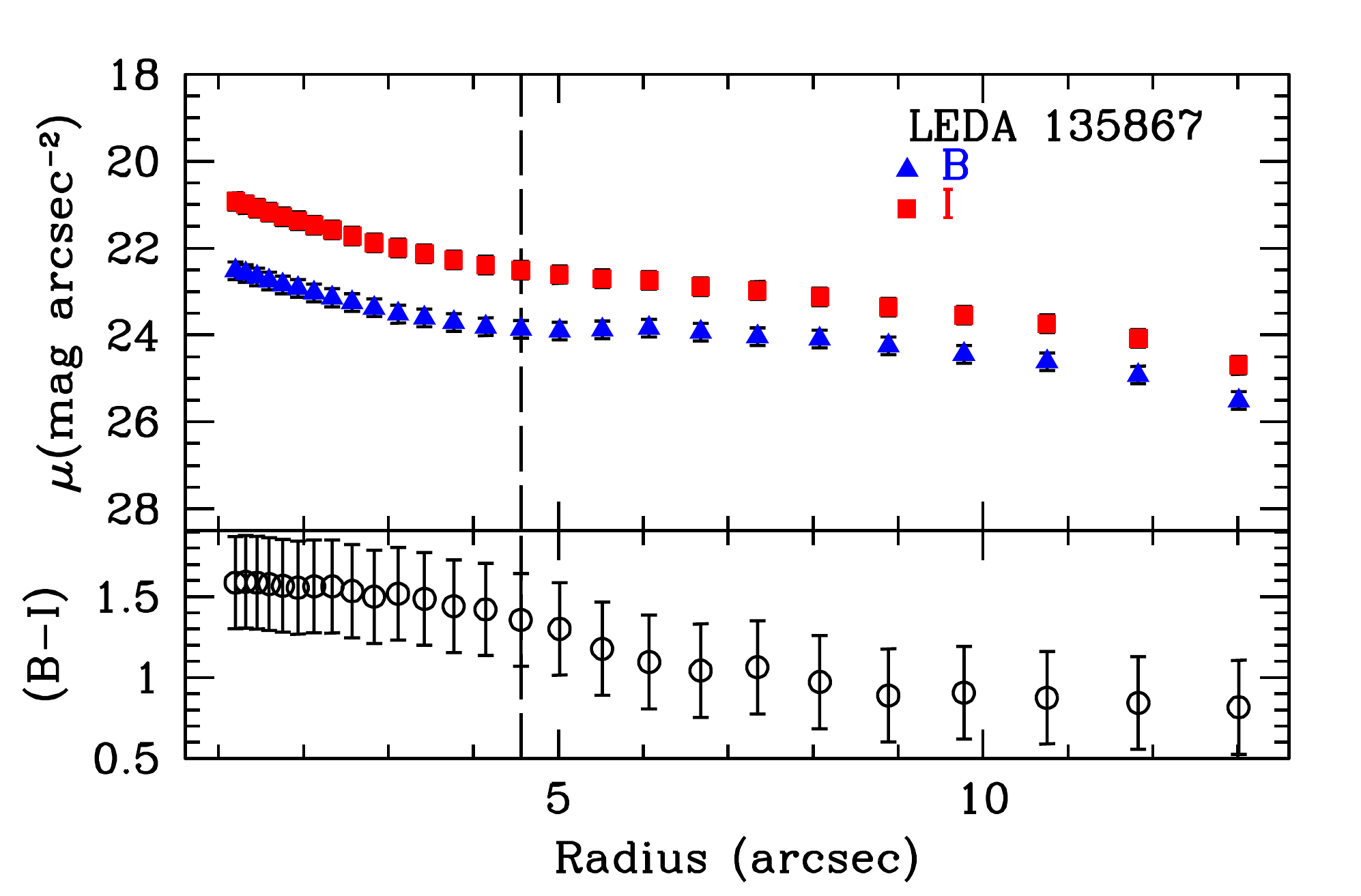} 
  \caption{\textit{B}-band (blue triangles) and \textit{I}-band (red squares) surface brightness profiles (top panels), and ($B-I$) radial colour profiles (bottom panels). The vertical dashed lines indicate the bar length from Table~\ref{barprops}.}
  \label{surfBright}
\end{figure*}

\addtocounter{figure}{-1}

\begin{figure*}
  \centering
  \includegraphics[scale=0.38]{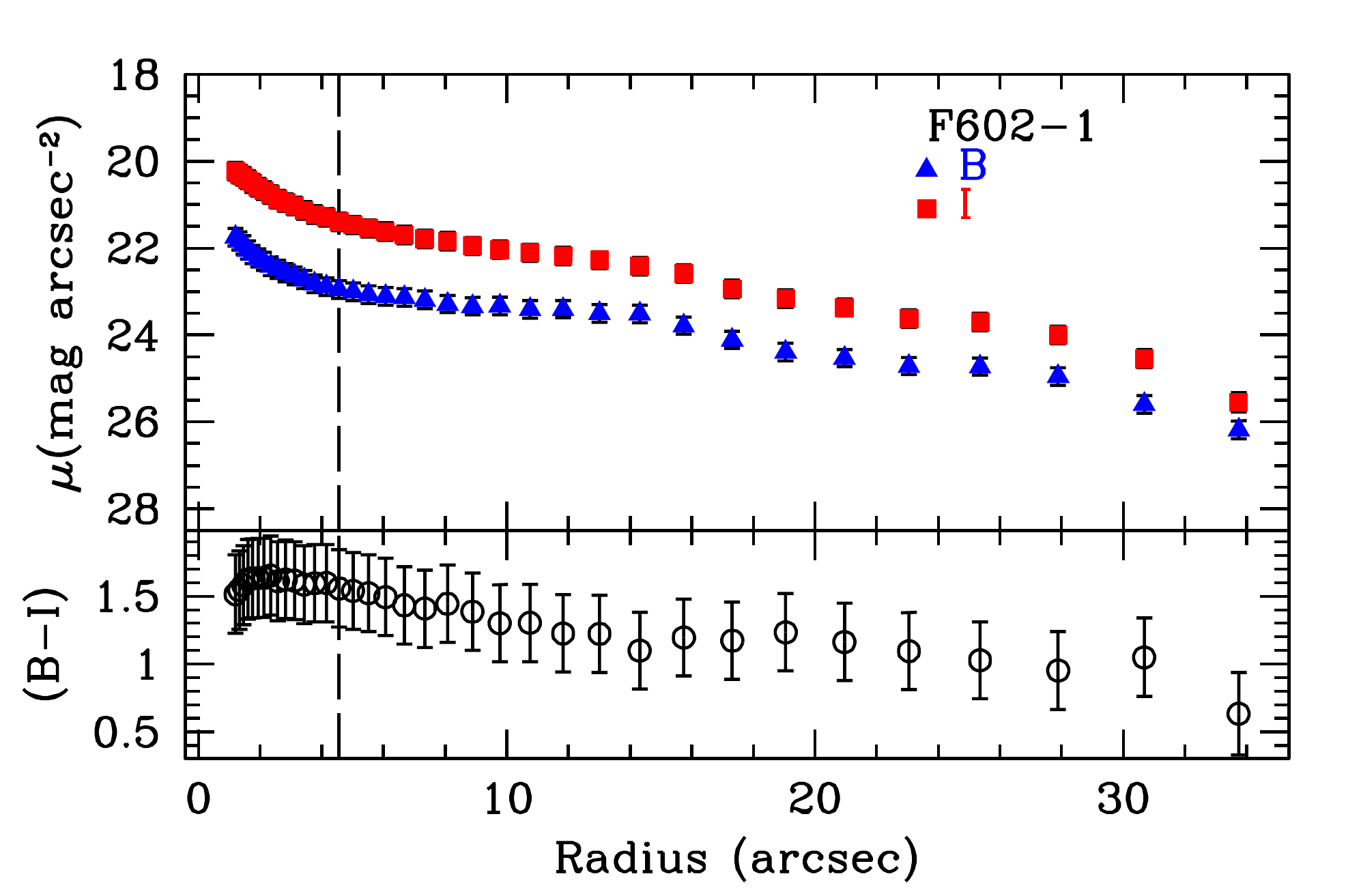} \includegraphics[scale=0.38]{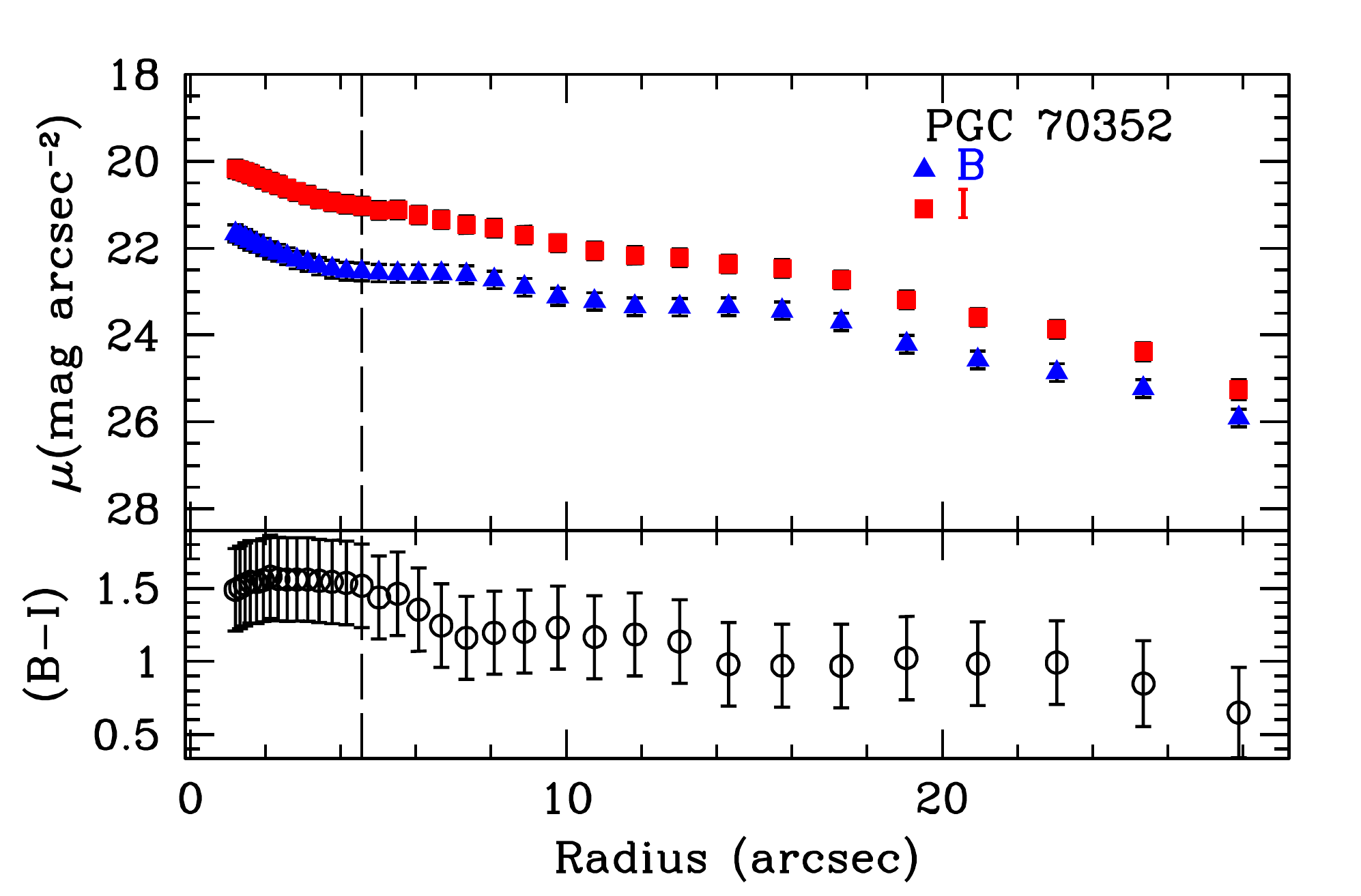}
  \\
  \includegraphics[scale=0.38]{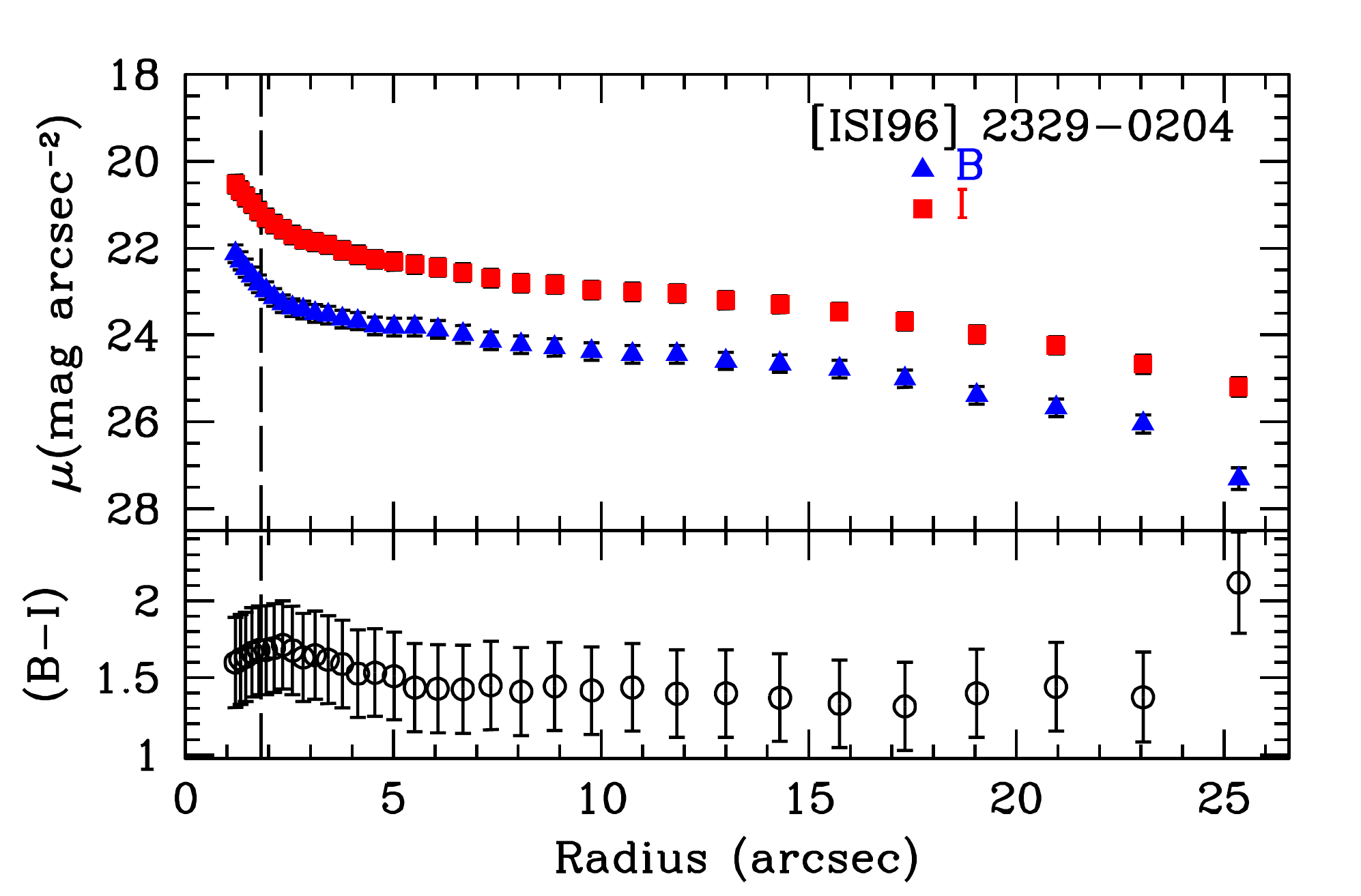} \includegraphics[scale=0.38]{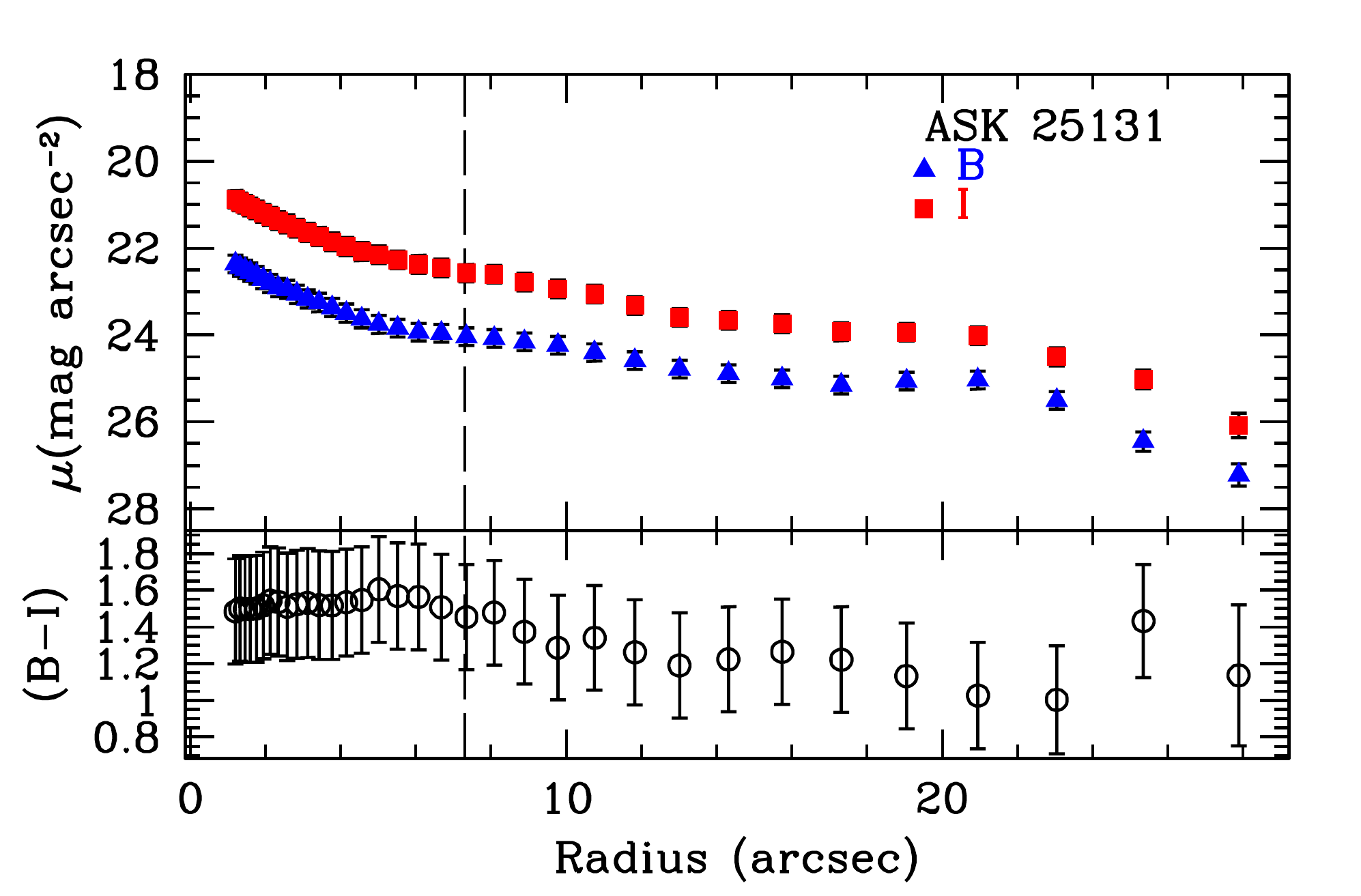}
  \caption{\textit{continued}}
\end{figure*}

\begin{figure}
  \centering
  \includegraphics[scale=0.38]{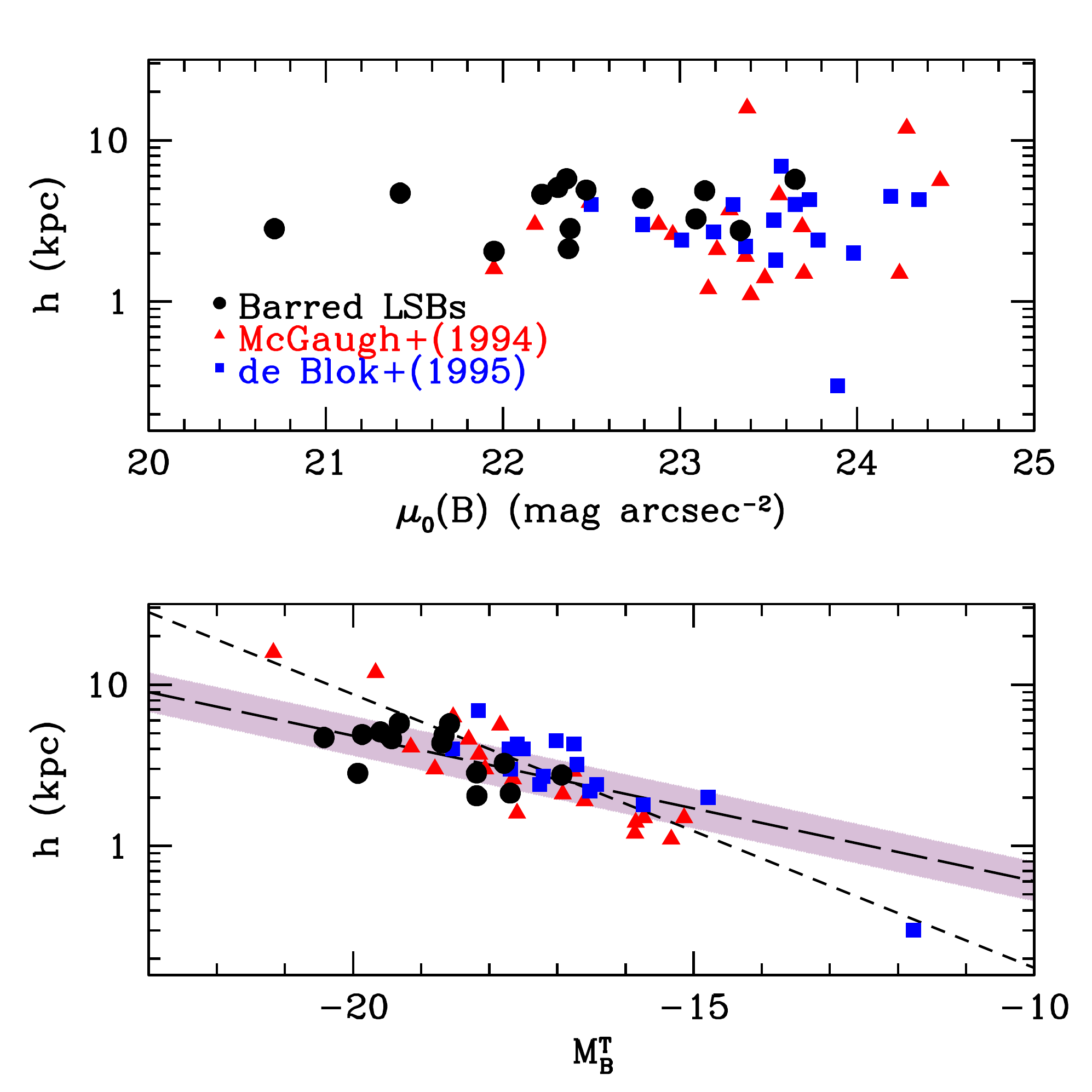}
  \caption{Comparison of our disc scale length (kpc) and $B$-band central surface brightness ($\mu_{0}(B)$) and absolute $B$-band magnitude (M$_{B}^{T}$): our data (black circles), \citet{mcgaugh1994} (red triangles), and \citet{deblok1995} (blue squares). The galaxies plotted here from \citet{mcgaugh1994} and \citet{deblok1995} are unbarred. The long dash line in the bottom panel shows the fit $\log{h} = -0.09M_{B} - 1.44$. The gray shaded region denotes a scatter of $\sigma = 0.13$\ about the fit. The short dashed line shows the fit to the unbarred LSBs from the red and blue points.}
  \label{scalePlot}
\end{figure}

\subsubsection{Magnitudes}
\label{sssec:magnitudes}

The mean absolute $B$-band and $I$-band magnitudes are $-18.89$\ and $-20.01$\ respectively. We find a relation between absolute magnitude and disc scale length, shown as the longer dashed line in the bottom panel of Fig.~\ref{scalePlot}. The relation is given by $\log{h} = -0.09M_{B} - 1.44$, with a scatter of $\sigma = 0.13$ (shown as the shaded gray region in Fig.~\ref{scalePlot}). This is slightly shallower than the relation for the general LSB population in \citet{zhong2008}. We also fit for the unbarred LSBs in Fig.~\ref{scalePlot} from \citet{mcgaugh1994} and \citet{deblok1995} and found the relation $\log{h} = -0.17M_{B} - 2.45$, shown as the shorter dashed line, indicating a significantly different slope between the two LSB populations.

\begin{figure}
  \centering
  \includegraphics[scale=0.38]{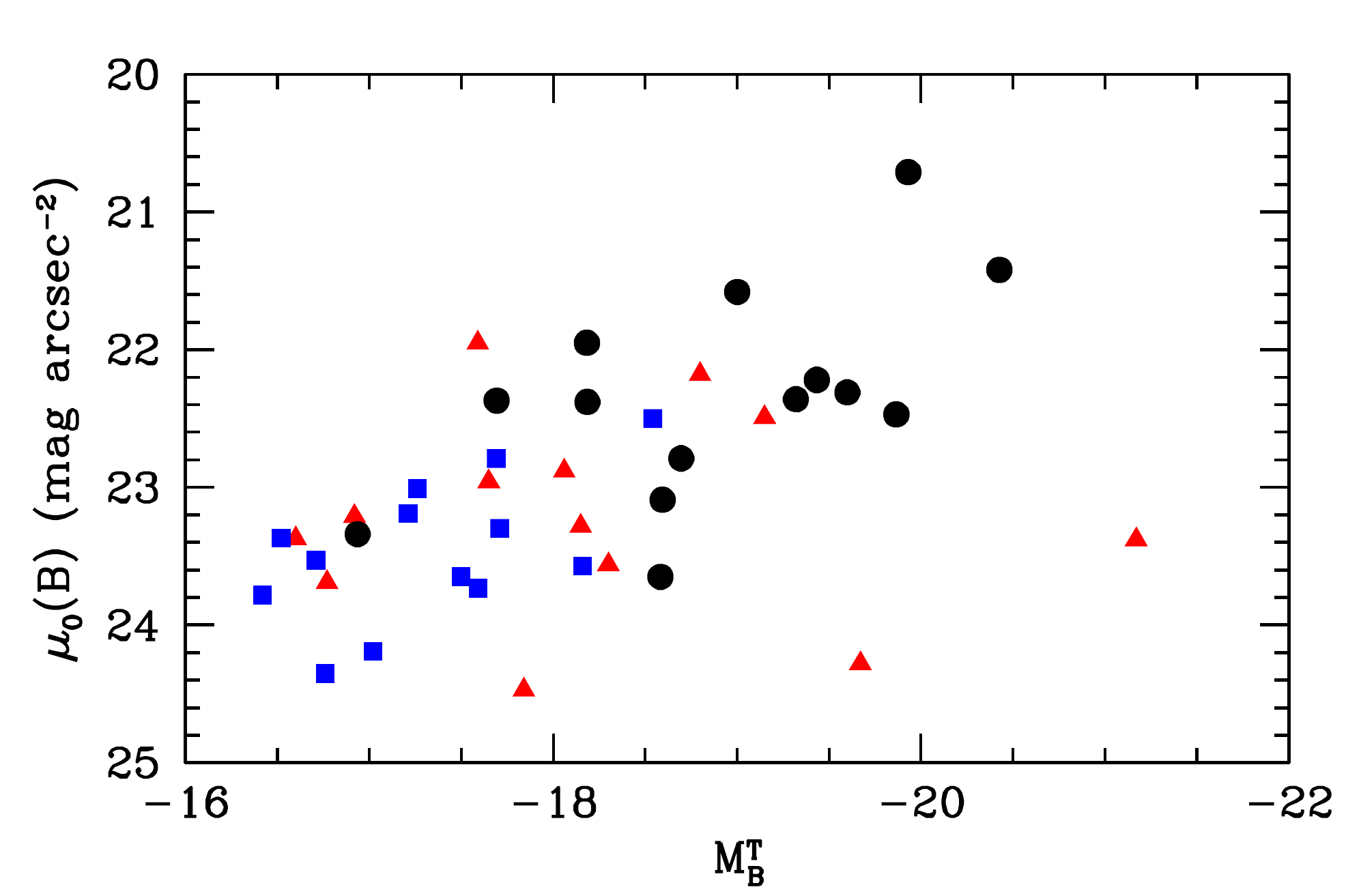}
  \caption{Comparison of \textit{B}-band central surface brightnesses ($\mu_{0}(B)$) and absolute magnitudes (M$_{B}^{T}$): same color scheme as in Fig.~\ref{scalePlot}.}
  \label{magComp}
\end{figure}

In the Fig.~\ref{magComp} we show a comparison of the \textit{B}-band central surface brightness ($\mu_{0}(B)$) and absolute magnitude ($M_{B}^{T}$) for our barred sample (black circles) with the unbarred LSBs in \citet{mcgaugh1994} (red triangles) and \citet{deblok1995} (blue squares). We find that our barred LSBs are noticeably brighter than their unbarred counterparts, extending off of the trend of the unbarred LSBs.

\subsubsection{Colours}
\label{sssec:colours}

Here we discuss our radial colour profiles, three different total colour measures, and our colour maps.

\paragraph{Radial Profiles}
\label{ssssec:colorProfs}

We show the radial ($B-I$) profiles in the bottom panels of Fig.~\ref{surfBright}. We can see that our galaxies are rather blue, with ($B-I$) values of roughly 1 in the disc region outside the bar. In addition, we find that most colour profiles are more red and constant within the bar region, consistent with a primarily stellar feature. All show a clear trend towards bluer values with increasing radius, as expected for LSBs \citep{deblok1995}.

\paragraph{Total Colours}
\label{ssssec:totColors}

Our three total colour measures (see Sec~\ref{sssec:methodsColors}) are shown in Table~\ref{colors}. Color data for F563-V2 are taken from \citet{mcgaugh1994} and data for F568-1 and F568-3 are taken from \citet{deblok1995} and \citet{mcgaugh1997} respectively. Average colours are also listed at the bottom of the table.

We find our bar colours (average value of 1.54$\pm$0.20) are comparable to the nuclear colours found in \citet{mcgaugh1994} (average value of 1.52) and \citet{deblok1995} (average value of 1.47). In these works, the nuclear region was defined to be the color within a 5$\arcsec$\ aperature for all galaxies. Here, our bar color is the color within the bar region for each galaxy, which is dependent on each individual galaxy. 

Our disc colours are noticeably bluer than the bar colours, with an average value of 1.01$\pm$0.31. We find that our area colour is often very close to the value of the disc colour, with an average value of 1.11$\pm$0.31. Since the bars in our sample are quite small, it is not surprising that the area colours are heavily weighted towards the larger area of the bluer disc. 

\begin{table}
  \centering
  \caption{Bar, disc, and area ($B-I$) colours. Data for F563-V2 are taken from \citet{mcgaugh1994}, and data for F568-1 and F568-3 are taken from \citet{deblok1995} and \citet{mcgaugh1997}. Average colours are listed at the bottom of the table.}
  \label{colors}
  \begin{tabular}{lccc}
    \hline
    Galaxy                  & bar  &  disc    &  area \\
    \hline
    UGC 628                 & 1.49 &   0.98   &  1.08 \\
    LEDA 135682             & 1.61 &   0.54   &  0.55 \\
    LEDA 135684             & 0.93 &   0.45   &  0.49 \\
    LEDA 135693             & 1.52 &   0.86   &  0.90 \\
    UGC 2925                & 1.94 &   1.53   &  1.55 \\
    F563-V2                 & 1.64 & $\dotso$ &  1.57 \\
    F568-1                  & 1.57 & $\dotso$ &  1.32 \\
    F568-3                  & 1.58 & $\dotso$ &  1.29 \\
    LEDA 135782             & 1.61 &   1.05   &  1.09 \\
    UGC 8066                & 1.52 &   1.04   &  1.06 \\
    LEDA 135867             & 1.46 &   0.87   &  0.96 \\
    F602-1                  & 1.57 &   1.13   &  1.14 \\
    PGC 70352               & 1.57 &   1.03   &  1.05 \\
    ASK 25131               & 1.53 &   1.20   &  1.24 \\
    $[$ISI96$]$ 2329-0204   & 1.64 &   1.41   &  1.42 \\
    \hline
    averages                & 1.54$\pm$0.20 &   1.01$\pm$0.31  &  1.11$\pm$0.31 \\
    \hline
  \end{tabular}
\end{table}

\paragraph{Colour Maps}
\label{ssssec:colorMaps}

\begin{figure*}
  \includegraphics[scale=0.28]{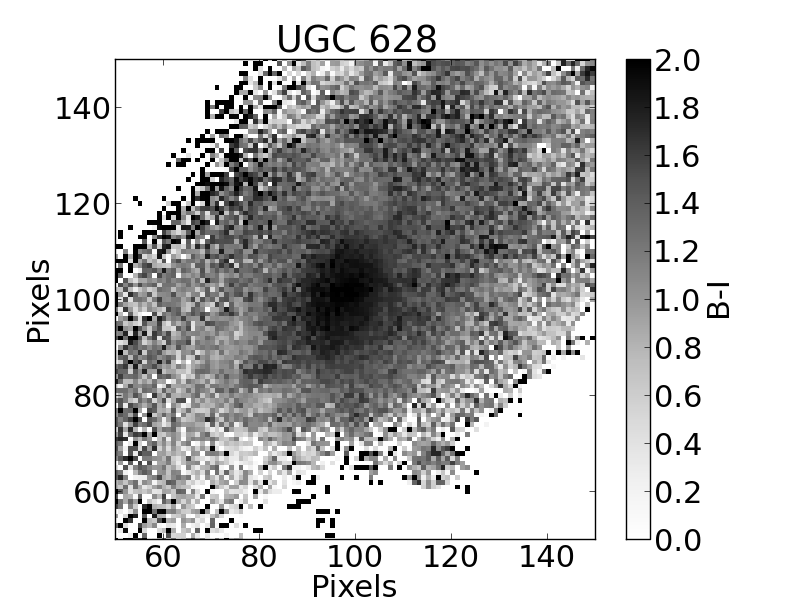} \includegraphics[scale=0.28]{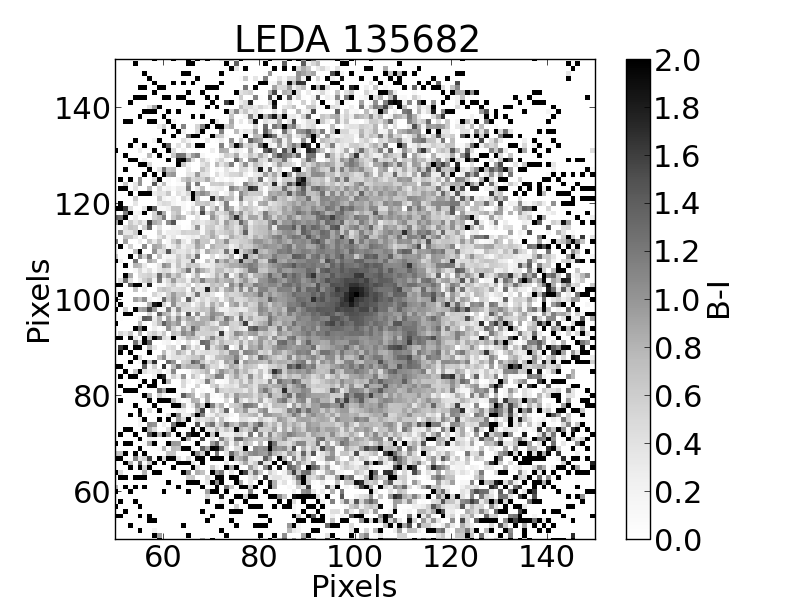} \includegraphics[scale=0.28]{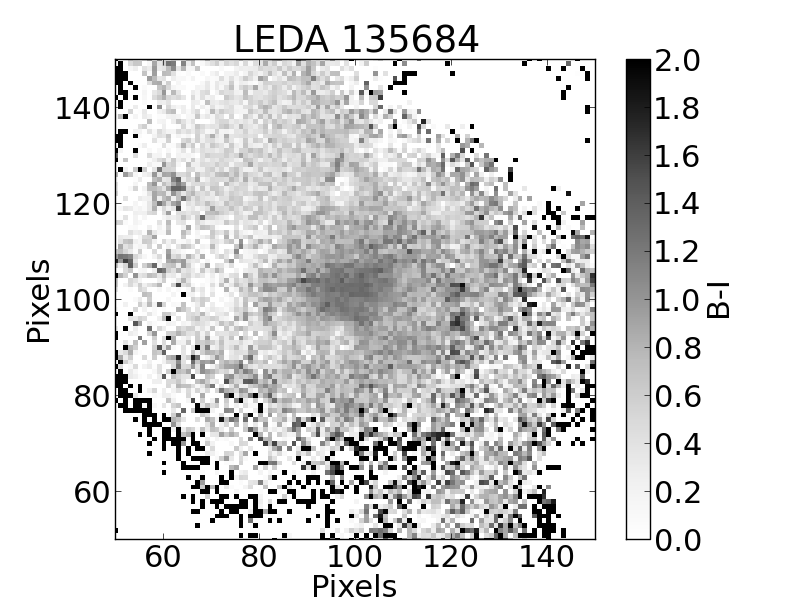} \\
  \includegraphics[scale=0.28]{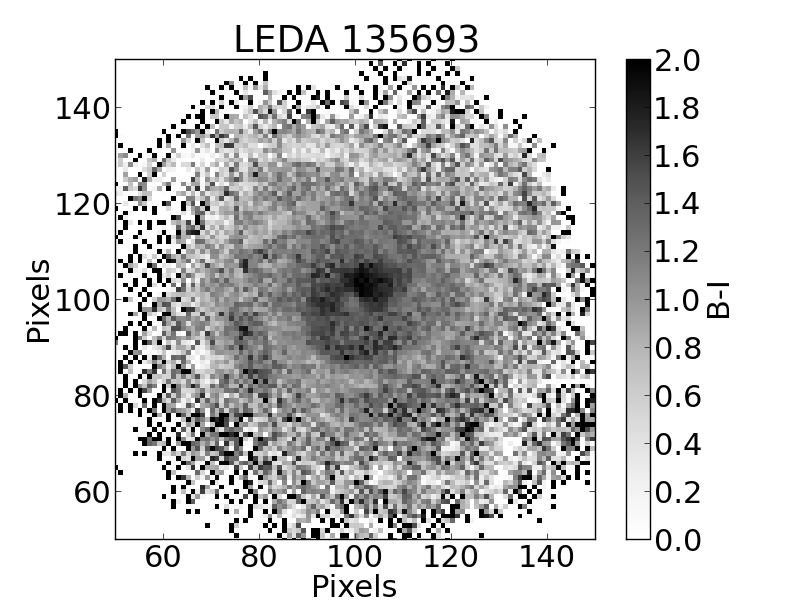} \includegraphics[scale=0.28]{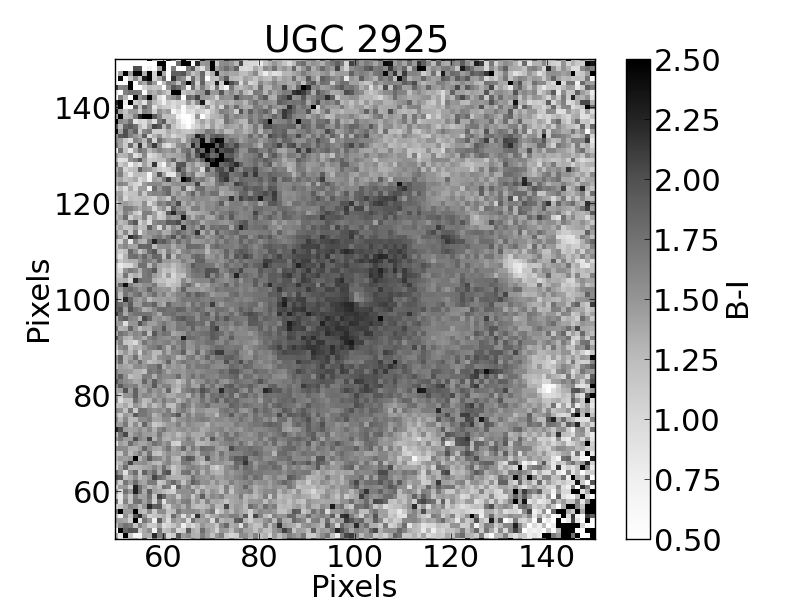} \includegraphics[scale=0.28]{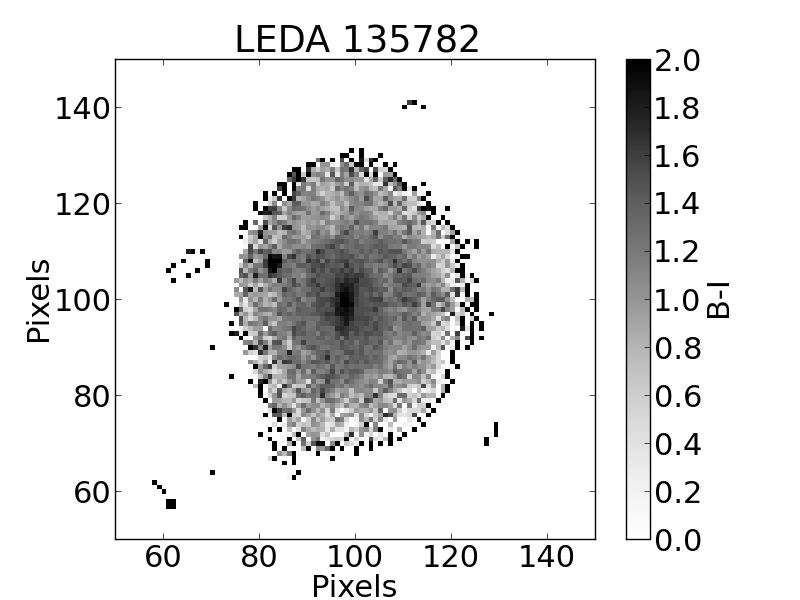} \\
  \includegraphics[scale=0.28]{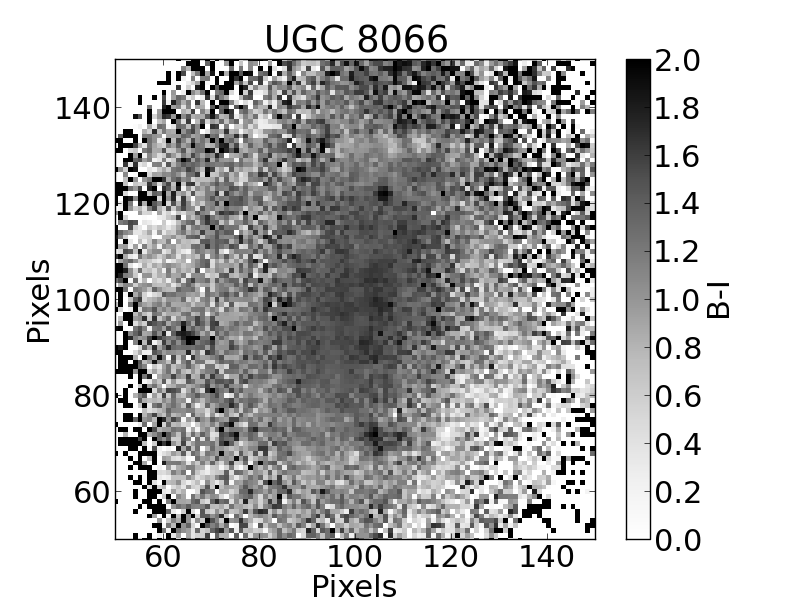} \includegraphics[scale=0.28]{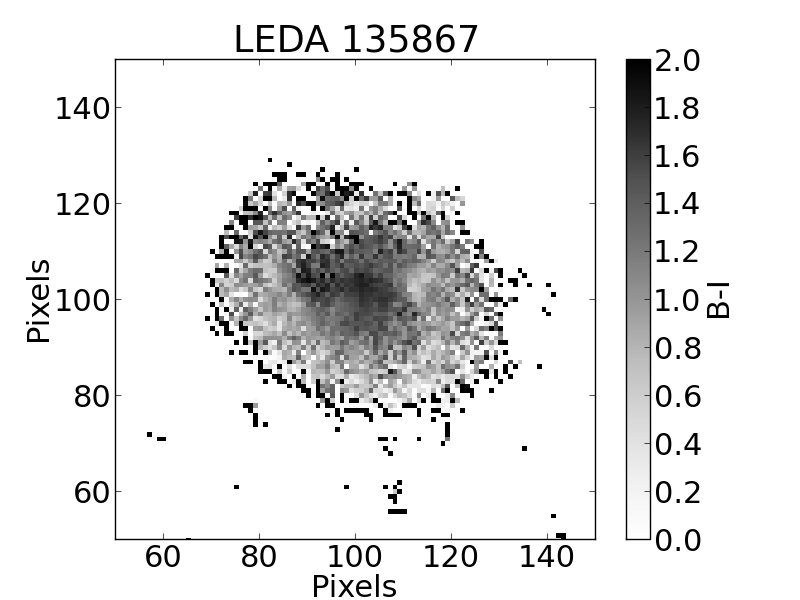} \includegraphics[scale=0.28]{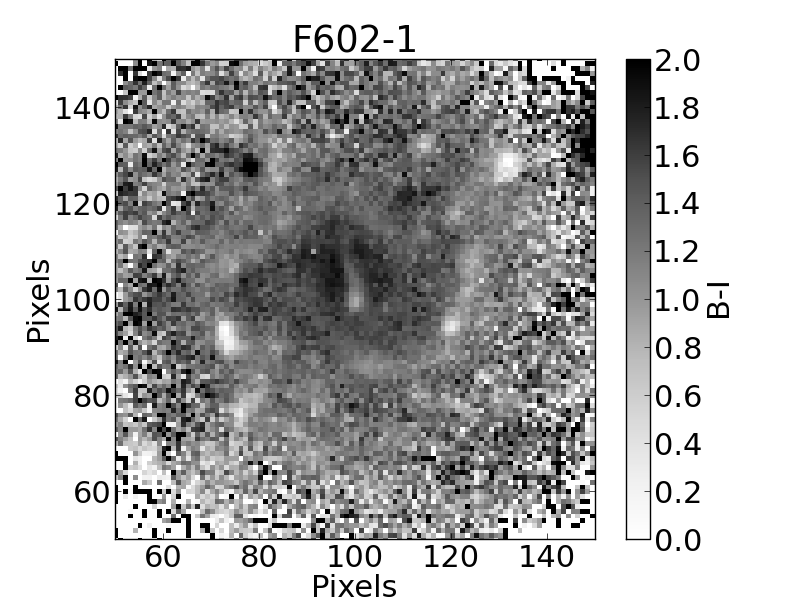} \\
  \includegraphics[scale=0.28]{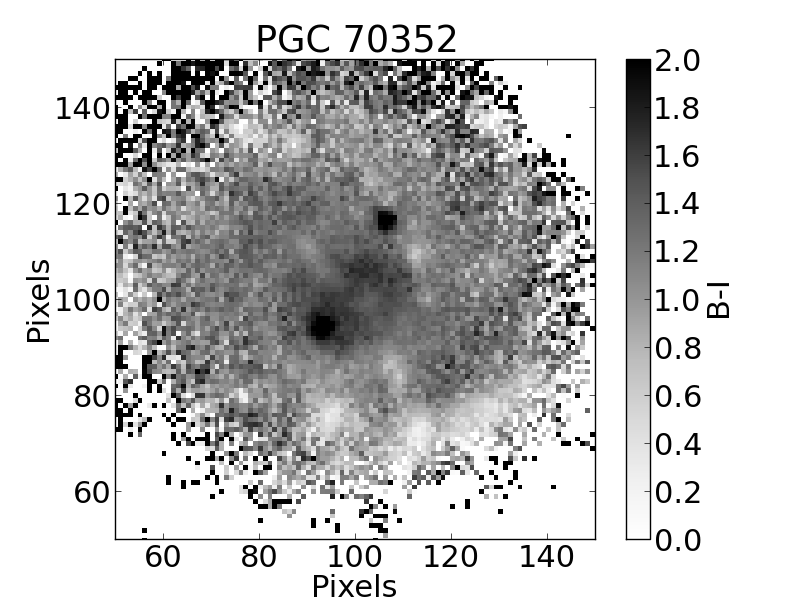} \includegraphics[scale=0.28]{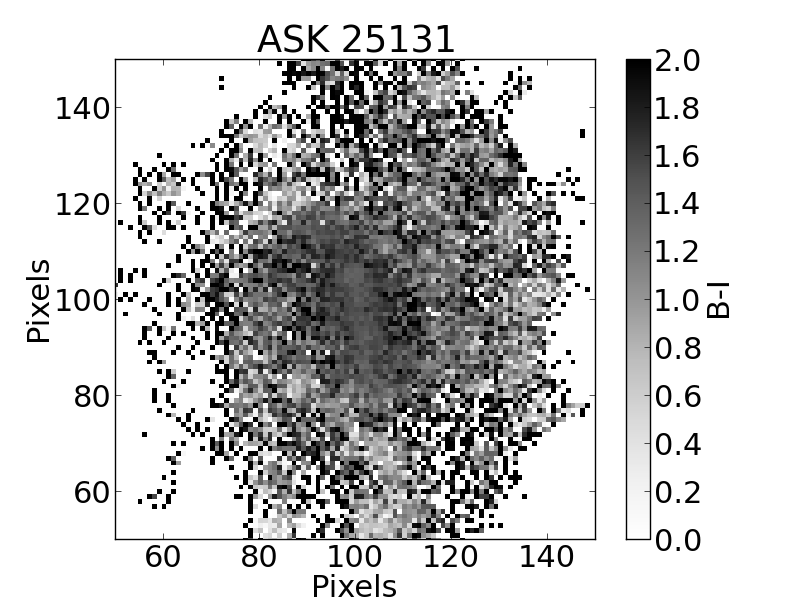} \includegraphics[scale=0.28]{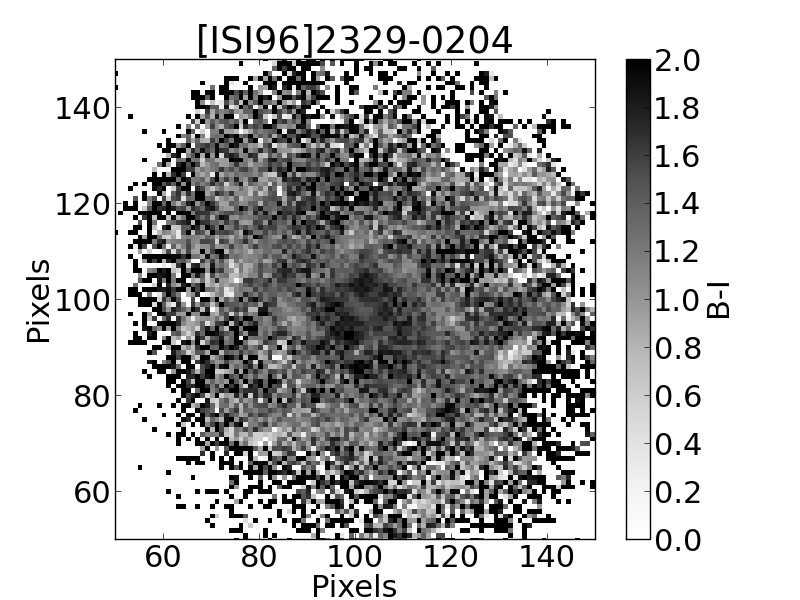} 
  \caption{($B-I$) maps of our barred LSBs, created from 2x2 rebinned $B$- and $I$-band images. North is up, and east is to the left in all maps, and each map is 0.84$\arcmin \times$0.84$\arcmin$. The two black dots in the map for PGC 70352 (bottom left) are two very bright stars in the $I$-band image that were not masked out to prevent distorting the bar. All maps use the same colourbar, with the exception of UGC 2925.}
  \label{colorImages}
\end{figure*}

We show the ($B-I$) maps of the galaxies in our sample with calibrated photometry used to determine the total colours in Sec.~\ref{ssssec:totColors} in Fig.~\ref{colorImages}. Here, we use a colour bar ranging $0 < (B - I) < 2$\ (with the exception of UGC 2925). Spiral arms are present in these maps as the slightly bluer (whiter) band in some of our maps (LEDA 135693 and F602-1, for example). In addition, a few galaxies show HII regions as white blobs in spiral arms (PGC 70352 and $[$ISI96$]$2329-0204, for example). The bars in our galaxies are clear as the redder (darker) regions in the centres of the maps.

All of our galaxies show a stark contrast between the bar and disc regions, with an almost immediate shift from redder colours to much bluer, consistent with our radial colour plots and total colours. For those galaxies with more tenuous disc structure (LEDA 135867 or ASK 25131, for example), the colour maps are not as clear or defined, quickly getting lost in the noise. 

Interestingly, some galaxies show bluer regions within the bar: noteably F602-1, ASK 25131, and PGC 70352. This is not a result of the images being misaligned when combining, but instead a real feature. When examining the $B$- and $I$-band images individually (Fig.~\ref{galimages}), it is noticeable that the bar in the $B$-band is noticeably different than in the $I$-band for these galaxies. Most noticeable is F602-1, which has a very narrow bar in $B$\ compared with $I$. 

\section{Gas Fraction}
\label{sec:gasFraction}

We have shown that bars in LSBs are characteristically shorter and weaker than those in HSBs (Sec.~\ref{ssec:barDiscuss}). It is thought that high gas content in galaxies prohibits bar formation, as well as forms shorter and weaker bars \citep[e.g.][]{mayer2004, cervantes2017b}. Therefore, we wish to examine the gas fractions of the galaxies in our sample in order to see if this explains the short and weak nature of the bars found.

To obtain gas fractions of our sample, we require estimates of both the gas and stellar mass. We use the \texttt{HyperLeda} online data base\footnote{http://leda.univ-lyon1.fr/} \citep{makarov2014} to obtain 21cm HI fluxes ($F_{\mathrm{HI}}$). Out of our whole sample, ten galaxies have published HI magnitudes, or $m_{21}$, which we convert into fluxes. We then convert these fluxes to HI masses via the relation from \citet{haynes1984}:

\begin{equation}
  M_{\mathrm{HI}} = 2.36 \times 10^{5} D^{2} F_{\mathrm{HI}}\ (M_{\odot}),
\end{equation}
where D is the distance in Mpc (Table~\ref{sample}). Although LSBs are thought to generally contain very little molecular gas \citep{mihos1999}, we obtain the gas mass by $M_{gas}=1.4M_{\mathrm{HI}}$.

With our $I$-band magnitudes and ($B-I$) colours it is possible to obtain a stellar mass via a mass-to-light ratio, which requires using stellar population models. We first use our $I$-band absolute magnitudes (Table~\ref{photometry}) to determine the $I$-band luminosity for each galaxy. We then use the ($B-I$) area colour (Table~\ref{colors}) to obtain an $I$-band stellar mass-to-light ratio using the appropriate equation from \citet{into2013}:

\begin{equation}
  \log{\Upsilon_{*}^{I}} = -0.997 + 0.641(B-I)
\end{equation}
Because we do not have an $I$-band absolute magnitude for F563-V2, we take the apparent magnitude from \texttt{HyperLeda}, and use the distance reported in Table~\ref{sample}. Finally, we obtain gas fractions via
\begin{equation}
  f_{gas} = \frac{M_{gas}}{M_{gas} + M_{*}}
\end{equation}

In Table~\ref{masses} we report the HI fluxes ($F_{\mathrm{HI}}$), HI masses ($\log{M_{\mathrm{HI}}}$), $I$-band luminosity ($\log{L_{I}}$), stellar mass-to-light ratios ($\Upsilon_{*}^{I}$), stellar masses ($\log{M_{*}}$) for our sample, and gas fractions ($f_{gas}$).

\begin{table*}
  \centering
  \caption{HI flux ($F_{\mathrm{HI}}$), HI mass ($\log{M_{\mathrm{HI}}}$), $I$-band luminosity ($\log{L_{I}}$), $I$-band mass-to-light ratio ($\Upsilon_{*}^{I}$), stellar mass ($\log{M_{*}}$), and gas fraction ($f_{gas} = M_{gas}/(M_{gas} + M_{*})$) for those galaxies in our sample with available HI magnitudes.}
  \label{masses}
  \begin{tabular}{lcccccc}
    \hline
    Galaxy      & $F_{\mathrm{HI}}$ & $\log{M_{\mathrm{HI}}}$ & $\log{L_{I}}$ & $\Upsilon_{*}^{I}$ & $\log{M_{*}}$  &  $f_{gas}$ \\
                &   (Jy km/s)    &     ($M_{\odot}$)      &  ($L_{\odot}$) &                  &   ($M_{\odot}$) &            \\
    \hline
    UGC 628     &        4.37    &             9.88      &      9.81    &      0.50         &      9.51     &   0.77   \\
    LEDA 135684 &        7.52    &             9.57      &      8.94    &      0.21         &      8.26     &   0.97   \\
    UGC 2925    &        4.66    &             9.56      &     10.28    &      1.01         &     10.28     &   0.21   \\
    F563-V2     &        3.57    &             9.49      &      8.75    &      1.02         &      8.76     &   0.88   \\
    F568-1      &        1.91    &             9.58      &      9.56    &      0.71         &      9.41     &   0.67   \\
    F568-3      &        2.31    &             9.58      &      9.60    &      0.68         &      9.43     &   0.66   \\
    LEDA 135782 &        1.45    &            10.17      &      9.87    &      0.50         &      9.57     &   0.85   \\
    UGC 8066    &        4.06    &             9.20      &      8.83    &      0.54         &      8.56     &   0.86   \\
    F602-1      &        1.94    &             9.70      &     10.03    &      0.54         &      9.76     &   0.55   \\
    PGC 70352   &        3.38    &             9.63      &      9.68    &      0.48         &      9.36     &   0.72   \\
    \hline
  \end{tabular}
\end{table*}

The mean HI mass for our galaxies is $\log{\left( M_{\mathrm{HI}} / M_{\odot} \right)} = 9.64 \pm 0.25$, consistent with the total LSB spiral population in \citet{honey2018}. The mean stellar mass for our galaxies is $\log{(M_{*})} = 9.29 \pm 0.60$. We find that the majority of our galaxies are gas rich ($f_{gas} > 0.5$), with only one galaxy being gas poor, UGC 2925. Specifically, the majority of our sample have $f_{gas} > 0.65$. This is consistent with  barred LSBs \citep{pahwa2018} and LSBs in general \citep{deblok1996}.

\begin{figure}
  \centering
  \includegraphics[scale=0.38]{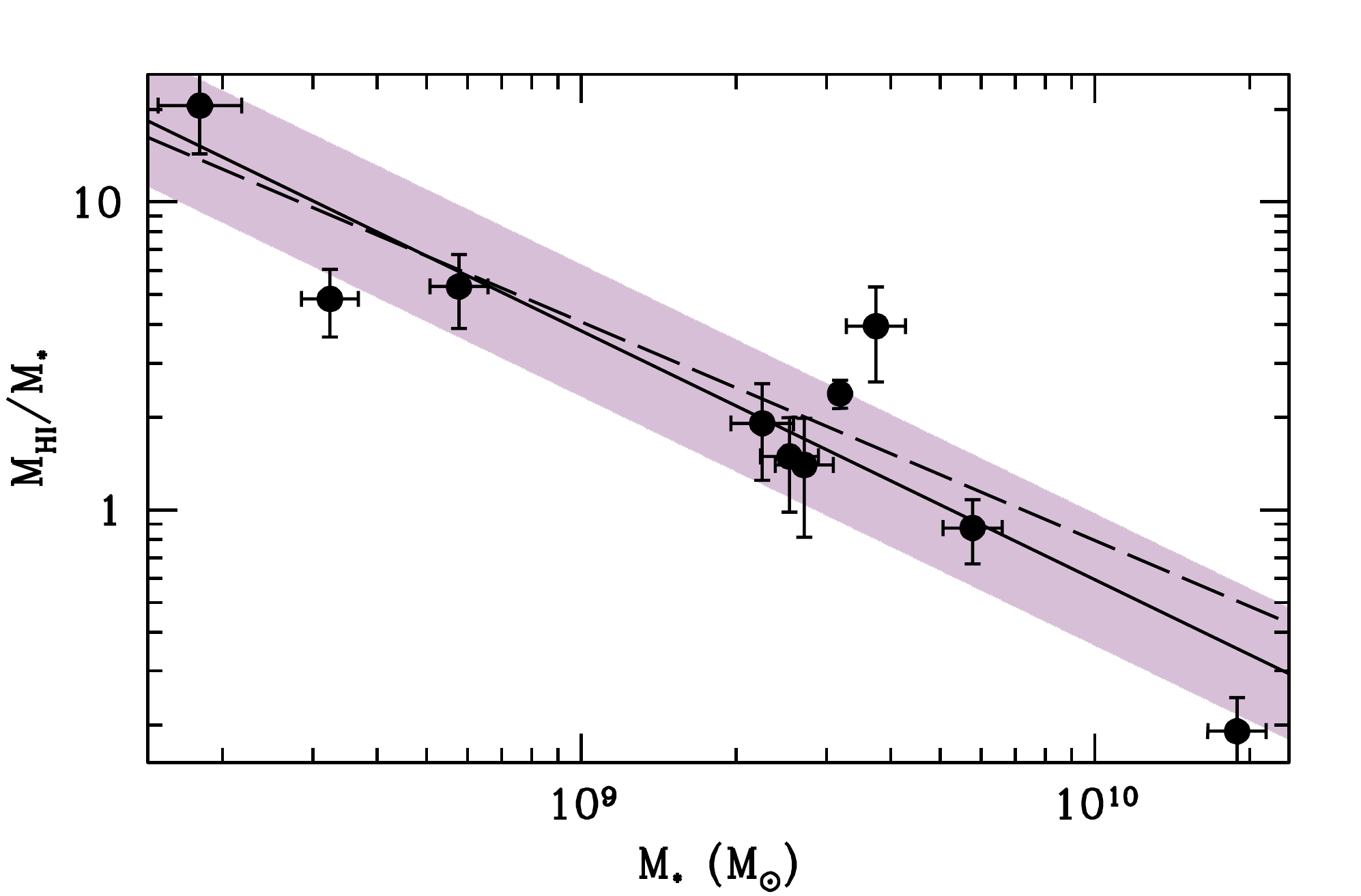}
  \caption{$\log{(M_{\mathrm{HI}}/M_{*})}$\ vs $\log{(M_{*})}$. The solid line denotes the fit $\log{(M_{\mathrm{HI}}/M_{*})} = -0.81\log(M_{*}) + 7.88$, and the shaded region denotes the scatter of 0.22. The dashed line shows the fit to general LSB spirals from \citet{honey2018}, $\log{M_{\mathrm{HI}}/M_{*}} = -0.71\log{M_{*}} + 7$.}
  \label{gas}
\end{figure}

In Fig.~\ref{gas} we show $\log{(M_{\mathrm{HI}}/M_{*})}$\ as a function $\log{(M_{*})}$. The solid line is the fit $\log{(M_{\mathrm{HI}}/M_{*})} = -0.81\log(M_{*}) + 7.88$, slightly steeper than the relation for general LSB spirals in \citet{honey2018}, $\log{M_{\mathrm{HI}}/M_{*}} = -0.71\log{M_{*}} + 7$\ (the dashed line in Fig.~\ref{gas}). The scatter about our relation, $\sigma = 0.22$, is shown as the shaded region in Fig.~\ref{gas}.

\section{Discussion and Conclusions}
\label{sec:discussion}

We have measured the photometric bar properties (length, strength, and corotation radius) of fifteen barred LSB disc galaxies assembled from \citet{schombert1992} and \citet{impey1996} using optical $B$- and $I$-band photometry. We have found that bars in LSBs are shorter and weaker than those in HSBs, consistent with numerical simulations. Our mean bar length and strength are 2.5 kpc and 0.19 respectively. In order to determine the best measure of the bar length, we have created and analyzed fake galaxy images consisting of an exponential disc and a bar. We have found that our azimuthal bar length technique performs best out of the four tested here, and that the $R_{\mathrm{P.A.}}$\ measure, while dependent on the inaccurate $R_{\epsilon}$\ measure, also does an excellent job at measuring the bar length.

Interestingly, we find that the corotation radius of the bar falls very close to the end of the bar for our sample, implying that bars in LSBs are fast rotators. In addition, the only \textit{very} slow bar is located in the only galaxy without any clear spiral structure, F563-V2. Recent mergers are thought to decrease the bar pattern speed \citep{gerin1990}, but LSBs are known to be more isolated than HSBs \citep{bothun1993, mo1994, rosenbaum2004, du2015, honey2018}. In addition, there are no known companions to F563-V2.

As discussed in Paper I, the dark matter halos should dynamically slow down any bar that may be present \citep{weinberg1985, debattista2000}, \textcolor{black}{although this only applies to nonrotating halos (see Sec.~\ref{ssec:barDiscuss} why this is likely not the case for LSBs)}. While there have been reports of slow bars found in dark matter dominated objects \citep[see][]{bureau1999, chemin2009, banerjee2013}, our results imply that bars in LSBs are generally fast. The reason for this remains unclear, and requires further detailed spectroscopic analysis in order to obtain bar pattern speeds. However, the faint nature and rough discs of LSBs makes application of the only direct bar pattern speed measurement \citep{tremaine1984} prohibatively difficult.

\textcolor{black}{In addition, it is possible that $\mathcal{R}$\ can give misleading information regarding the actual rotational speed of the bar. \citet{font2017} found that by examining the bar pattern speeds ($\Omega_{b}$) of a large sample of galaxies, some bars that were considered fast (i.e. $\mathcal{R} < 1.4$) were actually some of the slowest rotators. In order to determine this, they examined the `normalised' bar pattern speed which they defined as $\Gamma = \Omega_{b} / \Omega_{d}$\ with $\Omega_{d}$\ being the disc pattern speed. By setting a deliminator of $\Gamma = 2$\ as a means of separating fast and slow rotators, they found that bars with $\mathcal{R} < 1.4$\ could have $\Gamma < 2$, meaning the bar rotates at a comparable rate to the disc. Because of this, the authors conclude that bars have indeed been slowed down due to the dark matter halos, despite the prevalance of bars with $\mathcal{R} < 1.4$, and suggest $\Gamma$\ as a better means of separating fast and slow rotators.}

The galaxies in our sample have only been shallowly observed before. We have therefore presented $B$- and $I$-band surface brightness profiles, magnitudes, and colours of our sample. We have found that barred LSBs are brighter than unbarred LSBs, and fall on a shallower $\log{h}$\ vs $M_{B}$\ relation. The disc scale lengths of our sample are identical to those of unbarred LSBs. Our sample is also quite blue, having a mean ($B-I$) area colour of 1.11, while the bar region is significantly redder, with a mean colour of 1.54. Due to the small bars in our galaxies, it is not surprising that our area colours are heavily weighted towards the bluer disc.

Finally, we have used available 21cm HI fluxes to determine the HI masses and population synthesis models to get stellar masses of our sample and found our galaxies to be quite gas rich. Our sample falls on the same $\log{(M_{\mathrm{HI}})/M_{*}}$\ vs $\log{(M_{*})}$\ relation as the general LSB spiral population. \textcolor{black}{It is thought that gas rich, low luminosity galaxies with high halo spin $\lambda$\ form short and weak bars \citep{cervantes2013, cervantes2017a}, consistent with our findings here.}

We list our major conclusions based on our observed sample here.

\begin{enumerate}
  \item Bars in LSBs are shorter and weaker than those in HSBs, with an average length and strength of 2.5 kpc and 0.19 respectively.
  \item Bars in LSBs are fast rotators ($\mathcal{R} < 1.4$), with the corotation radius occuring close to the end of the bar. Barred LSBs also show multiple $B$- and $I$-band phase intersections, possibly indicating disc corotation radii.
  \item Barred LSBs are slightly brighter than unbarred LSBs, with average central surface brightnesses of $\mu_{0}(B) = 22.39$\ mag arcsec$^{-2}$\ and $\mu_{0}(I) = 20.28$\ mag arcsec$^{-2}$, and average absolute magnitudes of $M_{B} = -18.89$\ and $M_{I} = -20.01$.
  \item Barred LSBs fall on a slightly shallower $\log{h}$\ vs. $M_{B}$\ relation than unbarred LSBs.
  \item Barred LSBs have HI masses that are nearly identical to the general spiral LSB population, and are just as gas rich as unbarred LSBs ($f_{gas} > 0.5$). 
  \item \textcolor{black}{In order to fully probe the nature of the dark matter halos of barred LSBs, extensive spectroscopy is required in order to obtain bar pattern speeds and halo spin parameters ($\lambda$).}
\end{enumerate}

\section*{Acknowledgements}

\textcolor{black}{The authors would like to thank the anonymous referee for their very helpful comments in assisting with the discussions in this paper.} In addition, would like to thank Russet McMillan for her help taking images to create a master \textit{I}-band fringe pattern for the ARCTIC imager. We would also like to thank Dr. Stacy McGaugh for discussions on LSB photometry, especially morphological classifications.

This research has made use of the NASA/ IPAC Infrared Science Archive, which is operated by the Jet Propulsion Laboratory, California Institute of Technology, under contract with the National Aeronautics and Space Administration. This research made use of Astropy, a community-developed core Python package for Astronomy \citep{astropy2013, astropy2018}. We acknowledge the usage of the HyperLeda database (http://leda.univ-lyon1.fr).








\appendix

\begin{figure*}
  \centering
  \includegraphics[scale=0.25]{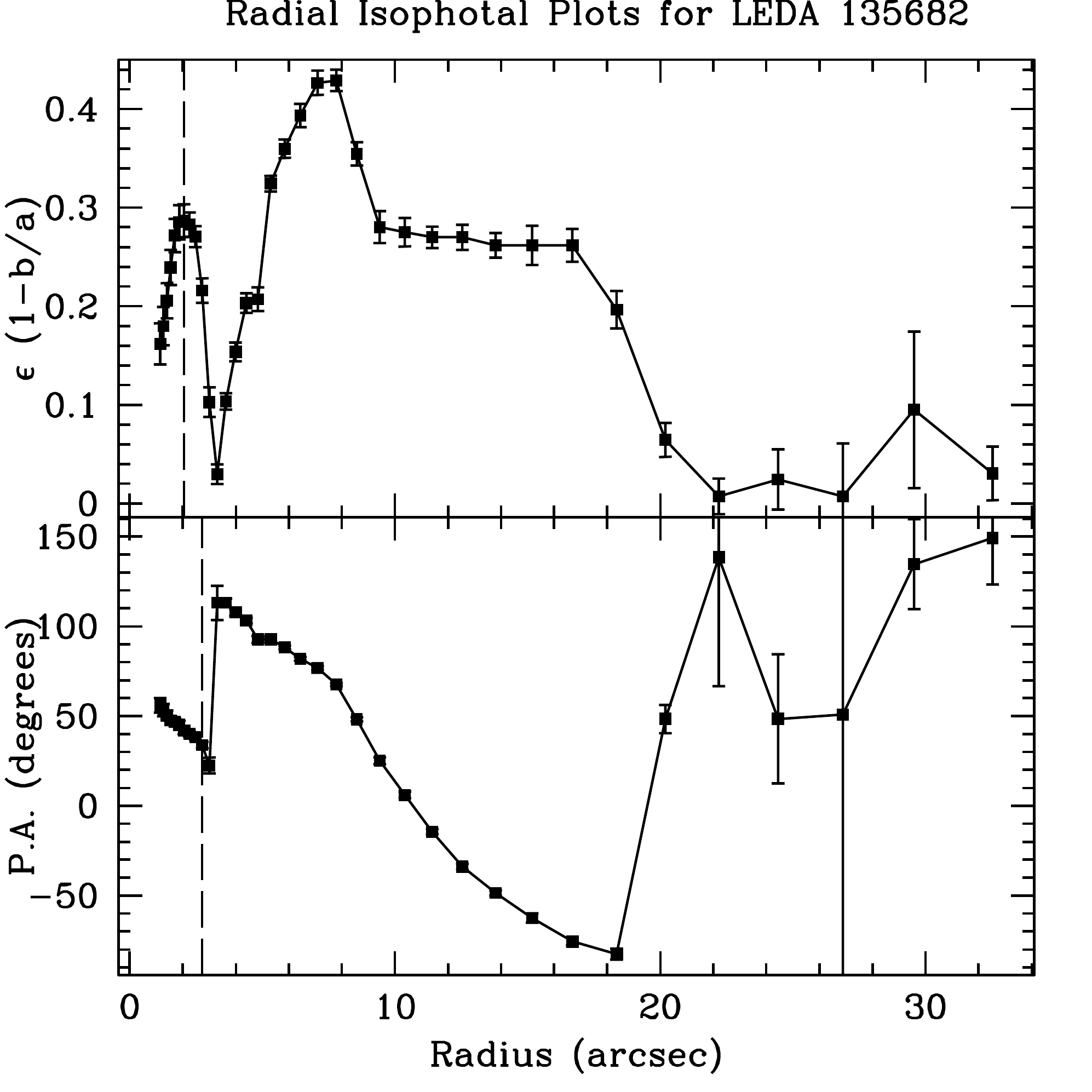} \includegraphics[scale=0.25]{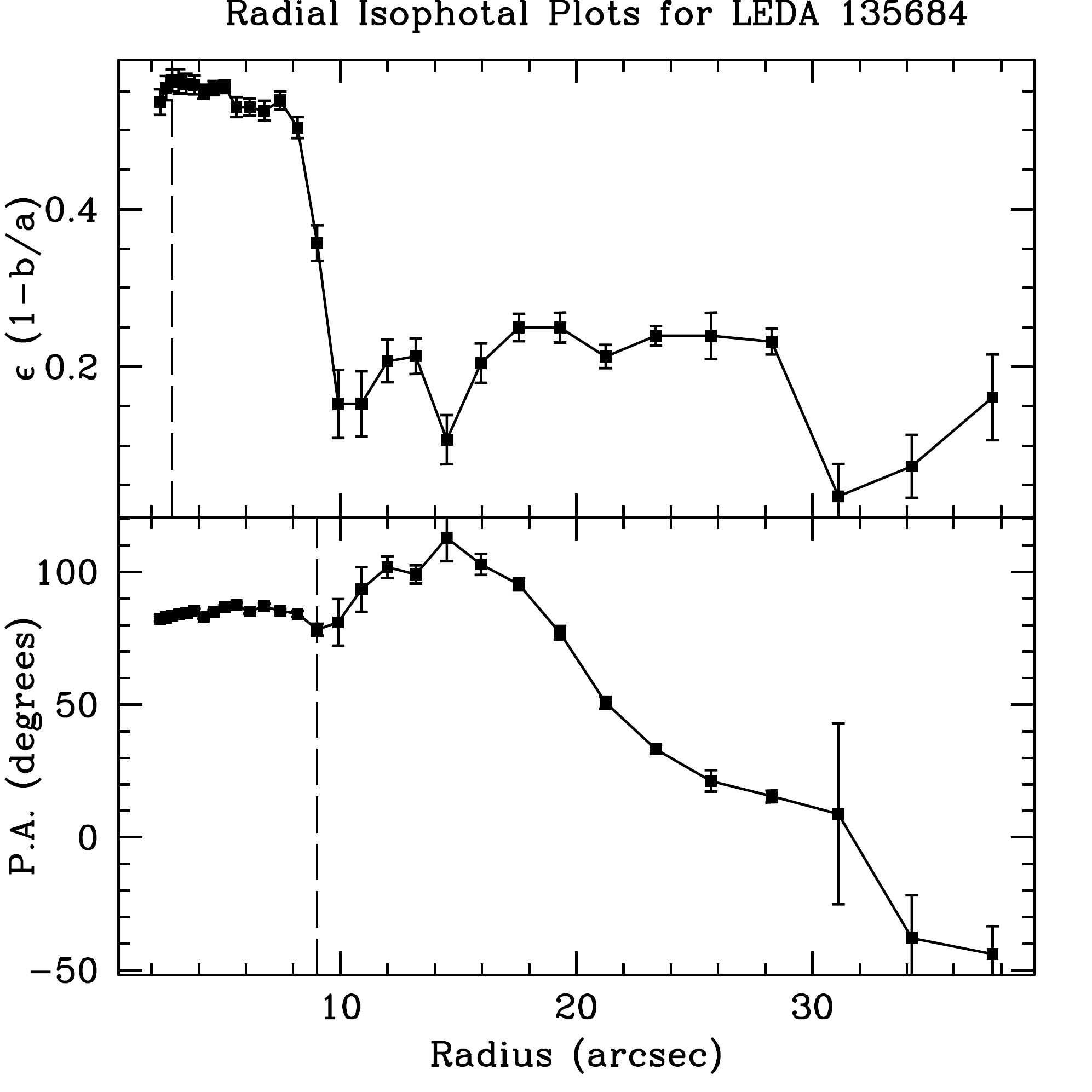} \includegraphics[scale=0.25]{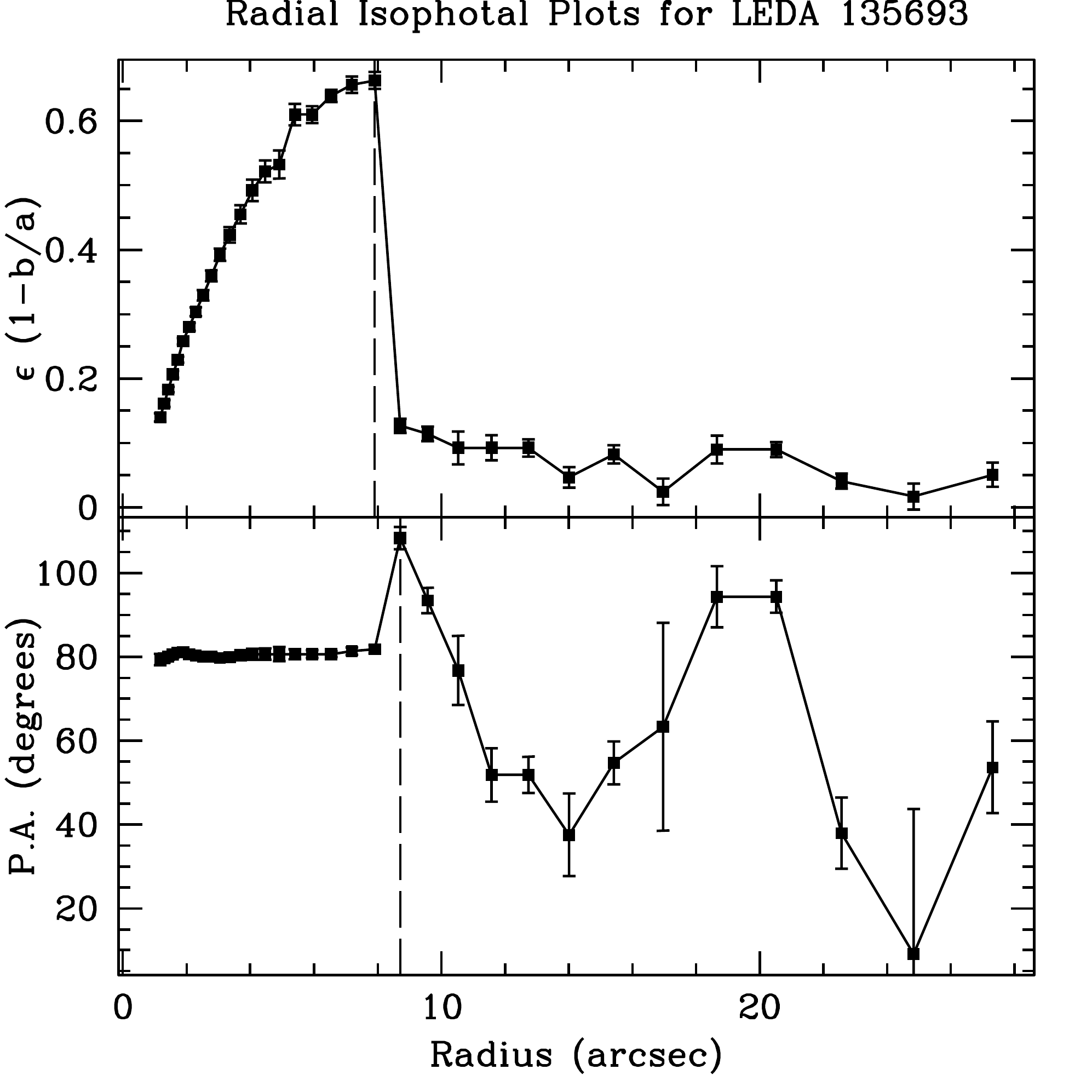} \\
  \includegraphics[scale=0.25]{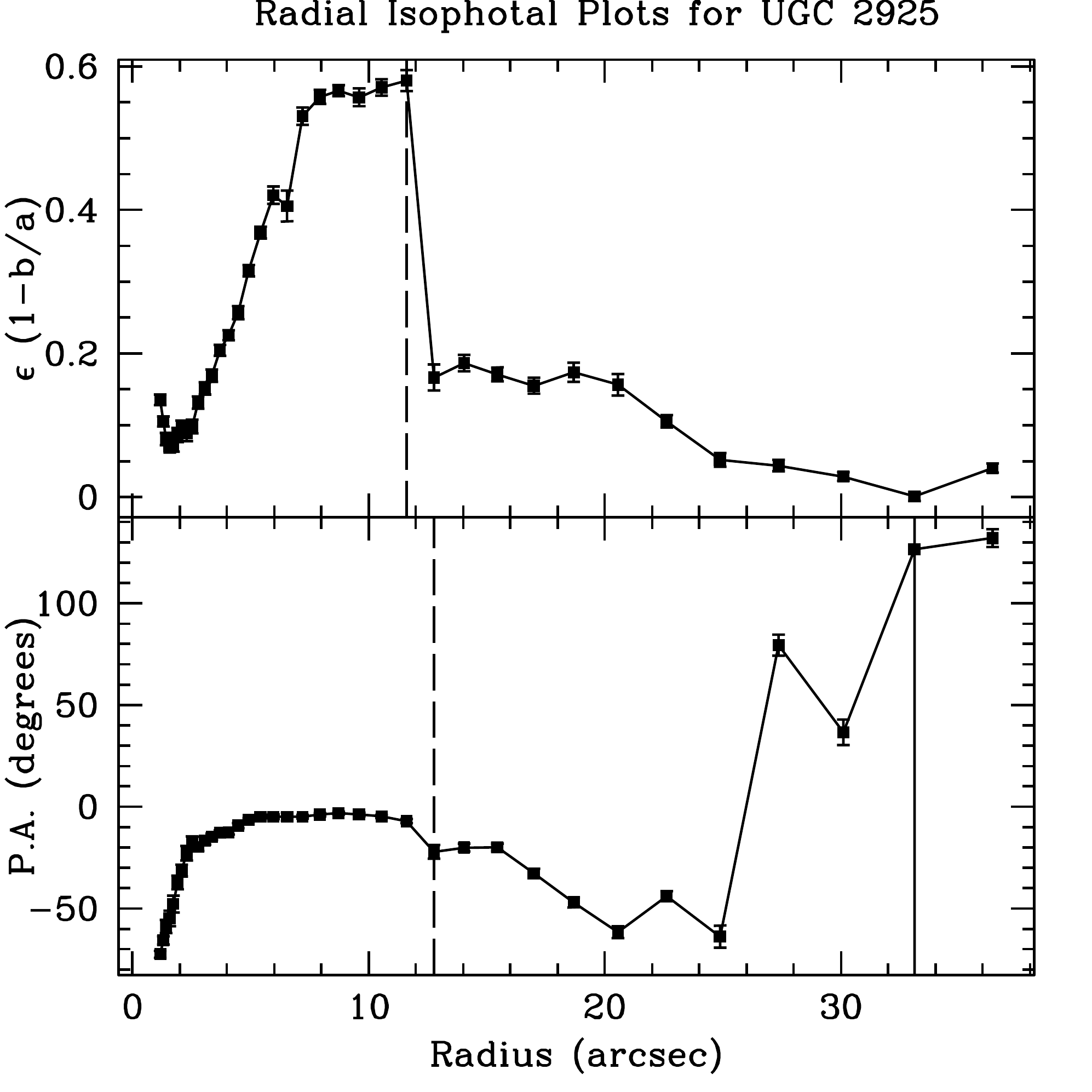} \includegraphics[scale=0.25]{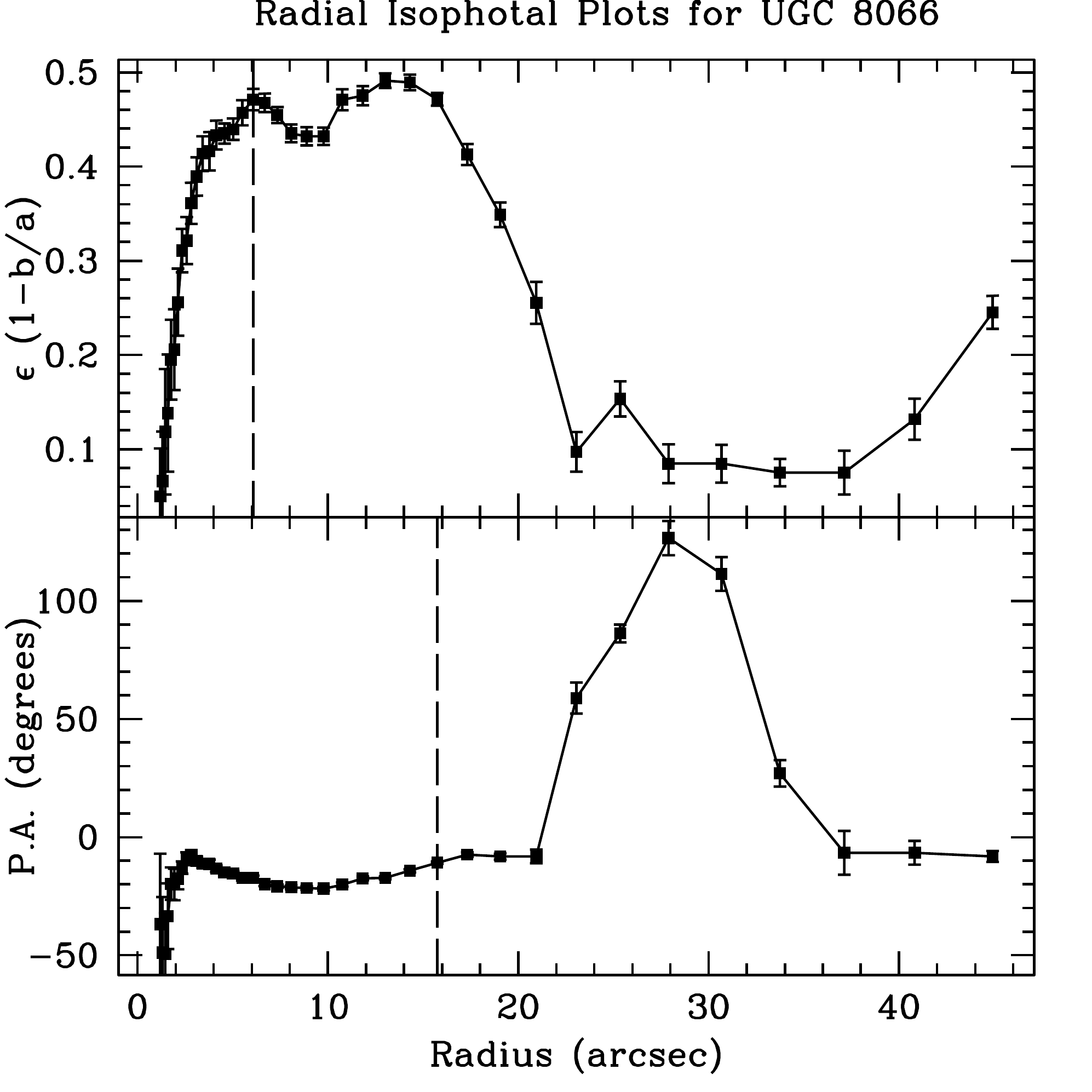} \includegraphics[scale=0.25]{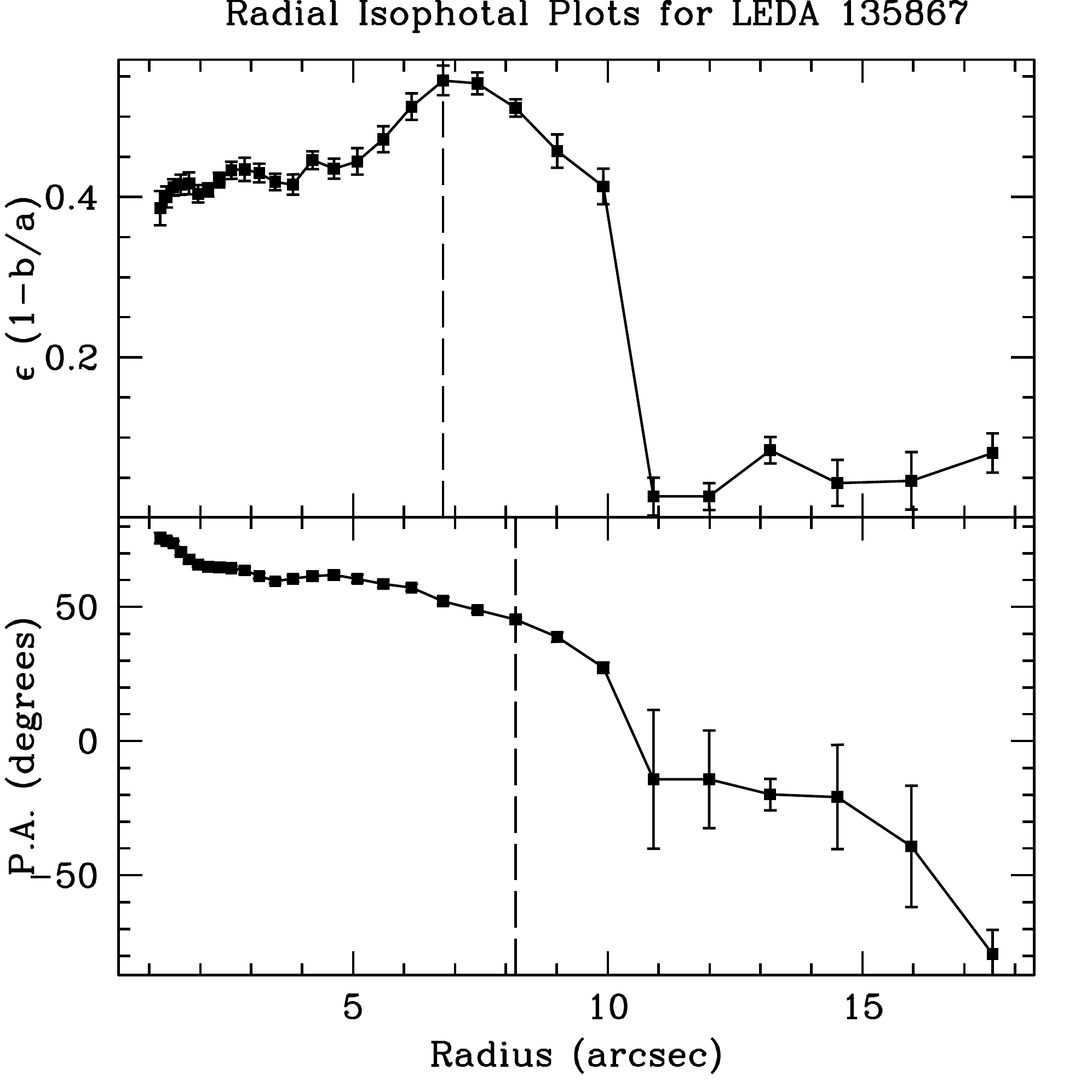} \\
  \includegraphics[scale=0.25]{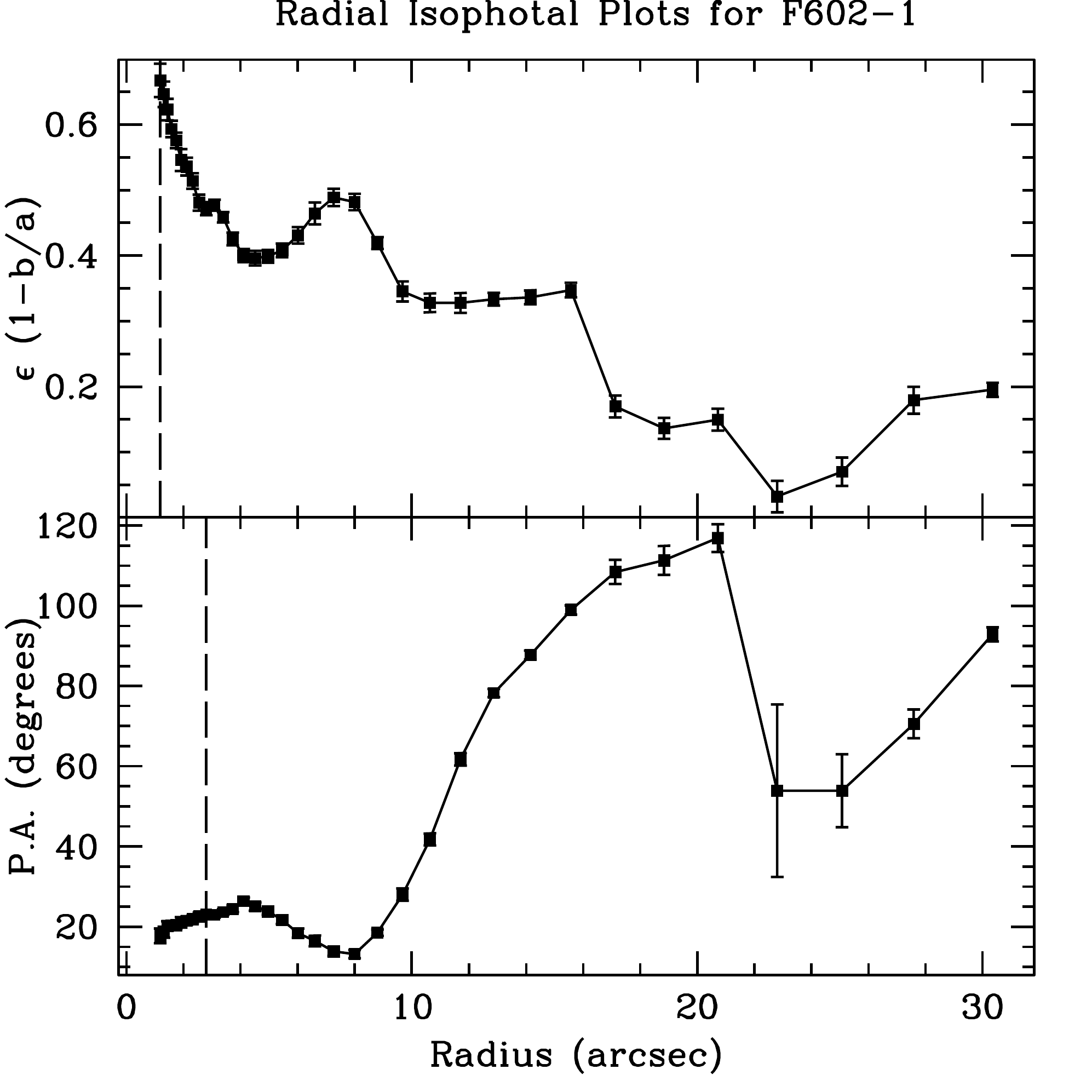} \includegraphics[scale=0.25]{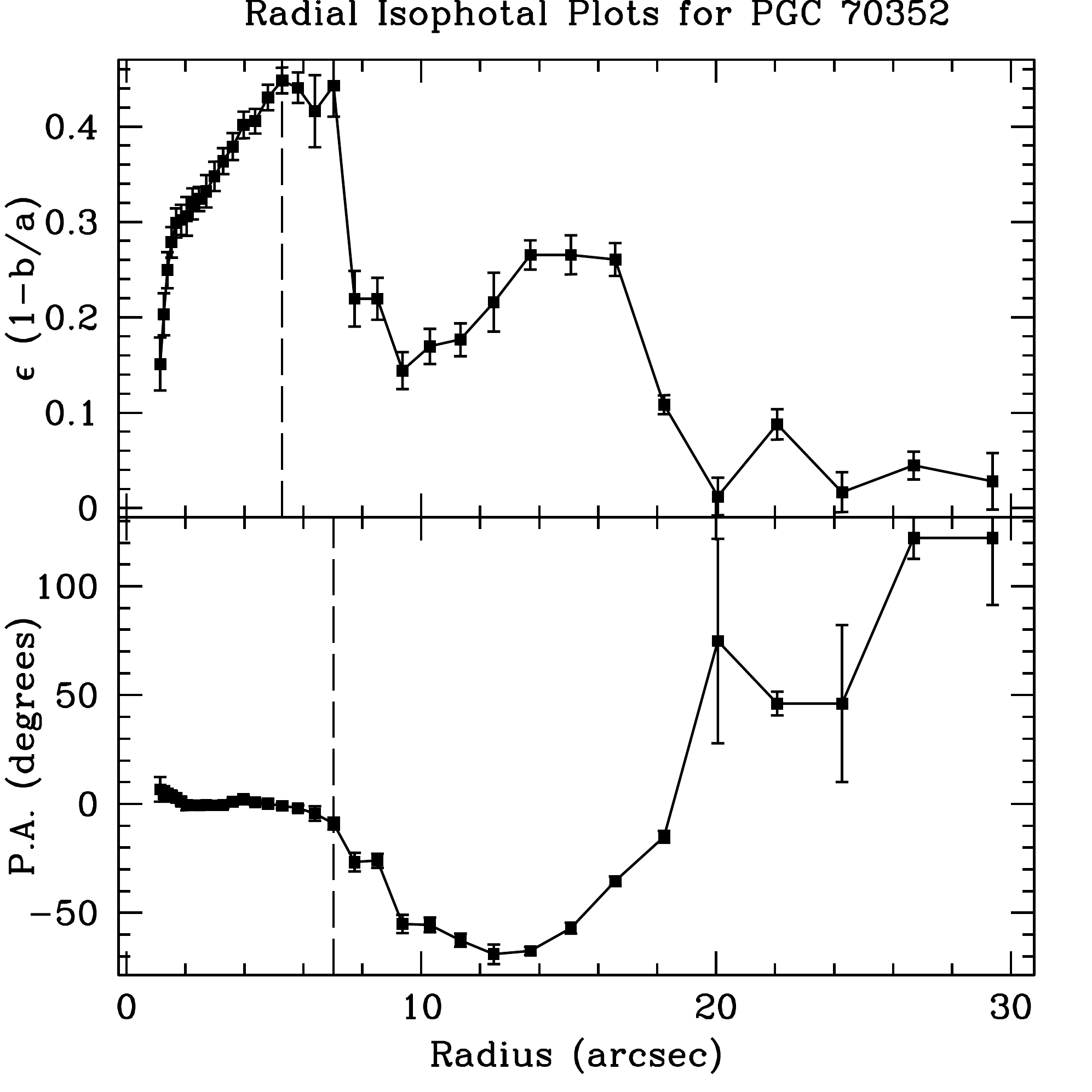} \includegraphics[scale=0.25]{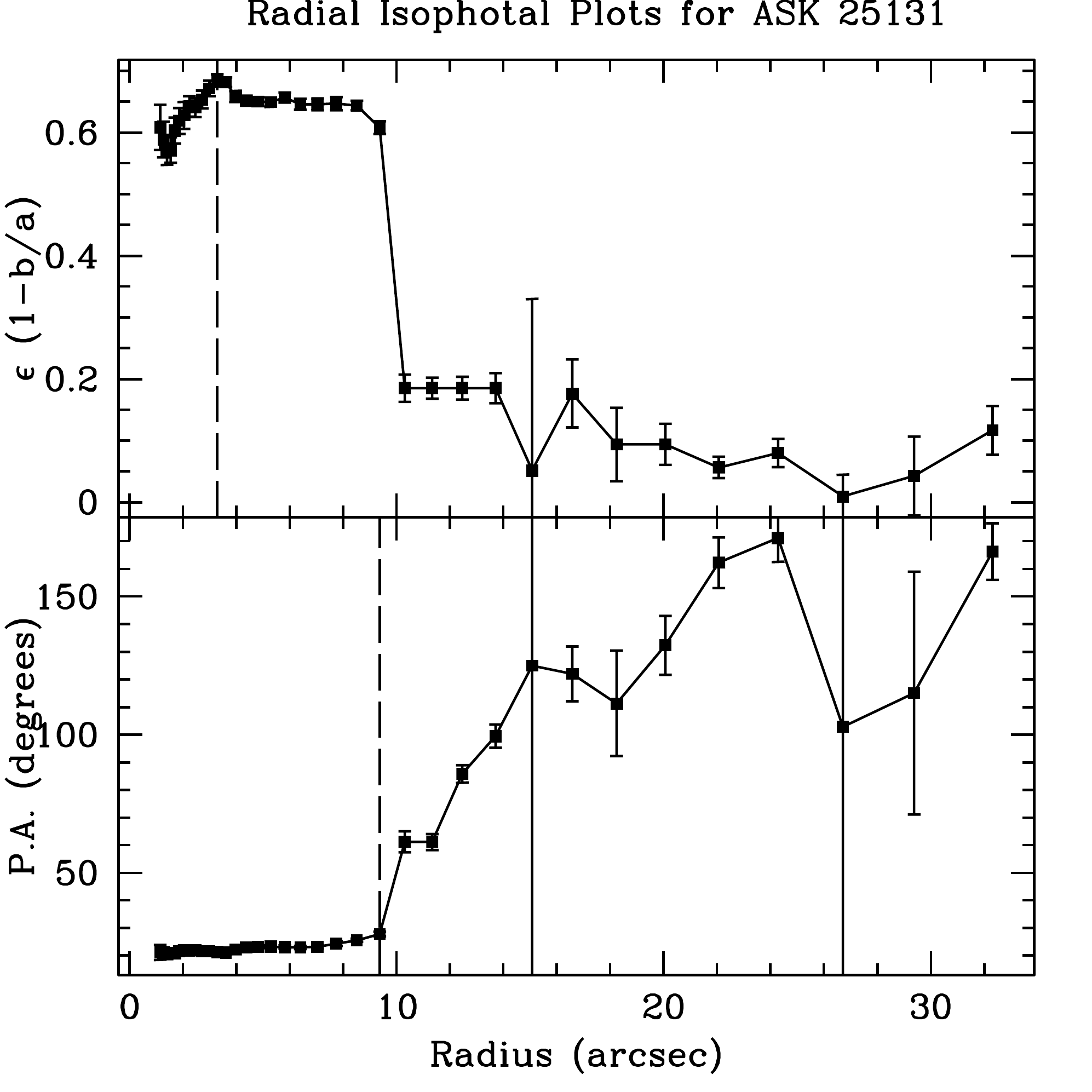} \\
  \includegraphics[scale=0.25]{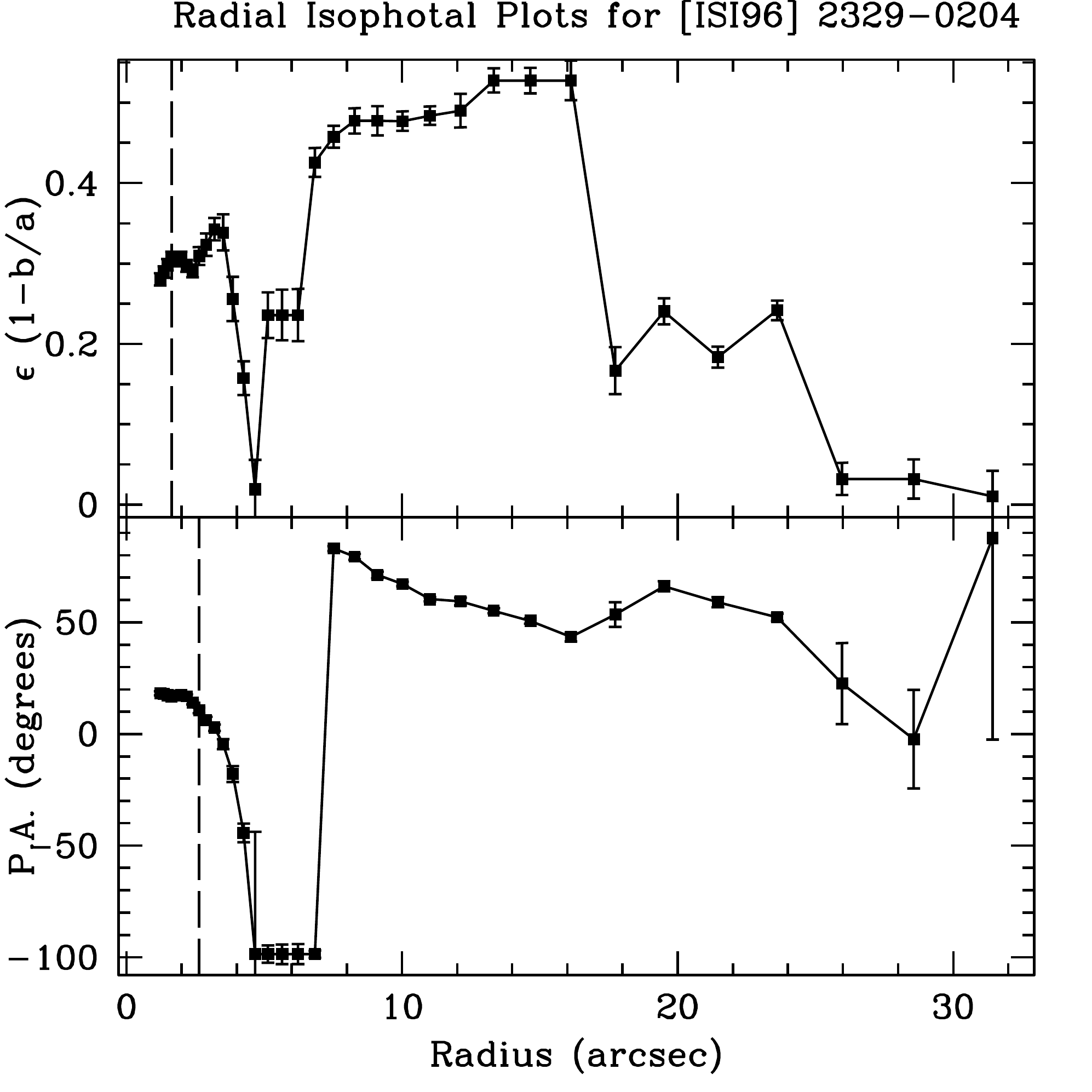} \\
  \caption{Radial plots of ellipticity (top panels) and position angle (bottom panels) for the remainder of our sample. The vertical lines in each panel denote the bar length measures $R_{e}$\ and $R_{\mathrm{PA}}$.}
  \label{elipBars}
\end{figure*}

\begin{figure*}
  \centering
  \includegraphics[scale=0.25]{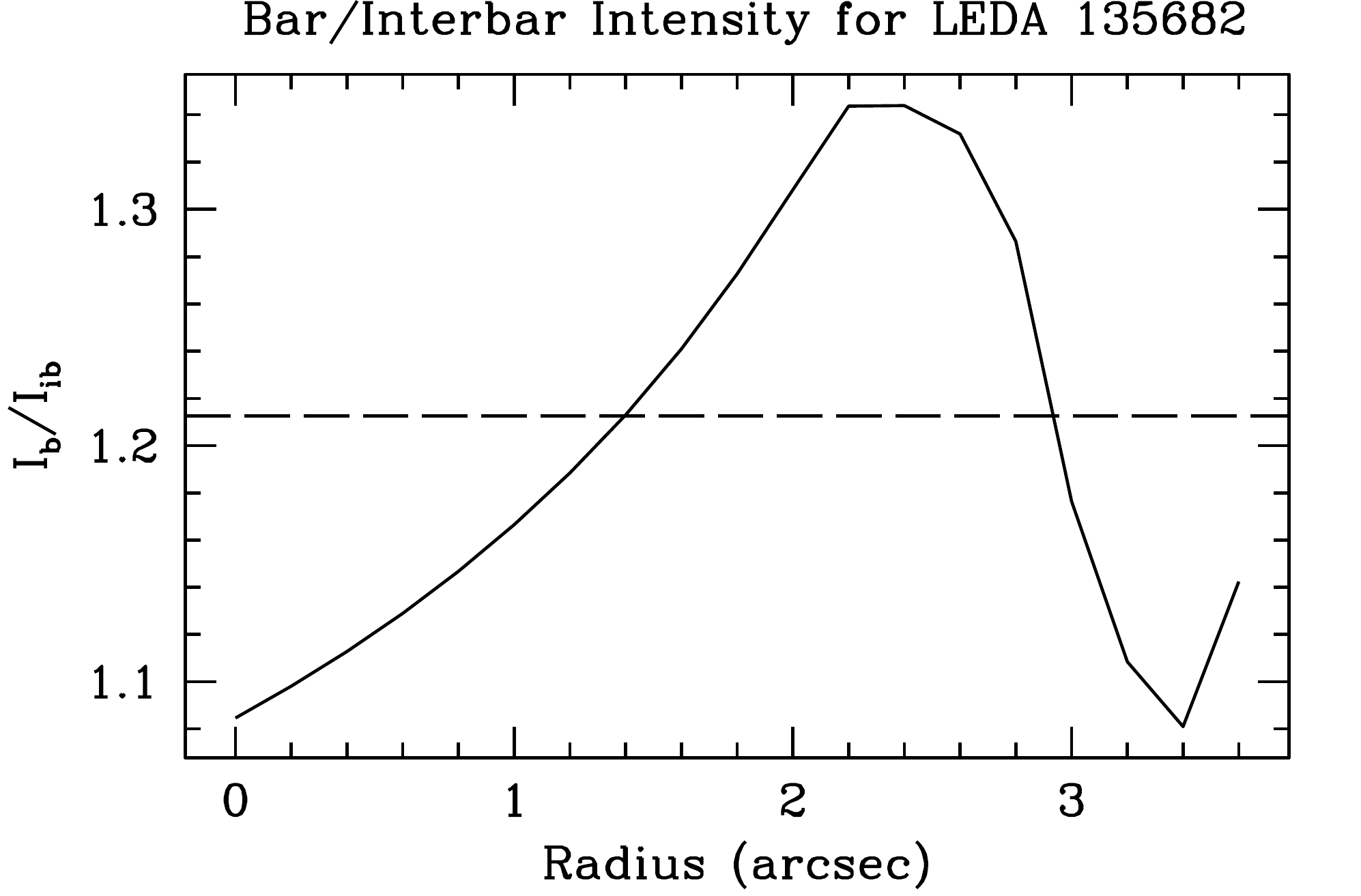} \includegraphics[scale=0.25]{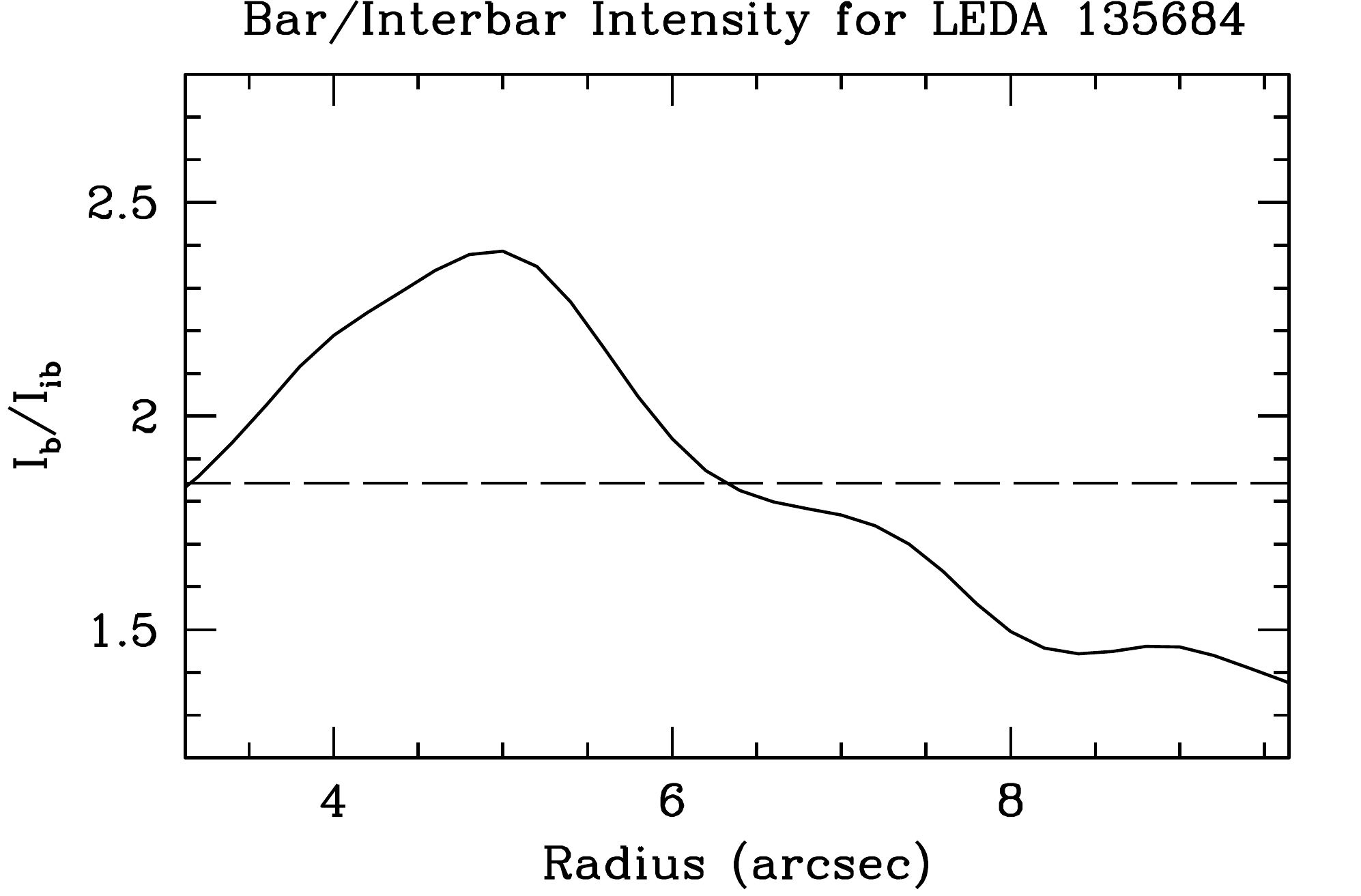} \includegraphics[scale=0.25]{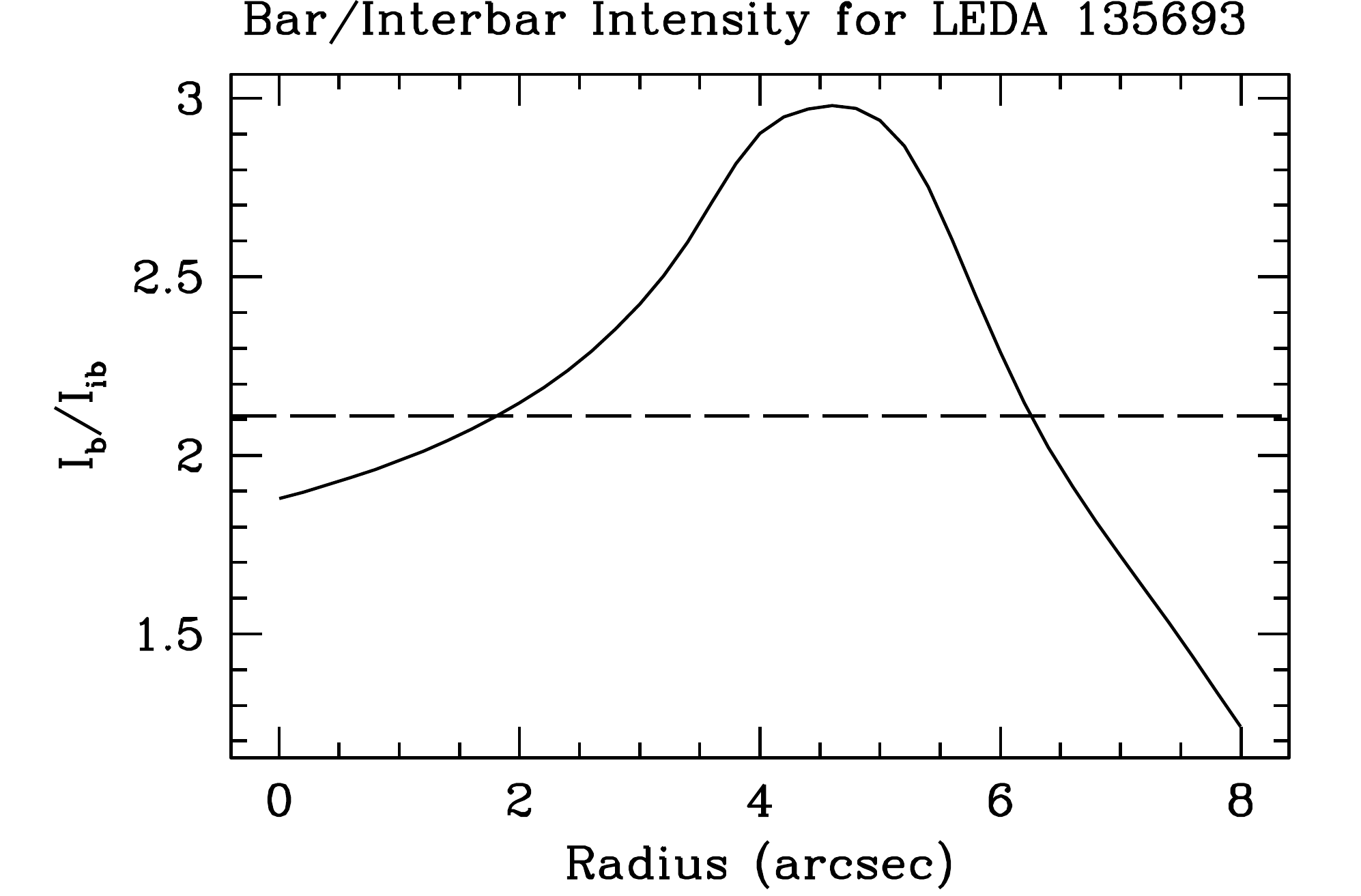} \\
  \vspace{0.05\textwidth}
  \includegraphics[scale=0.25]{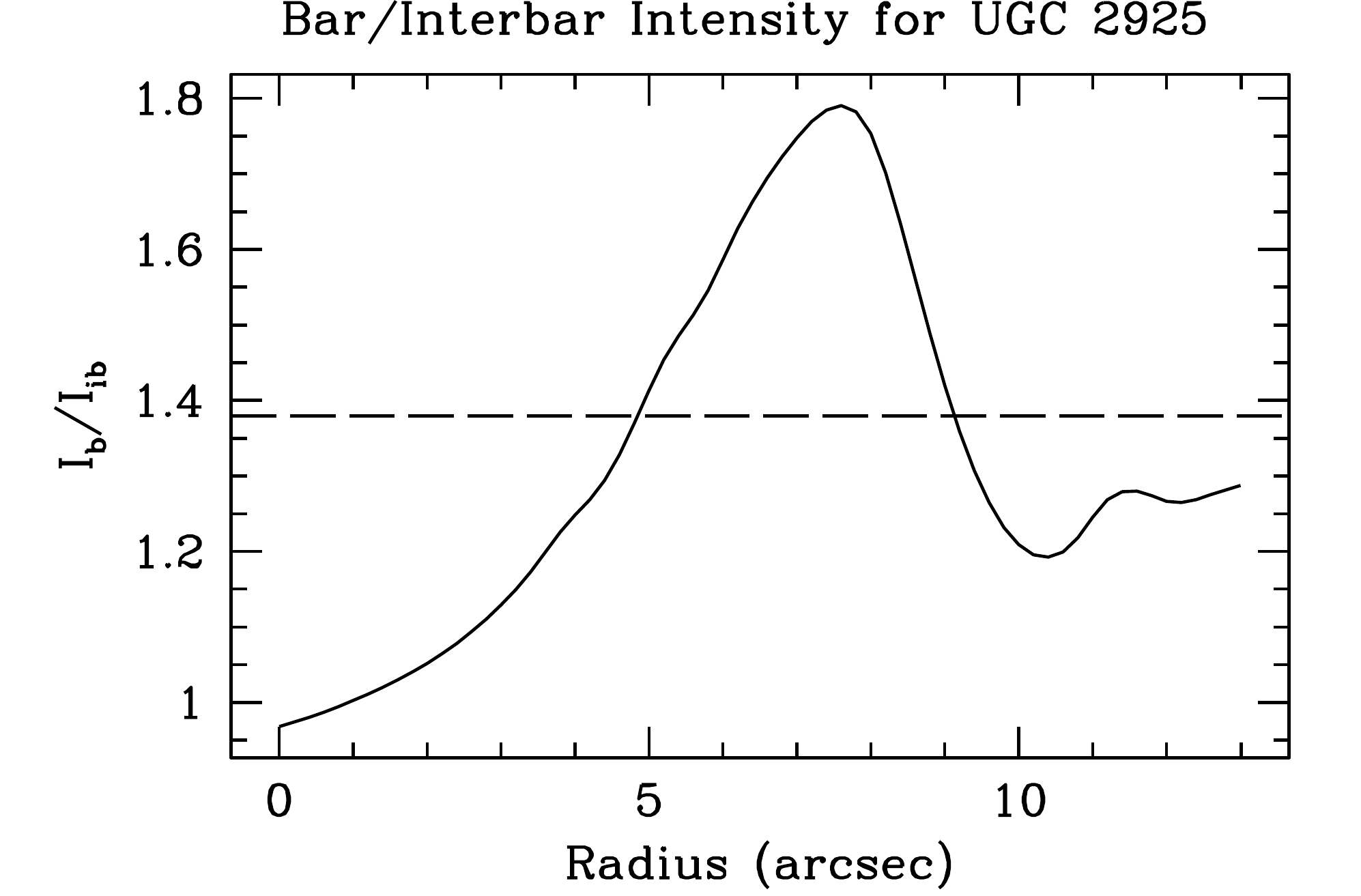} \includegraphics[scale=0.25]{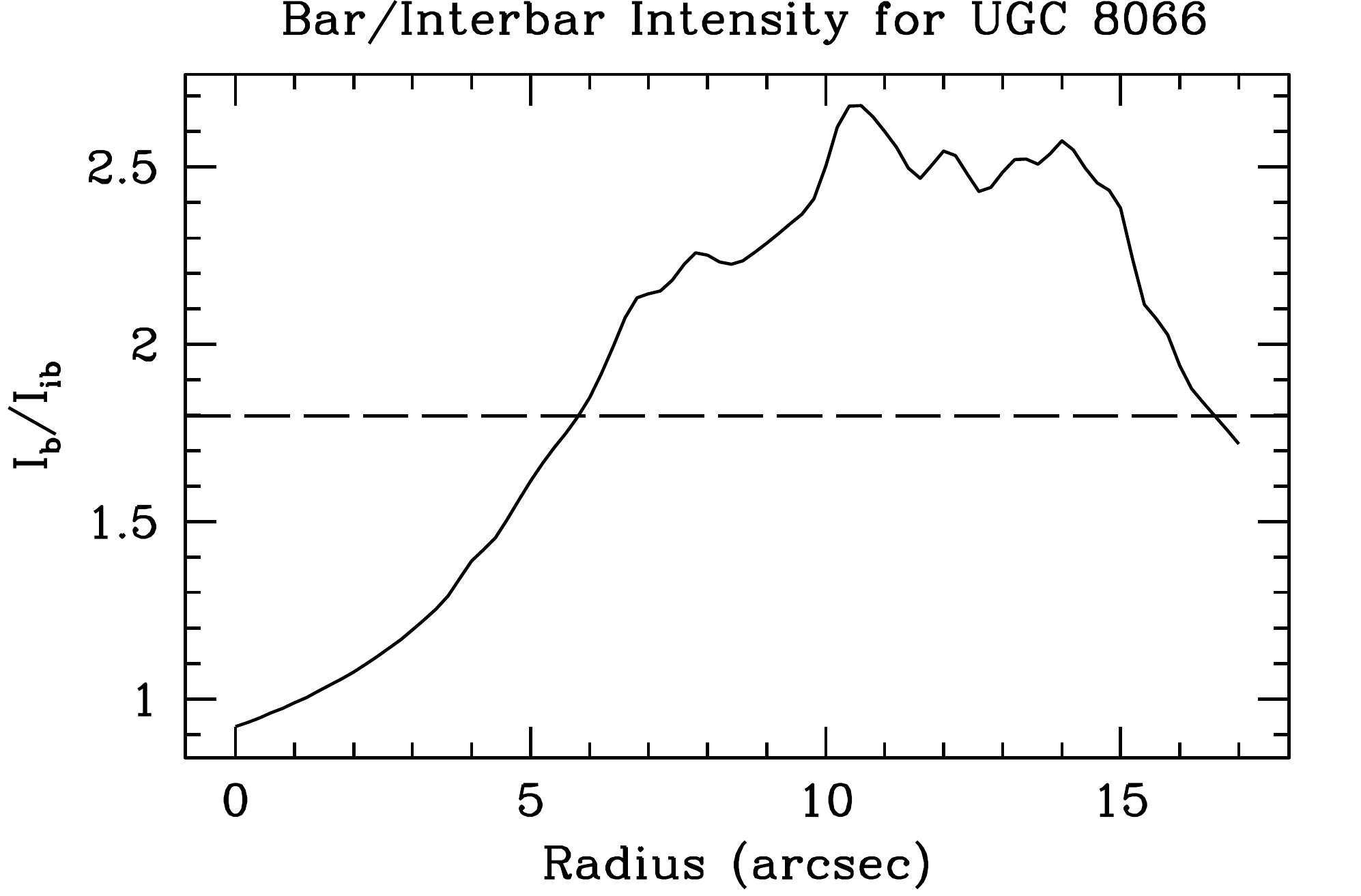} \includegraphics[scale=0.25]{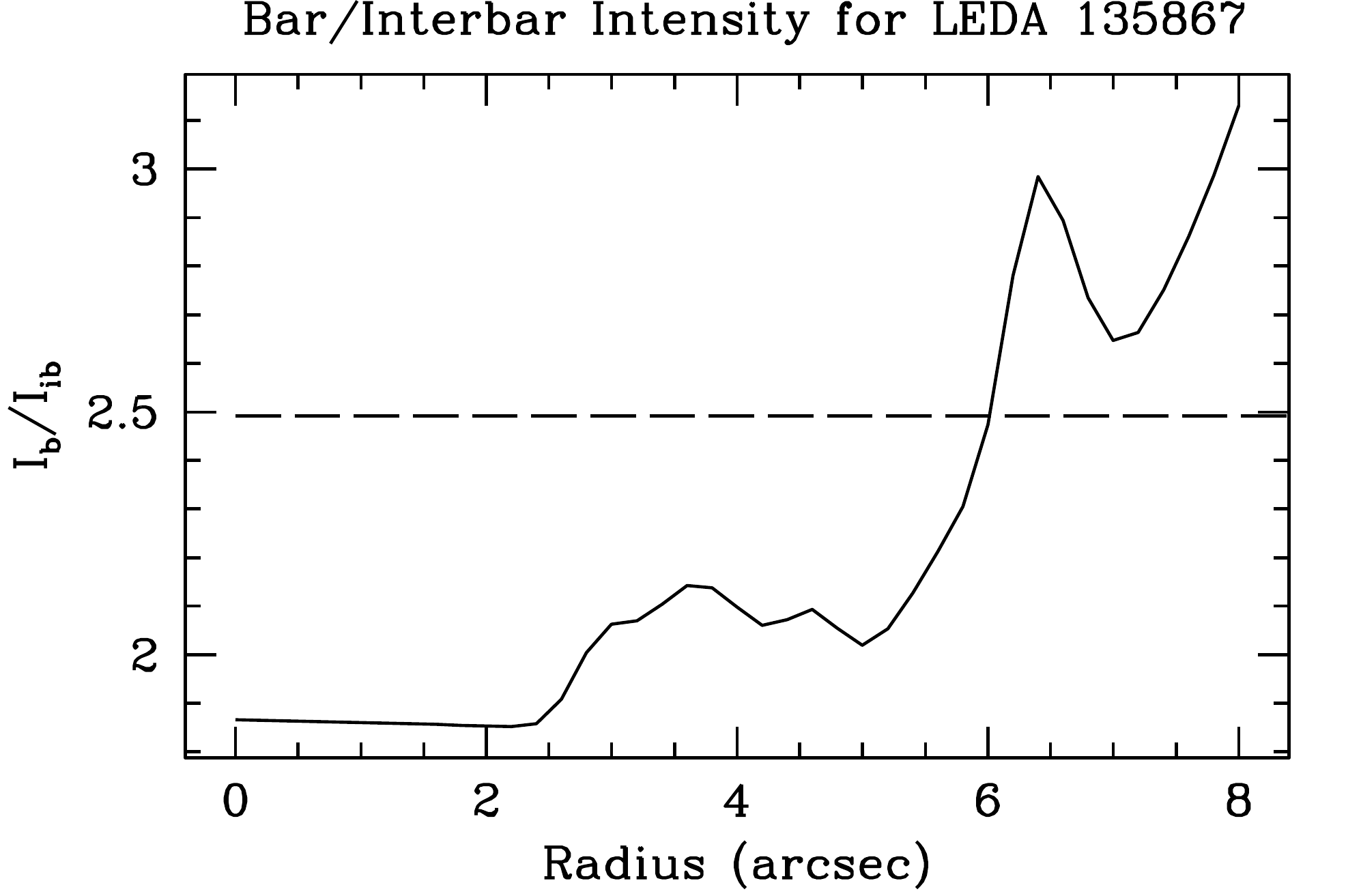} \\
  \vspace{0.05\textwidth}
  \includegraphics[scale=0.25]{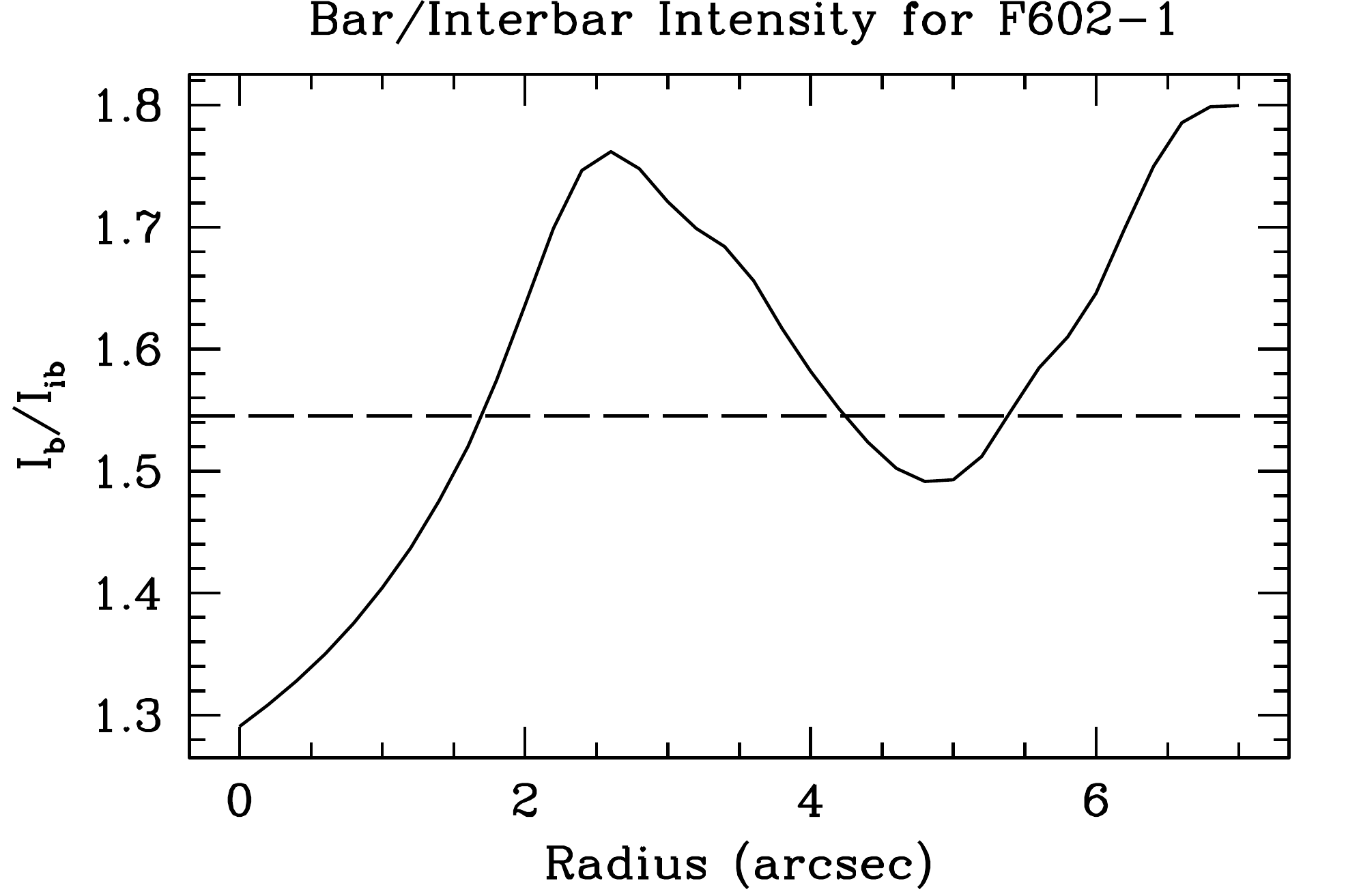} \includegraphics[scale=0.25]{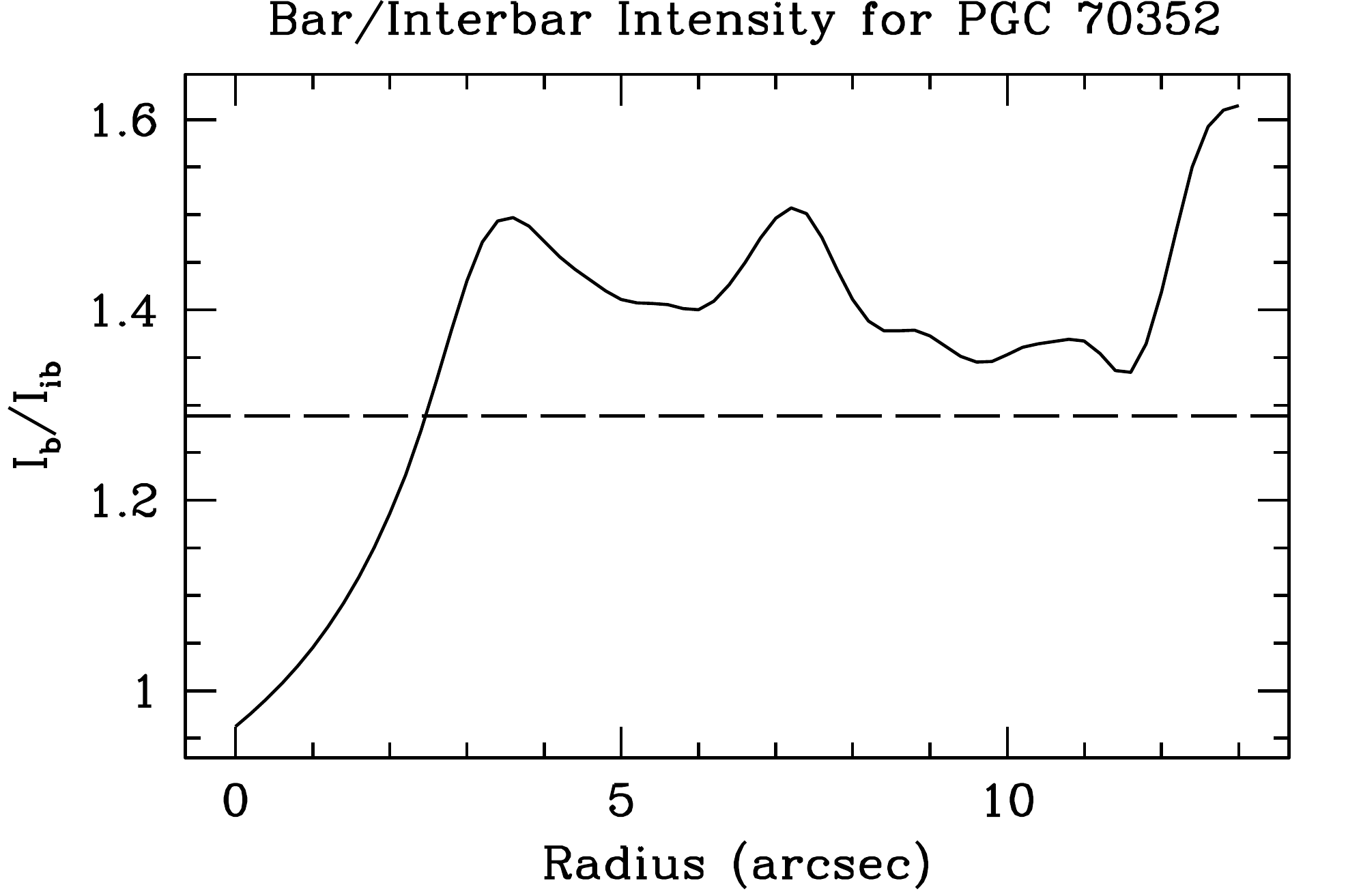} \includegraphics[scale=0.25]{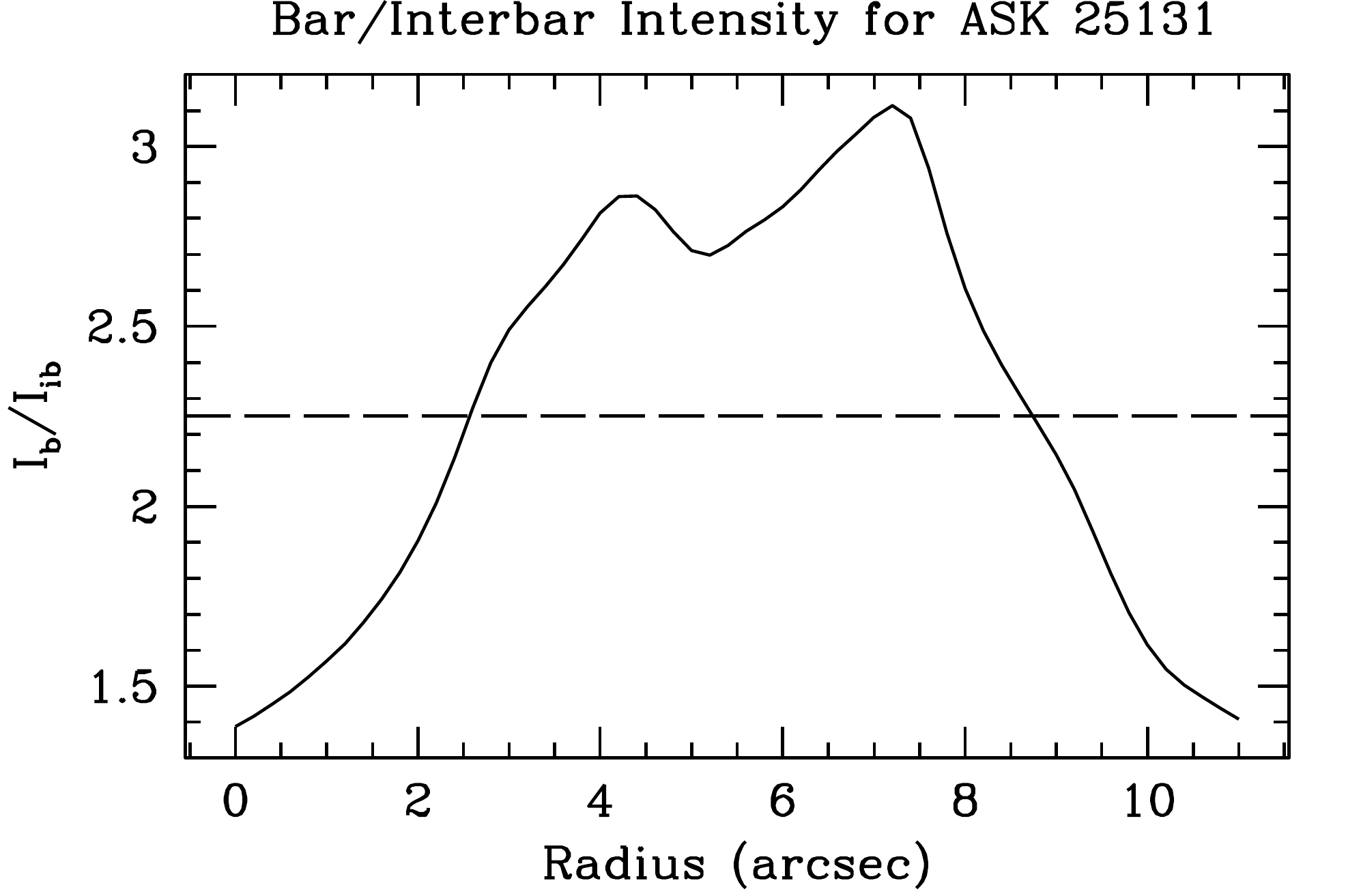} \\
  \vspace{0.05\textwidth}
  \includegraphics[scale=0.25]{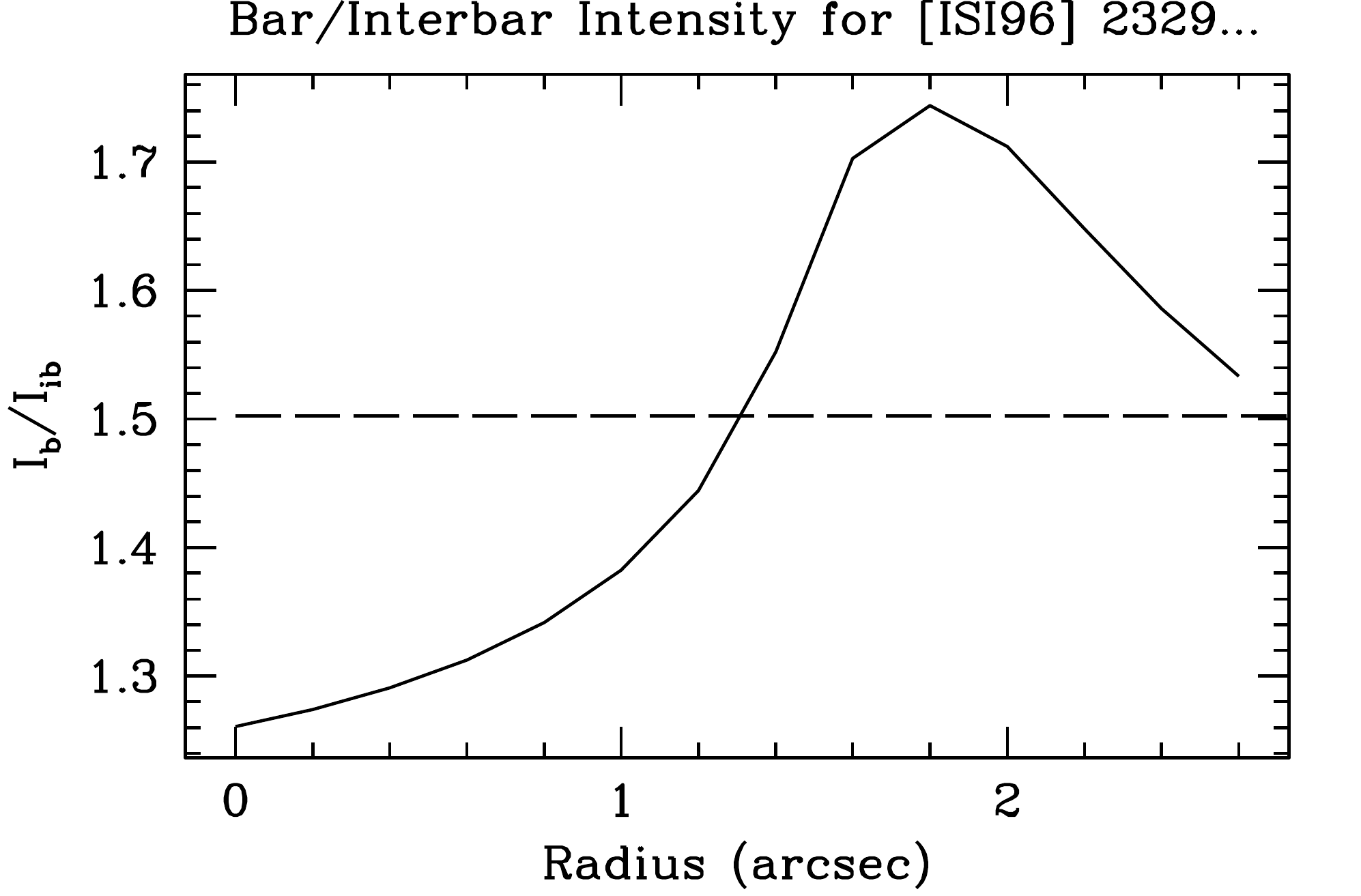} \\
  \caption{Radial plots of the Fourier bar ($I_{b}$) and interbar ($I_{ib}$) intensity ratio.}
  \label{fourierBars}
\end{figure*}

\begin{figure*}
  \centering
  \includegraphics[scale=0.25]{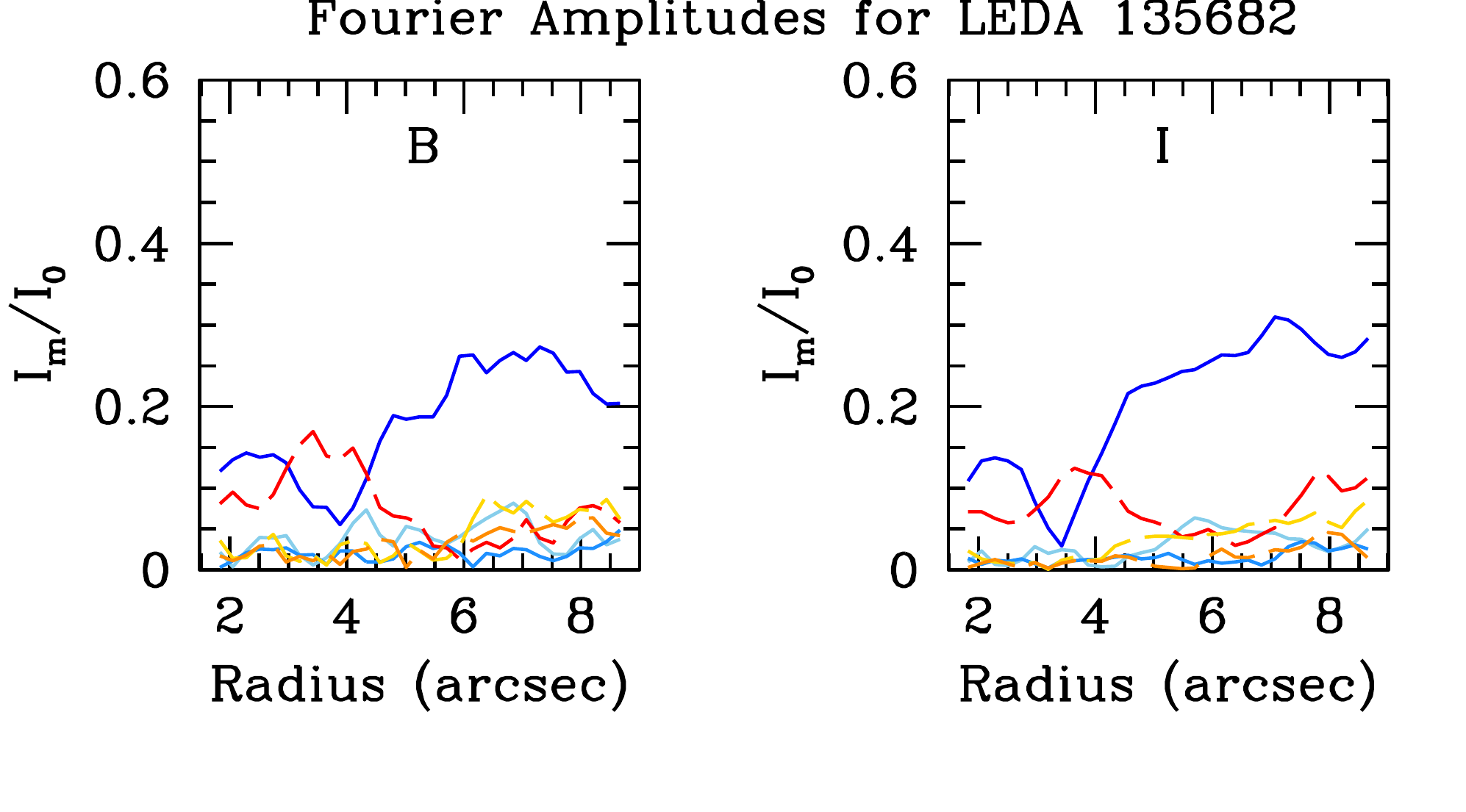} \hspace{0.01\textwidth} \includegraphics[scale=0.25]{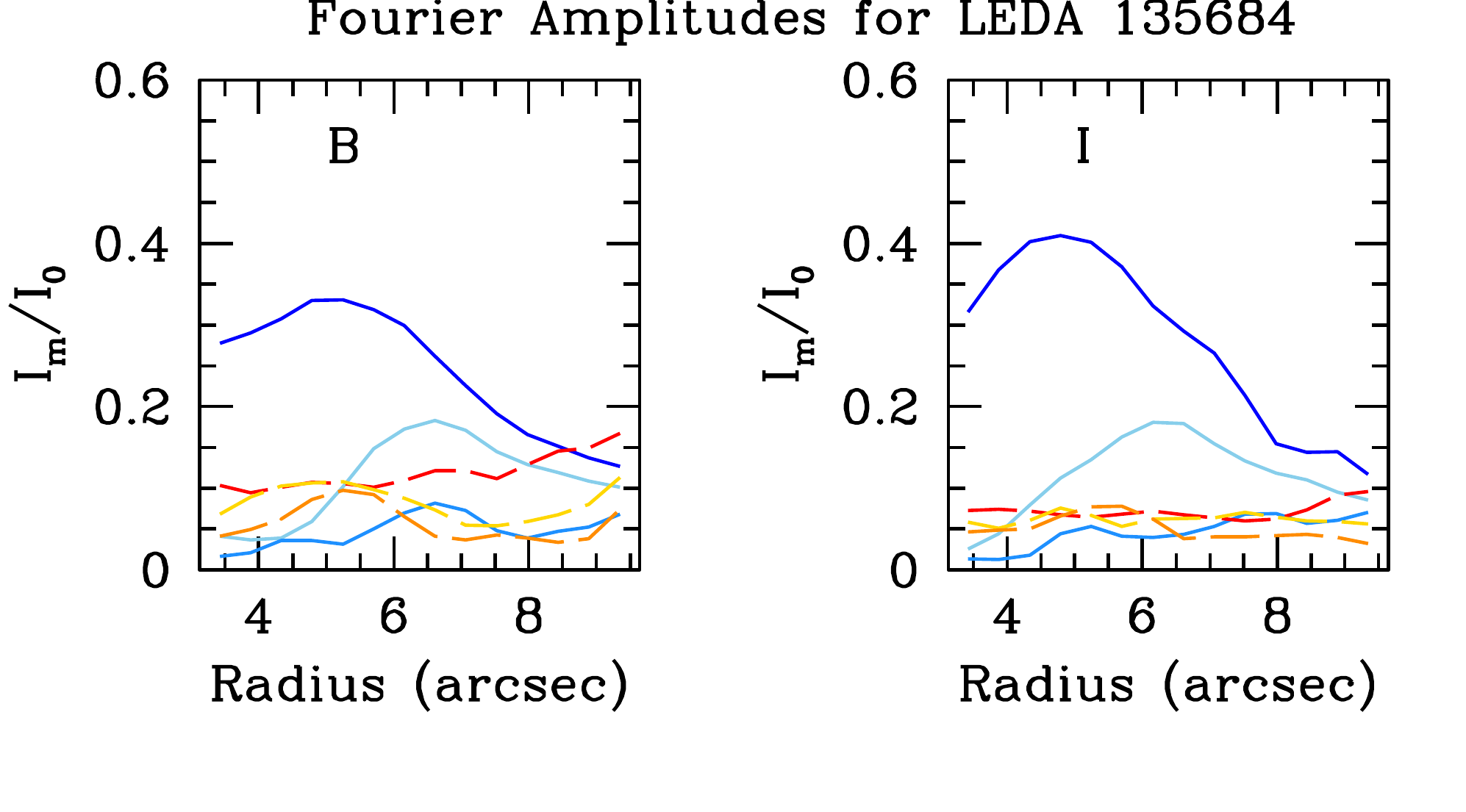} \hspace{0.01\textwidth} \includegraphics[scale=0.25]{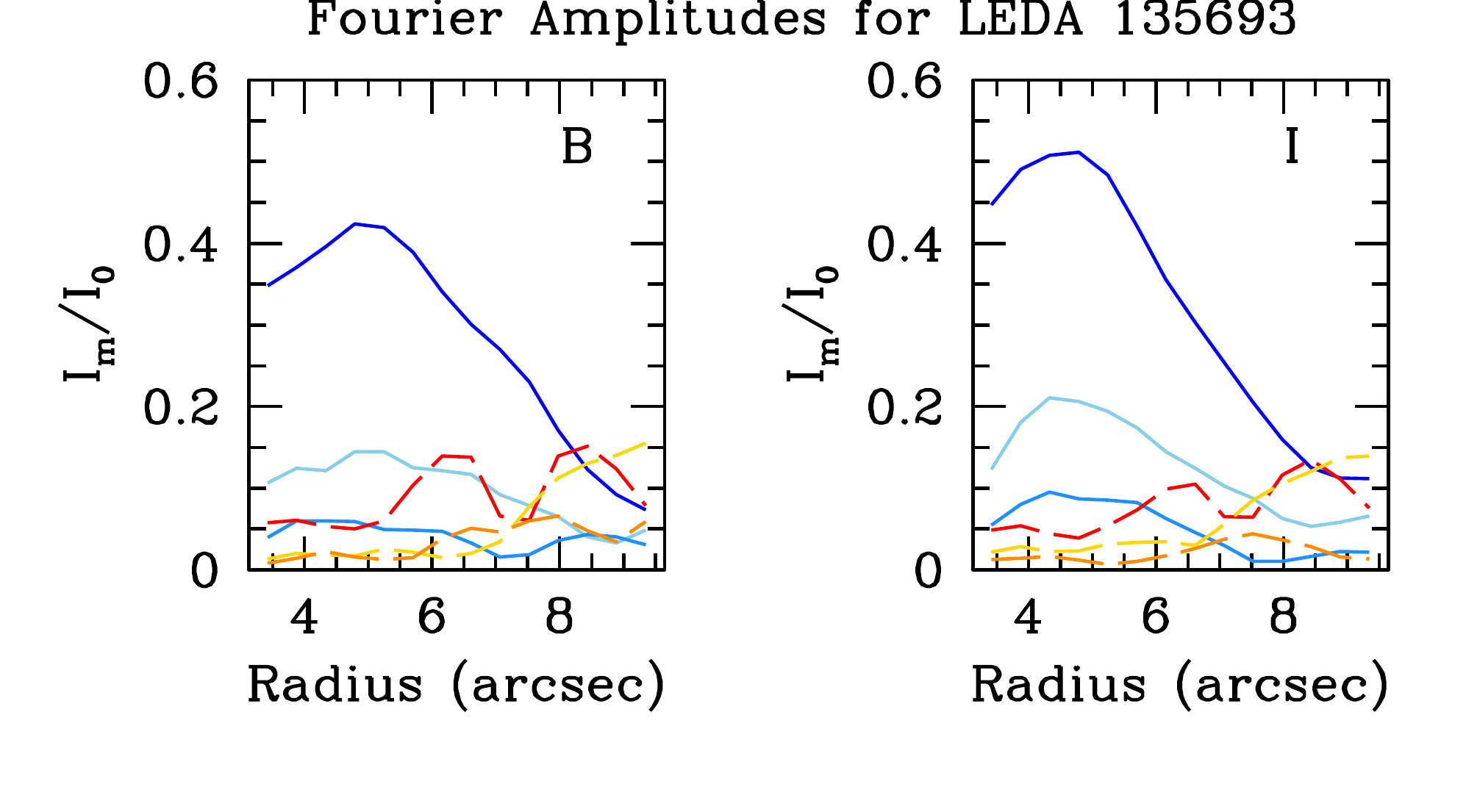} \\
  \vspace{0.05\textwidth}
  \includegraphics[scale=0.25]{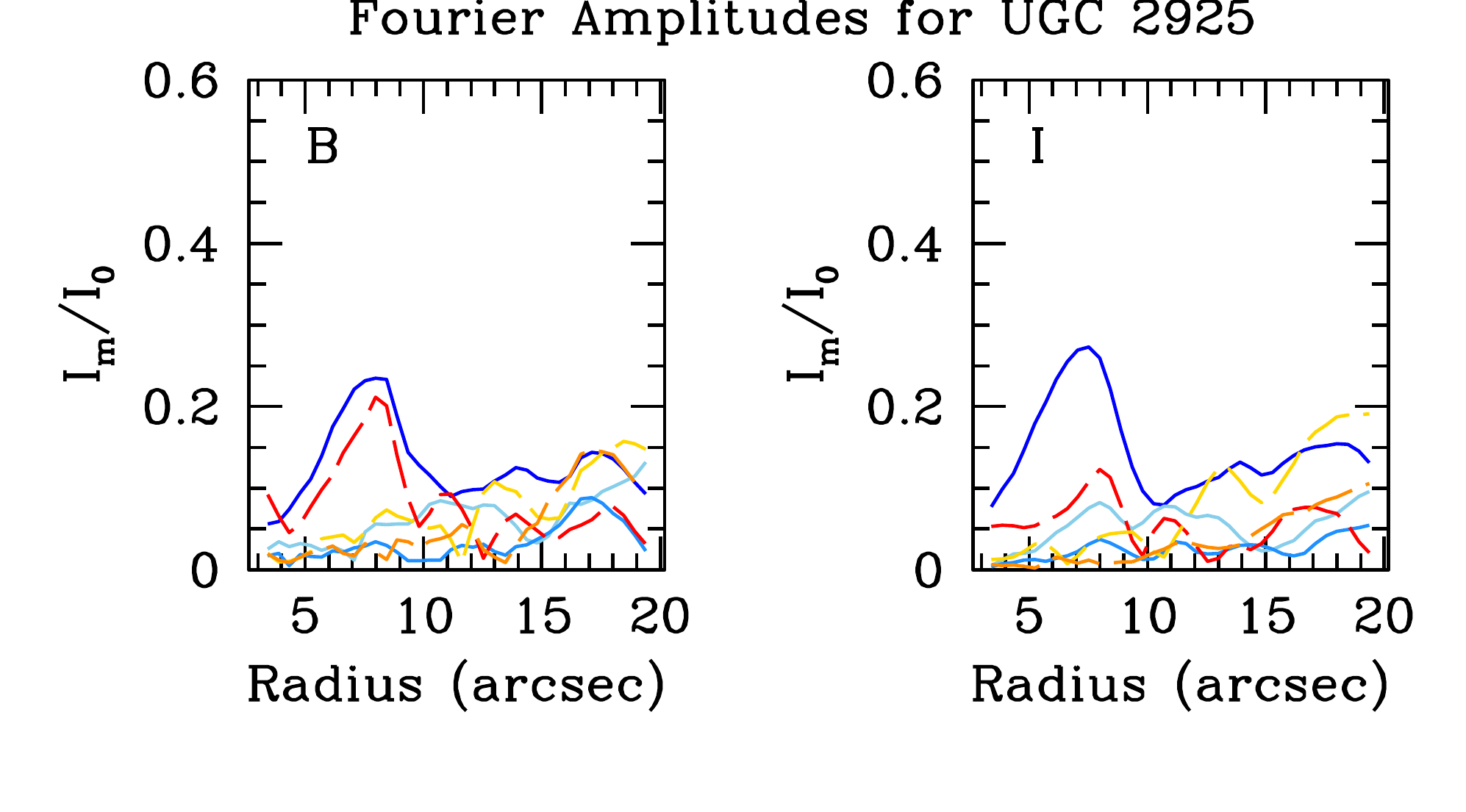} \hspace{0.01\textwidth} \includegraphics[scale=0.25]{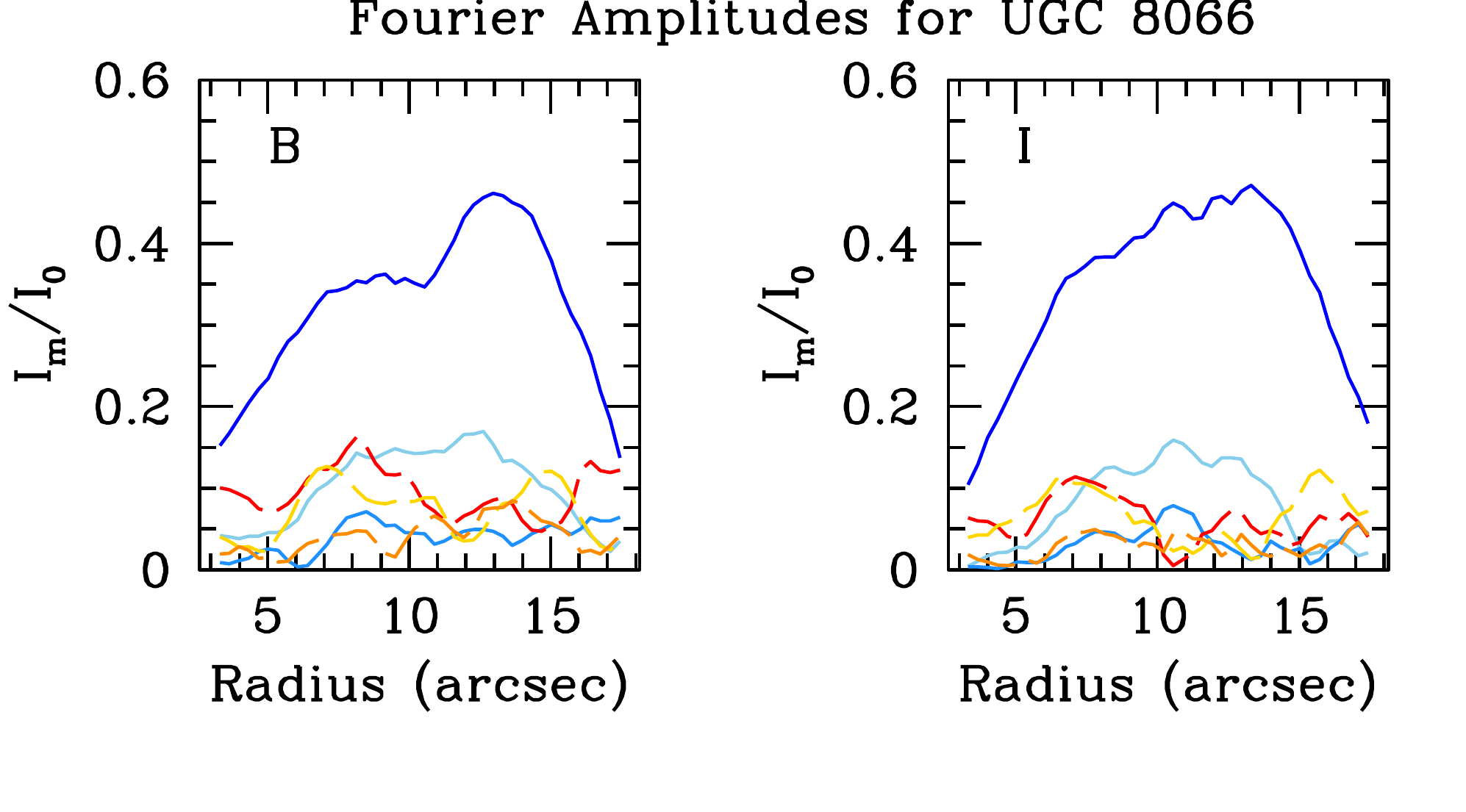} \hspace{0.01\textwidth} \includegraphics[scale=0.25]{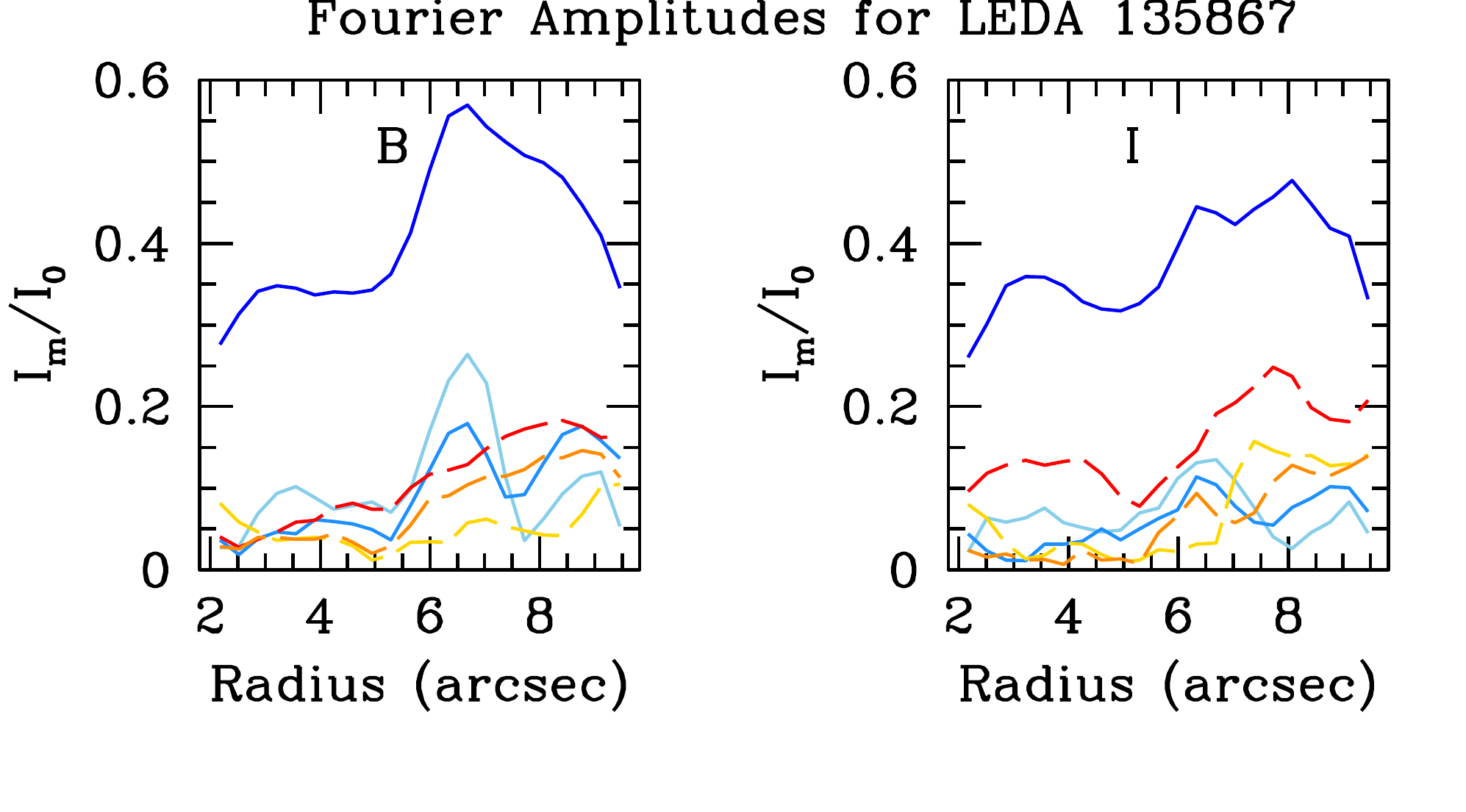} \\
  \vspace{0.05\textwidth}
  \includegraphics[scale=0.25]{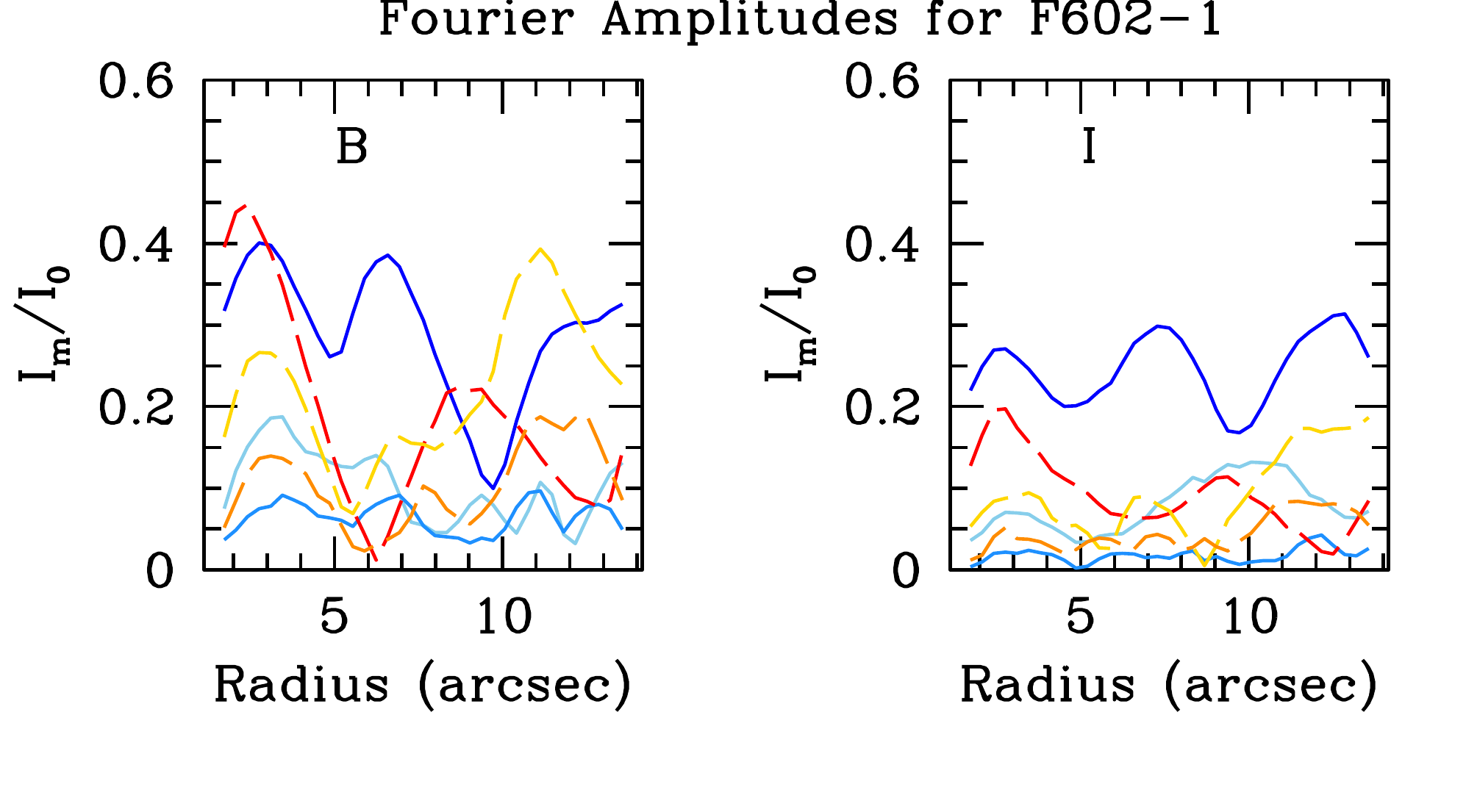} \hspace{0.01\textwidth} \includegraphics[scale=0.25]{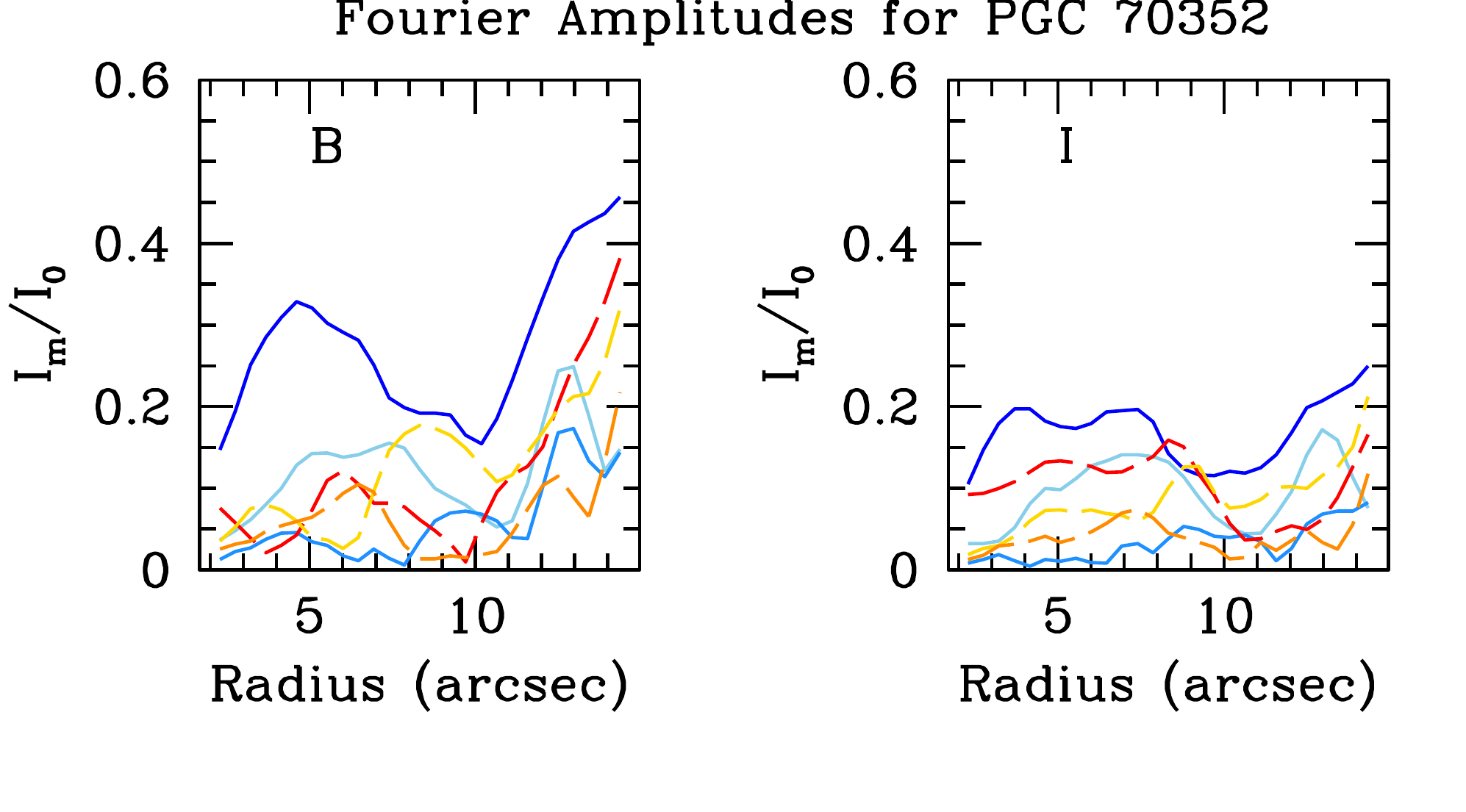} \hspace{0.01\textwidth} \includegraphics[scale=0.25]{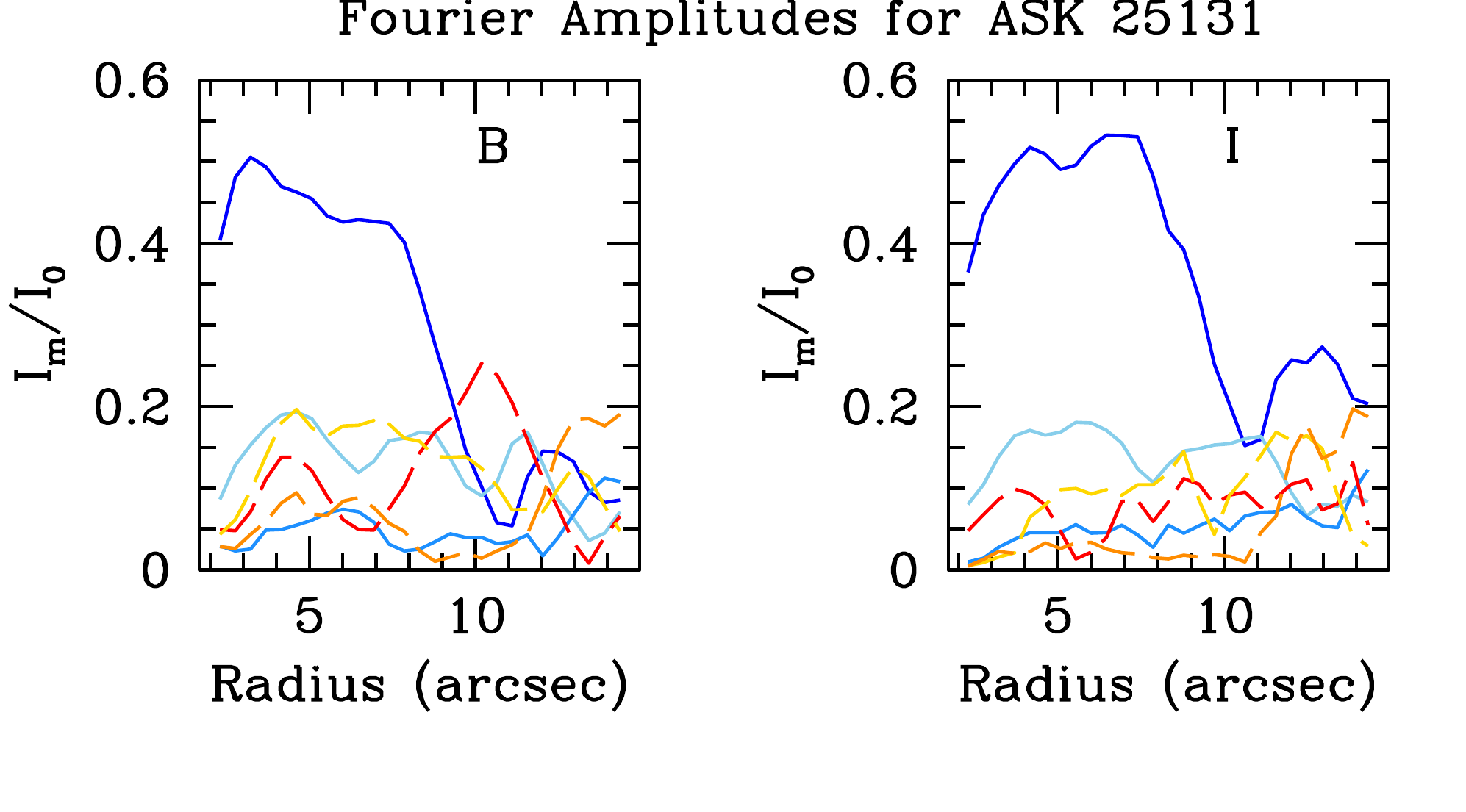} \\
  \vspace{0.05\textwidth}
  \includegraphics[scale=0.25]{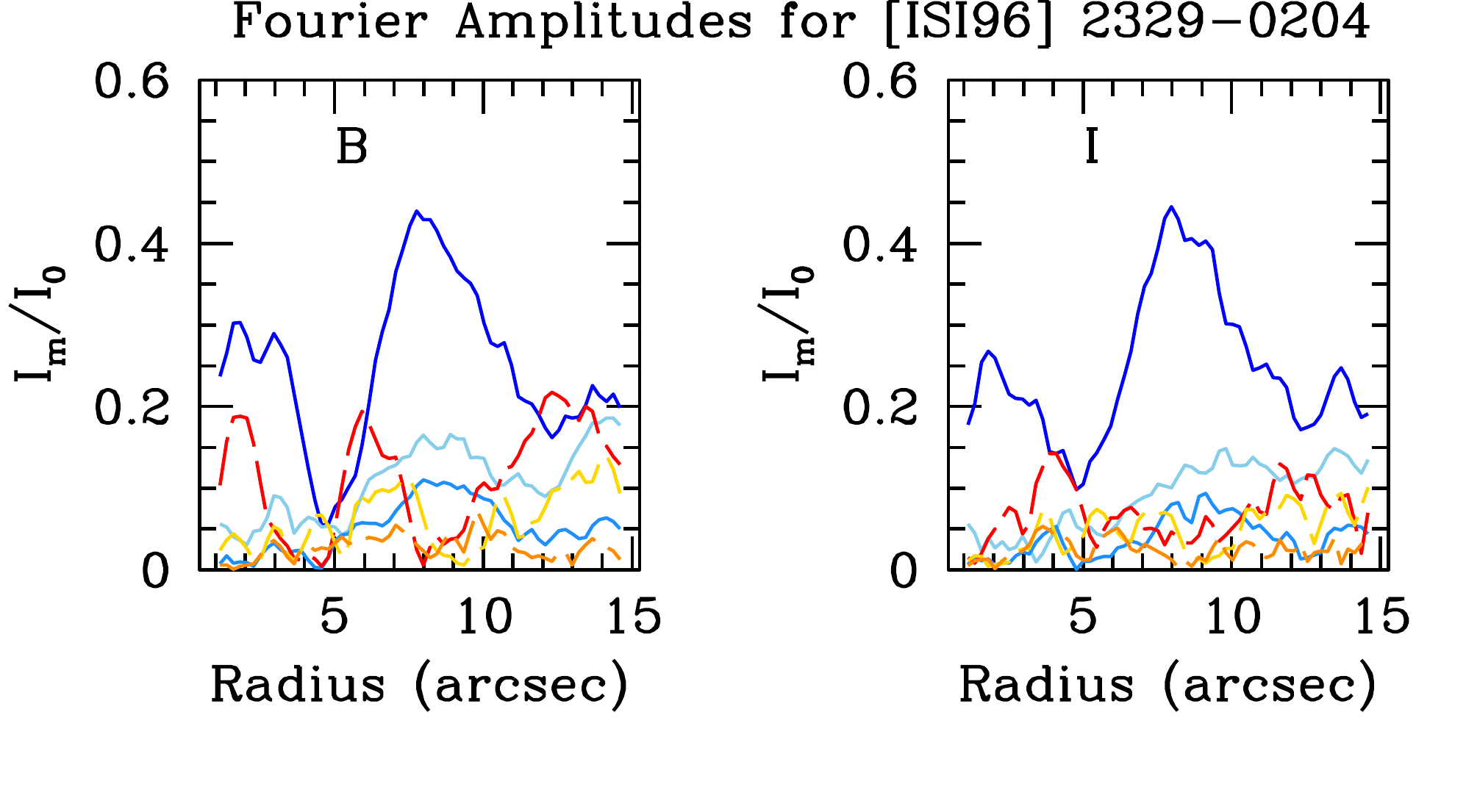} \\
  \caption{Fourier amplitudes for the remainder of our sample.}
  \label{fourAmps}
\end{figure*}

\begin{figure*}
  \centering
  \includegraphics[scale=0.25]{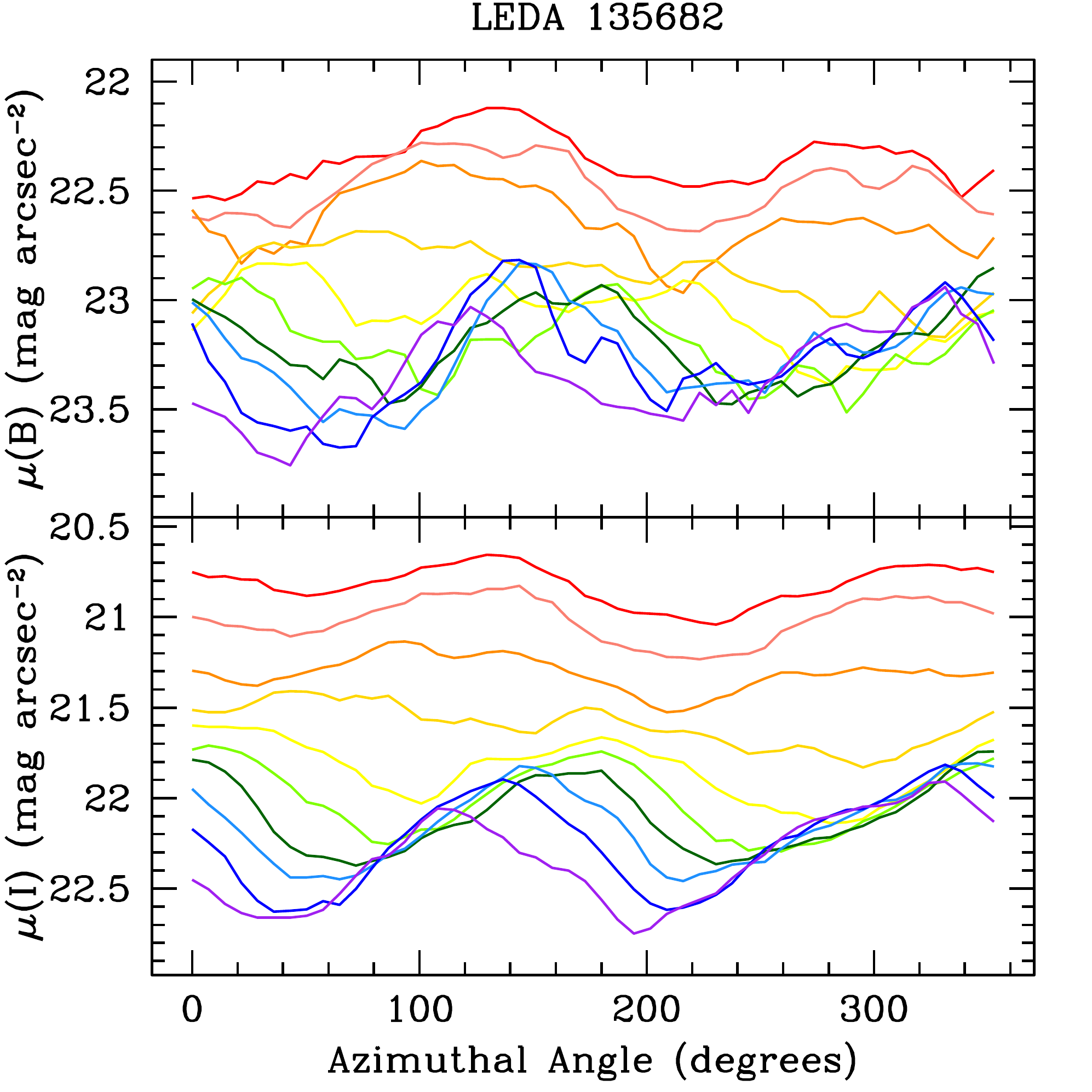} \includegraphics[scale=0.25]{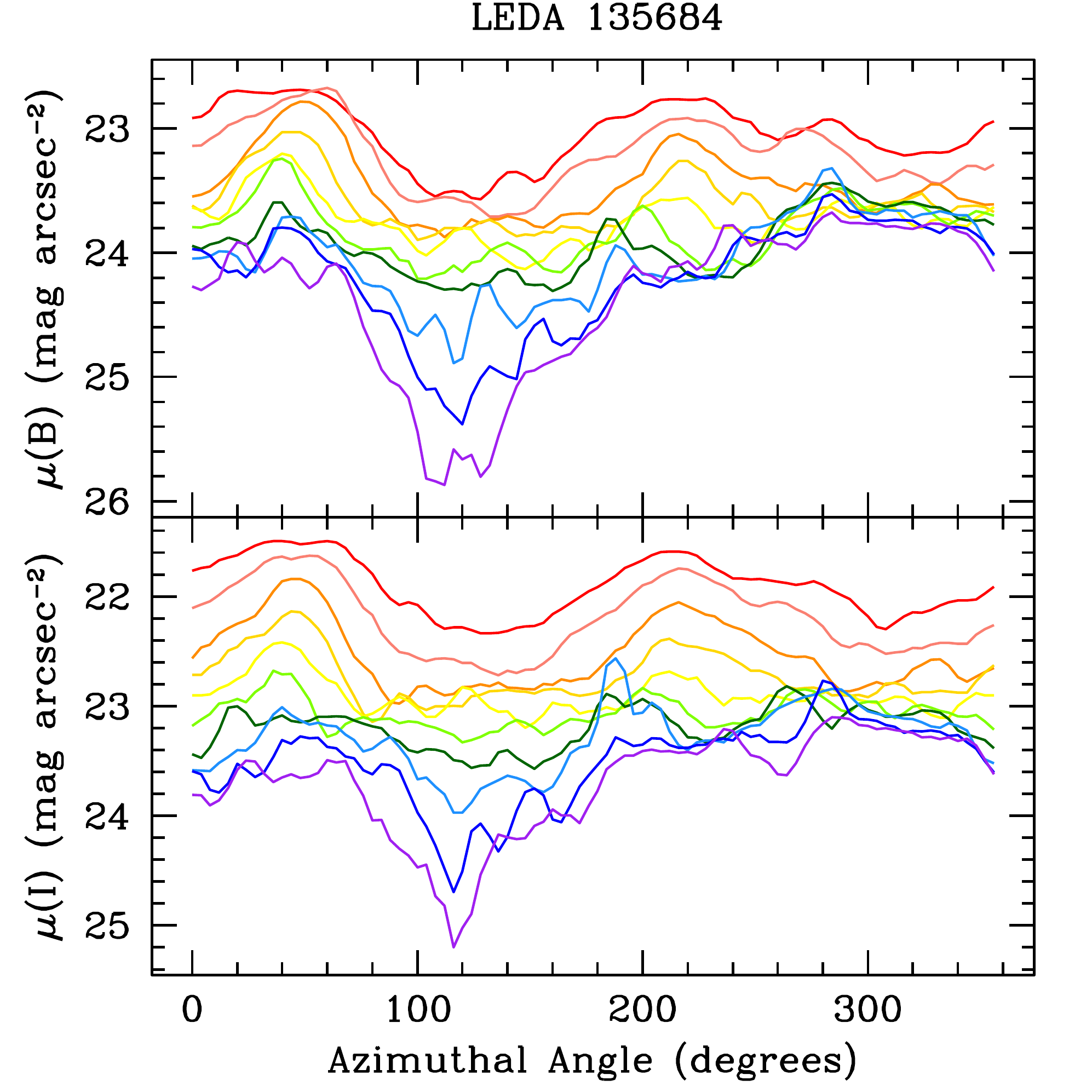} \includegraphics[scale=0.25]{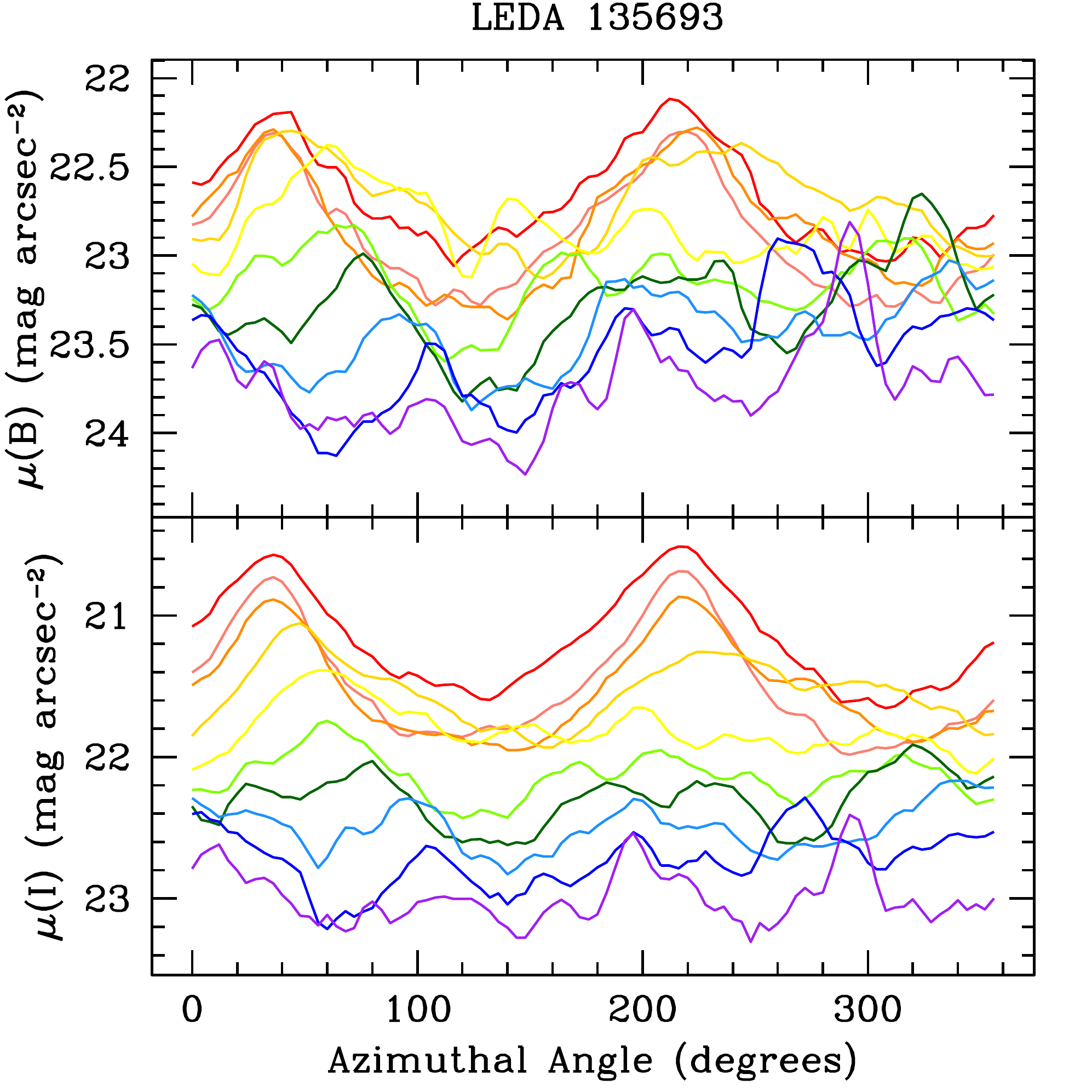} \\
  \includegraphics[scale=0.25]{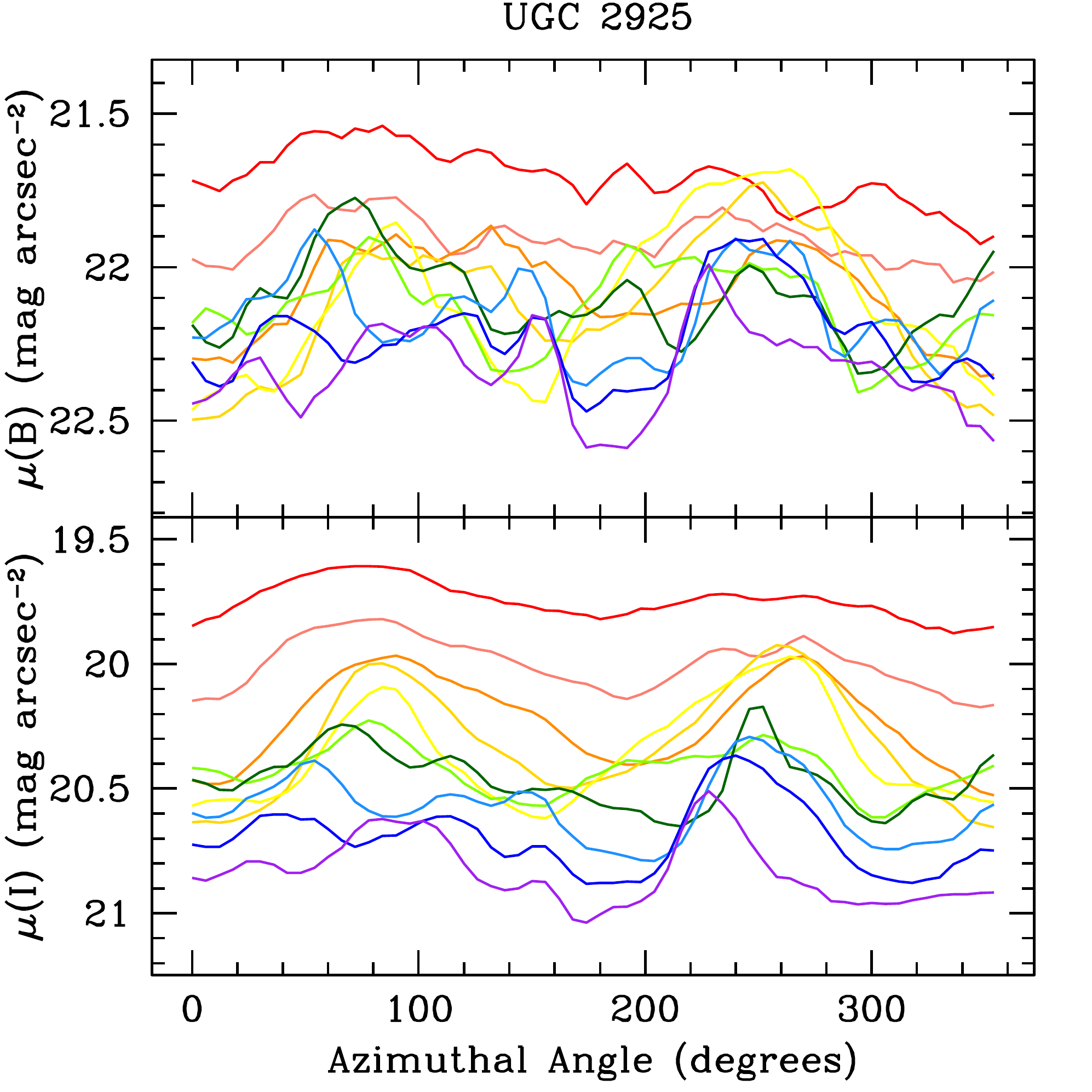} \includegraphics[scale=0.25]{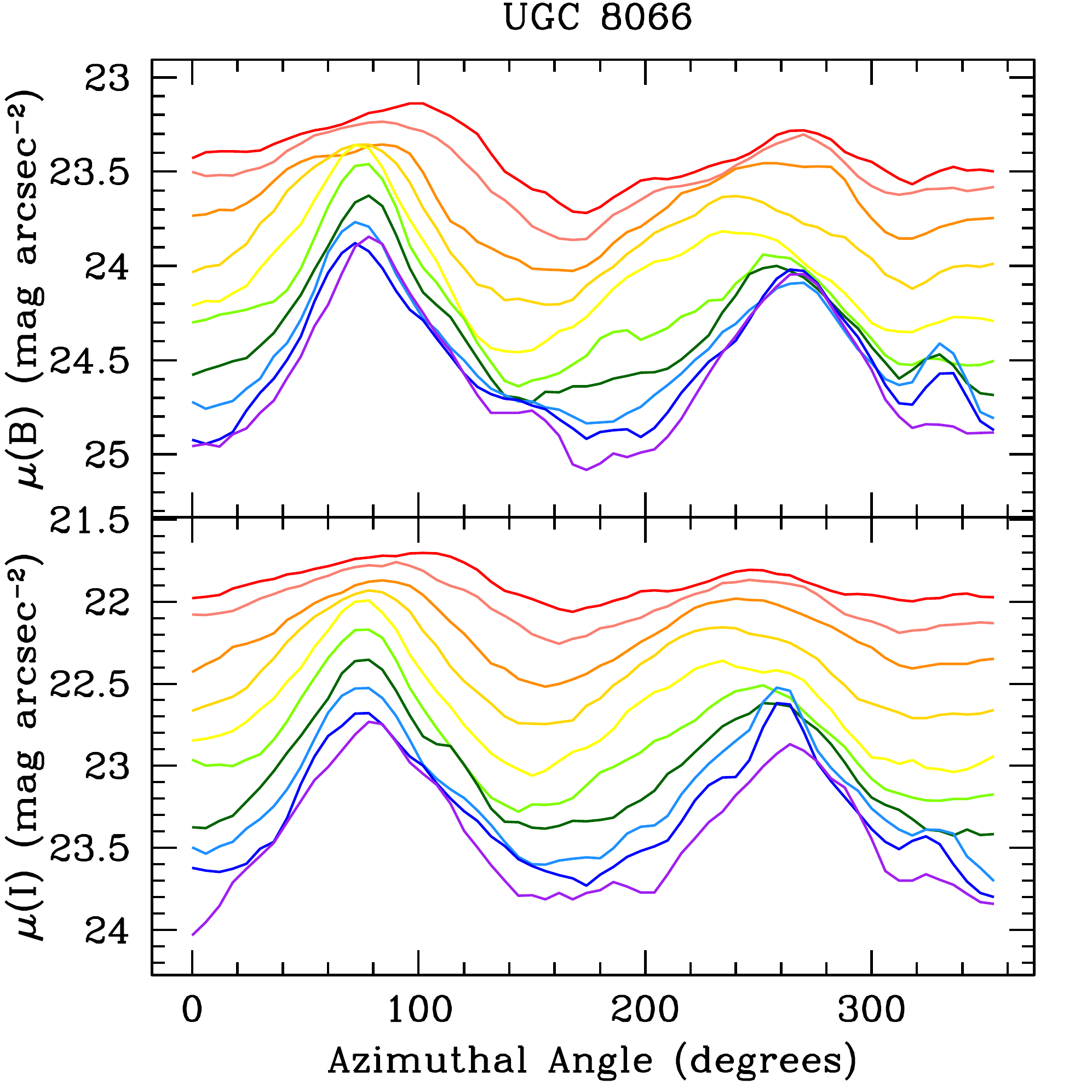} \includegraphics[scale=0.25]{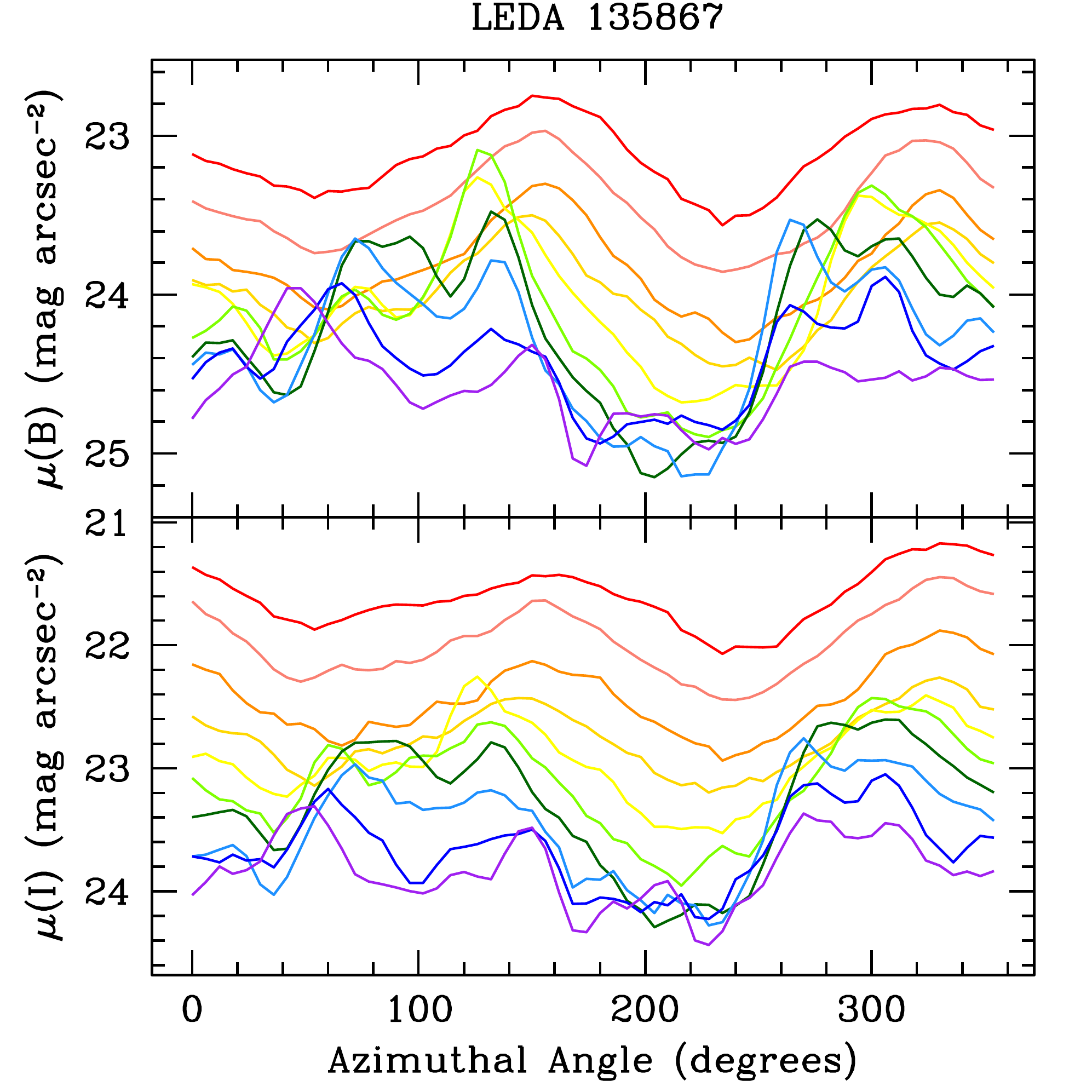} \\
  \includegraphics[scale=0.25]{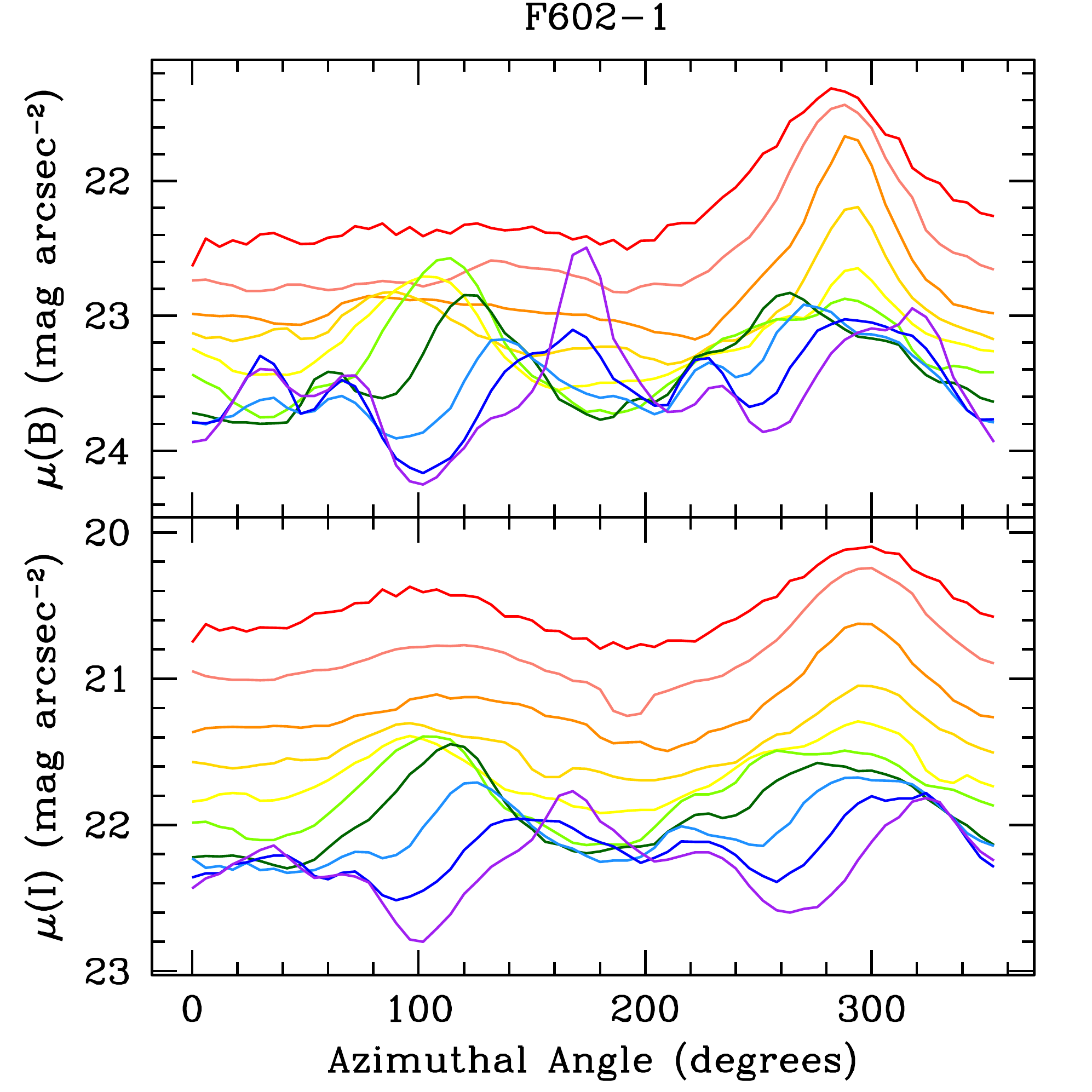} \includegraphics[scale=0.25]{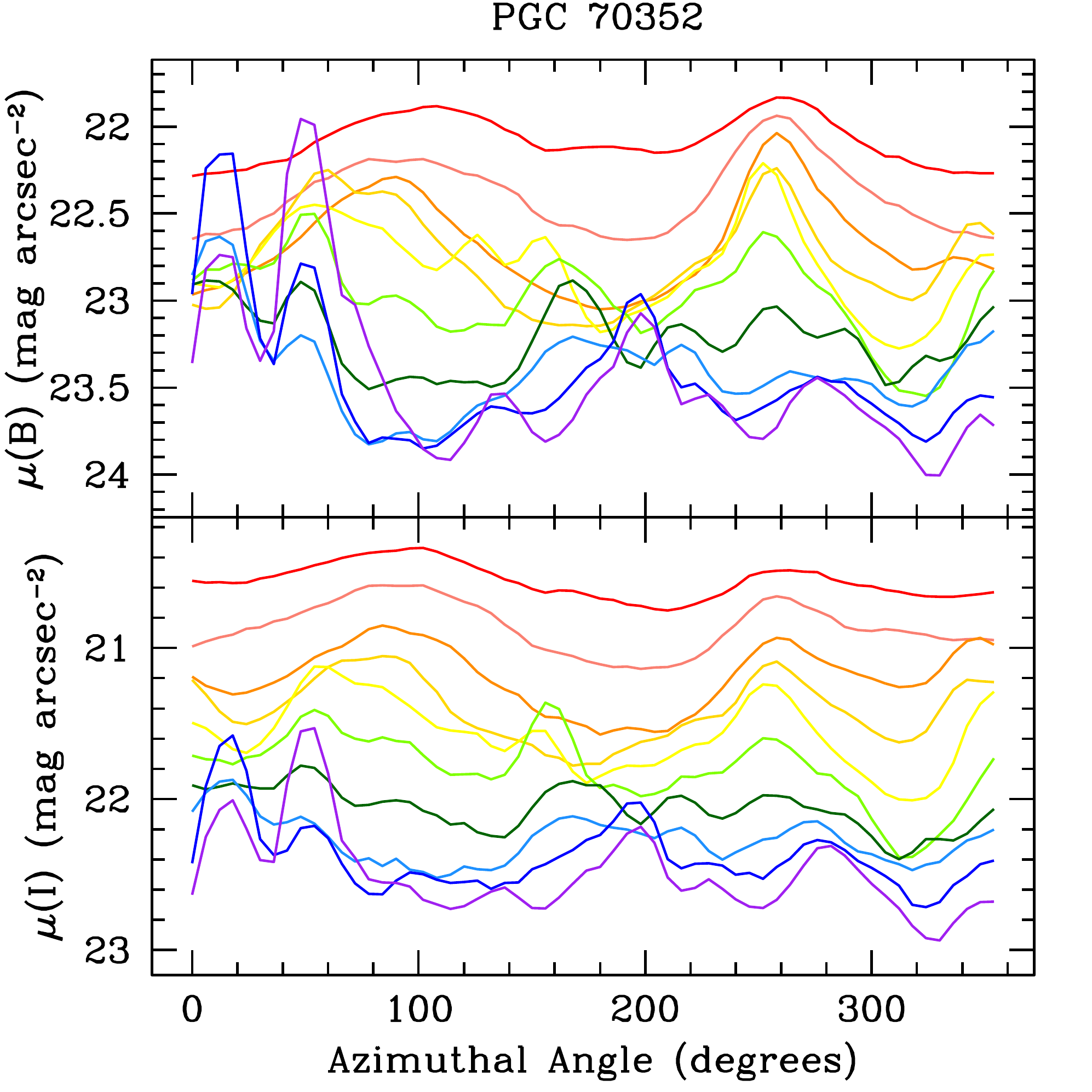} \includegraphics[scale=0.25]{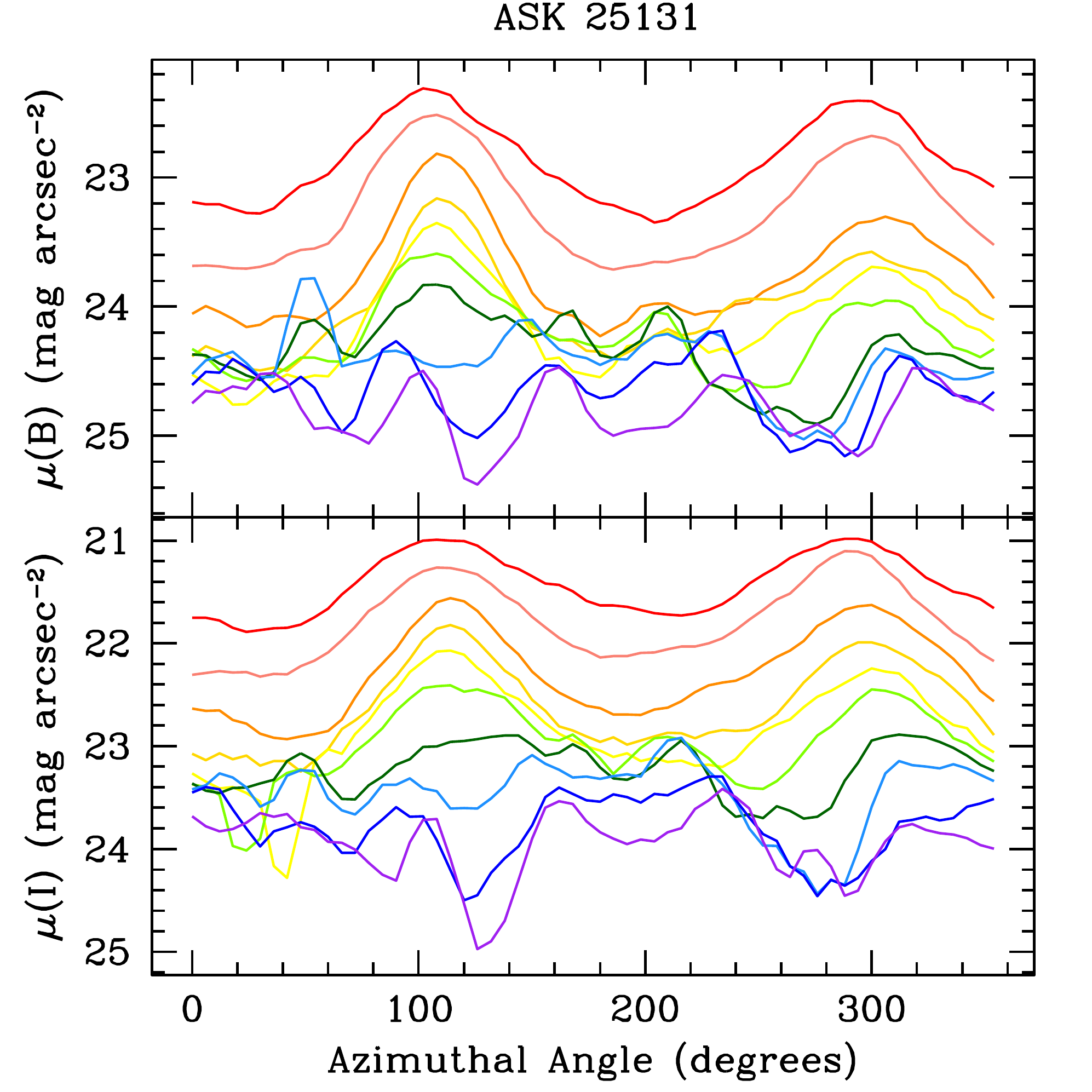} \\
  \includegraphics[scale=0.25]{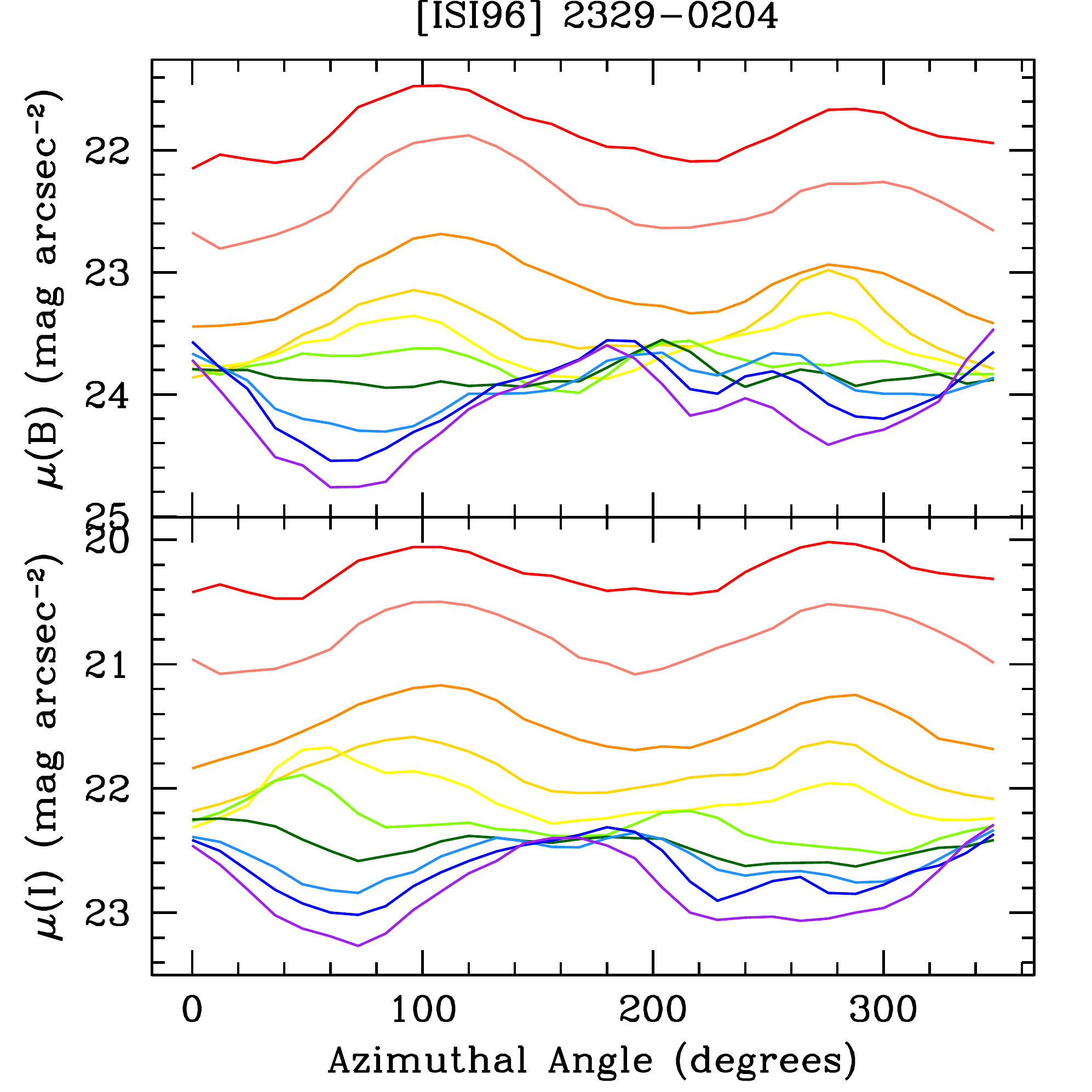} \\
  \caption{Azimuthal light profiles (in mag arcsec$^{-2}$) for the remainder of our sample; the $B$-band profiles are in the top panel and the $I$-band profiles are in the bottom panel for each subfigure.}
  \label{lightProfs}
\end{figure*}

\begin{figure*}
  \centering
  \includegraphics[scale=0.25]{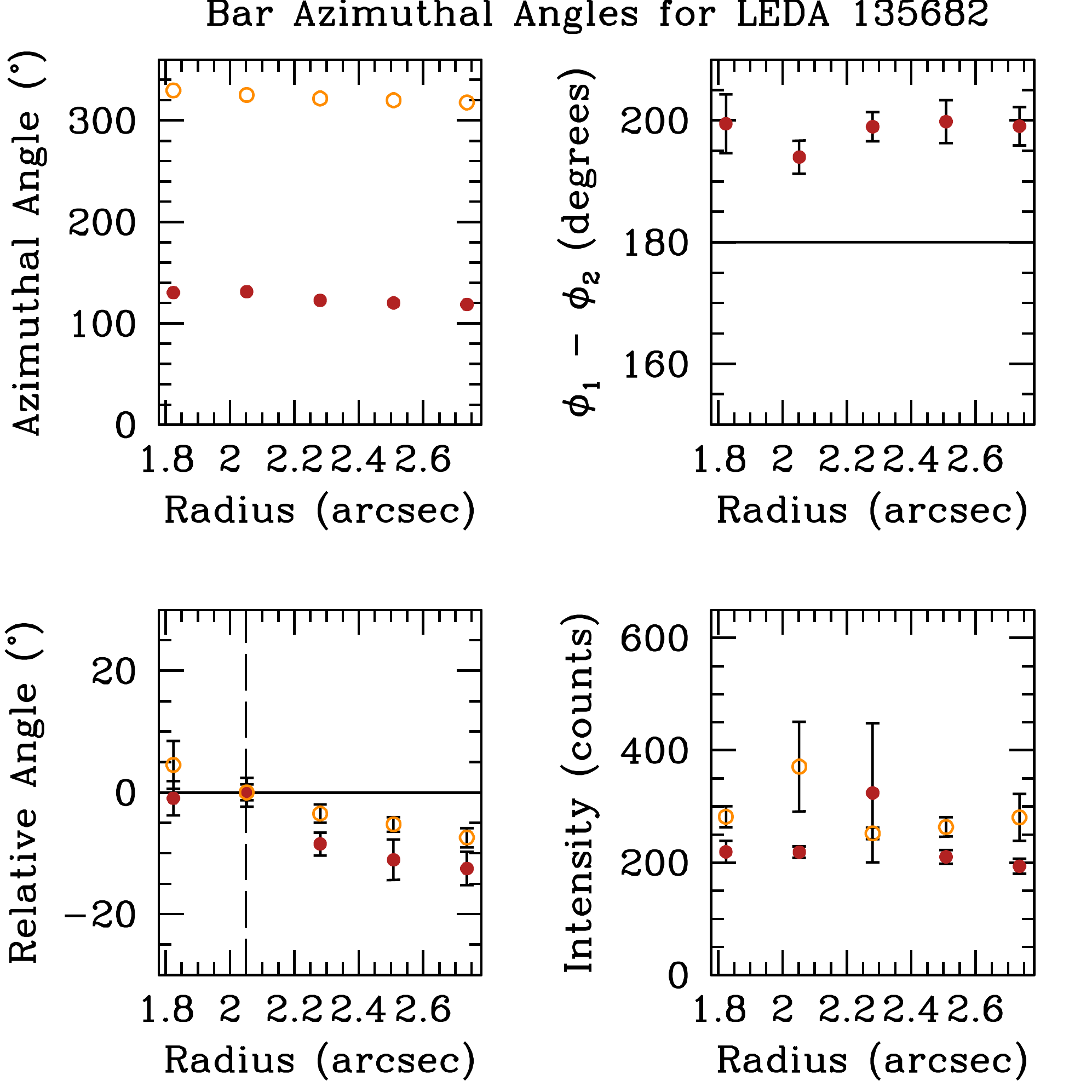} \hspace{0.01\textwidth} \includegraphics[scale=0.25]{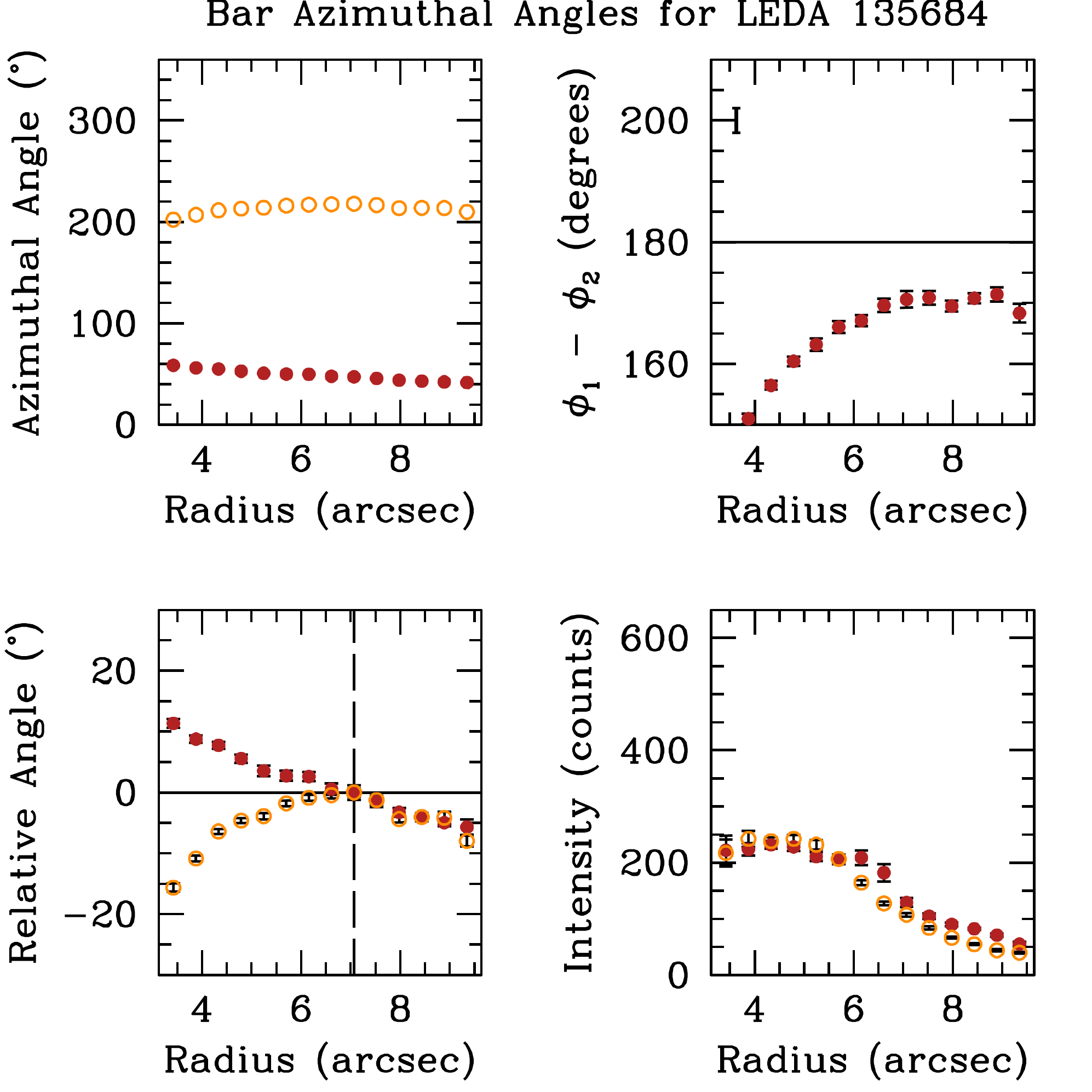} \hspace{0.01\textwidth} \includegraphics[scale=0.25]{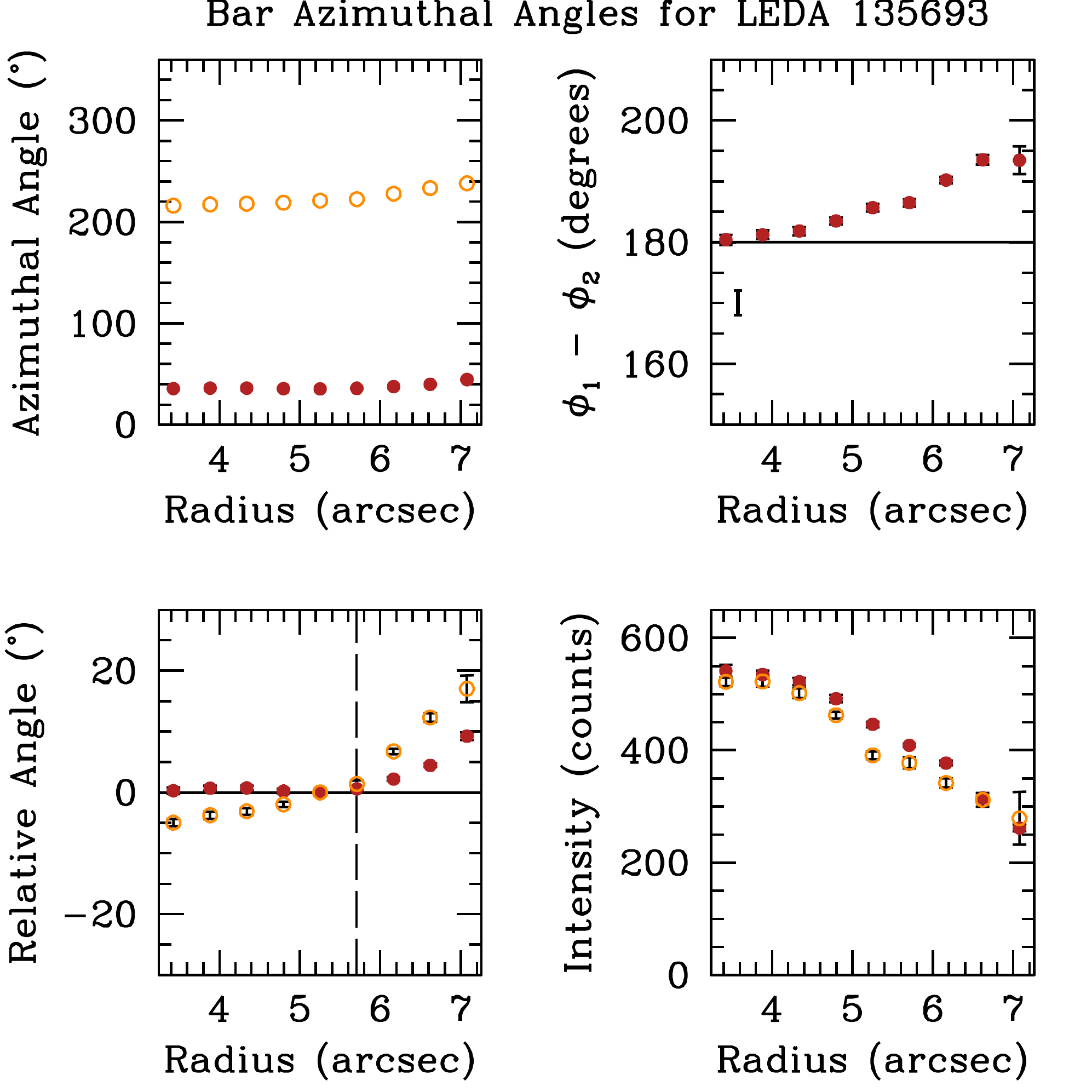} \\
  \vspace{0.05\textwidth}
  \includegraphics[scale=0.25]{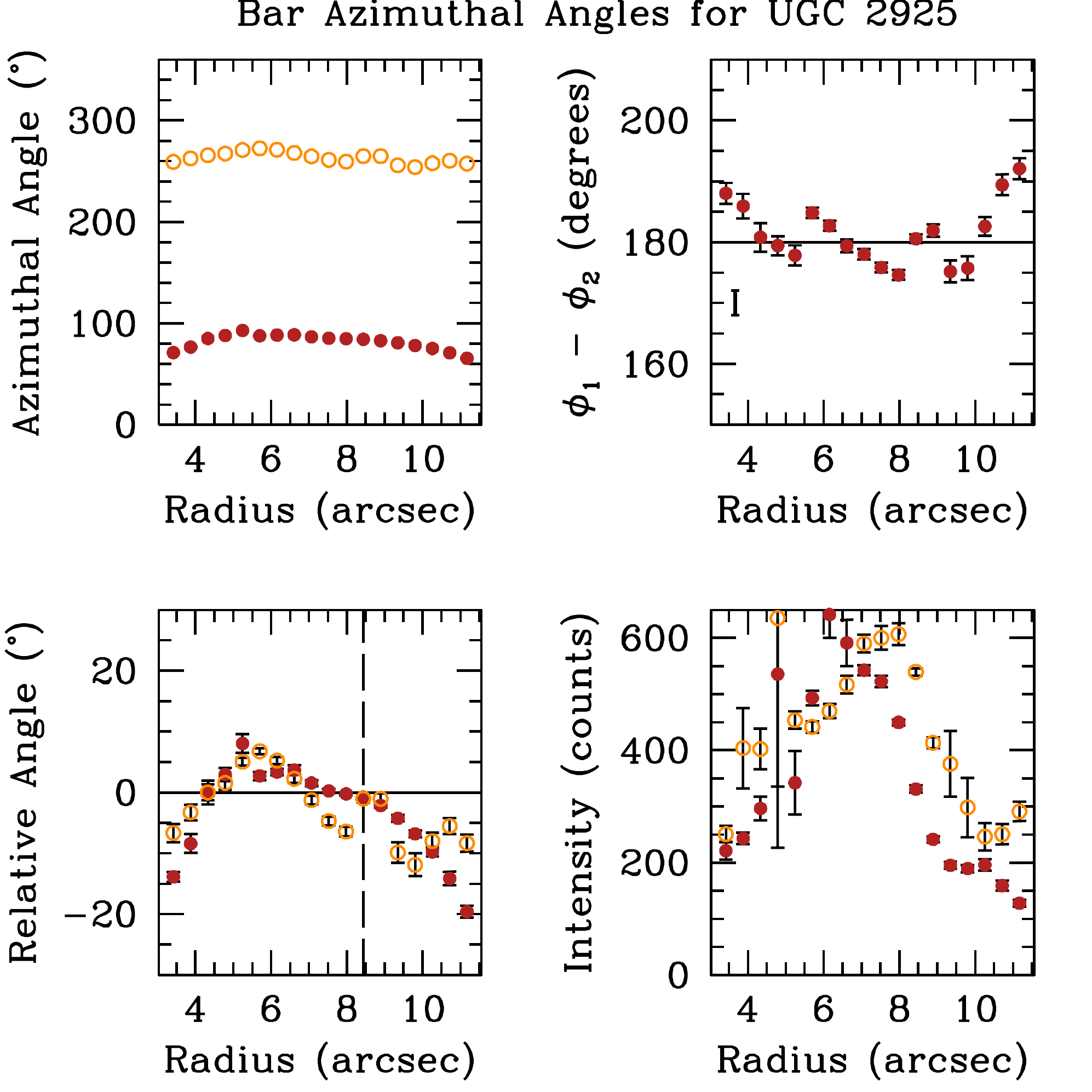} \hspace{0.01\textwidth} \includegraphics[scale=0.25]{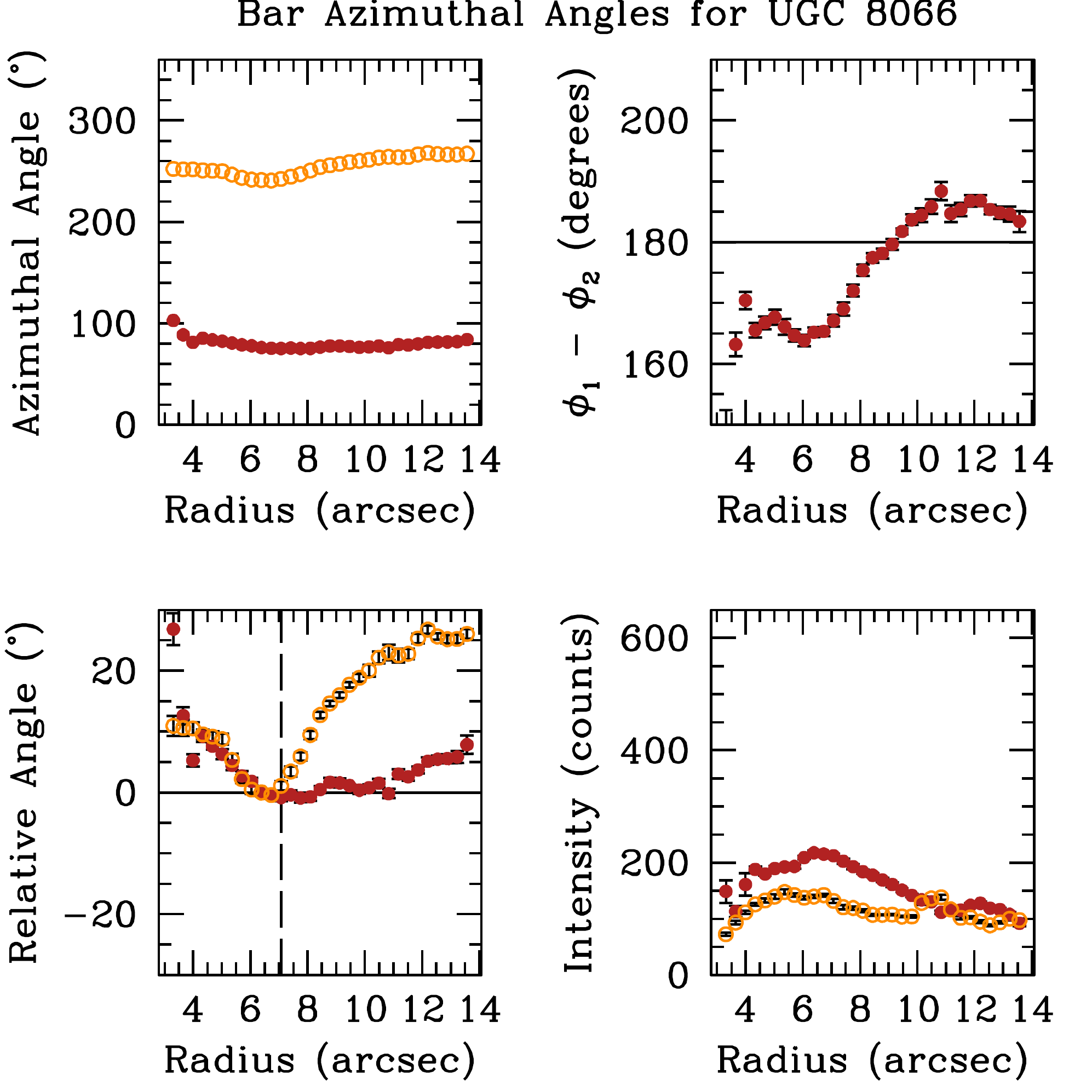} \hspace{0.01\textwidth} \includegraphics[scale=0.25]{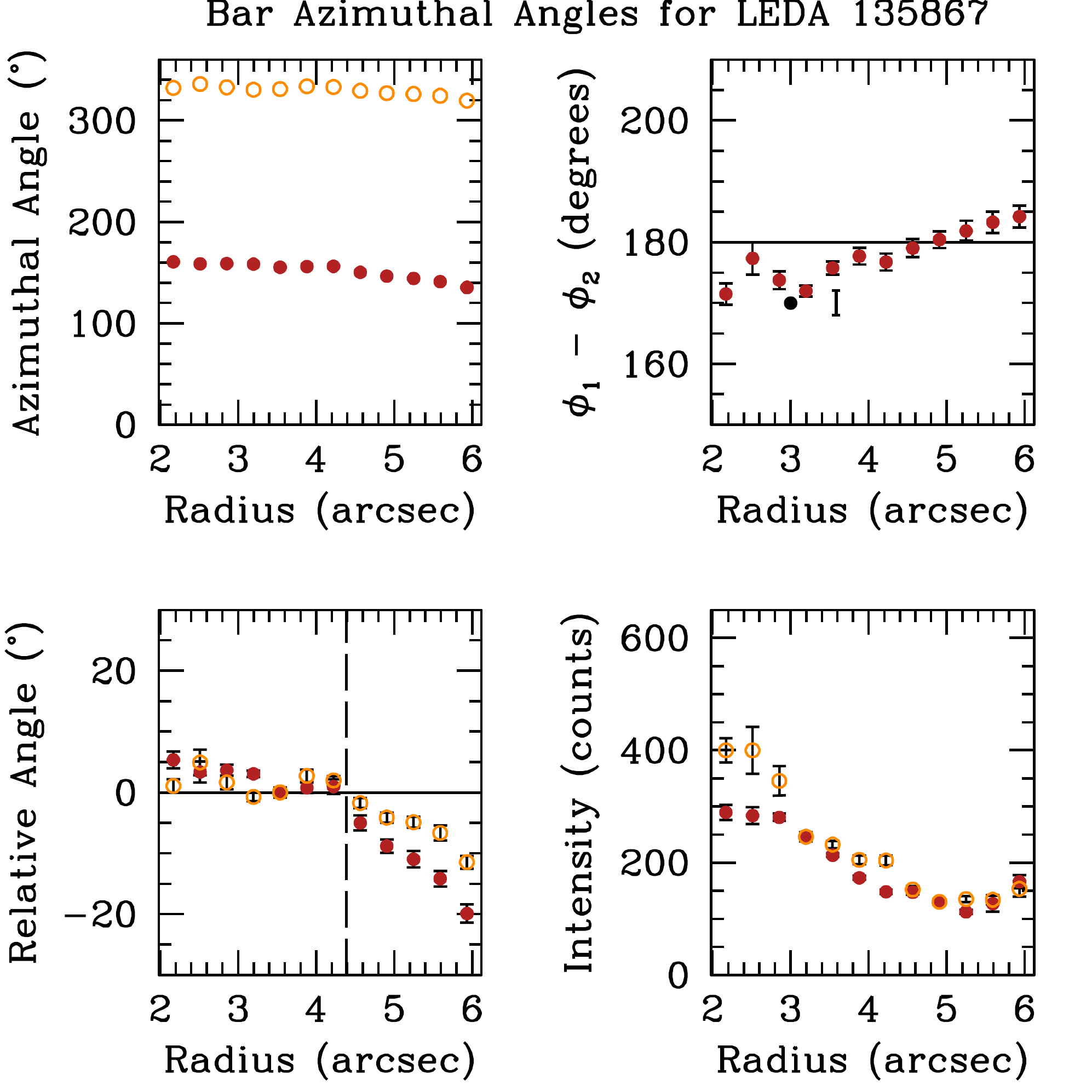} \\
  \vspace{0.05\textwidth}
  \includegraphics[scale=0.25]{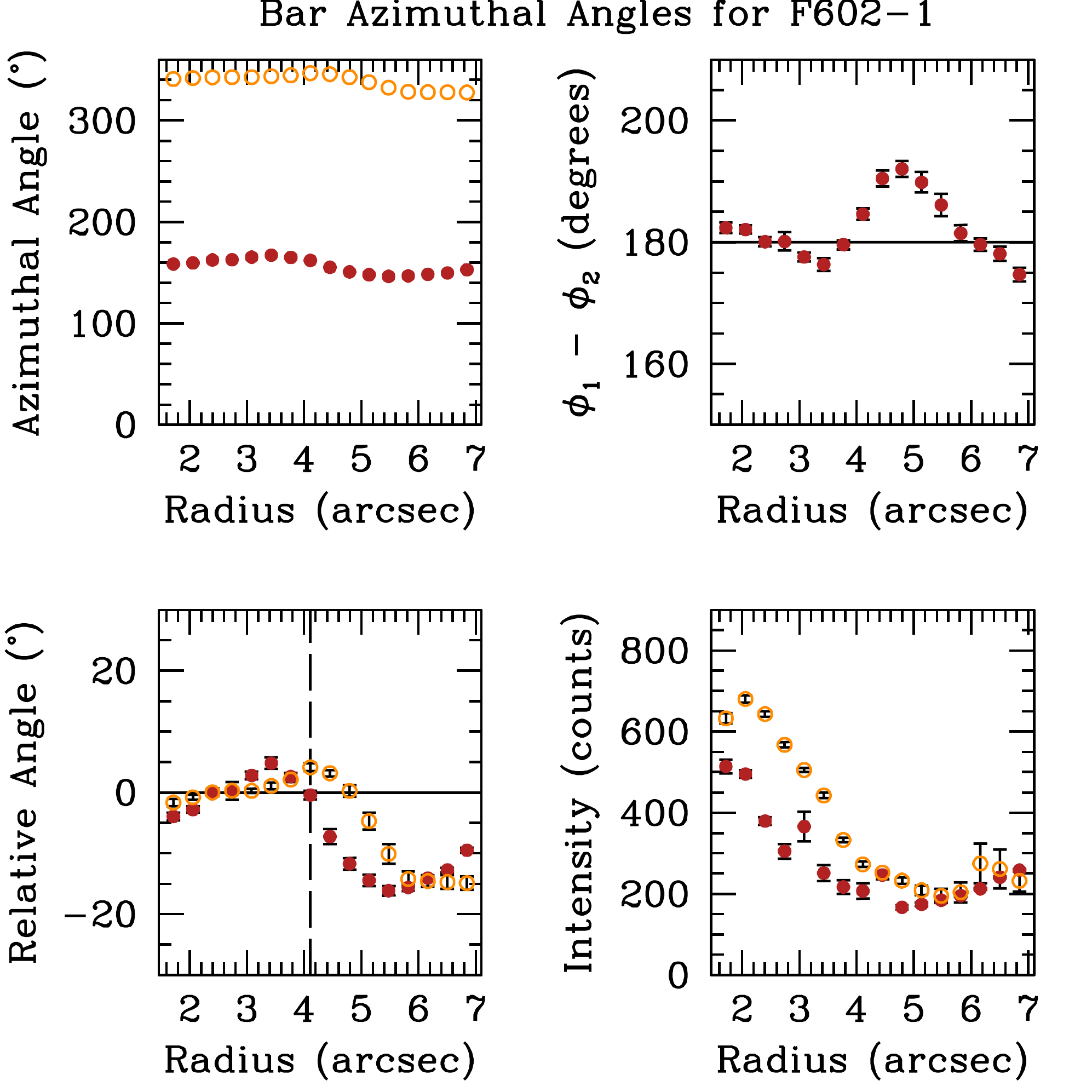} \hspace{0.01\textwidth} \includegraphics[scale=0.25]{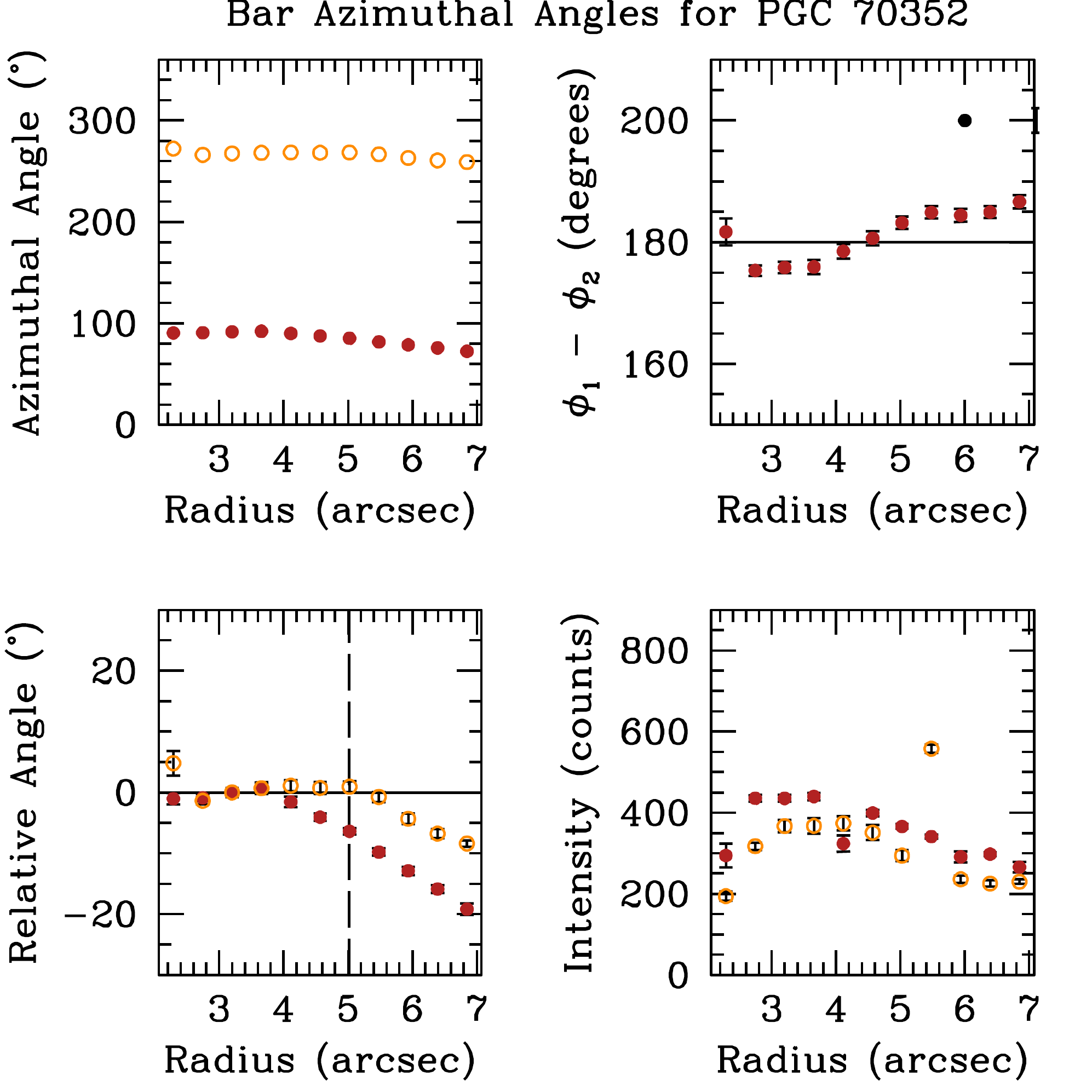} \hspace{0.01\textwidth} \includegraphics[scale=0.25]{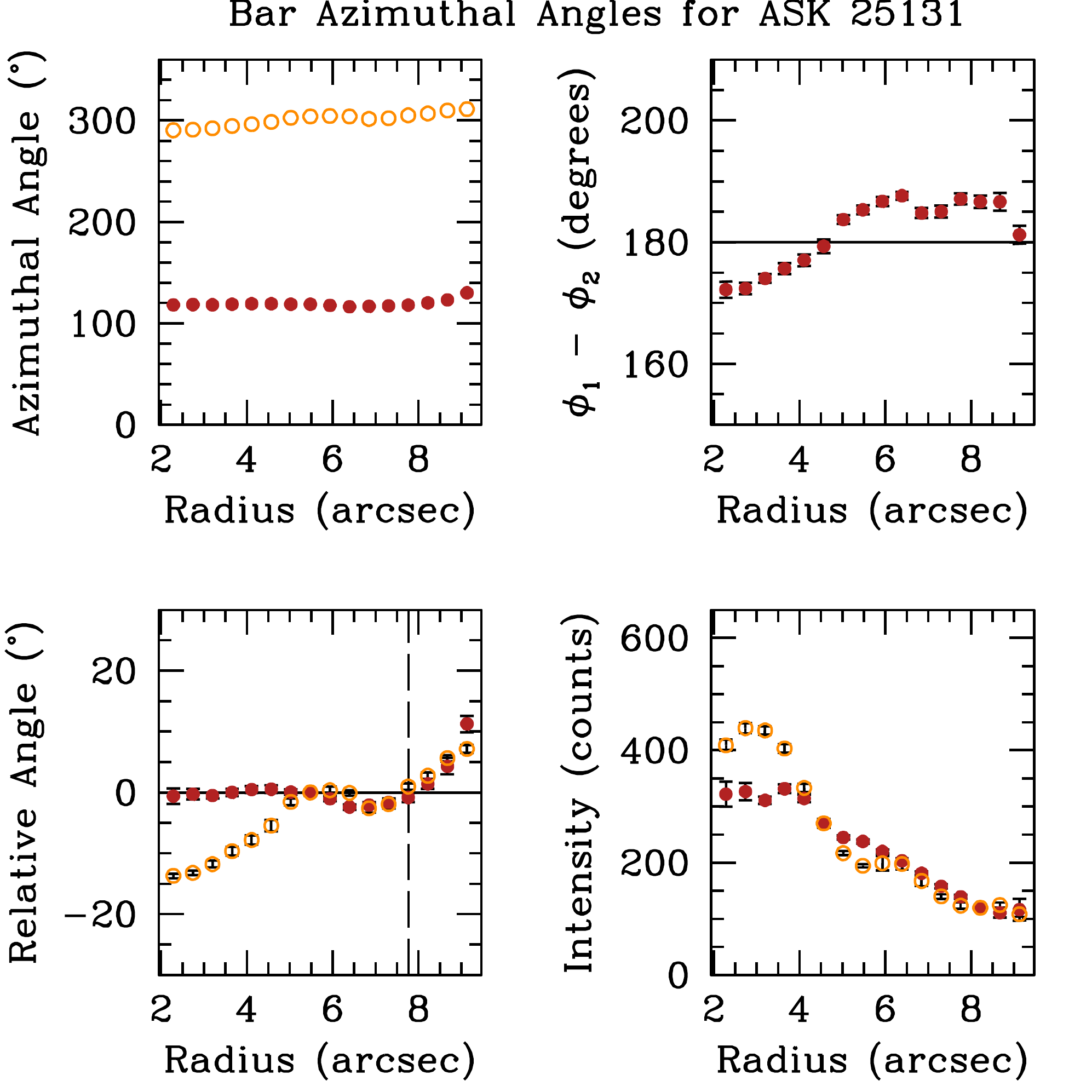} \\
  \vspace{0.05\textwidth}
  \includegraphics[scale=0.25]{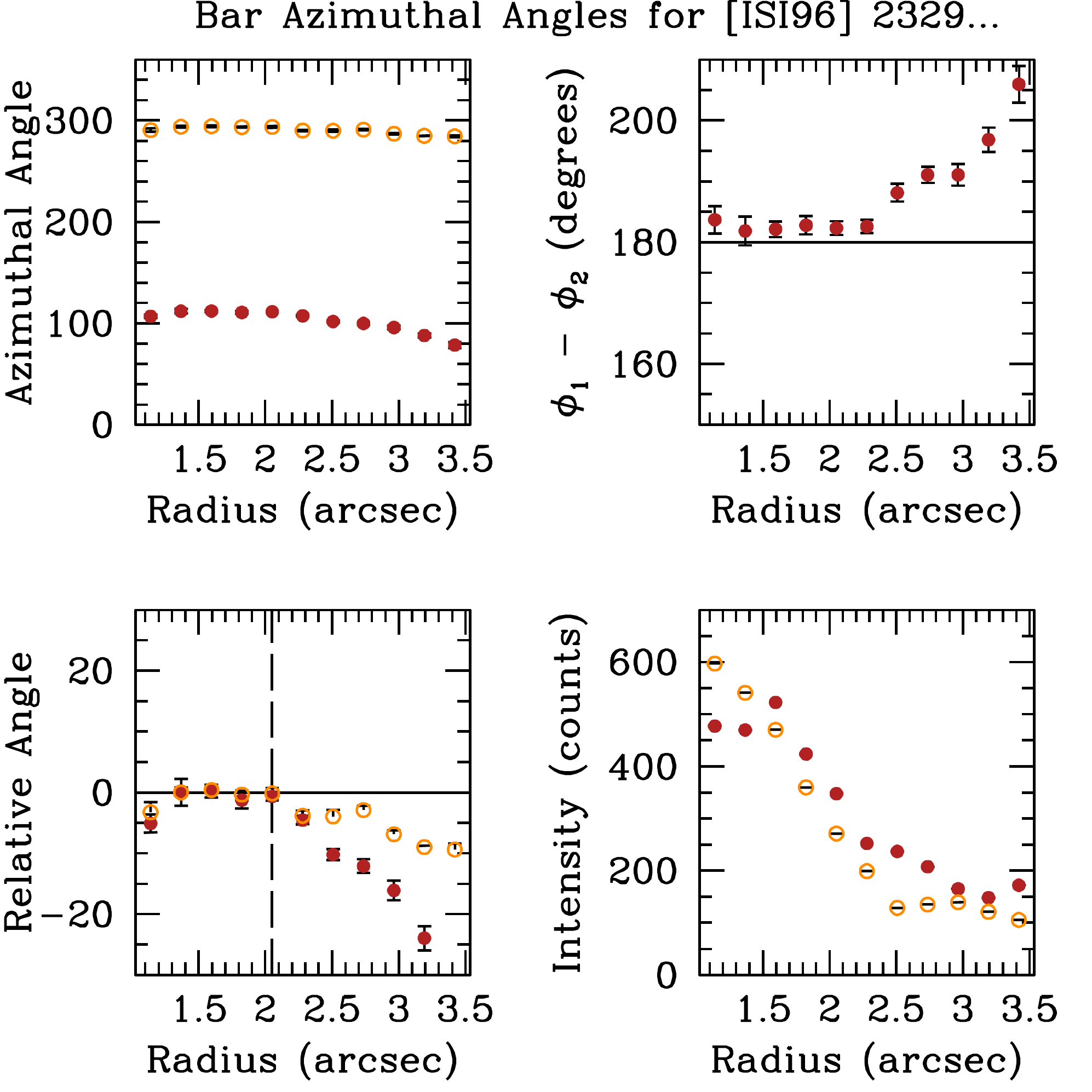} \\
  \caption{Azimuthal positions of the bars in our galaxies. Dashed vertical lines in the bottom left panels indicate the azimuthal bar length ($R_{az}$).}
  \label{azBars}
\end{figure*}

\begin{figure*}
  \centering
  \includegraphics[scale=0.25]{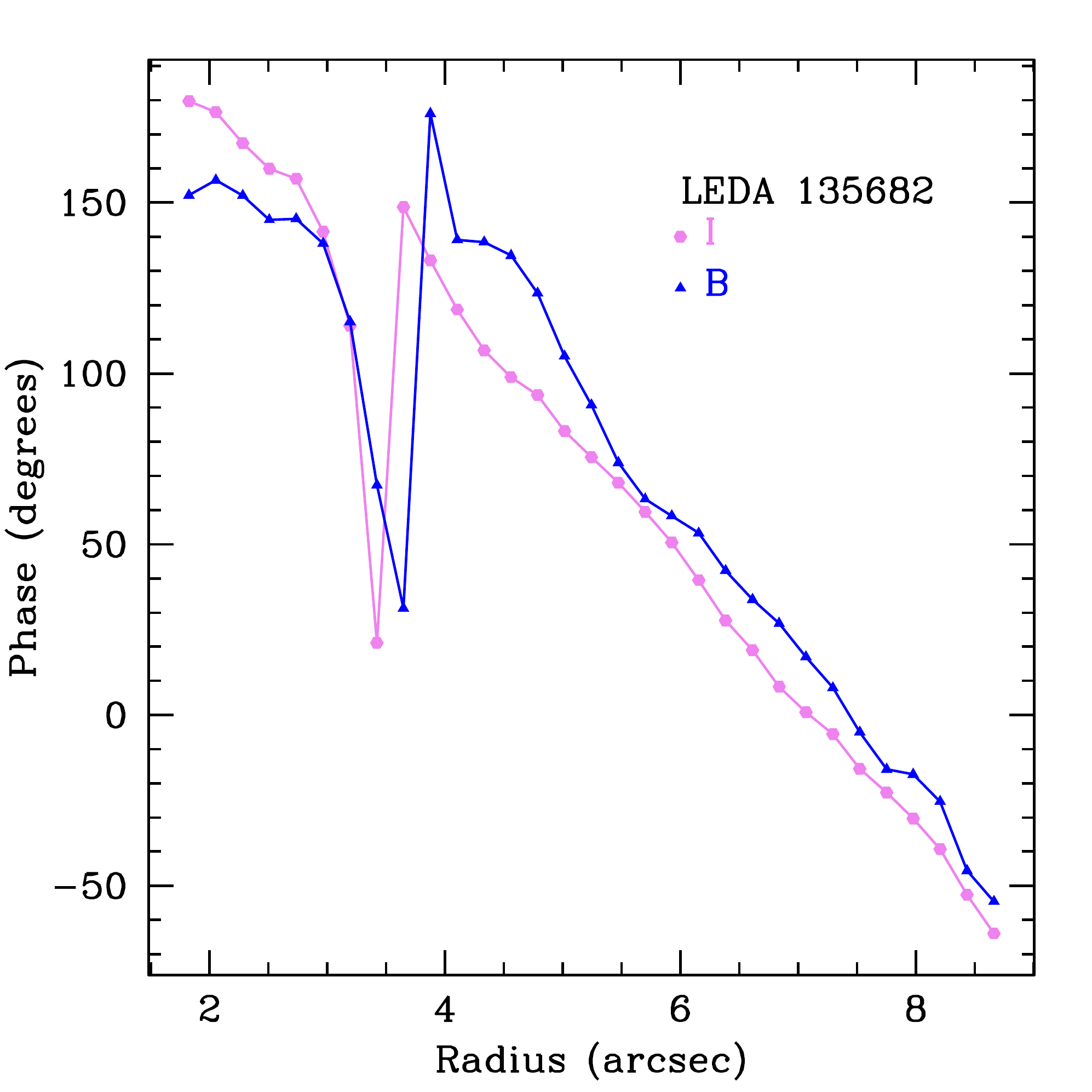} \includegraphics[scale=0.25]{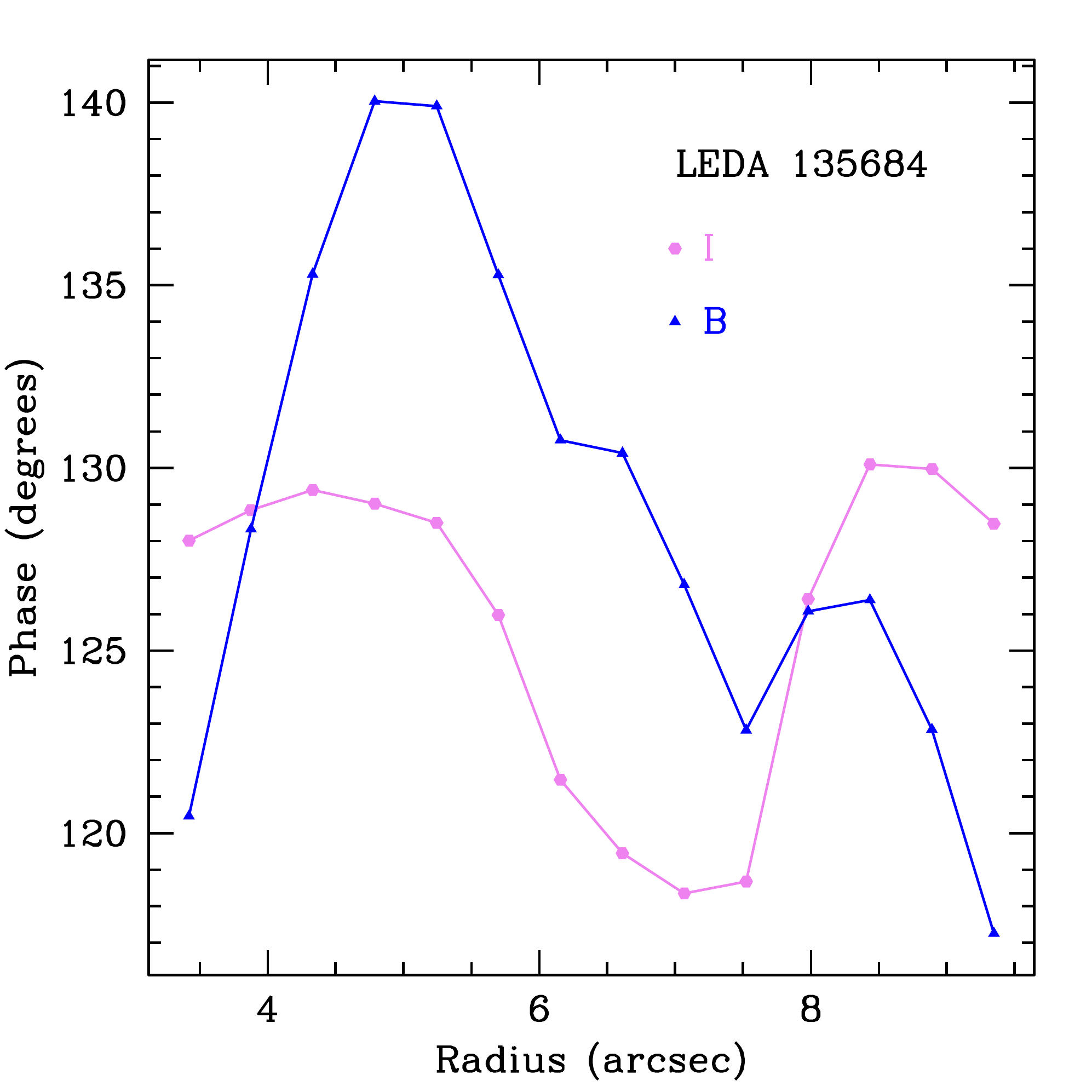} \includegraphics[scale=0.25]{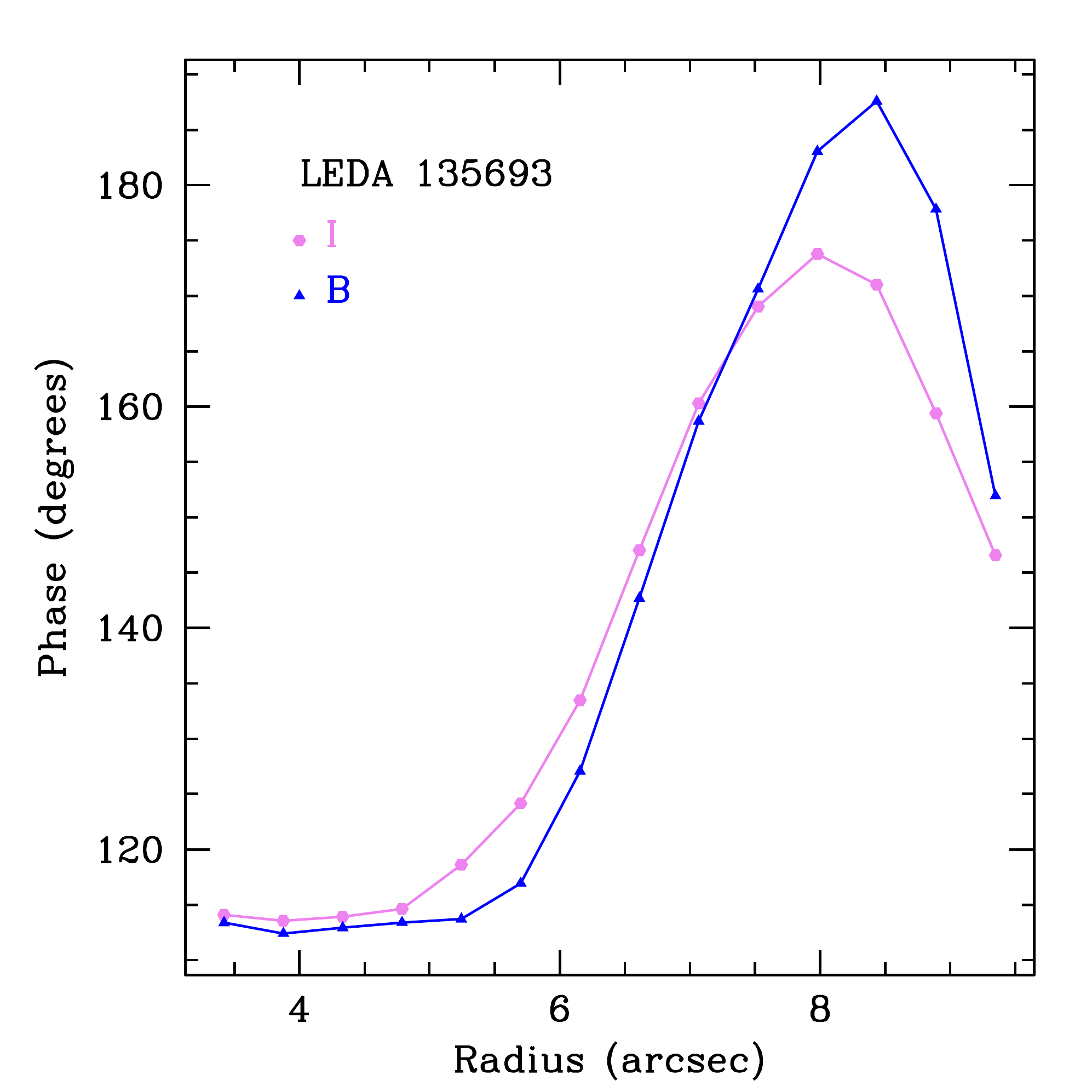} \\
  \includegraphics[scale=0.25]{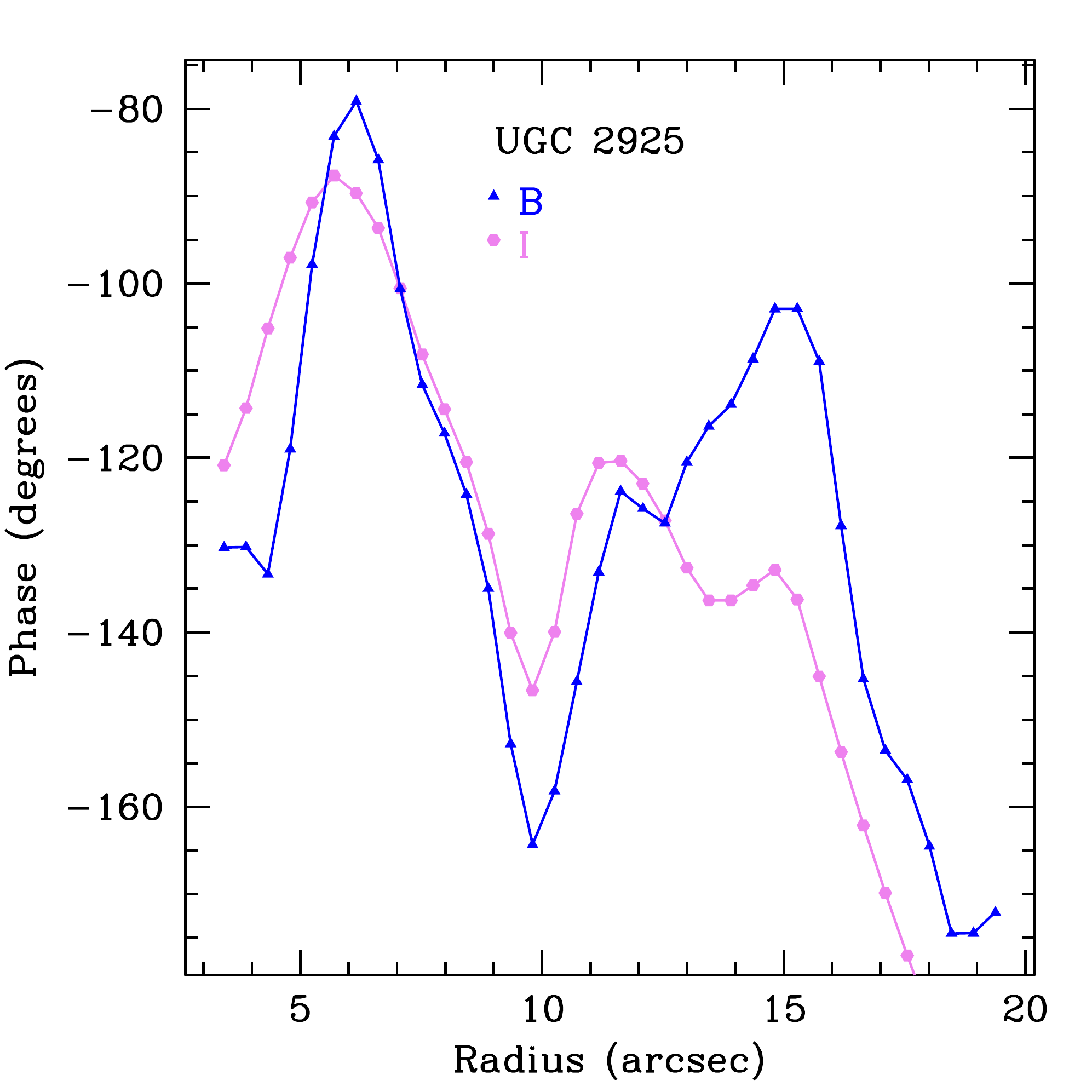} \includegraphics[scale=0.25]{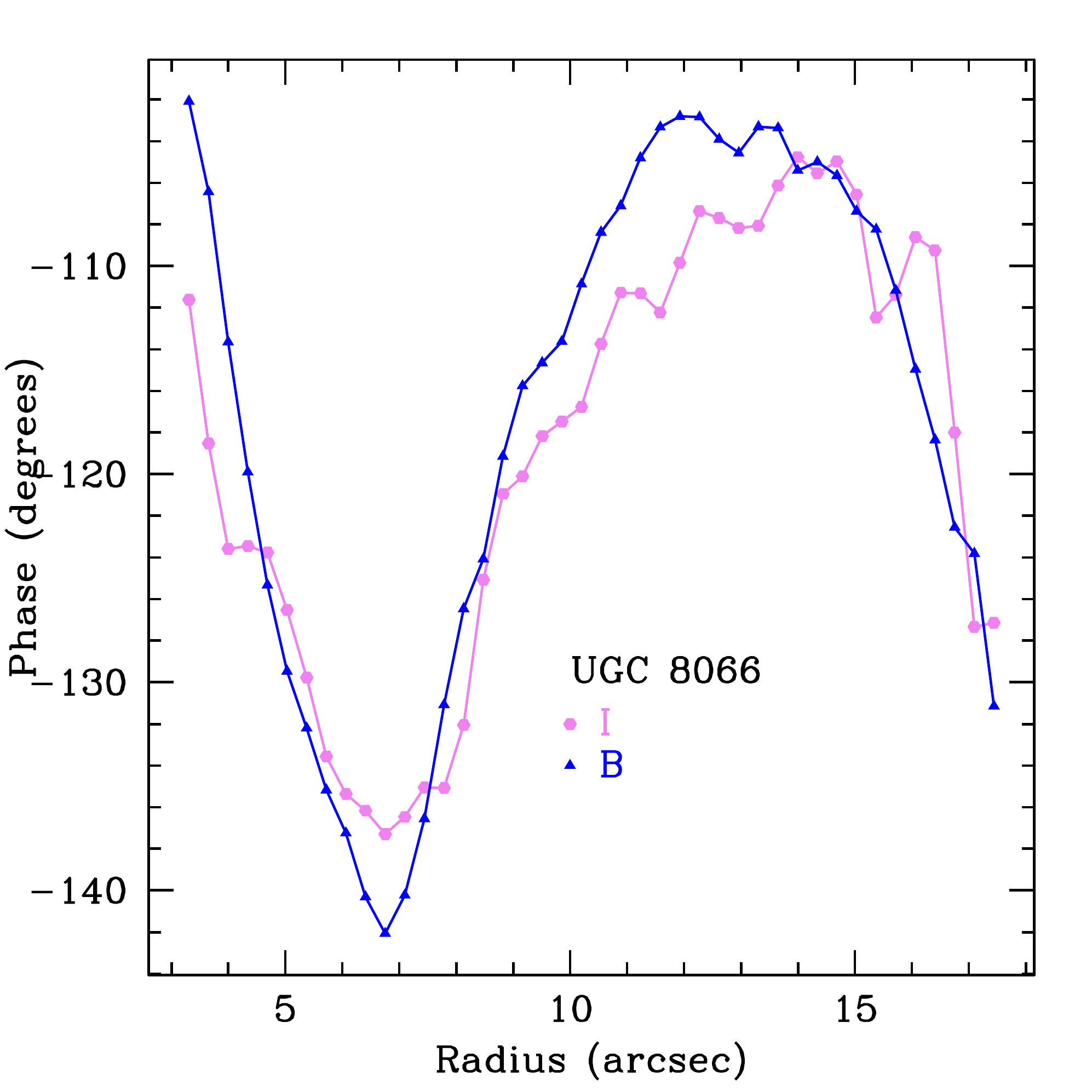} \includegraphics[scale=0.25]{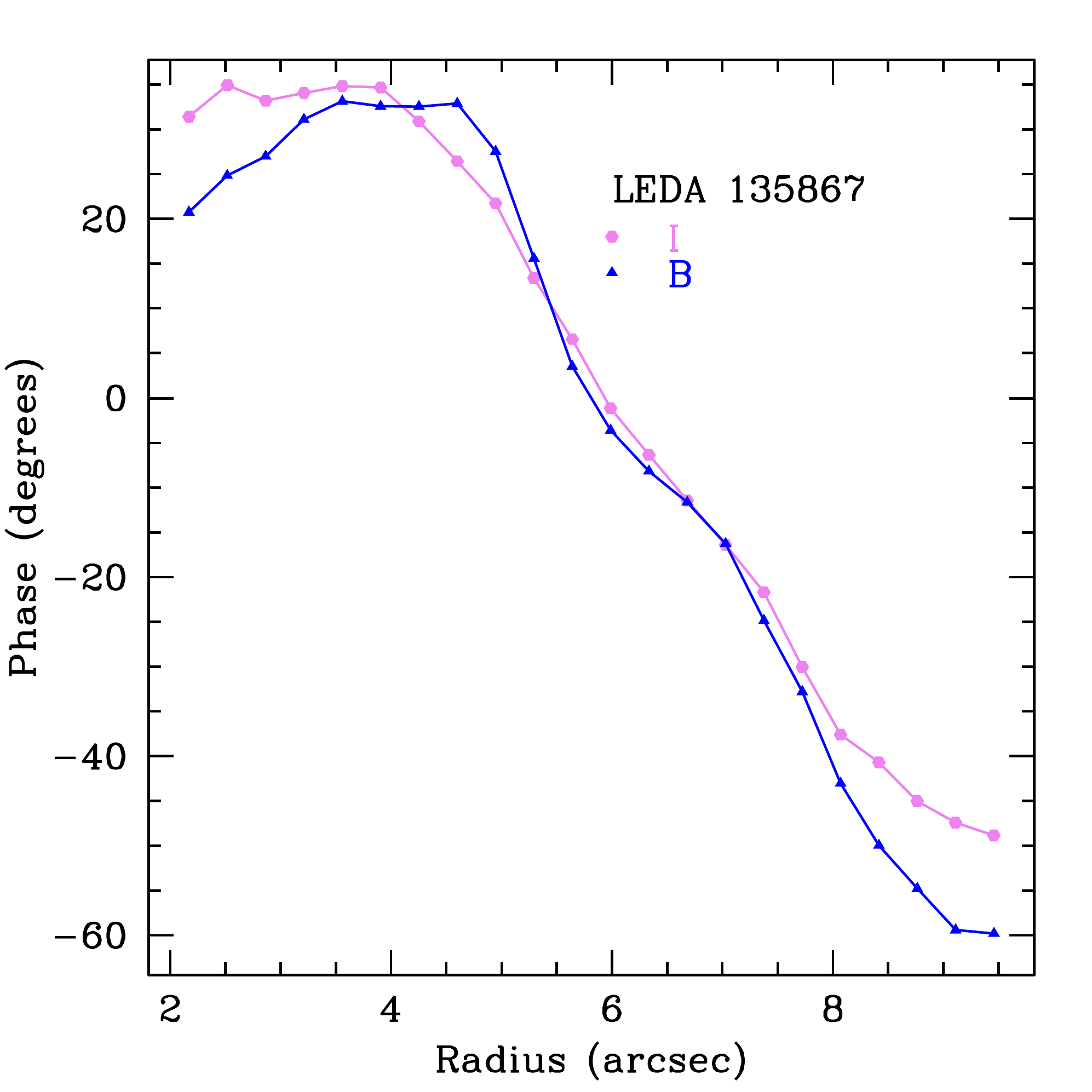} \\
  \includegraphics[scale=0.25]{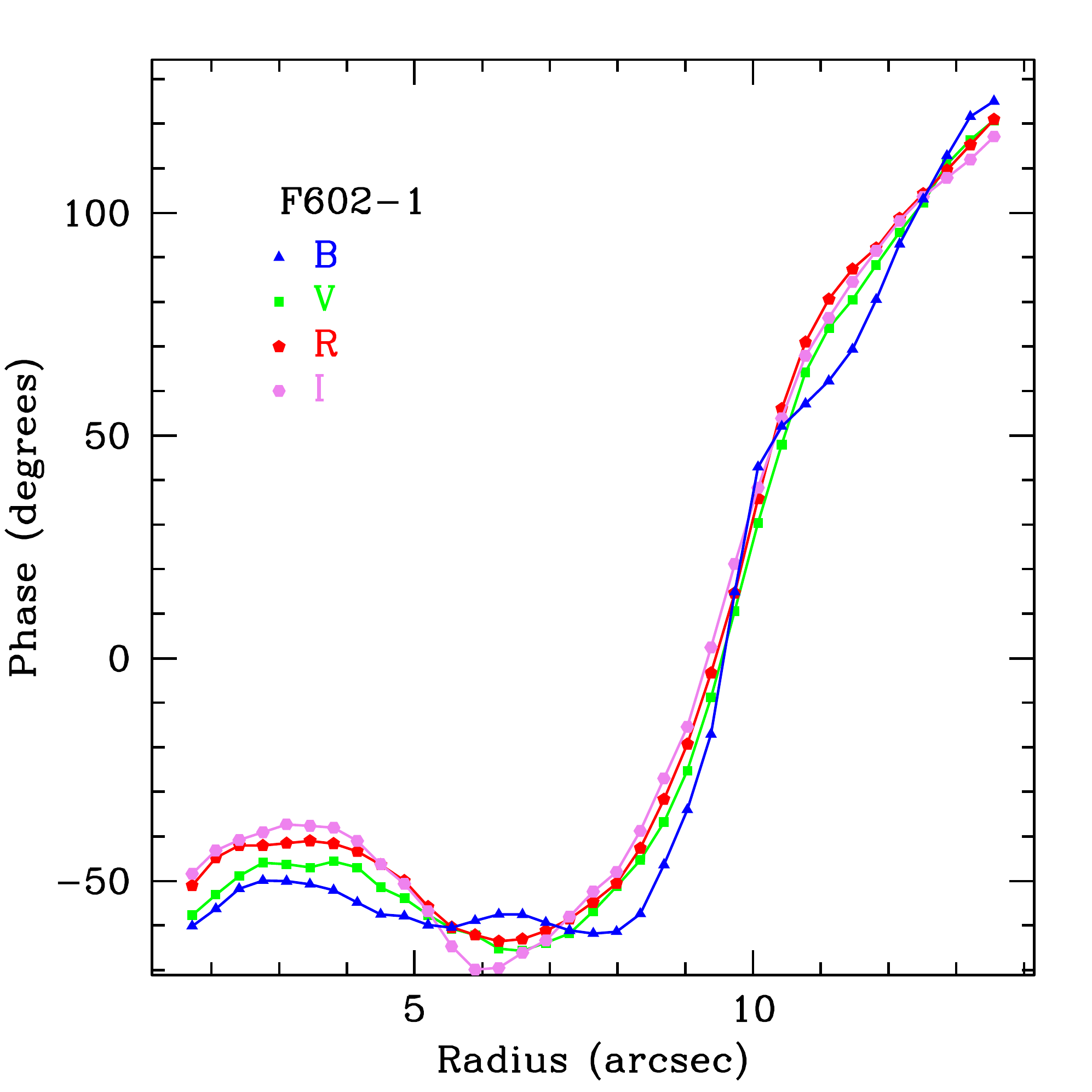} \includegraphics[scale=0.25]{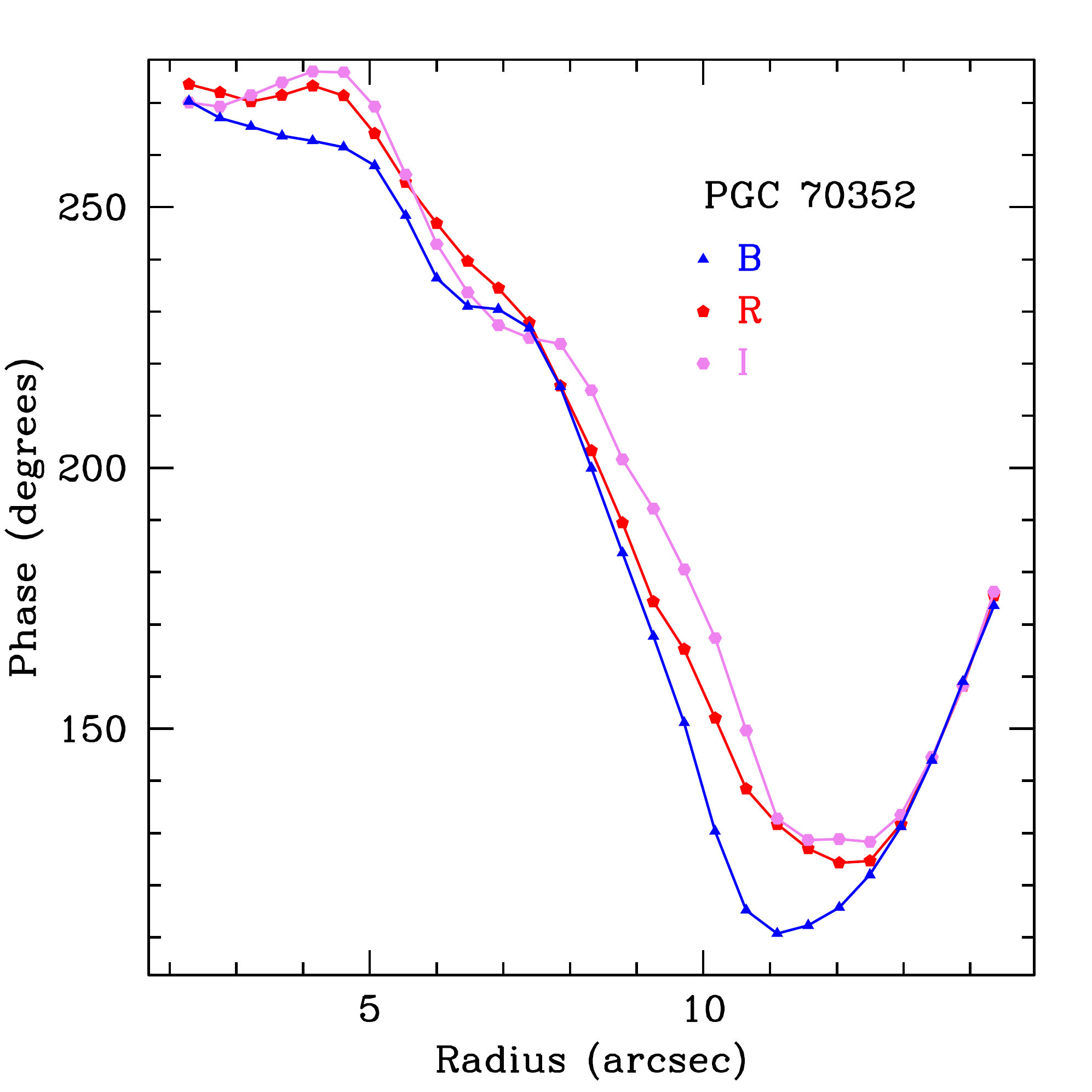} \includegraphics[scale=0.25]{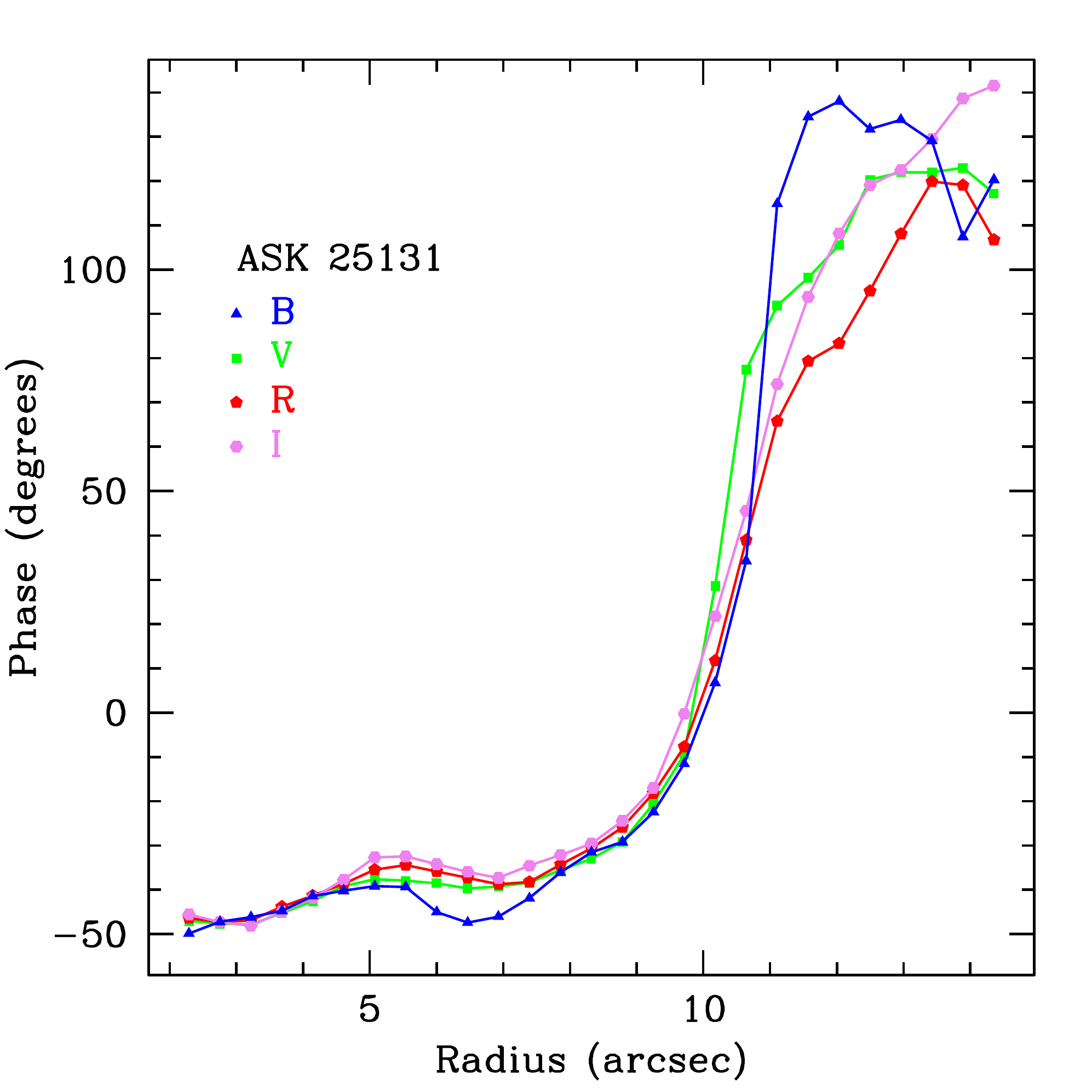} \\
  \includegraphics[scale=0.25]{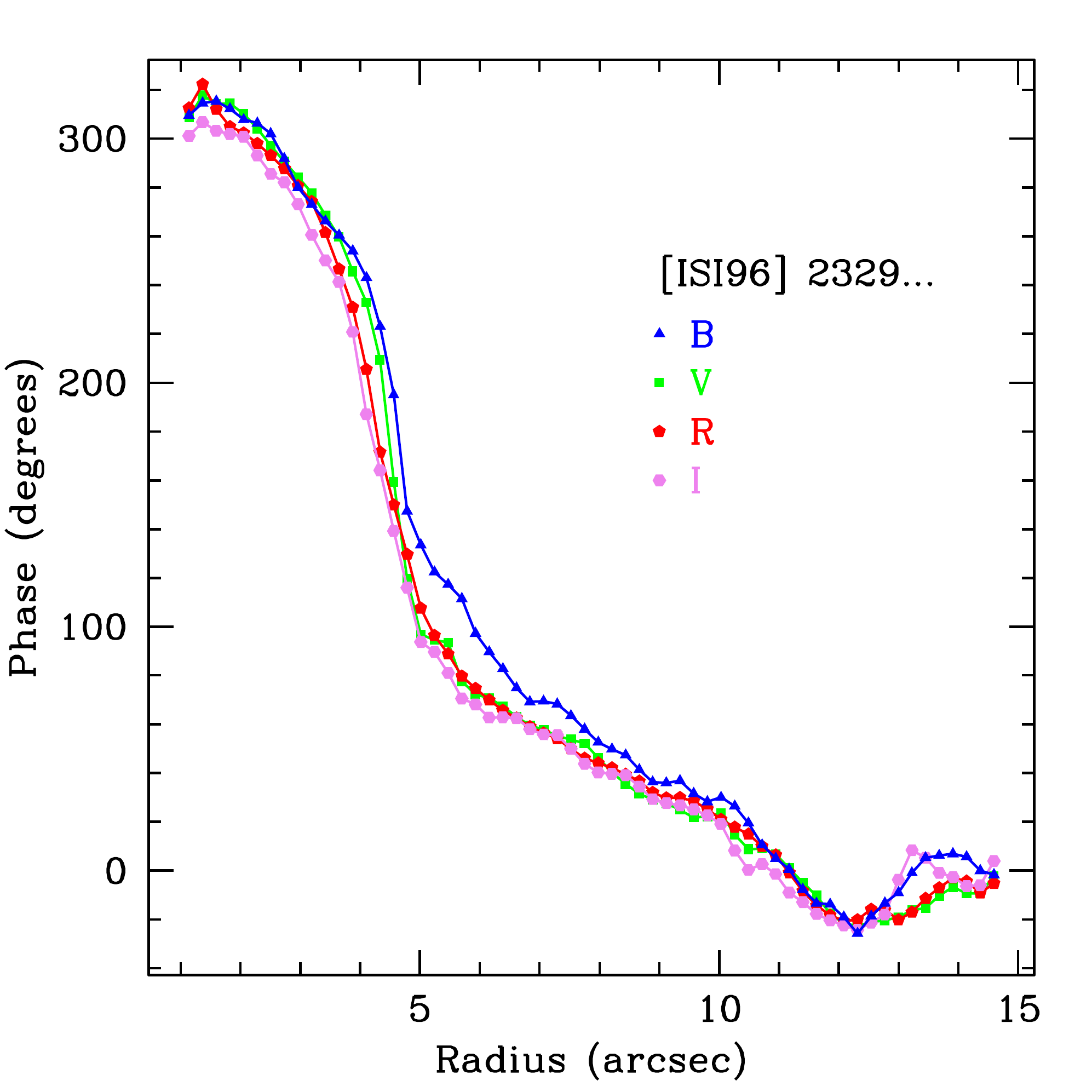} \\
  \caption{$B$- (blue) and $I$-band (violet) phase profiles for the remainder of our sample. Note that four galaxies have additional bands: $V$\ (green) and $R$\ (red).}
  \label{phases}
\end{figure*}

\begin{figure*}
  \centering
  \includegraphics[scale=0.11]{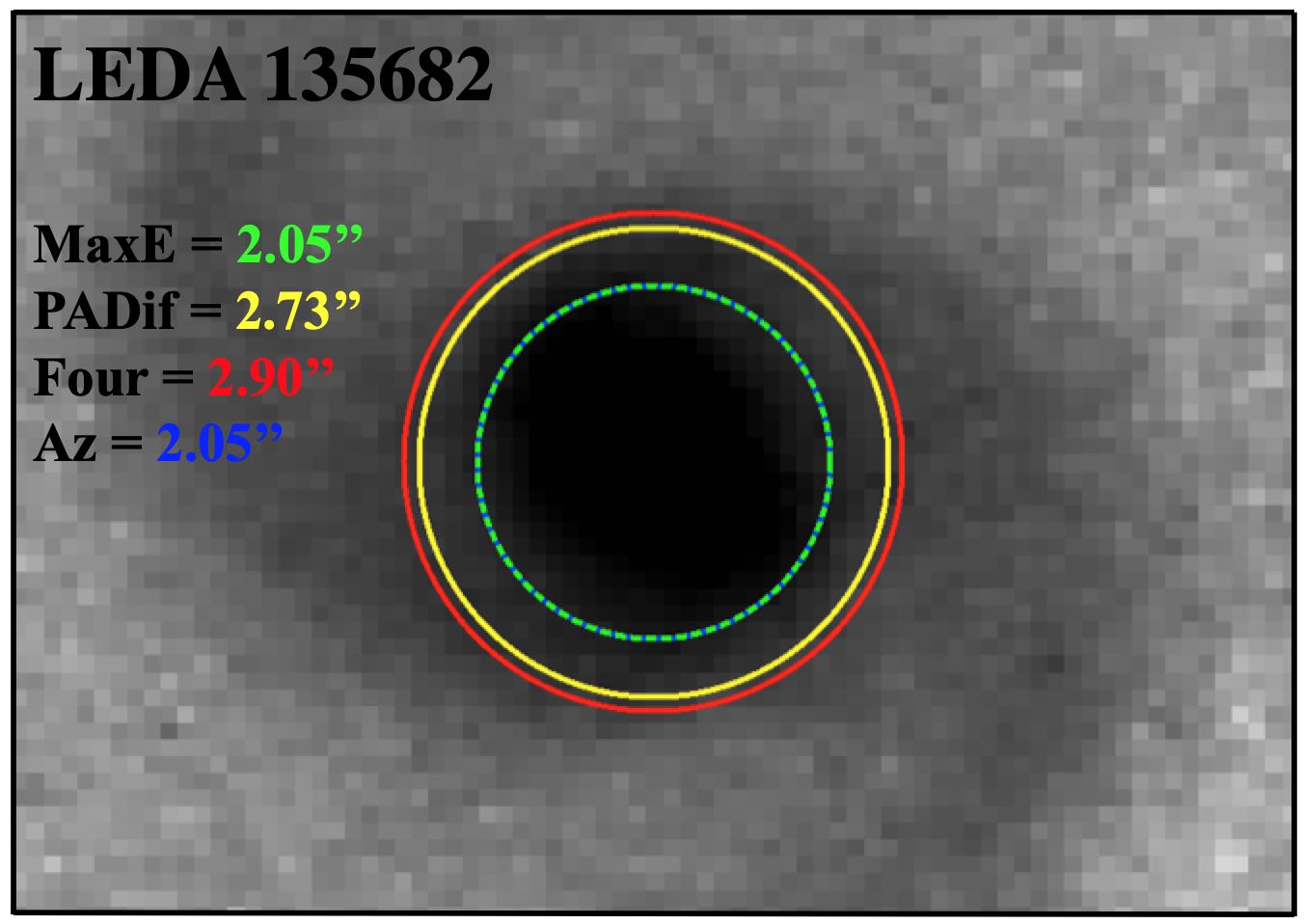} \includegraphics[scale=0.11]{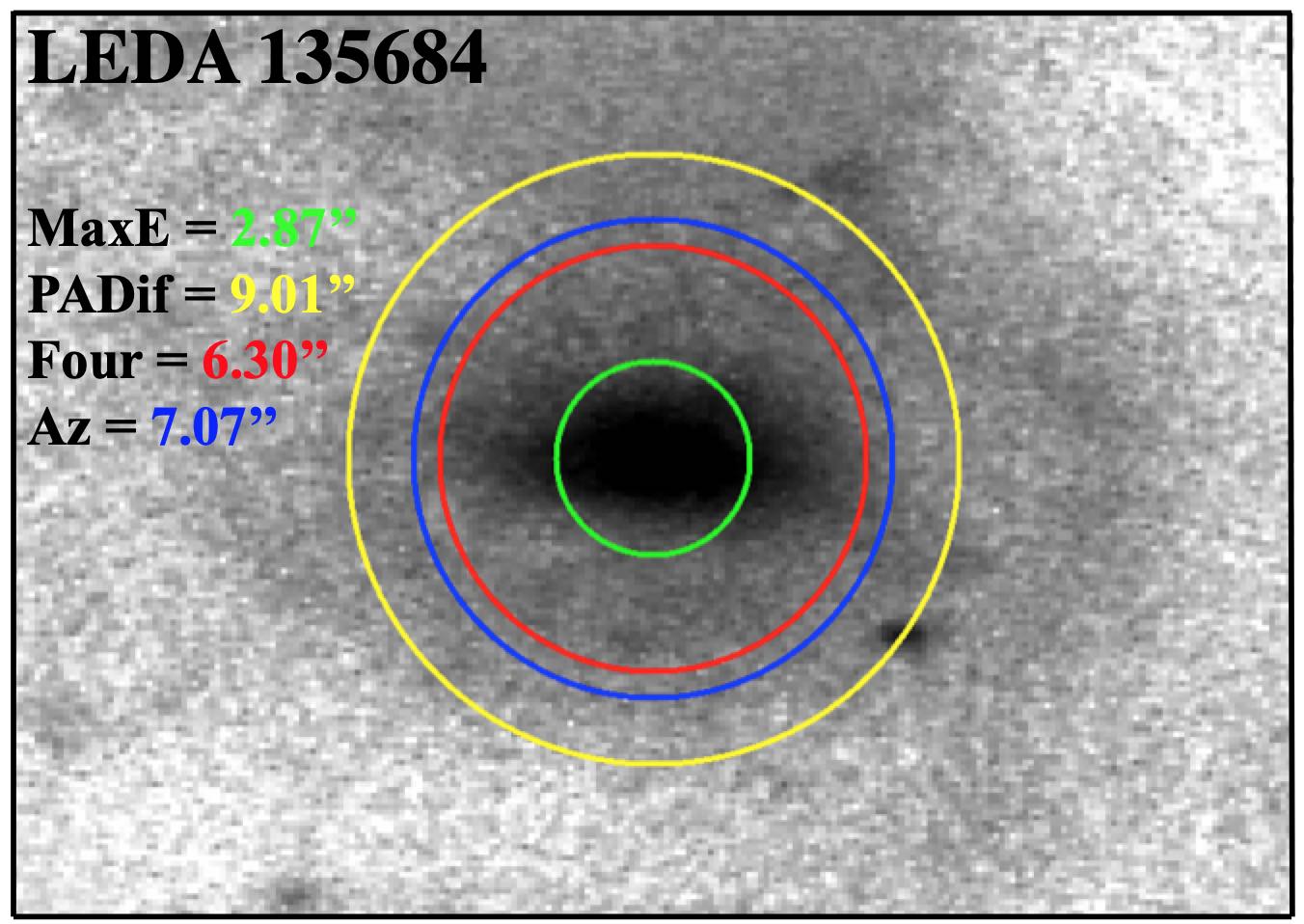} \includegraphics[scale=0.11]{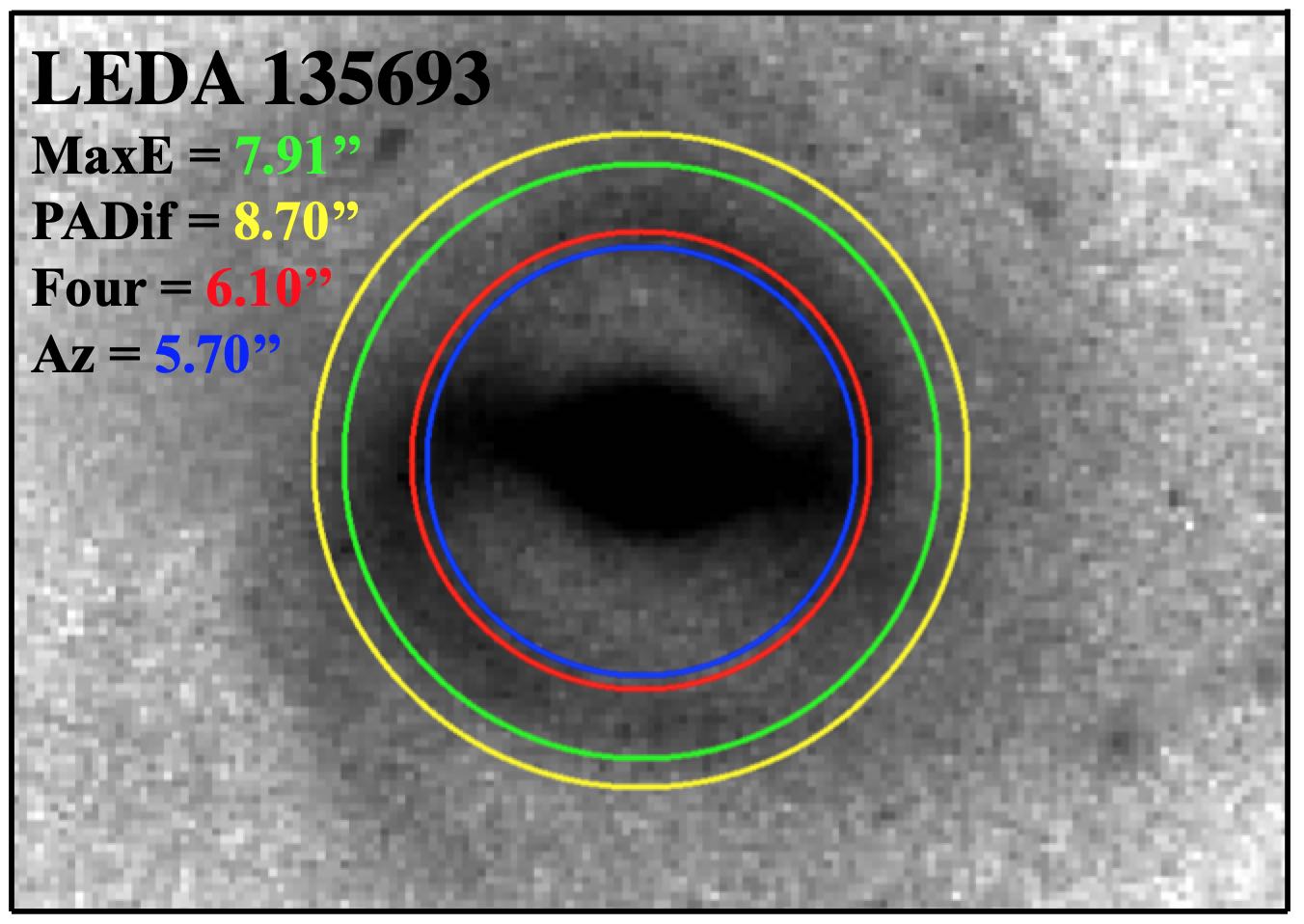} \\
  \includegraphics[scale=0.11]{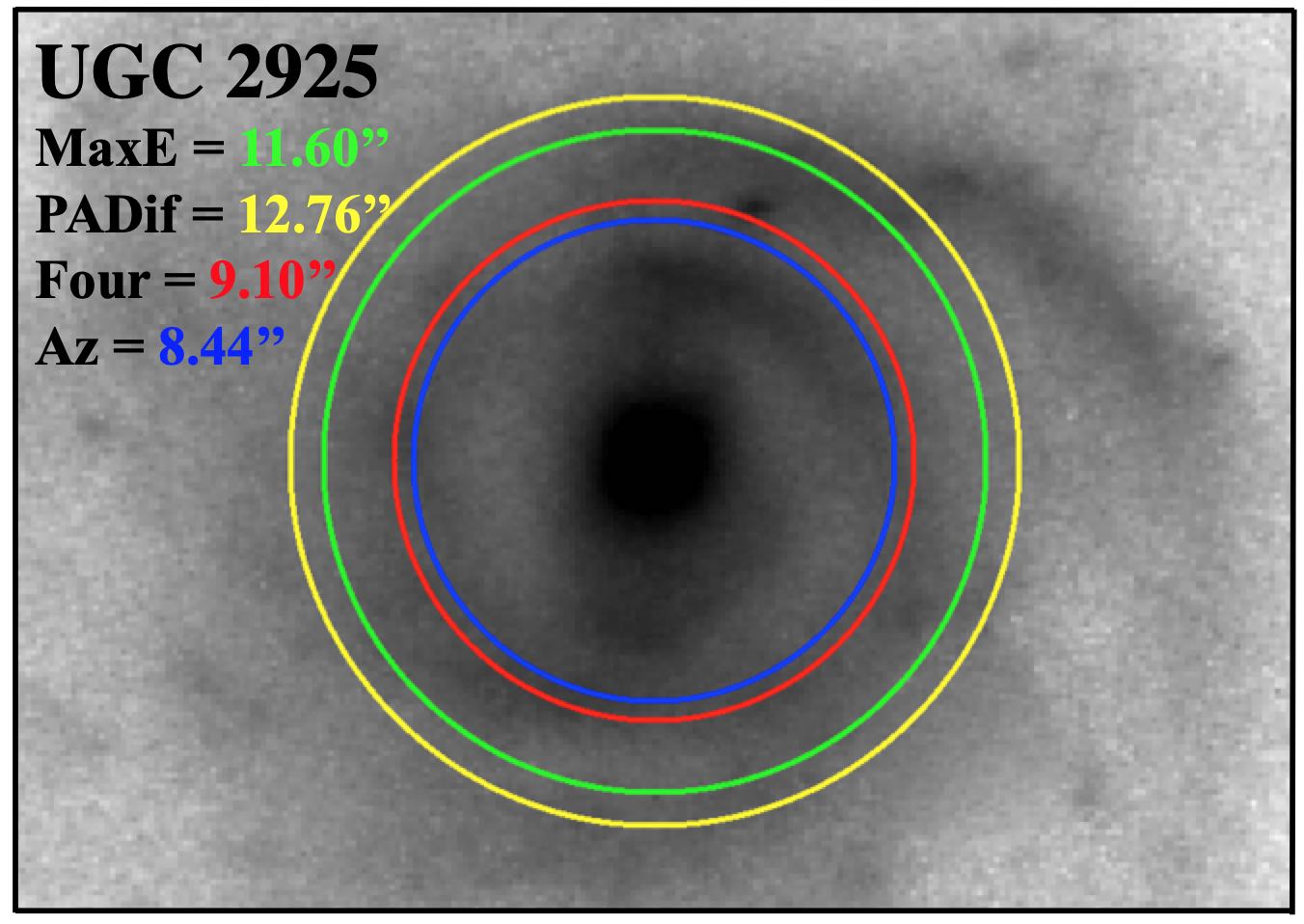} \includegraphics[scale=0.11]{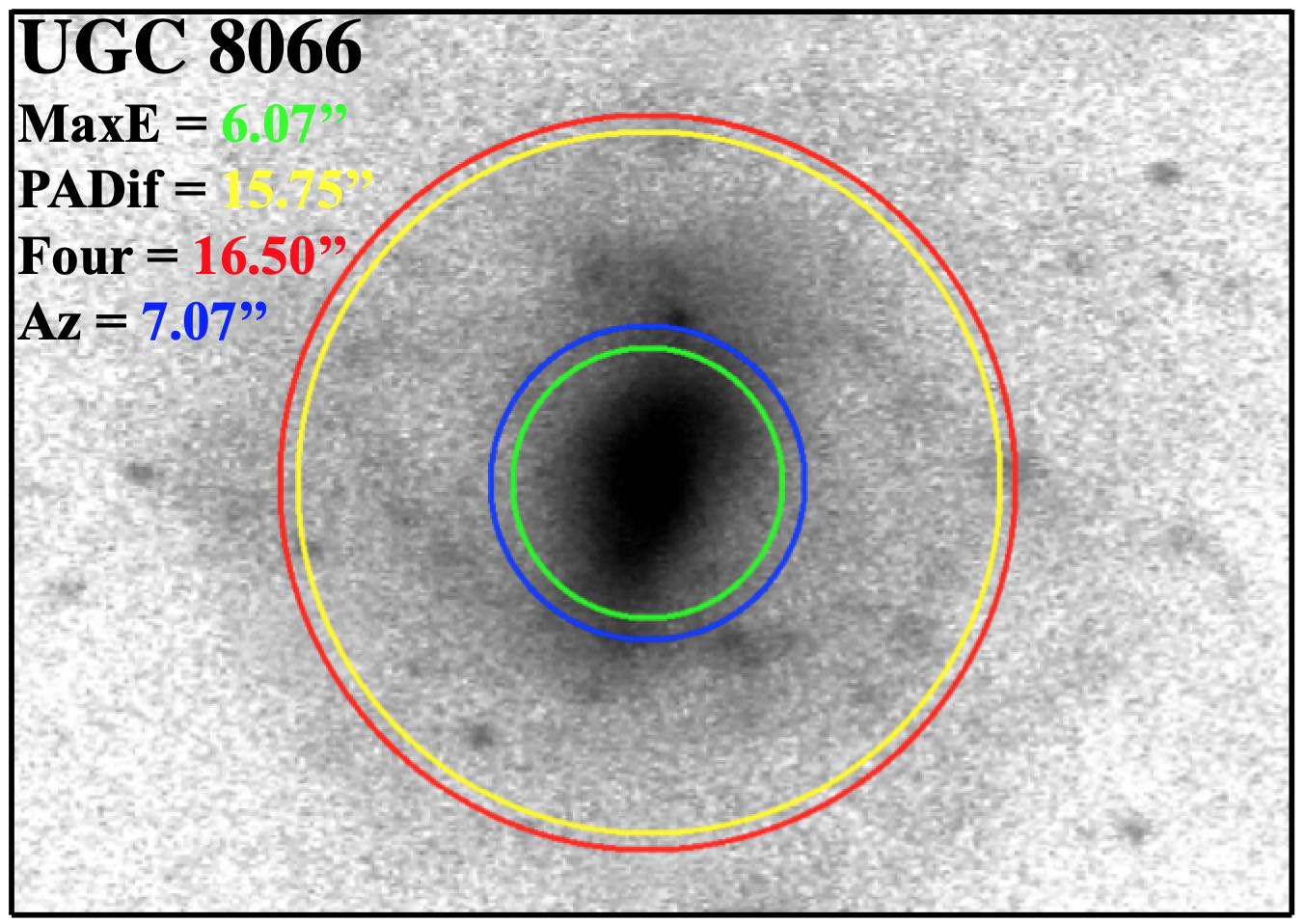} \includegraphics[scale=0.11]{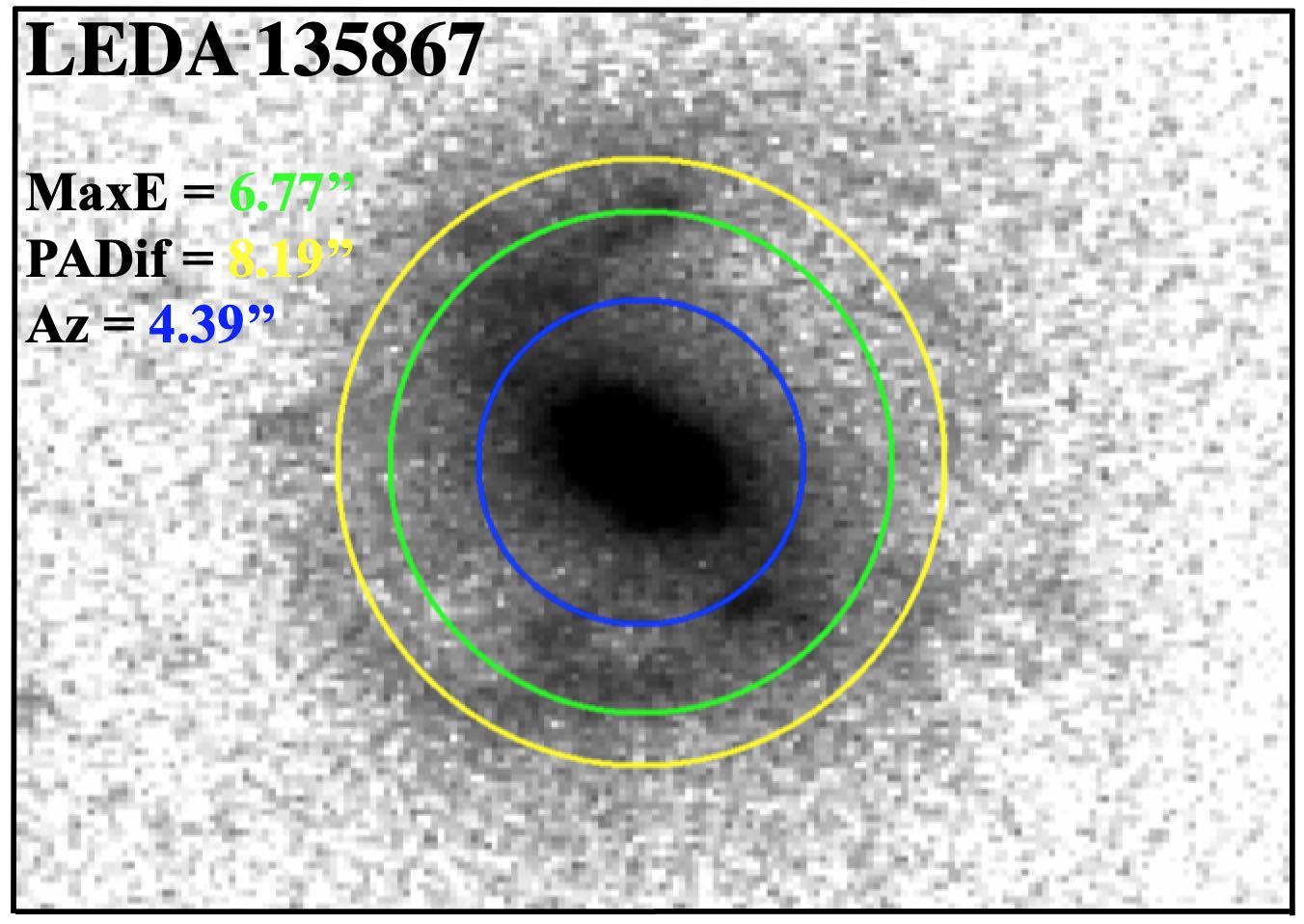} \\
  \includegraphics[scale=0.11]{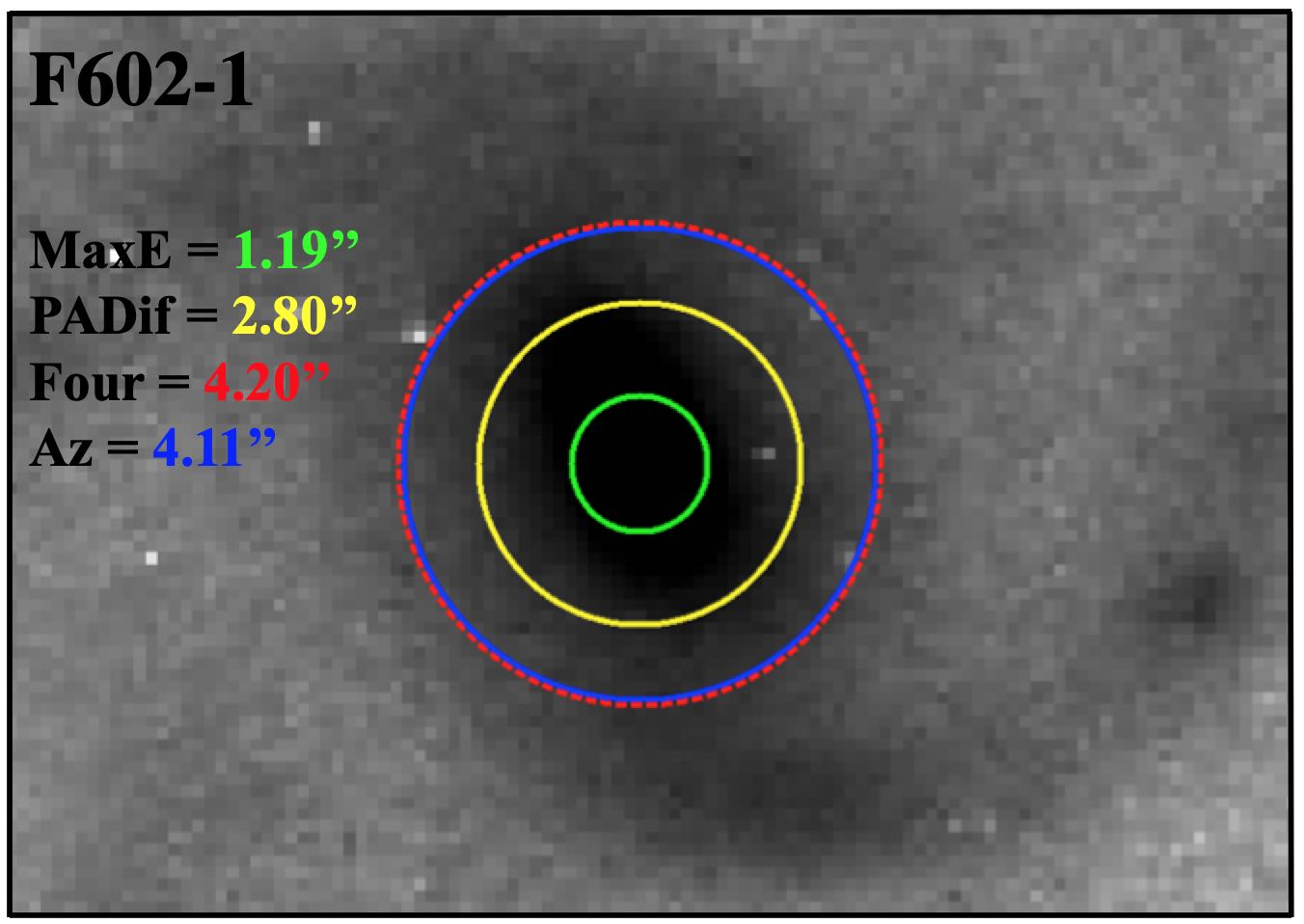} \includegraphics[scale=0.11]{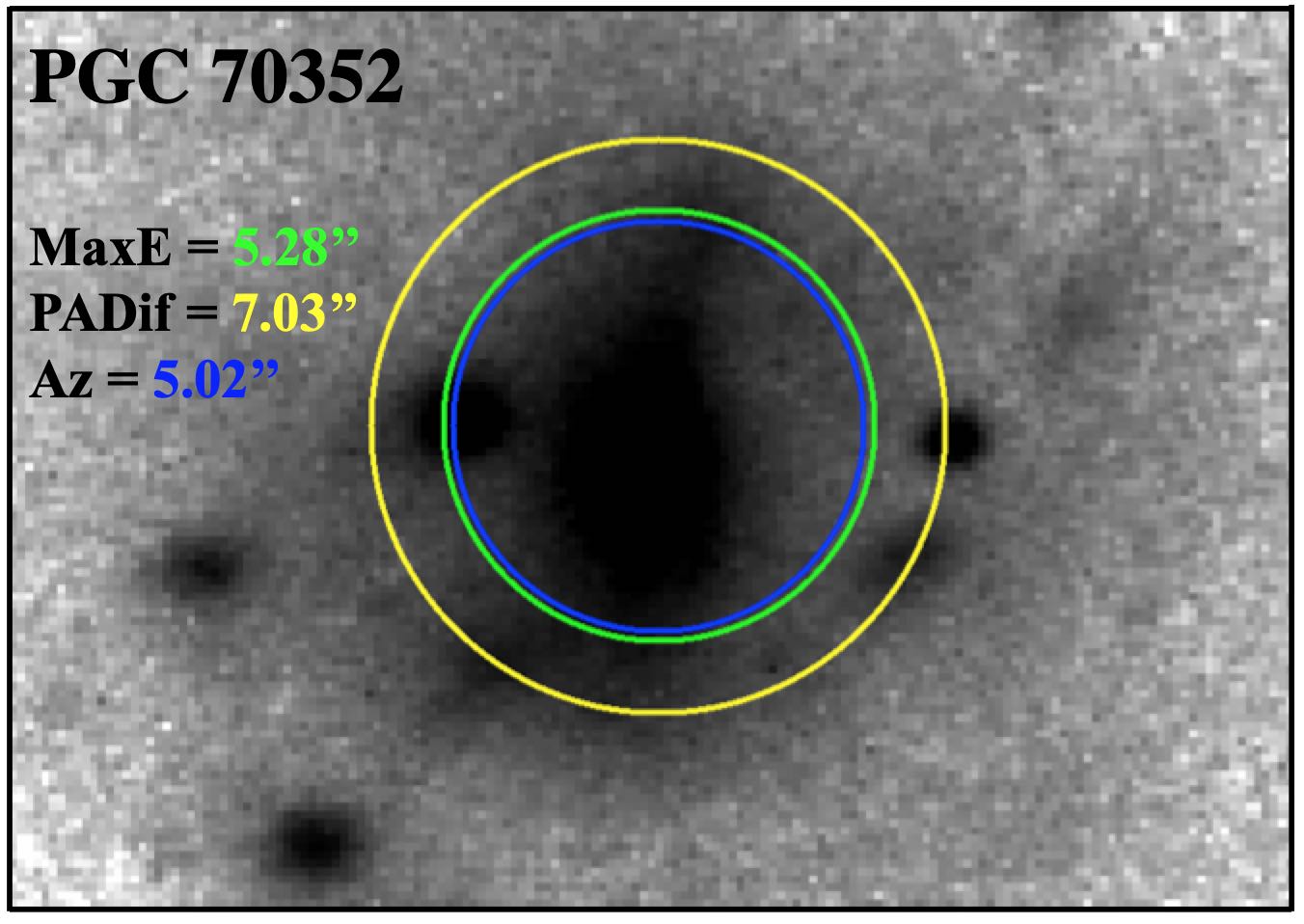} \includegraphics[scale=0.11]{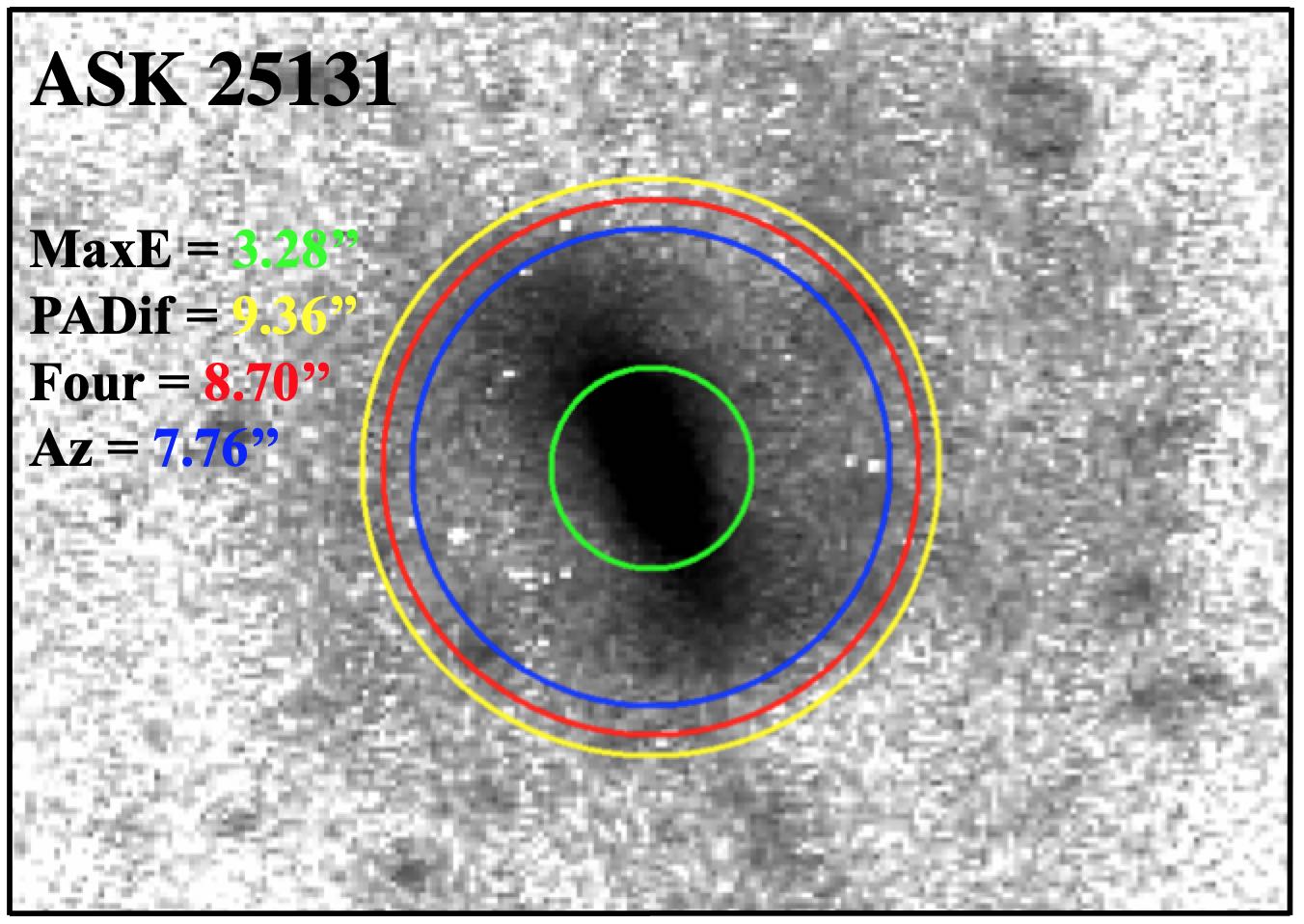} \\
  \includegraphics[scale=0.11]{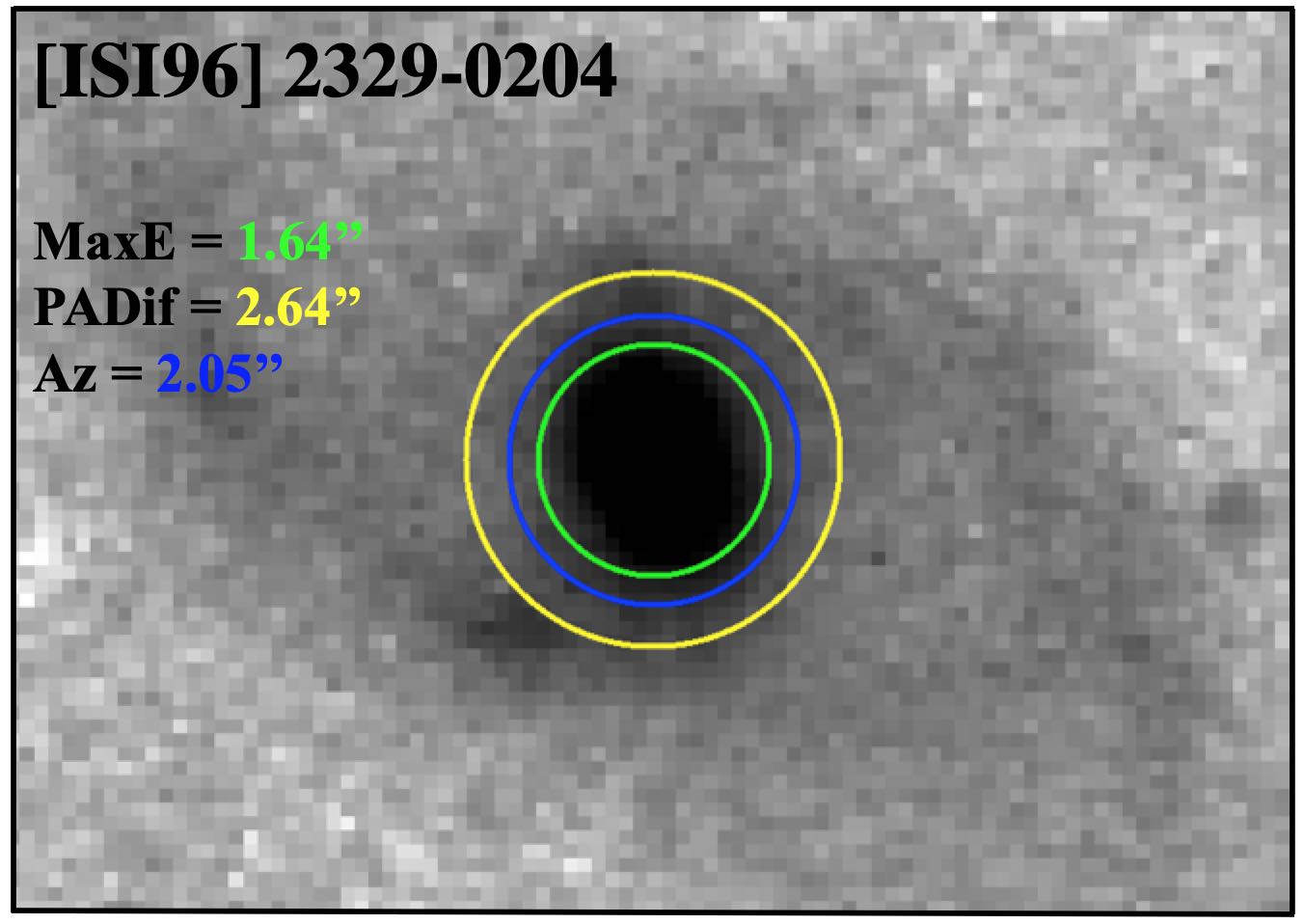} \\
  \caption{Comparison of the four bar length measures plotted over the deprojected $I$-band images for the remainder of our sample: $R_{\epsilon}$\ (green), $R_{\mathrm{P.A.}}$\ (yellow), $R_{\mathcal{F}}$\ (red), and $R_{az}$\ (blue).}
  \label{barPlots}
\end{figure*}


\bsp	
\label{lastpage}
\end{document}